\documentclass[final, 5p, 11pt,authoryear]{elsarticle}

\usepackage{graphicx}
\usepackage{multicol,multirow}
\usepackage[fleqn]{amsmath}
\usepackage{amssymb,amsfonts}
\usepackage{amsthm}
\usepackage[noend]{algpseudocode}
\usepackage{algcompatible}
\usepackage{algorithm}
\usepackage{listings}
\usepackage{lineno}
\usepackage[figuresright]{rotating}
\usepackage{appendix}
\usepackage{ifpdf}
\usepackage[T1]{fontenc}
\usepackage{textcomp}
\usepackage{xcolor}
\usepackage{float}    % For controlling figure placement [H]
\usepackage{booktabs} % For professional-looking tables
\usepackage{enumitem} % For customising lists
\usepackage{multicol,multirow}
\usepackage{array}
\usepackage{makecell}
\usepackage{subcaption} 
\usepackage{longtable}
\usepackage{colortbl}
\usepackage[colorlinks,allcolors=blue,bookmarks=false]{hyperref}
\usepackage[bottom]{footmisc}% places footnotes at page bottom
\usepackage{url}
\usepackage{graphics}
\usepackage{tabularray}
\usepackage{makecell} %for \xhline
\usepackage{svg}
\usepackage{indentfirst}
\usepackage{mathtools}
\usepackage{setspace}
\usepackage{wrapfig}
\usepackage{lscape}
\usepackage{rotating}
\usepackage{caption}  % For formatting caption and notes
\usepackage{epstopdf}
\usepackage{ctable} % for \specialrule command
\usepackage{lipsum}
\usepackage{colortbl}
\usepackage{tabularx} % Add this to your preamble if not already there
\usepackage[utf8]{inputenc}
\usepackage{enumitem}
\usepackage{bm} % For \textbf in math mode if needed, but mainly for bold

\emergencystretch=\maxdimen
\newcommand{\greyline}{\vspace{1.0em}{\color{lightgray}\hrule height 0.5pt}\vspace{1.0em}}

\journal{Computers in Industry}

\begin{document}

\begin{frontmatter}

%% Title, authors, and addresses

%% use the context command for the associated footnote;
\title{\Large{\textbf{Git for Sketches: An Intelligent Tracking System for Capturing Design Evolution}}}

\author{B. Sankar\fnref{label1}\corref{corauth}}
\ead{sankarb@iisc.ac.in}
\cortext[corauth]{Corresponding Author}

\affiliation[label1]{
            addressline={},
            organization={Department of Mechanical Engineering, Indian Institute of Science (IISc)},
            city={Bangalore},
            postcode={560012}, 
            state={Karnataka},
            country={India}}

\author{Amogh A S\fnref{label2}}
\affiliation[label2]{
           addressline={},
           organization={Department of Design and Manufacturing, Indian Institute of Science (IISc)},
           city={Bangalore},
           postcode={560012}, 
           state={Karnataka},
           country={India}}

\author{Sandhya Baranwal\fnref{label3}}
\affiliation[label3]{
           organization={Department of Design and Manufacturing, Indian Institute of Science (IISc)},
           addressline={}, 
           city={Bangalore},
           postcode={560012}, 
           state={Karnataka},
           country={India}}

\author{Dibakar Sen\fnref{label4}}
\affiliation[label4]{
           organization={Department of Design and Manufacturing, Indian Institute of Science (IISc)},
           addressline={}, 
           city={Bangalore},
           postcode={560012}, 
           state={Karnataka},
           country={India}}

% Note: Emails for co-authors can be added using \ead{} similar to the first author if needed.

%% Abstract
\begin{abstract}
During the conceptual phase of product development, the evolution of a design is as critical as the final output. Designers resort to product concept sketches (PCS) to represent their designs. However, traditional sketching tools fail to capture the non-linear history and cognitive intent behind design decisions, resulting in significant information loss. To address this, we present \textbf{DIMES} (Design Idea Management and Evolution capture System), a web-based sketching environment integrated with a novel custom-developed version control architecture named \textbf{sGIT} (Sketching Ground with Intelligent Tracking or SketchGit) embedded with a Generative AI system. Unlike existing software-centric Git, sGIT is tailored for the visual domain, along with a specialized module named \textbf{AEGIS} (AI Enhanced Gathering and Interpretation of Strokes). AEGIS records stroke dynamics and employs a hybrid classification pipeline that combines image-based Deep Learning models and feature-based Machine Learning models to automatically classify six distinct stroke types as identified by us: Constraining, Defining, Detailing, Shading, Shadow, and Annotation. The system maps Git primitives to design actions, enabling implicit branching and multi-modal commits (stroke data + voice intent). We validated the system through a comparative study involving expert generators and novice replicators. Results indicate that the sGIT-DIMES system significantly enhances the design process: expert designers using DIMES demonstrated a \textbf{160\% increase in concept exploration breadth} compared to traditional methods. Furthermore, Generative AI modules integrated into the system produced automated narrative summaries of the design evolution. These AI summaries facilitated superior knowledge transfer, with novice designers achieving significantly higher replication fidelity (\textbf{Neural Transparency-based Cosine Similarity: 0.97 vs 0.73}) compared to those relying on manual human summaries. Additionally, the AI-generated product concept render(ing)s (PCR) elicited significantly higher user acceptance (\textbf{Purchase Likelihood Score: 4.2 vs 3.1}). This research demonstrates that automated, intelligent version control can bridge the gap between creative action and cognitive documentation, offering a new paradigm for design education and process management in industrial design.
\end{abstract}

\begin{graphicalabstract}
\includegraphics[width=\textwidth, height=\textheight, keepaspectratio]{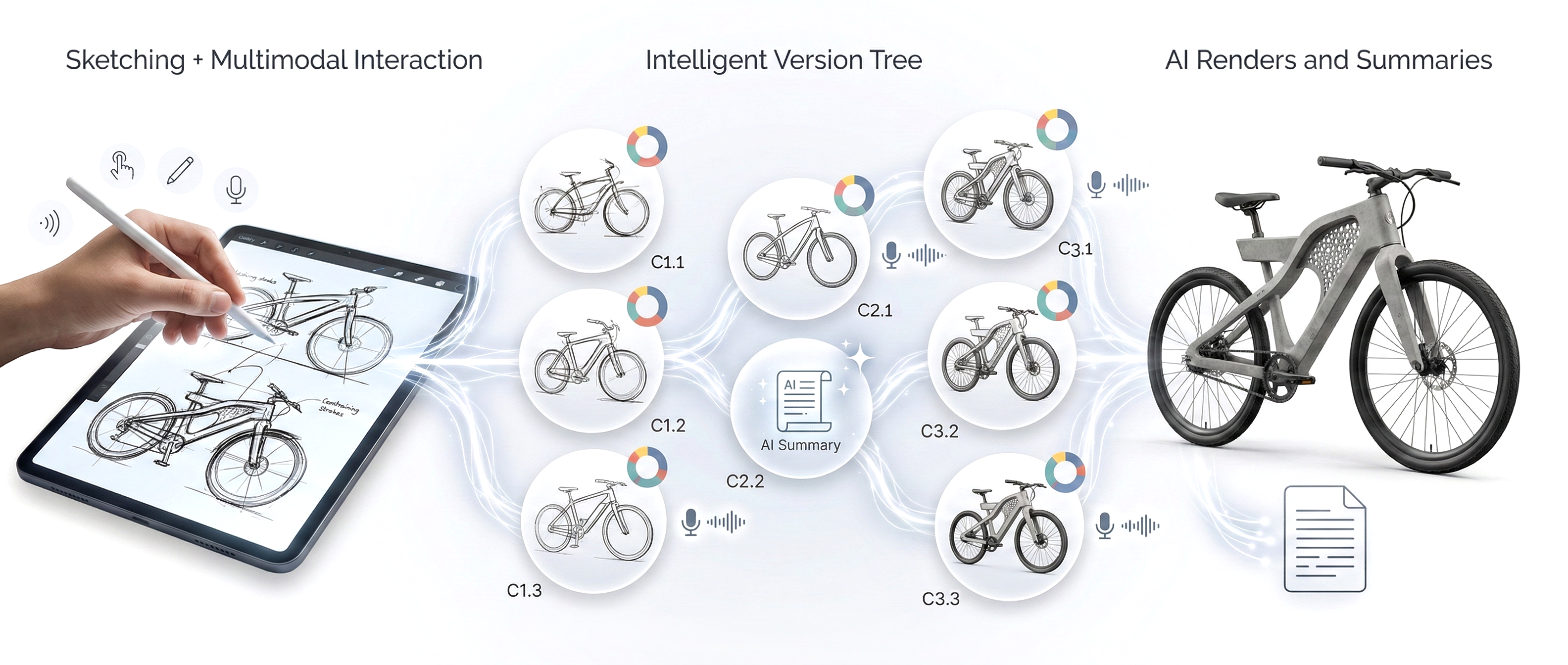}
\end{graphicalabstract}

\begin{highlights}
    \item We introduce \textbf{DIMES}, a web-based sketching environment integrated with \textbf{sGIT}, a novel version control architecture tailored for the visual domain that operates at the stroke level.
    \item \textbf{AEGIS}, an embedded AI module, automatically classifies six distinct stroke types (Constraining, Defining, Detailing, Shading, Shadow, Annotation) using hybrid Deep and Machine Learning models.
    \item Expert designers using DIMES demonstrated a \textbf{160\% increase in concept exploration breadth} and an \textbf{800\% increase in commit granularity} compared to traditional sketching methods.
    \item AI-generated design evolution narratives enabled novice replicators to achieve significantly higher fidelity (\textbf{Cosine Similarity: 0.97 vs 0.73}) compared to manual summaries using our \textbf{Neural transparency-based similarity analysis} technique.
    \item The system bridges the gap between creative action and cognitive documentation, offering a new paradigm for design education and \textbf{industrial design process management}.
\end{highlights}

%% Keywords

\end{frontmatter}

%\linenumbers


\clearpage
\section{Introduction}
\label{sec:introduction}

Product design is an intricate orchestration of creativity, engineering constraints, and user-centric problem solving. It transforms abstract user needs into tangible, functional solutions. Within this multifaceted process, the conceptual design phase is a crucial stage where a product is conceived, and its fundamental features are determined. It is during this phase that the vagueness of a problem statement is gradually crystallized into concrete geometric forms and functional architectures. 

Central to this phase is the cognitively intensive process of conceptualization, where ideas evolve into concepts. It is imperative to distinguish between an 'idea' and a 'concept' to appreciate the granularity of design evolution. An idea fundamentally addresses the question of \textit{what} to do; it is a seed of intent. In contrast, a concept addresses the question of \textit{how} to do it. When this 'how' describes the operational mechanism or utility, it is referred to as a functional concept. Conversely, when the 'how' describes the aesthetic, geometric, and tangible manifestation, it is referred to as a form concept. 

For decades, the primary medium for designers in externalizing these concepts has been Product Concept Sketching. From the rudimentary cave paintings of early humans to the sophisticated digital tablets of the modern era, sketching has remained the most natural, immediate, and high-bandwidth method for human communication and ideation. However, despite the ubiquity of sketching, the technological infrastructure supporting the \textit{management} of these sketches has lagged significantly behind the tools available for textual or code-based creation. 

\subsection{The Role of Conceptual Design and the Nature of PCS}
\begin{figure}[ht!]
    \centering
    \includegraphics[width=\linewidth]{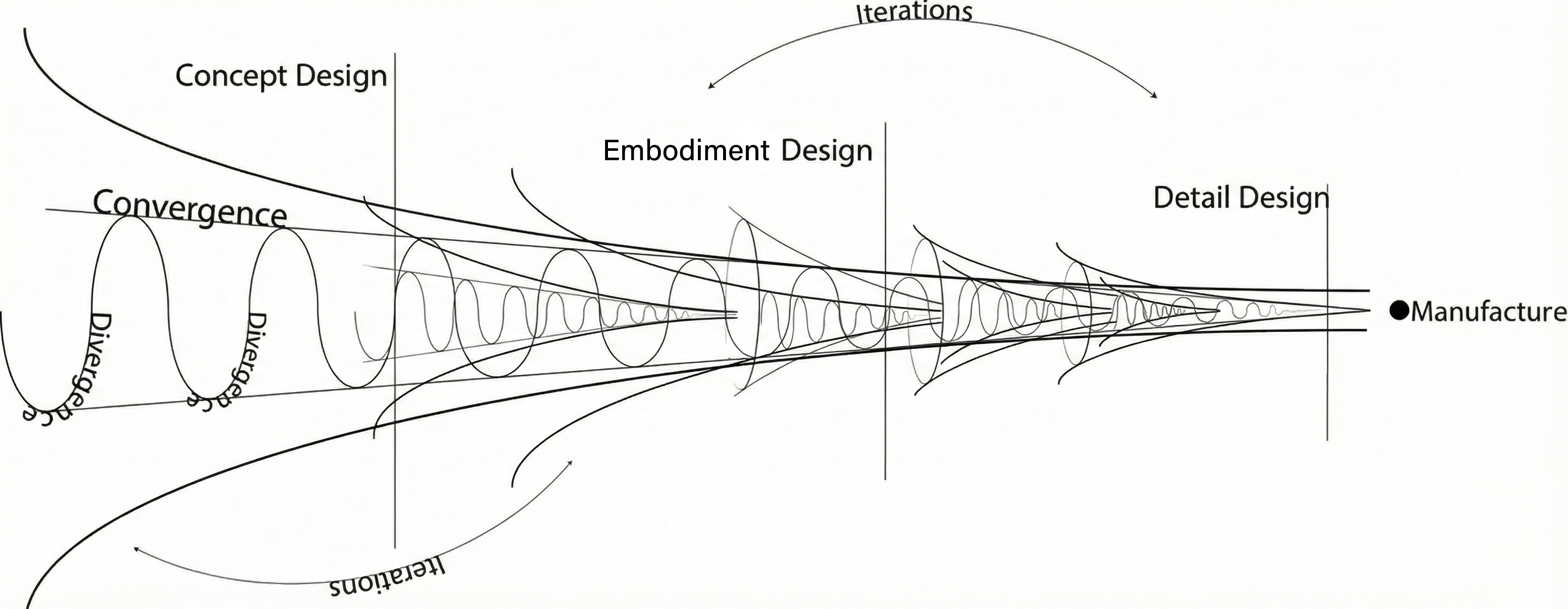}
    \caption{A diagram illustrating the design process as a converging funnel. It illustrates the stages of Concept, Development, and Detail Design, with cycles of divergence, convergence, and iteration culminating in manufacture.}
    \label{fig:design_funnel}
\end{figure}

Conceptual design is characterized as a chaotic exploration of the solution space (Figure~\ref{fig:design_funnel}). It is a non-linear process involving rapid iteration, divergence into multiple lines of thought, and convergence towards a feasible solution. The primary representation modality for this exploration is the Product Concept Sketch. Unlike technical drawings or artistic paintings, PCS occupies a unique position in the visual arts and design taxonomy.

\begin{figure}[ht!]
        \centering
        \includegraphics[width=\linewidth]{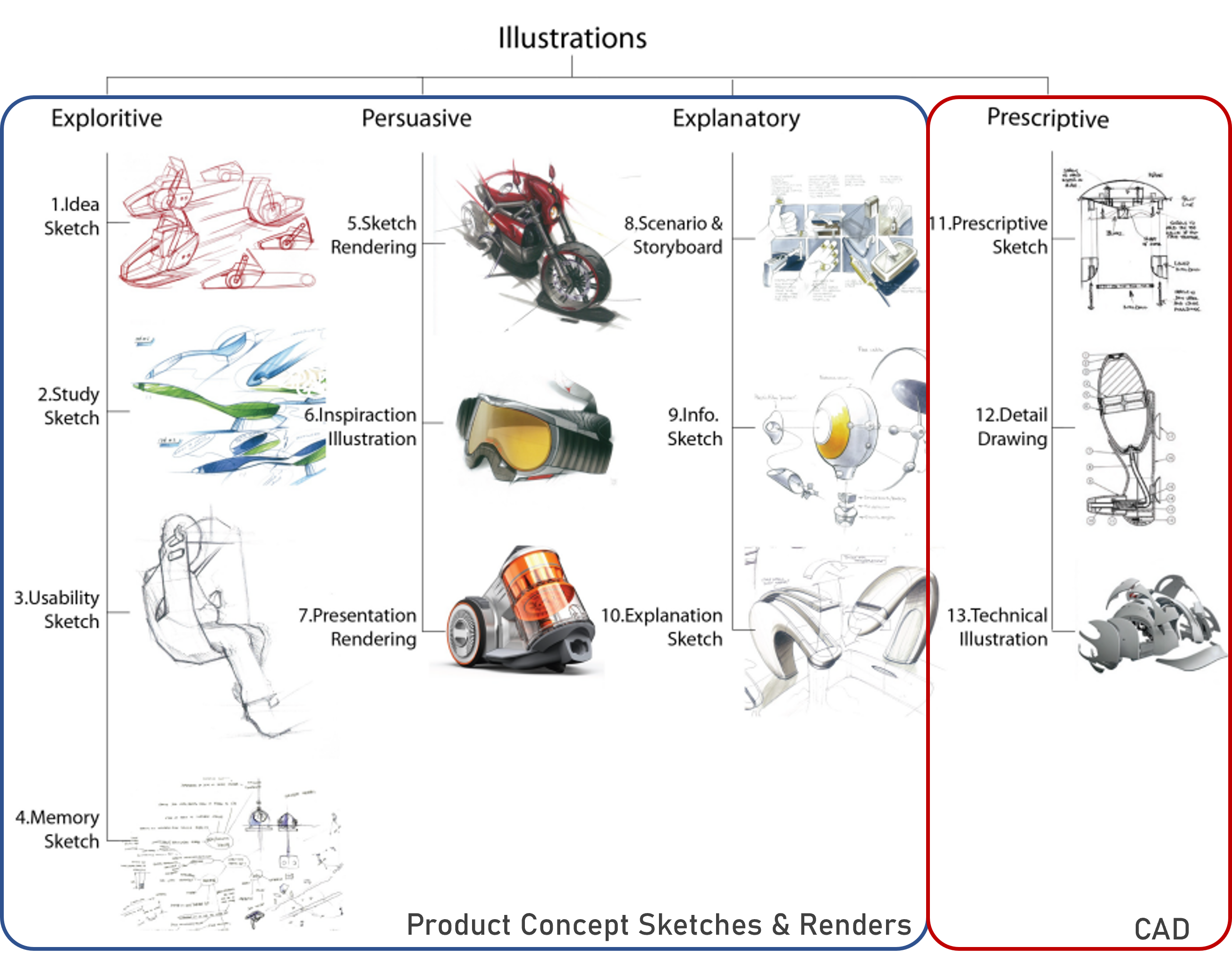}
        \caption{A classification chart of different types of illustrations used in design, categorized into Explorative, Persuasive, Explanatory, and Prescriptive, with the Prescriptive category highlighted and associated with CAD.}
        \label{fig:illustration_types}
\end{figure}

We differentiate PCS from other visual illustration activities to understand its specific requirements. Drawing is often a general term for mark-making; painting is associated with artistic expression, often devoid of functional constraints; and doodling is a subconscious act lacking directed intent. On the contrary, product concept sketches are a series of rapid explorations of ideas, creating impressions on paper with a pen, focused on a specific intent (shown in Figure~\ref{fig:illustration_types}). These sketches possess unique properties that distinguish them from doodles, paintings, technical drawings, CAD models, and/or renderings. (More details about the properties given in Section~\ref{subsubsec:properties_and_characteristics}).

\begin{figure*}[ht!]
    \centering
    \includegraphics[width=0.85\linewidth]{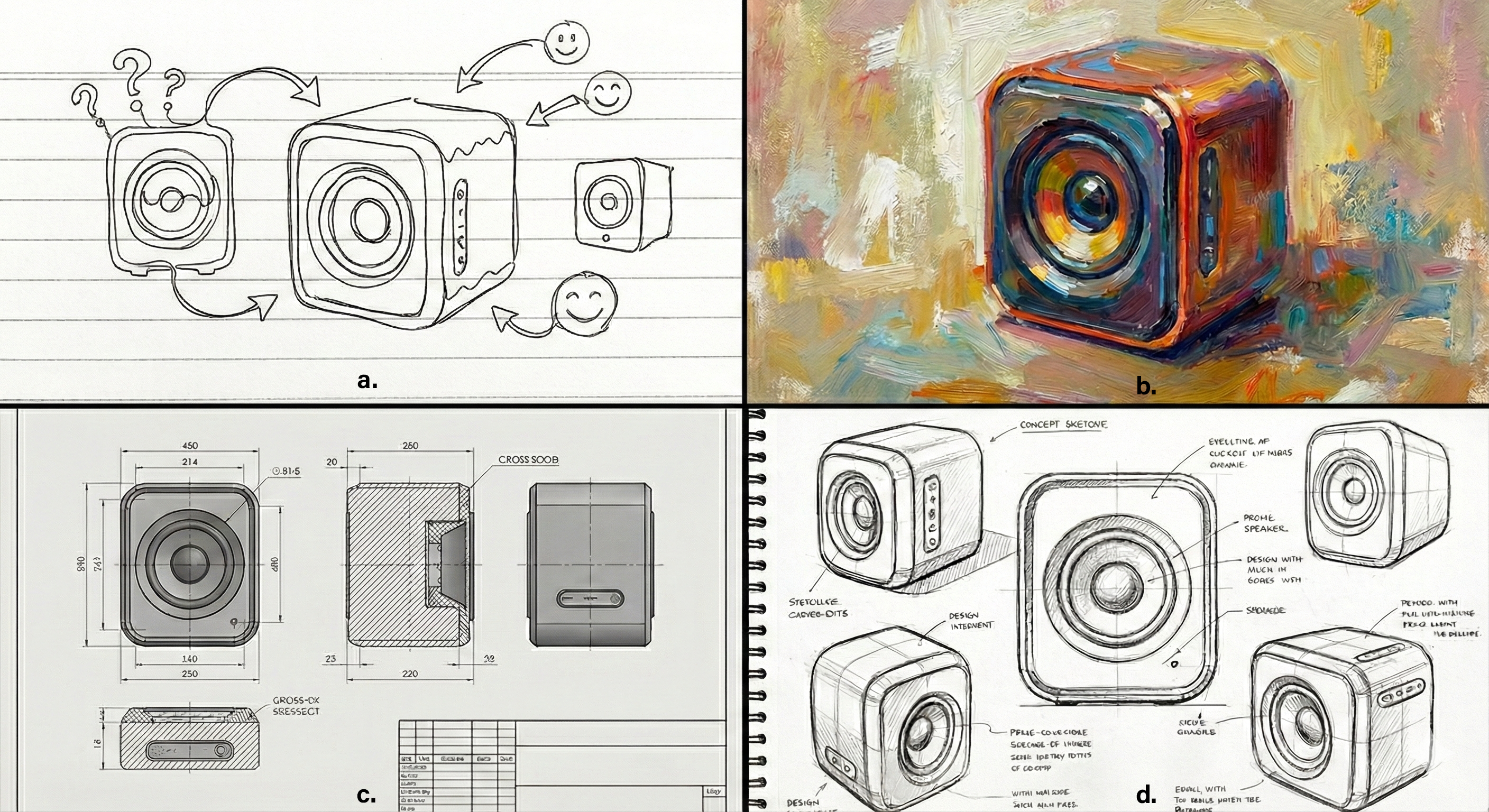}
    \caption{A comparative visual taxonomy illustrating the fundamental differences between four distinct modes of visual representation in industrial design practice: \\
    {a. Doodling} – rapid, low-fidelity ideation with minimal structure and high ambiguity; \\
    {b. Artistic Painting} – expressive, painterly rendering prioritizing aesthetic impact and emotional communication over precision; \\
    {c. Technical Drawing} – dimensioned orthographic projections and cross-sections adhering to engineering drawing standards for manufacturing clarity; and \\
    {d. Product Concept Sketch} – purposeful combination of expressive form exploration, annotated design intent, construction awareness, and visual hierarchy to communicate industrial design concepts effectively during the fuzzy front-end of product development.}
    \label{fig:visual-taxonomy-sketching}
    % \caption{Visual taxonomy of design representation techniques: Doodling (a), Artistic Painting (b), Technical Drawing (c), and Product Concept Sketch (d).}
\end{figure*}

The act of sketching involves a tight loop between the hand (action) and the mind (cognition), suggesting that the sketch is not merely a record of an idea but a medium through which the idea is actually constructed.

\subsection{The Current Landscape of Sketching Tools and Their Limitations}
Most contemporary designers have transitioned from paper to pixels by utilizing various software tools, including Autodesk Sketchbook, Concept, Alias Studio, Adobe Photoshop, and Procreate. These applications are incredibly robust in terms of image processing, offering advanced layer management, diverse brushes, and editing capabilities. However, we believe there is a fundamental misalignment between these tools and the conceptual design process.

These existing tools are predominantly developed for \textit{image editing} and artistic illustration. They excel at creating a single, polished "hero shot" or a final aesthetic output. They operate on the paradigm of 'Layers', thereby stacking visual elements on top of each other to build a final image. However, through the extensive use of these tools, we have found that they do not operate on the paradigm of 'Versions' or 'Time'. 

In the conceptual design phase, a designer does not merely create one image; they generate a branching tree of possibilities. They might start with a cylindrical form, iterate on it five times, reach a dead end, backtrack to the second iteration, and diverge into a cubic form. This non-sequential, non-linear evolution is the essence of creativity \cite{Sankar2023}. Currently, designers manage this by manually saving multiple files (e.g., \texttt{concept\_v1.jpeg}, \texttt{concept\_v1\_final.jpeg}, \texttt{concept\_v2\_new.jpeg}), or by duplicating layers endlessly within a single file until the system becomes too slow. This manual logistics is cumbersome, prone to error, and, most critically, disruptive to the "flow" state essential for creativity.

Furthermore, the overwhelming number of options in commercial software, such as hundreds of brushes, filters, and adjustment layers, often hinders the rapid, free-flowing nature of early-stage ideation. Our observations of expert designers reveal that early concept sketches are often monochromatic, utilizing a simple ballpoint pen aesthetic, and rarely involve the use of an eraser. The focus is on form exploration, not rendering perfection. The complexity of existing tools, combined with their lack of temporal management, creates a significant friction point in the design workflow.

\subsection{The Lost Dimension: Evolution and Design Intent}
Perhaps the most profound deficit in the current design workflow is the loss of process history. In the traditional paradigm, stakeholders, clients, and even design managers typically see only the final output, which is the polished render or the selected concept. The rollercoaster journey of how that solution evolved is simply lost. The hundreds of intermediate sketches, the rejected alternatives, the subtle refinements, and the branching decision paths are discarded or buried in undigested file dumps.

Understanding the evolution of concepts is paramount for two reasons: \textit{understanding the design} and \textit{understanding the designer}.
Firstly, the rationale behind a design decision is often embedded in the alternatives that were rejected. By knowing how a solution evolved, one gains insight into the constraints and trade-offs that shaped the final outcome. 
Secondly, capturing this evolution offers a window into the designer's cognitive style. Creativity increases with experience; an experienced designer navigates the solution space differently than a novice. A system that captures the "style" and "process" of an experienced designer could serve as an invaluable pedagogical tool for training novices. Currently, to the best of our knowledge, no system exists that automatically tracks, creates a genealogy of, and summarizes this evolutionary process for product concept sketching.

\subsection{Contributions}
This research presents the development and validation of DIMES (Design Idea Management and Evolution capture System), a web-based sketching application that integrates a custom-built version control system named sGIT, inspired by the existing GIT. The system is designed specifically to capture, track, and manage the evolution of product concept sketches (PCS), thereby bridging the gap between fleeting ideation and structured design management.

Our contribution to the field of Design Methodology and Artificial Intelligence in Design is fivefold:
\begin{enumerate}
\item \textbf{Development of sGIT:} We propose, to the best of our knowledge, the first dedicated Version Control System for Product Concept Sketching. We define a comprehensive mapping of Git primitives to design actions (e.g., \textit{Checkout} for exploring alternatives, \textit{Branch} for concept divergence, \textit{Diff} for visual comparison).
\item \textbf{Taxonomy of Strokes:} We expand upon existing literature to establish a robust classification of six stroke types: Constraining, Defining, Detailing, Shading, Shadow, and Annotation strokes. This classification serves as the foundational grammar for understanding the ``action'' component of sketching.
\item \textbf{AEGIS and Hybrid Stroke Classification:} Addressing the lack of labelled data, we developed \textbf{AEGIS} (AI Enhanced Gathering and Interpretation of Strokes) to record stroke dynamics and facilitate automated classification. We present a hybrid classification pipeline that employs \textbf{5 Image-Based Deep Learning models} and \textbf{10 Feature-Based Machine Learning models}. This dual approach leverages both visual morphology and kinematic features to accurately classify stroke types, enabling the machine to ``read'' the sketch construction.
\item \textbf{AI-Driven Intent Capture:} We integrate Generative AI capabilities to address the ``cognitive'' loss. The system utilizes Text-to-Speech (TTS) for capturing design intent during commits and employs Large Language Models (LLMs) to generate narrative summaries of the concept's evolution. Furthermore, we demonstrate the use of Generative AI to transform monochromatic product concept sketches into photorealistic renders based on textual style descriptions as directed by the designers.
\item \textbf{Neural Transparency-Based Similarity Analysis:} We introduce a novel methodological contribution for evaluating design knowledge transfer. We employ a \textbf{Neural Transparency} technique, utilizing the internal activation matrices of Vision-Language Models (LLaVA-NeXT), to quantify the semantic similarity between sketches. This objective metric surpasses traditional pixel-based comparisons, offering a robust tool for assessing the fidelity of design replication and education.
\end{enumerate}

% TODO: Insert Table 1: Summary of the gap analysis between traditional tools, software Git, and the proposed sGIT system.

This paper addresses the critical research gap: \textit{How can we capture the information-abundant, temporal evolution of product concept sketching in a way that preserves the designer's cognitive intent and facilitates learning?}
By answering this, we move closer to a future where machines not only generate images in a flash but also understand the nuanced, iterative, and deeply human process of design.

%%%%%%%%%%%%%%%%%%%%%%%%%%%%%%%%%%%%%%%%%%%%%%%%%%%%%%%%%%%%%%%%%%%%%%

\section{Literature Review}
\label{sec:literature_review}

The trajectory of product development is fundamentally established during the nascent stages of conceptual design. It is widely acknowledged in the engineering literature that, although this phase incurs a minimal fraction of the total project expenditure, it determines a significant portion, often estimated to be between 70\% and 80\%, of the product’s lifecycle costs. Consequently, the efficacy of the methods used to explore, represent, and manage ideas during this phase is crucial \cite{Relvas2021-yt}. Effective management of this "Fuzzy Front End" (FFE) requires structured methodologies to transform abstract intuition into tangible product specifications \cite{Uribe_Ocampo2023-aq}. Furthermore, as idea management becomes central to organizational innovation, the lack of a generalized model for corporate idea tracking remains a persistent challenge \cite{Gerlach2017}.

This review synthesizes the existing literature to provide a contextual understanding of the development of the sGIT-DIMES ecosystem. We examine the cognitive and physical aspects of design sketching, the evolution of digital support tools ranging from early digitizers to modern Generative Design algorithms, the application of Knowledge Graphs (KGs) for representing design rationale, and the potential of Version Control Systems (VCS) to manage design evolution. The review concludes by identifying a critical lacuna in the current state of the art: the absence of a dedicated, temporally-aware system capable of managing the non-linear evolution of sketch-based concept generation.

\subsection{The Cognitive and Physical Nature of Design Sketching}
To develop a system that supports design, one must first understand the fundamental behaviour of the designer and the conditions under which design practice occurs \cite{Kaszynska2025-wk}. Despite the proliferation of advanced CAD tools, freehand sketching remains the ubiquitous and primary method for externalizing ideas in the early phases of design \cite{McGown1998}.

\subsubsection{Sketching as Reflection-in-Action}
Schon’s seminal theory of the "Reflective Practitioner" posits that sketching is not merely a medium for recording an idea but a mechanism for thinking itself \cite{Schon1983}. It allows the designer to conduct a dialogue with themselves, constructing a 'virtual world' where ideas can be simulated, tested, and modified rapidly. This "reflection-in-action" relies on the spontaneity of the medium; the low fidelity and speed of sketching are crucial for capturing "fleeting ideas" before they evaporate from the designer's working memory \cite{Tovey1997}. \cite{Cross1982} defines these specific cognitive strategies as "designerly ways of knowing," distinct from scientific problem-solving. Furthermore, sketching serves as a vital knowledge management tool, physically instantiating the knowledge creation process \cite{Eppler2011}.

Research broadly categorizes sketches based on their intent. \cite{Ferguson1992} differentiate between the \textit{thinking sketch}, used to guide non-verbal private thought; the \textit{talking sketch}, utilized for peer communication; and the \textit{prescriptive sketch}, used for detailing instructions. For the purpose of our research, the focus is on the \textit{thinking sketch}. These artifacts are characterized by "vague knowledge" and shifting goals \cite{Goel1995}. They are inherently ambiguous, imprecise, and often unintelligible to an outside observer without context \cite{Goel1995}. They function not just as visual outputs but as tools for assessment and reflection, where the quality of the sketch often correlates with the understandability of the design concept \cite{Das2022-xa}.

\subsubsection{Taxonomy of Transformations and Strokes}
The evolution of a concept is defined by the movement between successive sketches. \cite{Goel1995} provided a critical framework for this evolution, identifying two distinct transformation types:
\begin{enumerate}
    \item \textbf{Lateral Transformations:} The movement from one idea to a slightly different idea, representing a divergence in the solution space. This associative extension allows designers to combine distinct concepts into novel solutions \cite{Kim2023}.
    \item \textbf{Vertical Transformations:} The movement from a vague idea to a more detailed version, representing convergence.
\end{enumerate}
Understanding these transformations is vital for any version control system, as they directly map to the concepts of 'branching' and 'committing'. \cite{Rodgers2000-dr} demonstrated that tracking these lateral and vertical movements can reveal the designer's thinking mode and progress.

\begin{figure*}[ht!]
    \centering
    \includegraphics[width=\textwidth]{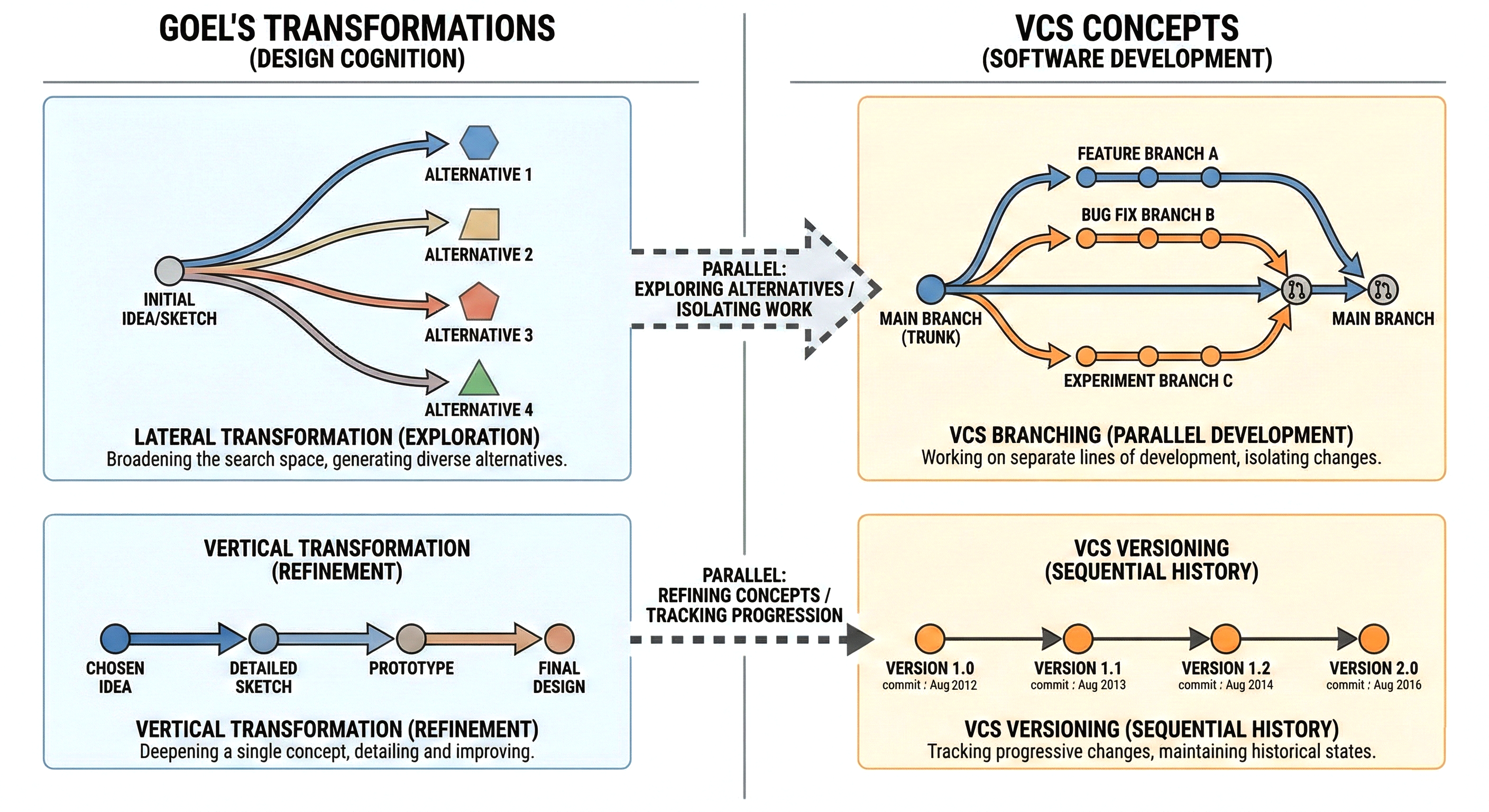}
    \caption{Diagram illustrating the parallel between Goel's \cite{Goel1995} Lateral and Vertical Transformations in design cognition and VCS Branching and Versioning in software development. The left side depicts the exploration of alternatives (Lateral) versus the refinement of a single concept (Vertical). The right side shows the corresponding VCS concepts of parallel development through Branching and sequential history through Versioning.}
    \label{fig:goel_vcs_parallel}
\end{figure*}

At the atomic level, the sketching activity is not a monolithic action but a sequence of distinct cognitive phases: \textit{Creation} (action-dominant generation), \textit{Evaluation} (cognition-dominant assessment), and \textit{Planning} (cognition-dominant strategy formulation) \cite{McGown1998, Rodgers2000-dr, Sankar2023}. This process is inherently non-sequential and iterative, involving frequent interruptions and temporal discontinuities where the designer shifts focus \cite{Sankar2023}. One researcher further decomposed the physical act into functional stroke segments, arguing that the temporal evolution of stroke pressure and trajectory encodes the designer's mental organization \cite{Onkar2010-dz}. Researchers have hypothesized that the language of sketching can enable early conceptual analysis if the system can perceive the 3D model within the 2D strokes \cite{Lipson2000-ug}. Furthermore, visualizing the collective generation of these ideas can reveal the underlying social and cognitive networks driving innovation \cite{Cao2023}. Therefore, a robust support system must capture not just the static image but the temporal data of these strokes to decode the design intent.

\subsection{Computational Support for Conceptual Design}
Historically, general-purpose Computer-Aided Design (CAD) systems have neglected the conceptual phase in favour of the detailed design phase \cite{Hwang1990}. The rigidity of CAD geometry definition stifles the ambiguity required for creative exploration. This gap precipitated the development of Sketch-Based Interfaces for Modeling (SBIM) and various digital sketching tools.

\subsubsection{From Electronic Napkins to Digital Styluses}
Early attempts to digitize the "back of the envelope" sketch included systems like Lakin’s VMACS \cite{Lakin1989} and Gross's "Electronic Cocktail Napkin" \cite{gross1996electronic}. These tools sought to process digital freehand input and recognize diagrams. Other approaches, such as the Design Capture System (DCS) \cite{Hwang1990}, focused on capturing the stroke data to preserve the "fluency" of the designer. 

Modern systems leverage digital graphical tablets (such as Wacom, XP-Pen, iPads, etc.) capable of capturing rich, non-geometric data, including pressure, azimuth, velocity, etc. \cite{Elsen2012} highlights that while digital tools offer benefits like storage and "undo" capabilities, they often fail to replicate the friction and tactile feedback of paper, which aids in cognitive processing. Recent efforts, such as the OpenSketch dataset \cite{Gryaditskaya2019-bw}, have focused on capturing professional design sketches to better understand stroke topology and construction lines, bridging the gap between 2D input and 3D inference \cite{Hahnlein2022-yj}.

\subsubsection{The Generative Design Era}
The field has witnessed the rise of Generative Design and AI-driven image synthesis. Tools described by \cite{Su2024} utilize generative models to transform sketches into high-fidelity images. Recent frameworks, such as CustomSketching \cite{Xiao2024}, enable fine-grained editing by extracting concepts from user sketches, while systems like "Block and Detail" \cite{Sarukkai2024} support iterative refinement from coarse blocking strokes. However, while these "sketch-to-image" tools facilitate visual refinement (Vertical Transformation) and aesthetic rendering, they predominantly treat the sketch as a static input for a generative algorithm rather than a dynamic, evolving historical artifact. They do not inherently capture or organize the multi-branching version history or the cognitive rationale driving the sequence of iterations, leaving a gap in managing the \textit{process behind the generation of the image}.

\subsection{Knowledge Graphs: Structuring Design Information}
Parallel to sketching tools, research has focused on structuring design knowledge using Knowledge Graphs (KGs). KGs provide a semantic network connecting entities and relationships, making complex data machine-interpretable \cite{Gutierrez2021}.

\subsubsection{Applications in Engineering}
In the design domain, KGs have been effectively deployed to formalize product assemblies and configure engineering equipment \cite{Hao2021, Huang2025}. \cite{Peng2017} and \cite{Huang2024} explored the integration of KGs to manage collaborative design knowledge, combining formal CAD data with tacit knowledge (rationale). These graphs are increasingly used to capture data from large model repositories \cite{Bharadwaj2022} and to mine prototyping knowledge for process insights \cite{Gopsill2025}.

Specific applications include using KGs to drive User Experience (UX) improvements \cite{Li2023}, support the development of Smart Product-Service Systems (PSS) \cite{Li2020-pu}, and ensure scalability in smart manufacturing contexts \cite{Hu2024, Jawad2023-sc}. Moreover, Graph Models have been utilized to evaluate model fidelity and encoding in early design stages \cite{Coatanea2023-mm}. Recent work has even proposed interpretable conceptual design KGs to enhance heuristic ideation \cite{Cong2025} and DKG-aided approaches for joint entity extraction \cite{Huang2022-ud}.

\subsubsection{Limitations in Temporal Dynamics}
Despite their organizational strengths, KGs are structurally ill-suited to capture the dynamic, temporal evolution of concept sketching. KGs excel at representing \textit{static relationships} of what a final concept \textit{is}. They struggle to manage the volatile, spontaneous process of \textit{becoming}. The core challenge of conceptual design is managing uncertainty and non-linear evolution. While KGs can store metadata, they lack a native mechanism to handle the incremental creation and deletion of graphical entities (strokes) and the concurrent management of multiple diverging versions (Lateral Transformations) in the manner of a time-series log \cite{Cong2025, Gopsill2025}. None of the reviewed KG approaches provides a specialized, integrated version management framework necessary to track design evolution at the stroke level.

\subsection{Version Control Systems: Managing Evolution}
To manage non-linear evolution, we turn to the domain of Software Engineering and the concept of Version Control Systems (VCS), also earlier known as Revision Control Systems (RCS).

\subsubsection{The VCS Paradigm}
The problem of managing evolution and versions is not new; it has been solved definitively in the domain of software engineering. The Git version control system, developed by Linus Torvalds in 2005, allows software developers to track changes, branch into experimental features, merge code, and revert to previous states with granular precision. As reviewed by \cite{Zolkifli2018} and \cite{Koc2017}, VCS are essential for managing evolving codebases, operating on `checkout`, `modify`, and `commit` cycles. Crucially, they support `branching`, allowing multiple parallel lines of thought to evolve simultaneously without interference. Modern Distributed Version Control Systems (DVCS) like Git have decentralized this process \cite{Schwabe2021}. The utility of version control extends to asynchronous collaboration in Building Information Modeling (BIM), where graph analysis is used for model merging \cite{Esser2023}. Similarly, novel frameworks for collaborative conceptual-embodiment design are being proposed to bridge the gap between individual ideation and group development \cite{Hasby2024-ic}.

\subsubsection{VCS in the Creative Domain}
The parallels between software and product design, such as the incremental improvement, iterative testing, and collaborative divergence, have prompted investigations into "Creative Version Control" \cite{Sterman2022-rk}. However, adoption is slow because text-oriented VCS paradigms are poorly suited for binary graphical data \cite{Zolkifli2018} and the inherent ambiguity of concept sketches. Designers also face challenges in self-tracking and regulating their creative state, similar to mood-tracking in personal informatics \cite{Overdijk2022-qs}, which a VCS could potentially alleviate by externalizing progress.

\subsubsection{The Need for a Dedicated Version Control System for Sketching}
The most pertinent prior work is that of \cite{Sankar2023}, "A Novel Version Control Scheme for Supporting Interrupted Product Concept Sketching". This study explicitly attempted to leverage off-the-shelf VCS, Git combined with a digital sketching tool (Autodesk Sketchbook) for interrupted product sketching. The authors mapped the different Git features to the sketching process. Through a pilot study, they were able to establish that a VCS was highly effective in facilitating the exploration of diverse concepts.

While the authors concluded that the VCS logic successfully supported divergent exploration and helped manage interruptions, they identified a critical failure point: the "Cognitive-Logistical Friction." The Git application is designed for text-based code, not visual data. Designers found the command-line interface (CLI) and even standard Git GUIs (like GitHub Desktop/GitKraken) to be counterintuitive and disconnected from their sketching canvas. Using two separate software applications, one for sketching and one for versioning, forced designers to constantly switch contexts, obstructing their creative flow. They reported that the logistical burden of managing commits disrupted the benefits of versioning. Yet, the designers unanimously recognized the potential value of such a system. They expressed a strong desire for a tool that could capture not just the final output but the intermediate "work-in-progress" states, facilitating a diverse range of concept generation unlike traditional linear methods. Furthermore, standard Git could not "read" the sketch content, treating it as a binary blob, thus preventing any intelligent analysis of the design progress. We identified that this necessitates an \textit{Integrated} Git-cum-Sketching system—a dedicated environment where version control is transparent, visual, and embedded directly into the sketching workflow.

\subsection{Research Gap}
Synthesizing the literature reveals a clear convergence of needs but a divergence of solutions. Table \ref{tab:tool_comparison} illustrates the functional gap in existing tools regarding the specific requirements of conceptual design tracking.

\begin{table*}[ht!]
\centering
\caption{Comparative analysis of existing tools against requirements for conceptual design tracking.}
\label{tab:tool_comparison}
% Increase row height slightly for better readability
\renewcommand{\arraystretch}{1.3}

% Using X columns with multipliers ensures the table always fits \textwidth exactly.
% The multipliers determine relative width (1.5 = 1.5x standard X width).
\begin{tabularx}{\textwidth}{@{}
    >{\raggedright\arraybackslash\hsize=1.5\hsize}X 
    >{\centering\arraybackslash\hsize=0.8\hsize}X 
    >{\centering\arraybackslash\hsize=0.8\hsize}X 
    >{\centering\arraybackslash\hsize=0.8\hsize}X 
    >{\raggedright\arraybackslash\hsize=1.1\hsize}X 
@{}}
\toprule
\textbf{Tool Category} & \textbf{Temporal Capture (History)} & \textbf{Branching Support (Lateral Transform)} & \textbf{Stroke-Level Awareness} & \textbf{Design-Intent Integration} \\ 
\midrule
\textbf{Digital Sketching Tools} \cite{Elsen2012, Gryaditskaya2019-bw} & Low (Undo stack) & None & High (Pixels/Vector) & None \\ 
\addlinespace
\textbf{Generative AI Tools} \cite{Sankar2023, Xiao2024} & Low (Input/Output) & Low (Variations) & Medium (Concept Extraction) & Low \\ 
\addlinespace
\textbf{Knowledge Graphs} \cite{Bharadwaj2022, Cong2025} & Low (Static Relations) & None & N/A & High (Metadata) \\ 
\addlinespace
\textbf{Standard VCS (Git)} \cite{Elsen2012, Zolkifli2018} & High & High & Low (Binary Blob) & Low (Commit Msg) \\ 
\addlinespace
\textbf{Proposed sGIT} & \textbf{High} & \textbf{High} & \textbf{High} & \textbf{High} \\ 
\bottomrule
\end{tabularx}
\end{table*}

As highlighted in Table \ref{tab:tool_comparison}, a critical gap exists for a system that unifies these capabilities:
\begin{enumerate}
    \item \textbf{The Need:} Conceptual design is a non-linear process \cite{Goel1995, Sankar2023} requiring the capture of both physical action (strokes) and cognitive intent (rationale) to be truly useful \cite{Schon1983, Cross1982}.
    \item \textbf{The Gap:} Sketching tools capture pixels but forget history \cite{Su2024, Elsen2012}; KGs capture static relationships but miss temporal fluidity of sketching \cite{Bharadwaj2022, Cong2025}; and standard VCS captures history but is designed for text, creating high friction for visual thinkers \cite{Sankar2023, Sterman2022-rk}. Therefore, a critical research gap exists for a dedicated \textit{Graph-based Version Control System} specifically for conceptual sketching, capable of integrating the diverse data streams identified in this review. 
    \item \textbf{The Requirements:} A novel solution is required that creates a "dedicated VCS for PCS". This system must move beyond simply storing versions of image files. It must integrate the fine-grained capture of drawing actions (strokes, pressure, timestamps, etc.) with the explicit recording of cognitive intent (via multi-modal inputs) into a cohesive branching structure. This integration would support the designer's natural, non-sequential workflow, enabling coherent concept reconstruction and traceability in the evolving landscape of Generative Design. The system should free the designers from logistical tasks involved during sketching and free up their cognitive memory for pure creative generation.
\end{enumerate}

%%%%%%%%%%%%%%%%%%%%%%%%%%%%%%%%%%%%%%%%%%%%%%%%%%%%%%%%%%%%%%%%%%%%%%

\section{Theoretical Framework: Understanding Product Concept Sketching}
\label{sec:understanding_pcs}

To develop a computational system capable of managing the evolution of design concepts, one must first establish a theoretical understanding of the domain itself. This section deconstructs the ontology of PCS, analyses its constituent atomic units (strokes), and presents an empirical study of expert sketching behaviours to derive the functional requirements for our proposed system.

\subsection{Definition and Ontology of PCS}
Sketching, in its most fundamental sense, is a well-established mode of communication that has been known to humanity since the early cave paintings. However, in the context of industrial design, it transcends mere representation. We define \textit{Product Concept Sketching as the manual action of creating impressions in the form of strokes on a medium, enabled through a cognitive process focused on satisfying a set of requirements for a given problem.}

Product Concept Sketches distinguish themselves from other visual media illustrations, such as technical drafting, artistic painting, or casual doodling, through specific, unique characteristics. We define \textit{PCS as a class of unfinished, rapidly created, multi-stroke freehand drawings generated by product or industrial designers to explore the different ideas present in their minds}. These representations differ from the final "product renderings" commonly seen in marketing materials.

\begin{figure}[ht!]
        \centering
        \includegraphics[width=\linewidth]{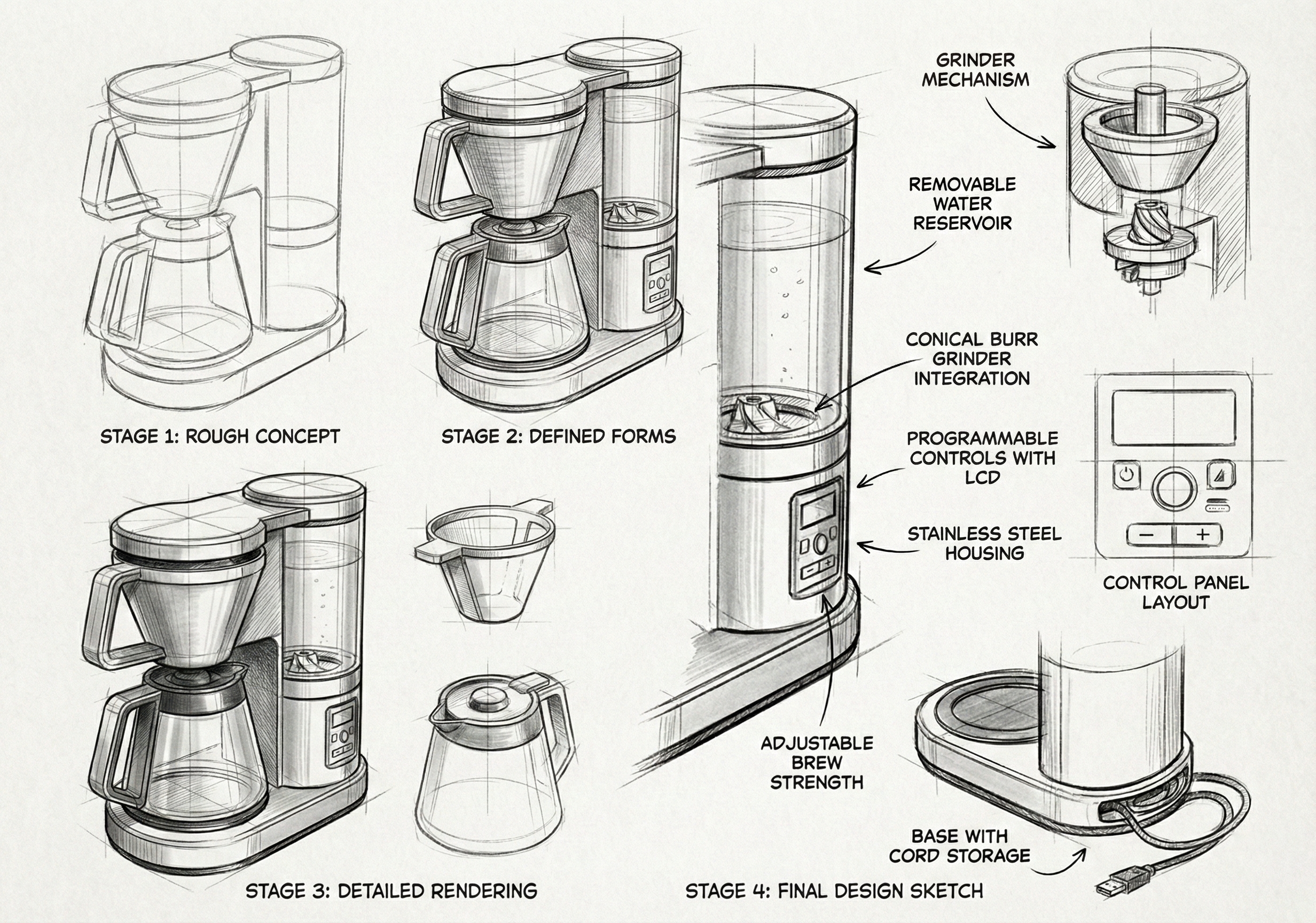}
        \caption{The progression of a coffee maker design from a rough concept (Stage 1) to a final detailed sketch (Stage 4), including specific component details like the grinder mechanism and control panel layout.}
        \label{fig:coffeemaker_progression}
\end{figure}

\subsubsection{Properties and Characteristics}
\label{subsubsec:properties_and_characteristics}
Based on our synthesis of design literature and observation, PCS exhibits the following properties that any support system must accommodate:
\begin{enumerate}
    \item \textbf{Vagueness:} The sketch is rarely a definitive declaration of form. It is represented with multiple, often overlapping strokes. This vagueness is functional; it allows the designer to delay commitment to a specific geometry while exploring the general topology.
    \item \textbf{Information Richness:} Despite their rough appearance, PCS are informationally complete regarding the designer's intent. They often contain non-geometric data such as annotations, force vectors, and operational logic.
    \item \textbf{Freeform Nature:} PCS is non-regimental. Unlike CAD, which demands precise dimensioning and constraints, PCS thrives on the fluidity of the hand motion, allowing for rapid iteration without the cognitive load of precision.
    \item \textbf{Ambiguity:} The sketch is subjective. A circle might represent a wheel, a hole, or a sphere. This ambiguity encourages "inferences and rediscovery," where the designer or a peer might perceive a new solution in a previously drawn ambiguous shape.
    \item \textbf{Imprecision:} The representation is qualitative. A line representing a curve need not be mathematically perfect; it only needs to convey the \textit{character} of the curve (e.g., tense, relaxed, organic).
    \item \textbf{Aesthetic Completeness:} Even in its rough state, a concept sketch by a skilled designer possesses a holistic aesthetic balance that conveys the "feel" of the product.
    \item \textbf{Entirety:} Through the use of multiple views (orthographic, isometric, perspective) on a single 2D canvas, the sketch represents the 3D product in its entirety.
\end{enumerate}

\subsubsection{The Cognitive-Action Duality}
Sketching is an activity where the cognitive aspect (thought) and the action aspect (drawing) go hand in hand. The free flow of sketching externalizes the designer's thought process. We posit that ideas manifest into multiple concepts through three interwoven, repetitive phases:
\begin{enumerate}
    \item \textbf{Creation (Action Dominant):} The physical production of strokes on the paper. This is a high-velocity phase where the hand moves rapidly to capture fleeting ideas.
    \item \textbf{Evaluation (Cognition Dominant):} The designer pauses to assess the present state of the sketch against the mental model or the requirements.
    \item \textbf{Planning (Cognition Dominant):} The formulation of the subsequent course of action—whether to modify the current form, spawn a new concept, or abandon the current path.
\end{enumerate}

The sketching process begins with a broad plan, but typically does not reach the last level of detail. As the process continues through these phases, ideas take shape, and the plans for adding other elements become more detailed.

Crucially, this process is discontinuous. It is punctuated by halts, moments when the physical sketching comes to a stop. These halts are not merely breaks; they are moments of critical thinking when new ideas emerge, or existing concepts receive quick assessments. The causes for these halts can be either \textit{Internal} (Pauses for thought) or \textit{External} (Gaps or disturbances). This temporal discontinuity is referred to as \textbf{Interrupted Product Concept Sketching (IPCS)}. In our previous work \cite{Sankar2023}, we classified these interruptions, noting that they often signify a branching point in the decision tree, a logic we later map to the `commit` structure in our VCS.

\subsection{Taxonomy of Sketching Strokes}
\label{sec:taxonomy_sketching_strokes}
To digitize and track the evolution of a sketch, one must first define its atomic units. While a sketch appears as a cohesive image, it is composed of thousands of individual strokes. Building upon the foundational classification established by \cite{Onkar2010-dz}, we have refined and expanded the taxonomy to classify strokes into six distinct types based on their intent and execution. This classification forms the basis for the AI recognition module in our system.

% TODO: Insert Figure 2: Visual examples of the six stroke types: Constraining, Defining, Detailing, Shading, Shadow, and Annotation.
\begin{figure}[ht!]
\centering
\includegraphics[width=\linewidth]{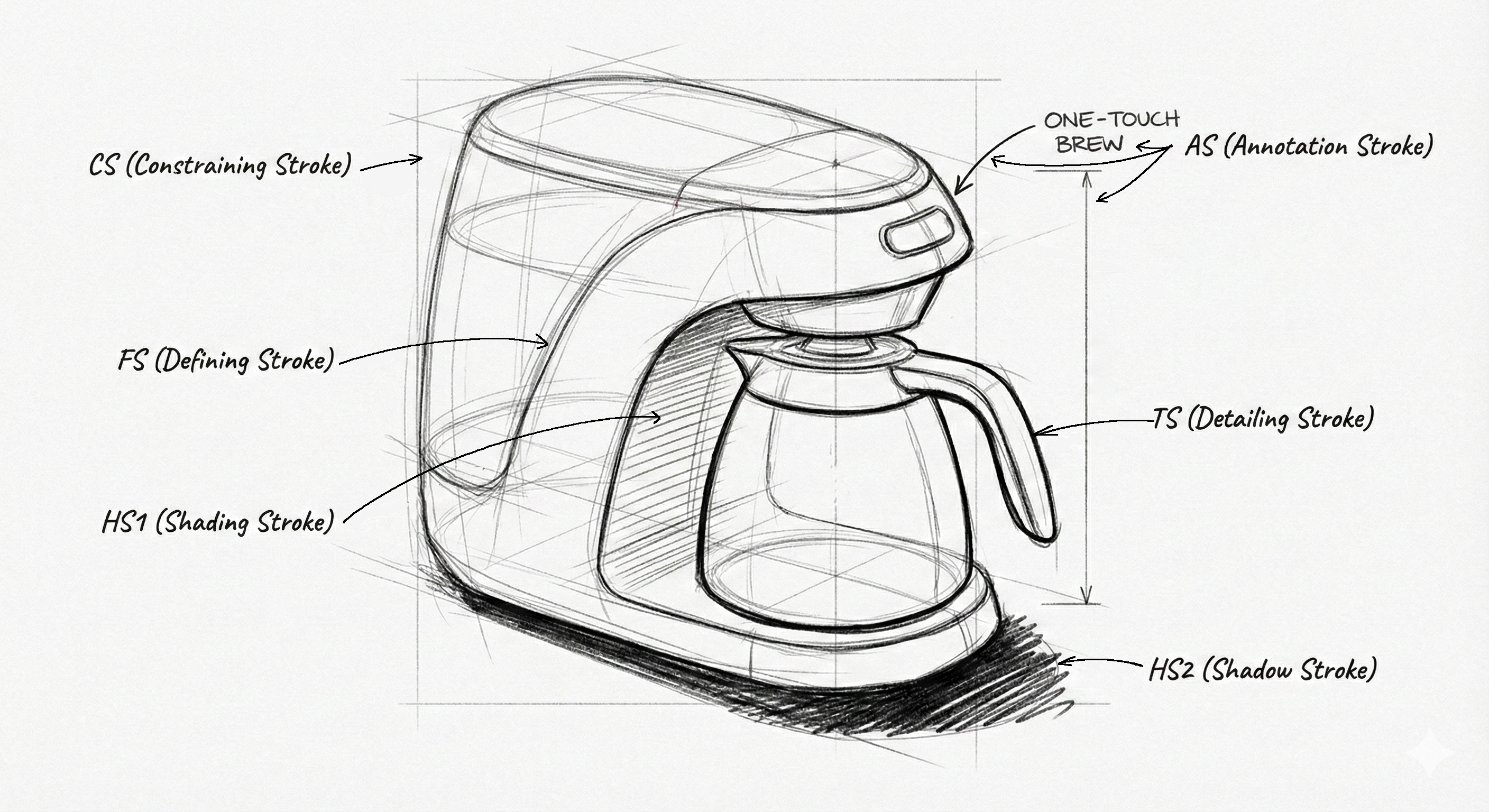}
\caption{Taxonomy of strokes used in product concept sketching shown on a drip coffee maker design. 
CS = Constraining Stroke, FS = Defining Stroke, AS = Annotation Stroke, 
TS = Detailing Stroke, HS1/HS2 = Shading/Shadow Strokes.}
\label{fig:concept-sketch-stroke-types}
\end{figure}

\subsubsection{Constraining Strokes (CS)}
The genesis of a sketch often begins with Constraining Strokes. These are generally light, straight lines drawn with high velocity. Their primary function is to establish a rough boundary or "bounding box" within which the sketch will be contained. They serve to establish the aspect ratio and the proportionality of the different elements, ensuring that the subsequent geometry adheres to the intended scale.

\subsubsection{Defining Strokes (FS)}
Once the constraints are set, the designer employs Defining Strokes to explore the form. These are rapid, light curves drawn across the canvas to suggest the silhouette and internal features. This is an exploratory phase; the designer is "searching" for the shape. Multiple strokes might be laid down in the same area to define a single curve, representing the evolution of the shape in real-time.

\subsubsection{Detailing Strokes (TS)}
As the exploration converges, the form emerges from the fuzzy definitions to something concrete. Detailing Strokes are used to concretize each edge. These are generally represented by slower, deliberate, and darker curves. They confirm which among the multiple Defining Strokes represents the final intent. This transition from Defining to Detailing represents a micro-evolution from ambiguity to certainty.

\subsubsection{Shading Strokes (HS1)}
Form is not merely an outline; it is a surface. Shading Strokes are employed to define the surface topology and texture. They indicate how light interacts with the product, whether a surface is planar or curved, and whether it is matte or glossy. For instance, a sharp white region surrounded by dark hatching might indicate a metallic finish, while a gradual gradient suggests a cylindrical form. These manifest as hatches, cross-hatches, or streaks.

\subsubsection{Shadow Strokes (HS2)}
To ground the object in reality, Shadow Strokes are applied. These provide a sense of "3D-ness" and depth perception. They are typically dark, dense regions representing the cast shadow of the object on the ground plane or the occlusion shadows between parts. They convert a floating 2D shape into a grounded 3D object.

\subsubsection{Annotation Strokes (AS)}
Finally, the "Information Richness" of PCS is often conveyed through annotation strokes. These are non-geometric graphical elements, including vectors, arrows, call-outs, and text. They are used to represent behavioural aspects (e.g., an arrow showing rotation), assembly logic, or material specifications that cannot be conveyed through geometry alone.

\subsection{Empirical Analysis of Expert Sketching Behaviour}
\label{subsec:empirical_analysis}

While the taxonomy defines \textit{what} is drawn, it is equally important to understand \textit{how} it is drawn, i.e., the temporal and procedural action of the designer. To facilitate this understanding, we conducted a rigorous video analysis of expert sketching sessions.

\subsubsection{Methodology}
We curated a dataset of 20 first-person concept sketching video tutorials from YouTube (YT), featuring professional industrial designers. The selection criteria ensured diversity in the products sketched, which included an Armchair, Clock, Coffee Maker, Gaming Mouse, Hair Dryer, Home Speaker, Lamp, Lawn Mower, LED Mood Lamp, Pottery, Toolbar, USB Hub, Washing Machine, and Wristwatch. The YouTube video ID for each of the product along with the video duration is given in Table~\ref{tab:sketching_videos}.

\begin{table}[ht!]
    \centering
    \caption{Dataset of First-Person Concept Sketching Videos}
    \label{tab:sketching_videos}
    \setlength{\tabcolsep}{5pt}
    \begin{tabularx}{\columnwidth}{@{} c >{\raggedright\arraybackslash}X c c @{}}
        \toprule
        PID & P. Name & VID & Dur. (HH:MM) \\
        \midrule
        P1 & Armchair & 7u-q6PrsfYg & 04:35 \\
        P2 & Clock & FUXuU8v3omI & 01:59 \\
        P3 & Clock & PdXRdyOzLeE & 02:58 \\
        P4 & Coffee Maker & 4ZlrvGRsz5M & 04:13 \\
        P5 & Coffee Maker & BMS-NbBfL6o & 01:43 \\
        P6 & Furniture & xpE8qy8lOXI & 04:22 \\
        P7 & Gaming Mouse & 6wpMzzowpXw & 03:59 \\
        P8 & Hair Dryer & tIBkucLlvkY & 02:12 \\
        P9 & Home Speaker & ZJYlmu09AGU & 05:02 \\
        P10 & Lamp & EJAGRJj7bpI & 04:41 \\
        P11 & Lawn Mower & HRlZgDwH2SI & 05:52 \\
        P12 & Lawn Mower & r5rcORybSVw & 05:30 \\
        P13 & LEDMood Lamp & ECrcWzAeN3E & 04:15 \\
        P14 & Mouse & 0RdjF2zF0F8 & 01:26 \\
        P15 & New Clock & 7VUcZX4cULk & 02:30 \\
        P16 & Pottery & gHyCA1MlZfg & 02:18 \\
        P17 & Tool bar & eS158QYQBGw & 03:26 \\
        P18 & USB Hub & wB86PtYFOH4 & 04:38 \\
        P19 & Washing Machine & s-AWyRNruWM & 08:13 \\
        P20 & Watch & 9aZDk7EzC5Y & 11:11 \\
        \bottomrule
    \end{tabularx}

    \vspace{10pt}
    
    \begin{minipage}{\columnwidth}
        \footnotesize
        PID: Product ID, P.Name: Product Name, VID: YouTube Video ID, Dur.: Duration
        \textbf{Note:} To access the specific video, append the ID listed in the table to the base URL.\\
        The full format is: \url{https://www.youtube.com/watch?v=ID} \\
        Example: \url{https://www.youtube.com/watch?v=7u-q6PrsfYg}
    \end{minipage}
\end{table}

\begin{figure*}[ht!]
    \centering
    \begin{subfigure}[b]{0.48\textwidth}
        \centering
        \includegraphics[width=\linewidth]{Figures/VideoFrames/Armchair_v=7u-q6PrsfYg.png}
        \caption{Armchair (7u-q6PrsfYg)}
        \label{subfig:armchair-7u-q6PrsfYg}
    \end{subfigure}
    \hfill
    \begin{subfigure}[b]{0.4\textwidth}
        \centering
        \includegraphics[width=\linewidth]{Figures/VideoFrames/Clock_v=PdXRdyOzLeE.png}
        \caption{Clock (PdXRdyOzLeE)}
        \label{subfig:clock-PdXRdyOzLeE }
    \end{subfigure}
    \vfill
    \begin{subfigure}[b]{0.48\textwidth}
        \centering
        \includegraphics[width=\linewidth]{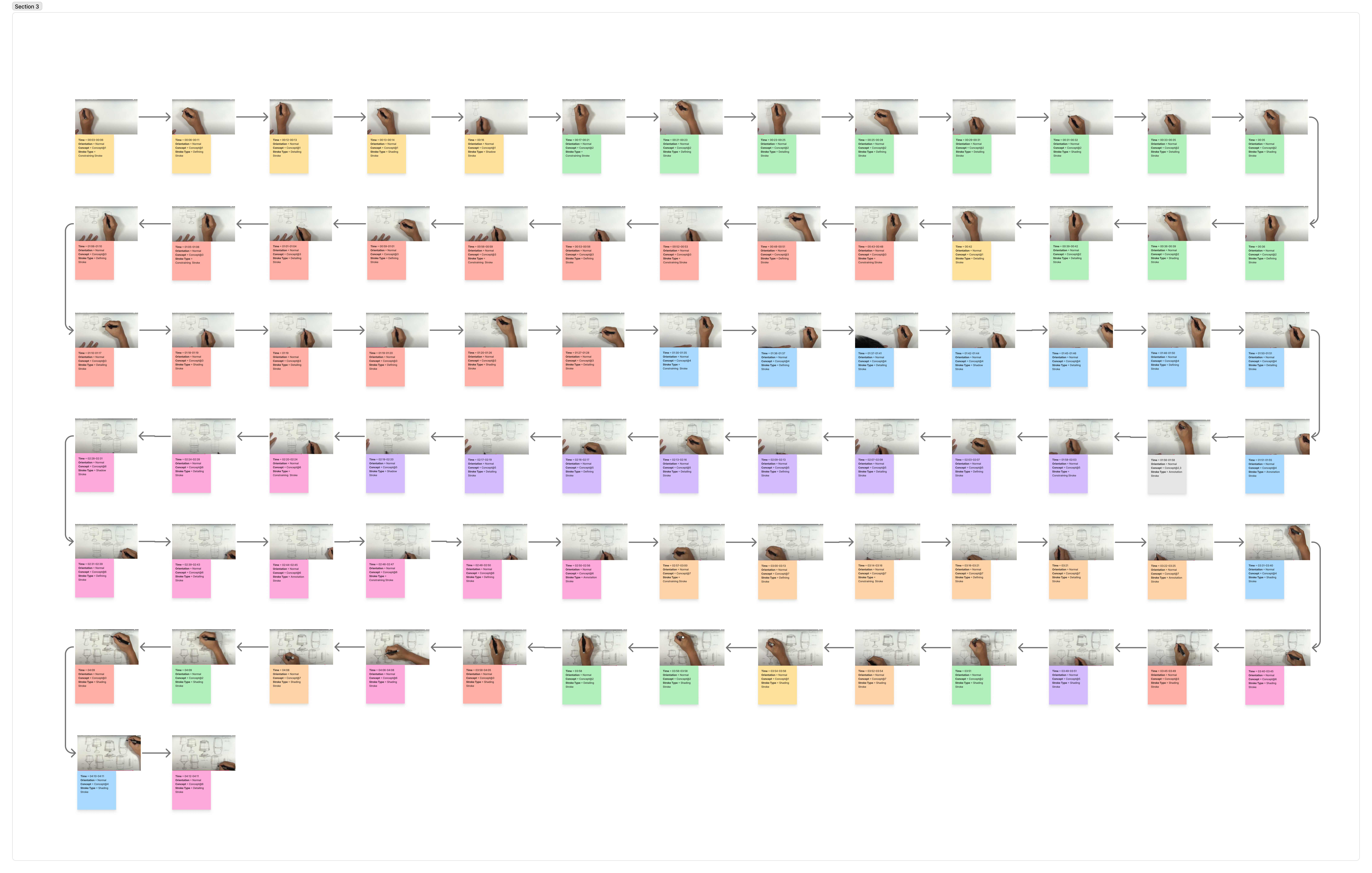}
        \caption{Lamp (EJAGRJj7bpI)}
        \label{subfig:lamp-EJAGRJj7bpI}
    \end{subfigure}
    \hfill
    \begin{subfigure}[b]{0.44\textwidth}
        \centering
        \includegraphics[width=\linewidth]{Figures/VideoFrames/Pottery_v=gHyCA1MlZfg.png}
        \caption{Pottery (gHyCA1MlZfg)}
        \label{subfig:pottery-gHyCA1MlZfg}
    \end{subfigure}
    
    \caption{Visual representations of frame sequences with annotated labels and metadata. The subfigures display the temporal progression of sketching actions, where individual frames are tagged with coloured metadata (e.g., time, operation type, or state) indicating the workflow stages for the Armchair, Clock, Lamp, and Pottery concepts.}
    \label{fig:sketch_metadata}
\end{figure*}

\begin{figure*}[ht!]
    \centering
    \begin{subfigure}[b]{0.48\textwidth}
        \centering
        \includegraphics[width=\linewidth]{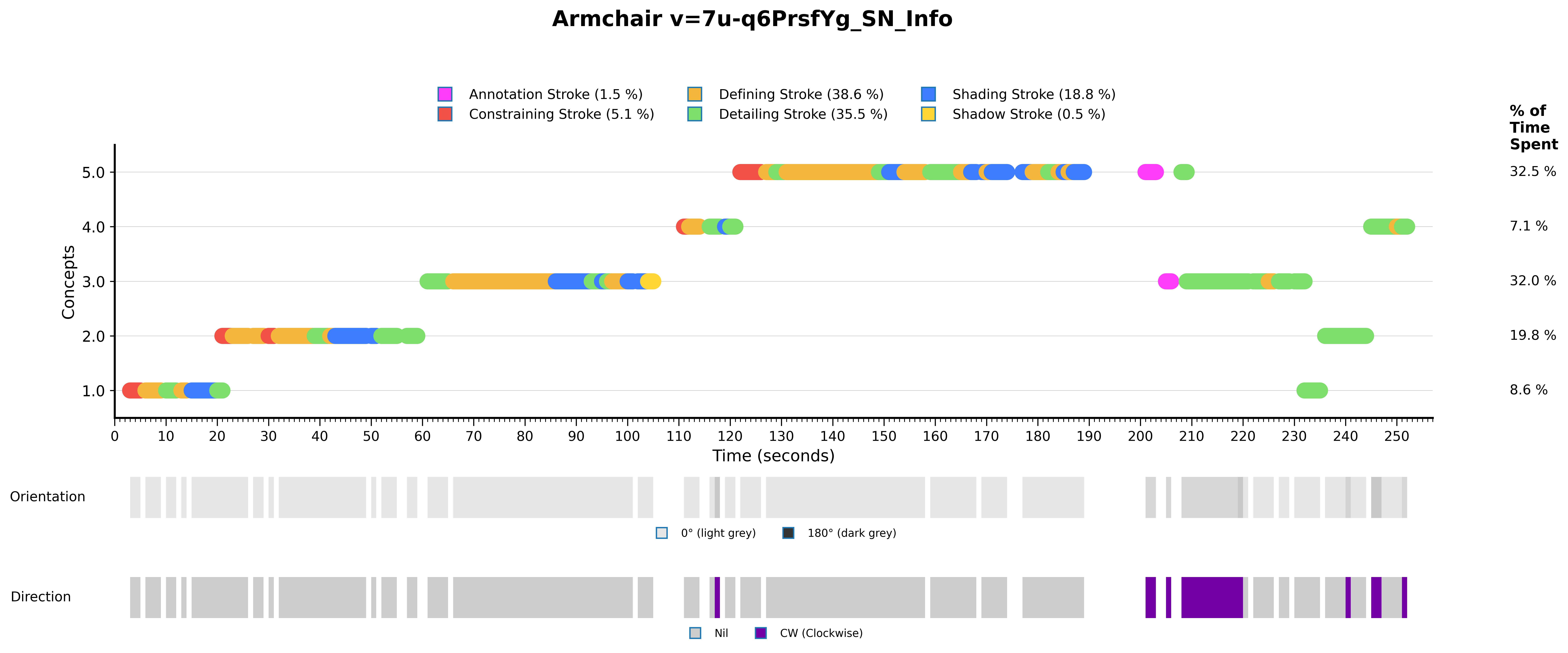}
        \caption{Armchair (7u-q6PrsfYg)}
        \label{subfig:armchair-7u-q6PrsfYg_plot}
    \end{subfigure}
    \hfill
    \begin{subfigure}[b]{0.4\textwidth}
        \centering
        \includegraphics[width=\linewidth]{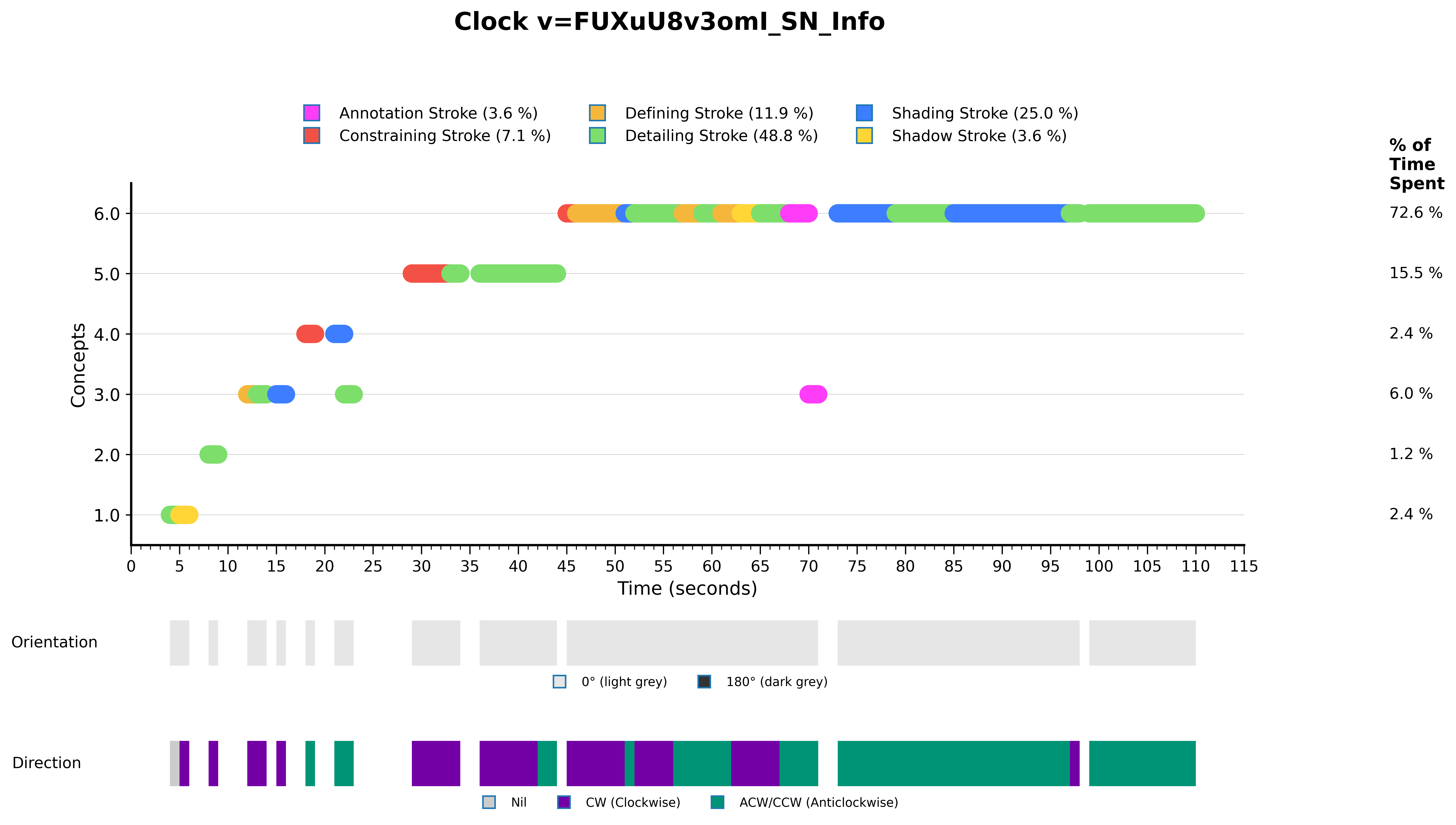}
        \caption{Clock (FUXuU8v3omI)}
        \label{subfig:clock-PdXRdyOzLeE_plot}
    \end{subfigure}
    \vfill
    \begin{subfigure}[b]{0.48\textwidth}
        \centering
        \includegraphics[width=\linewidth]{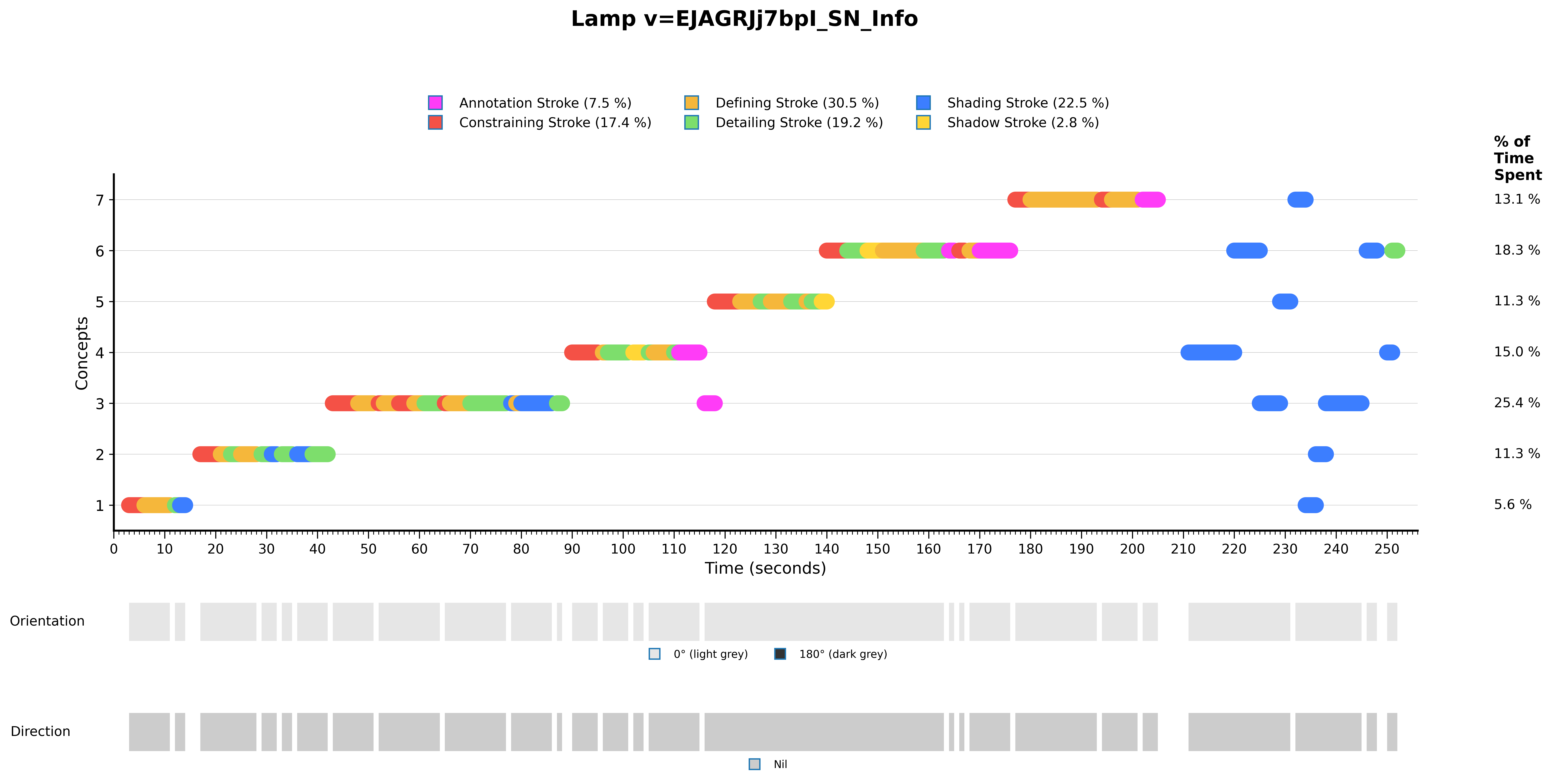}
        \caption{Lamp (EJAGRJj7bpI)}
        \label{subfig:lamp-EJAGRJj7bpI_plot}
    \end{subfigure}
    \hfill
    \begin{subfigure}[b]{0.44\textwidth}
        \centering
        \includegraphics[width=\linewidth]{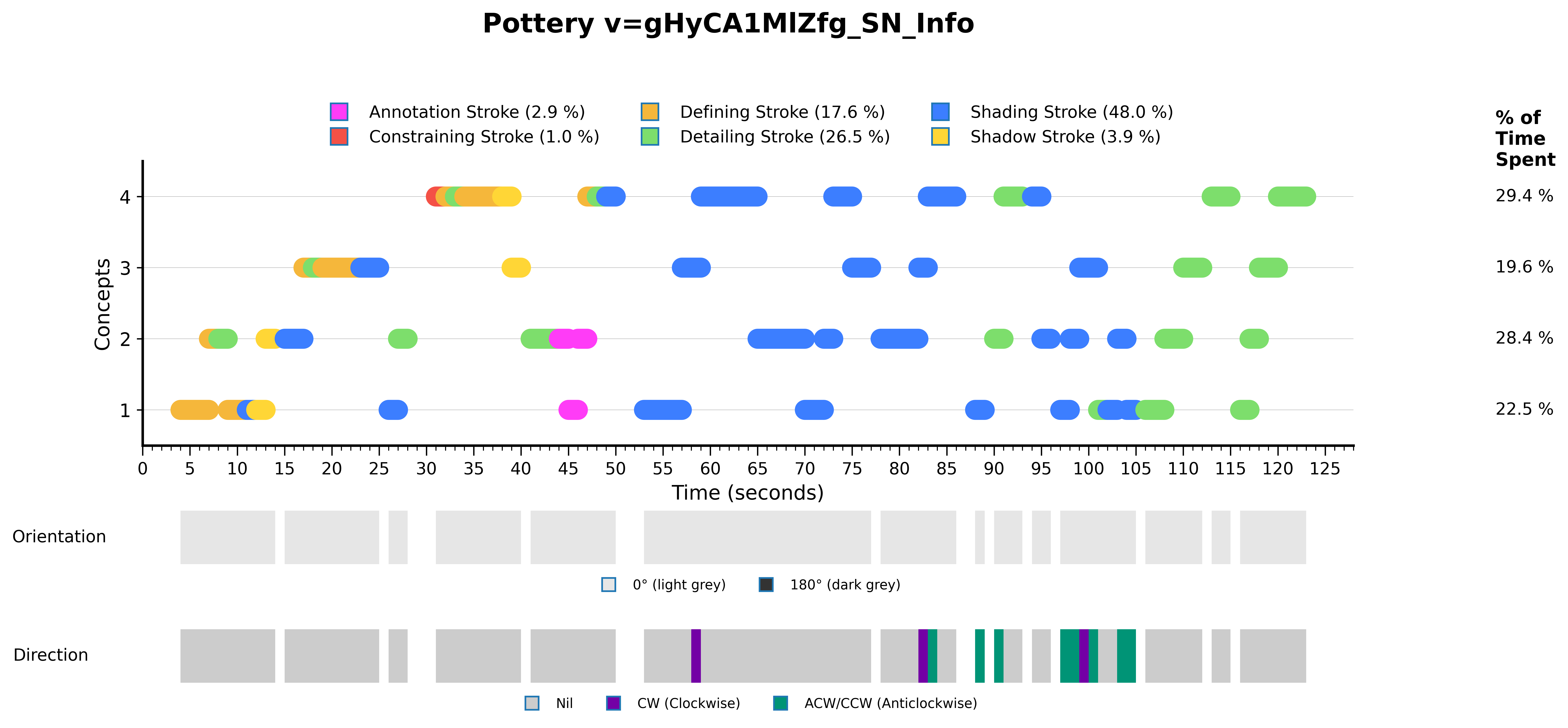}
        \caption{Pottery (gHyCA1MlZfg)}
        \label{subfig:pottery-gHyCA1MlZfg_plot}
    \end{subfigure}
    
    \caption{Time-series concept-wise stroke distribution plots for selected products. Each plot visualizes the temporal distribution of different stroke types (e.g., Defining, Shading, Detailing) across the various sketched concepts, highlighting the duration and sequence of actions.}
    \label{fig:stroke_distribution_plot}
\end{figure*}

A custom Python programme was developed to extract frames from these videos at one-second intervals and a similarity score was assigned between subsequent frames. Based on the score, duplicated frames were removed. Also any frames where the sketching is not depicted like the credit frame, intro frame, extro frame, end-card frame, ads frame were also removed. A product design student (pursuing an M.Des) skilled in the domain of PCS analysed each frame to identify and label the strokes according to the six categories defined in Section~\ref{sec:taxonomy_sketching_strokes}. Additionally, metadata regarding timestamp, concept version number, duration, stroke type and canvas orientation was also recorded. Figure~\ref{fig:sketch_metadata} presents a sample of frames and metadata for four products. Another python programme was written to create a \textit{Time-Series Stroke Distribution Plot}. For each video, a single plot was generated displaying the number of concepts, the percentage distribution of strokes and time across concepts, and canvas orientation and rotation data. For reference, the stroke distribution plots corresponding to the four products shown in Figure~\ref{fig:sketch_metadata} are displayed in Figure~\ref{fig:stroke_distribution_plot}. The frames and metadata diagrams, along with the time-series stroke distribution plots for the remaining products, are provided in the \textcolor{blue}{\href{https://drive.google.com/drive/folders/1Oo-dW_yXXP5wF87bBm_wSQaMFVSw91wk?usp=sharing}{Google Drive folder}}.

\subsubsection{Aggregate Analysis of Sketching Dynamics}
\label{subsubsec:aggregate_analysis}

\begin{figure*}[ht!]
    \centering
    \begin{subfigure}[b]{0.48\textwidth}
        \centering
        \includegraphics[width=\linewidth]{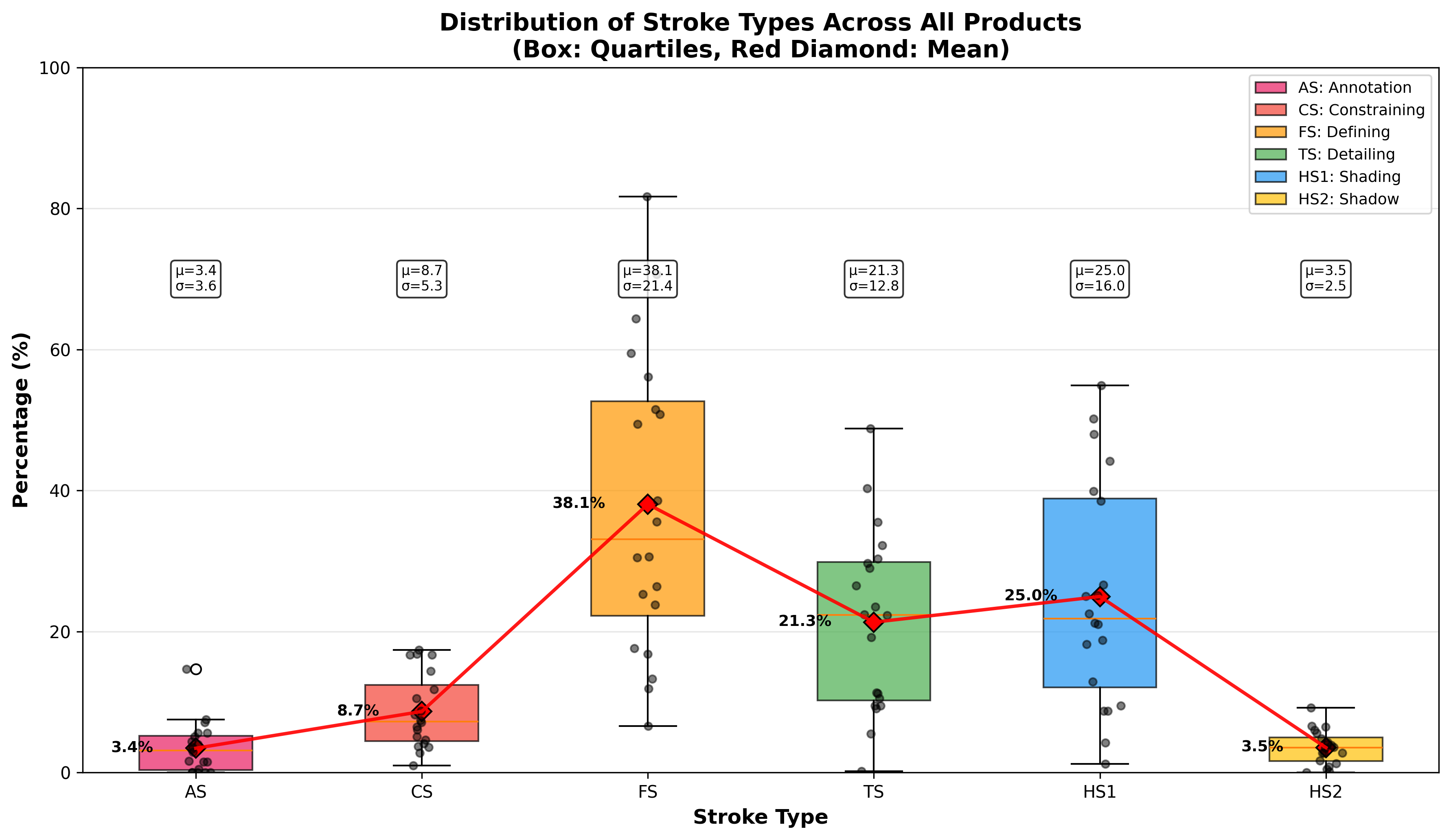}
        \caption{Distribution of Stroke Types Across All Products}
        \label{subfig:box_whisker_strokes}
    \end{subfigure}
    \hfill
    \begin{subfigure}[b]{0.48\textwidth}
        \centering
        \includegraphics[width=\linewidth]{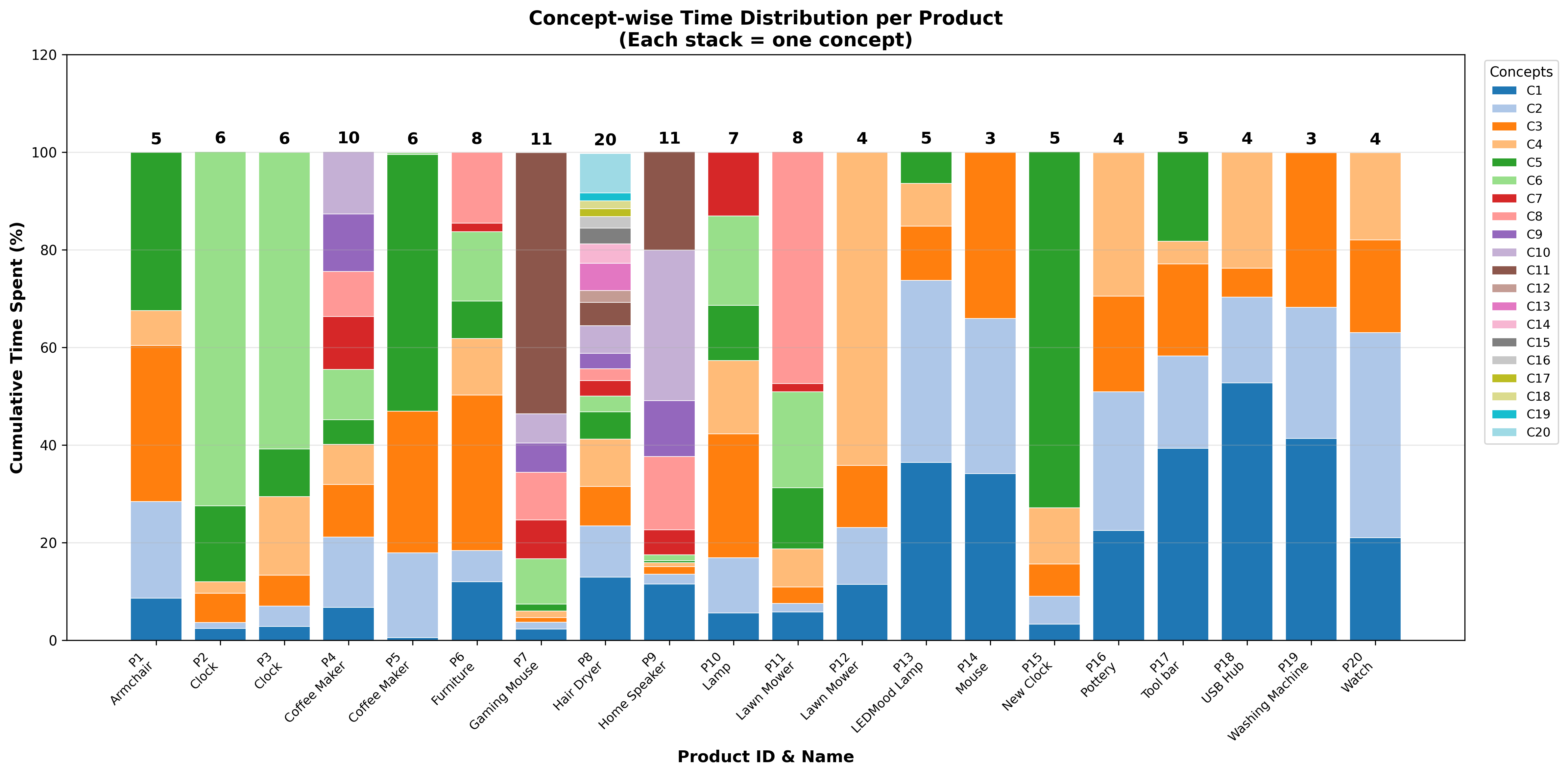}
        \caption{Concept-wise Time Distribution per Product}
        \label{subfig:stacked_bar_concepts}
    \end{subfigure}
    \vfill
    \begin{subfigure}[b]{0.48\textwidth}
        \centering
        \includegraphics[width=\linewidth]{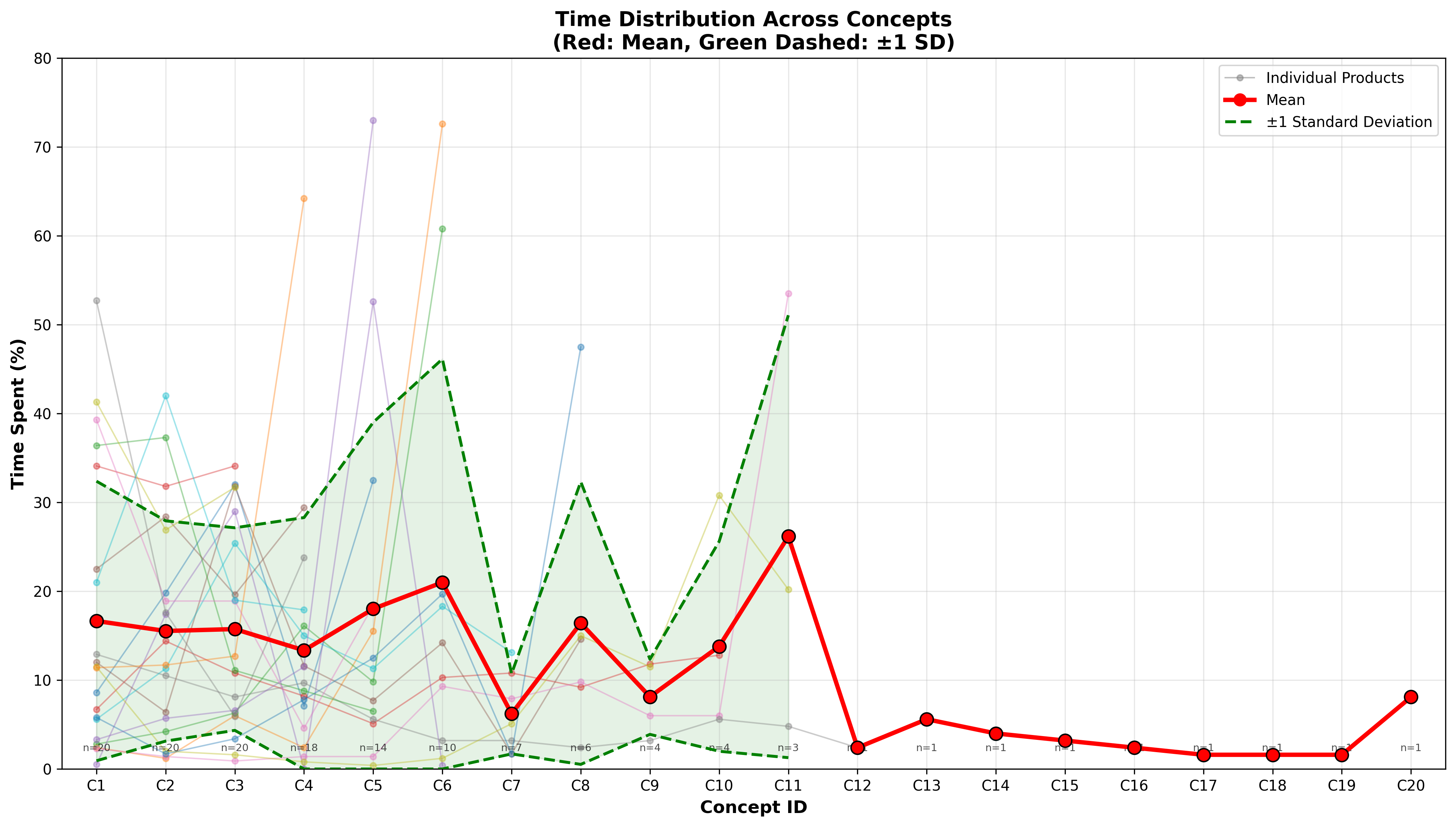}
        \caption{Time Distribution Across Concepts}
        \label{subfig:line_plot_concepts}
    \end{subfigure}
    \hfill
    \begin{subfigure}[b]{0.48\textwidth}
        \centering
        \includegraphics[width=\linewidth]{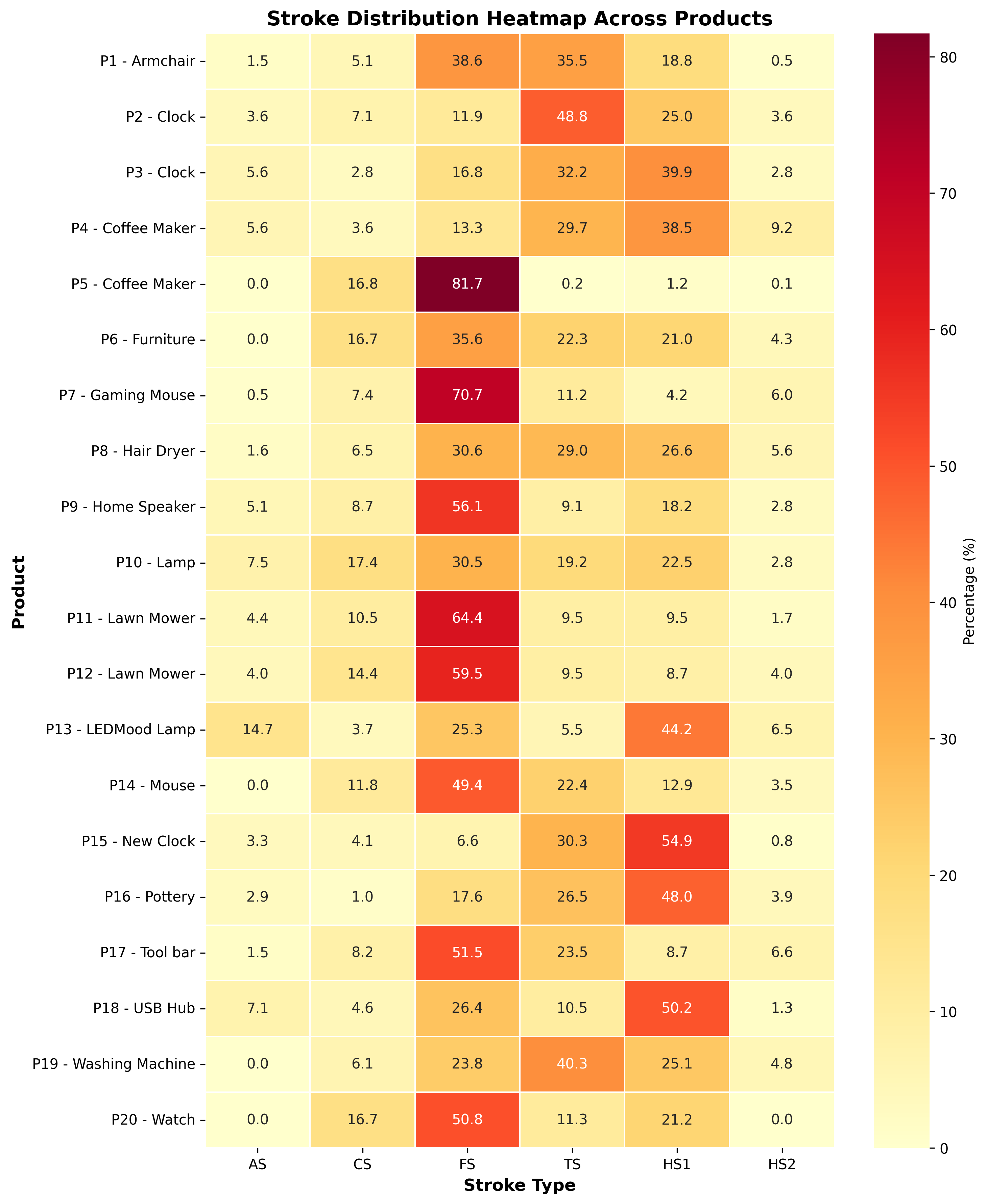}
        \caption{Stroke Distribution Heatmap Across Products}
        \label{subfig:heatmap_strokes}
    \end{subfigure}
    
    \caption{Aggregate analytical plots derived from the dataset. (a) Box plot showing the dominance of Defining and Shading strokes. (b) Stacked bar chart illustrating the variability in concept count (3 to 20) and time allocation. (c) Line plot depicting the trend of time expenditure across sequential concepts. (d) Heatmap visualizing the product-specific intensity of stroke usage.}
    \label{fig:aggregate_plots}
\end{figure*}

To derive generalizable insights beyond individual case studies, we performed an aggregate analysis across all 20 products. This involved generating composite visualizations to examine the global distribution of stroke types and the temporal allocation of effort across concepts. Figure~\ref{fig:aggregate_plots} presents a $2 \times 2$ matrix of these analytical plots.

\paragraph{Dominance of Form and Shading}
The Box and Whisker plot (Figure~\ref{subfig:box_whisker_strokes}) reveals a clear hierarchy in stroke usage. \textbf{Defining Strokes (FS)} are the most predominant, with a mean usage of approximately 38.1\% across all products. This aligns with the exploratory nature of conceptual design, where the search for form is paramount. Interestingly, \textbf{Shading Strokes (HS1)} occupy the second largest share ($\mu = 25.0\%$), followed closely by \textbf{Detailing Strokes (TS)} ($\mu = 21.3\%$). This indicates that designers spend nearly a quarter of their effort on rendering surface qualities to communicate 3D geometry. Conversely, Constraining (CS), Shadow (HS2), and Annotation (AS) strokes act as support elements, each constituting less than 10\% of the total stroke volume on average.

\paragraph{Variability in Concept Exploration}
The Stacked Bar Chart (Figure~\ref{subfig:stacked_bar_concepts}) highlights significant variability in the "width" of exploration. The number of concepts generated per session ranged from a minimum of 3 (e.g., P14, P19) to a maximum of 20 (P8 Hair Dryer). This variance suggests that the complexity of the product and the designer's individual process heavily influence the divergence factor.

\paragraph{Temporal Decay of Effort}
The Line Plot (Figure~\ref{subfig:line_plot_concepts}) demonstrates a temporal trend in effort allocation. The mean time spent (red line) is highest for the initial concepts (C1--C6), typically hovering between 15\% and 20\% per concept. As the concept count progresses beyond C10, the time spent per concept stabilizes or decreases, although the high standard deviation (green shaded region) indicates that some designers engage in "late-stage deep dives," spending significant time on later concepts (e.g., C11). This supports the notion that early phases are for broad exploration, while later phases may involve rapid iterations or specific refinements.

\paragraph{Product-Specific Signatures}
The Heatmap (Figure~\ref{subfig:heatmap_strokes}) provides a granular view of how stroke distribution shifts based on the product. For instance, the "Coffee Maker (P5)" and "Gaming Mouse (P7)" rely heavily on Defining Strokes ($>70\%$), suggesting a focus on silhouette and ergonomics. In contrast, products like "New Clock (P15)" and "USB Hub (P18)" show a dominance of Shading Strokes ($>50\%$), implying that surface finish and planar transitions were the primary design drivers. This variance validates the need for a flexible classification system that can adapt to different design intents.

\subsubsection{Observations on the "Action" of Sketching}
The analysis of the plots shown in Figure~\ref{fig:stroke_distribution_plot} and Figure~\ref{fig:aggregate_plots} revealed several consistent patterns in expert behaviour that directly informed the design of our sGIT system:
\begin{itemize}
    \item \textbf{Tool Simplicity:} In all observed sessions, designers utilized simple tools, predominantly ballpoint pens or fine-liners on paper. There was a notable absence of complex rendering markers or paints during the initial conceptualization.
    \item \textbf{The "No-Eraser" Paradigm:} Experts almost never used erasers. Mistakes or deviations were integrated into the design or simply ignored as the focus shifted to a new iteration. This supports our decision to de-prioritize "undo" in favour of "versioning."
    \item \textbf{Monochromatic Ideation:} The sketches were universally black and white. Colour was treated as a separate, later stage of rendering. This observation guided our decision to separate the sketching module (B\&W) from the rendering module which could be supported by an AI system.
    \item \textbf{Canvas Orientation:} Designers frequently rotated the canvas to facilitate comfortable hand mechanics for drawing specific curves.
    \item \textbf{Multiplicity and Branching:} Designers rarely drew a single concept. As evidenced by the concept counts in Figure~\ref{subfig:stacked_bar_concepts}, they drew multiple concepts (avg $\approx$ 6-8) on the same canvas or across pages, switching between them as needed. A typical session involved starting Concept A, pausing to sketch Concept B, and then returning to refine Concept A. This non-sequential workflow confirms that concept sketching is a non-linear tree structure, making a Git-based architecture theoretically the perfect fit.
\end{itemize}

The analysis provided a robust understanding of the \textit{action} aspect of sketching, the "what" and "how." However, it also highlighted a critical limitation: the \textit{cognition} or the thought process behind these actions was lost. For example, the video demonstrated that a designer transitioned from a round button to a square button, but it could not explain \textit{why}. This "Cognitive Gap" validated the need for the second pillar of our system: capturing design intent through simultaneous narration and AI summarization, which will be detailed in the following section.

In summary, the theoretical framework establishes PCS as a complex, multi-layered activity. It is informationally rich yet vague, action-heavy yet cognitively driven, and linear in time but non-linear in logic. These characteristics dictate that a Version Control System for design cannot simply copy the text-based logic of software engineering; it must be re-architected to accommodate the visual, distinct, and interrupted nature of the
sketching process.

%%%%%%%%%%%%%%%%%%%%%%%%%%%%%%%%%%%%%%%%%%%%%%%%%%%%%%%%%%%%%%%%%%%%%%
\section{Automated Stroke Classification using Hybrid AI Approaches}
\label{sec:stroke_classification}

The taxonomy of sketching strokes, as delineated in the preceding section, provides a necessary semantic framework for understanding design intent. However, for a computational system like DIMES to effectively assist a designer, this semantic understanding must be transitioned from a manual theoretical construct to an automated computational process. The machine must possess the capability to transition from merely recording pixel data to interpreting the cognitive intent behind each stroke. It must answer the fundamental question: \textit{Is this specific curve intended to define a boundary, or is it detailing an edge?}

To achieve this, we developed a comprehensive classification pipeline employing both Deep Learning (DL) and classical Machine Learning (ML) techniques. This section details the development of the data collection infrastructure, the creation of a bespoke dataset, and the comparative analysis of multiple AI architectures designed to classify strokes into four primary functional categories: Constraining, Defining, Detailing, and Hatches (Shading).

\subsection{The Data Scarcity Challenge and the AEGIS Platform}
\label{subsec:aegis_platform}

In the realm of Computer Vision (CV), the efficacy of a model is inextricably linked to the quality and volume of its training data. While massive datasets exist for general object detection (e.g., ImageNet~\cite{deng2009imagenet}, COCO~\cite{lin2014coco}) or handwriting recognition (e.g., MNIST~\cite{lecun1998gradient}), the domain of Product Concept Sketching (PCS) suffers from a severe paucity of labelled data. 

A single concept sketch drawn by a professional designer is a complex aggregate of tens of thousands of individual strokes. These strokes are often overlapping, intersecting, and varying in opacity. Asking human designers to manually segment and label each stroke within a completed sketch is a Herculean task, prone to fatigue and inconsistency. Furthermore, existing sketch datasets, such as the Google QuickDraw dataset, focus on object categories (e.g., "cat", "bicycle") rather than the \textit{type of stroke} used to construct them. Consequently, to the best of our knowledge, there exists no pre-existing dataset that classifies strokes according to the cognitive-action taxonomy required for this research.

\subsubsection{AEGIS: AI Enhanced Gathering and Interpretation of Strokes}
To address this logistical bottleneck, we developed a dedicated web-based application named \textbf{AEGIS} (AI Enhanced Gathering and Interpretation of Strokes). The application was engineered using ReactJS and deployed on the Vercel platform to ensure accessibility and low-latency performance.

AEGIS was designed not merely as a drawing tool, but as a scientific instrument for data acquisition. It features granular controls over stroke parameters, allowing us to mimic the diverse range of tools used in physical sketching. The interface provides configurable sliders for:
\begin{itemize}
    \item \textbf{Thickness:} Adjustable from $0.5$ to $20.0$ pixels, allowing for fine lines and broad markers.
    \item \textbf{Pressure Sensitivity:} Configurable variance from $-1$ to $1$, simulating the response of digital styluses.
    \item \textbf{Smoothing and Streamline:} Controlled from $0$ to $1$, enabling the stabilisation of hand tremors which is crucial for distinguishing between rapid (low smoothing) and deliberate (high smoothing) strokes.
    \item \textbf{Colour and Opacity:} Grayscale values ranging from $0.0$ (completely white) to $1.0$ (black), along with variable opacity.
\end{itemize}

By configuring these parameters, the user can accurately simulate the four distinct types of strokes identified for our classification task:
\begin{enumerate}
    \item \textbf{Constraining Strokes:} Characterised by smooth, light, straight lines used to define boundaries.
    \item \textbf{Defining Strokes:} Light strokes with curvature, used to provide the initial shape.
    \item \textbf{Detailing Strokes:} Darker, more emphatic strokes used for completion and emphasis.
    \item \textbf{Hatches and Shadows:} Continuous strokes created by rapid up-and-down or left-to-right hand movements, constituting the shading component of the sketch.
\end{enumerate}

The interface allows designers to draw these strokes in isolation or in sequence, providing options to clear the canvas and a counter to track the volume of data generated.

% TODO: Insert Figure 4.1: The User Interface of the AEGIS application showing the canvas, parameter sliders (Thickness, Pressure, Smoothing), and the stroke category selection buttons.

\subsubsection{Data Collection and Serialization Protocol}
The core innovation of AEGIS lies in its dual-modal recording capability. As the designer sketches, the system performs two simultaneous save operations to capture both the raster (image) and vector (metadata) information.

\paragraph{JSON Metadata Logging}
For every stroke created, AEGIS automatically generates a JSON object capturing high-fidelity temporal and spatial data. This includes the User ID, the ground-truth category (selected by the user prior to drawing), and a timestamp. Crucially, it records arrays for $X$ and $Y$ coordinates, pressure values, and thickness values throughout the entire trajectory of the stroke. It also logs the specific brush parameters (thinning, smoothing, streamline) active at the moment of creation.

\paragraph{Image Normalization}
Simultaneously, the visual representation of the stroke is captured. To ensure consistency for the Deep Learning models, a down-scaling algorithm was implemented. Before saving the screenshot of the stroke, the system ensures that the larger dimension (length or width) is resized to 256 pixels, with the other dimension scaled proportionately to maintain the aspect ratio. The filename of this processed image is linked within the JSON file, ensuring a perfect mapping between the visual artifact and its mathematical description.

This rigorous recording process was repeated extensively to validate that the interface faithfully records all stroke properties and that the stored JSON information is sufficient to mathematically reconstruct the stroke if necessary. An exemplar of the captured JSON structure is provided below:
{
\small
\begin{verbatim}
[
  {
   "_id": "693703646ba3764e63fbe10b",
   "username": "User",
   "category": "constraining",
   "stroke_number": 1,
   "timestamp": "2025-12-08T16:57:08.290000",
   "x_coordinates": [328, 329, 330, ..., 1334],
   "y_coordinates": [765, 764, 763, ..., 226],
   "pressure_values": [0.2, 0.2, 0.2, ..., 0.2],
   "thickness_values": [1, 1, 1, ..., 1],
   "color": "#333333",
   "grayscale_value": 0.8,
   "opacity": 0.8,
   "stroke_parameters": {
     "size": 1,
     "thinning": 0,
     "smoothing": 0.5,
     "streamline": 0.5,
     "simulatePressure": false
    },
   "path": "M 327.65 764.65 L 328.80 763.49 ...",
   "image_filename": "Constraining_Strokes_1.png"
  }
]
\end{verbatim}
}

% TODO: Insert Figure 4.2: Visualisation of the Data Acquisition Pipeline showing the User Input -> AEGIS Interface -> Splitting into JSON Logs and 256x256 PNG Images.

\subsection{Deep Learning (DL) Approach}
\label{subsec:dl_approach}

To enable the automated recognition of stroke types based on their visual morphology, we implemented a Deep Learning pipeline. This approach treats stroke classification as a supervised image classification problem, leveraging Convolutional Neural Networks (CNNs) to extract hierarchical features from the rasterized stroke images captured by the AEGIS platform.

\subsubsection{Dataset Expansion and Augmentation}
Following the validation of the AEGIS recording protocol, we initiated a rigorous data collection phase. A cohort of 10 master's level students in product design were recruited to generate the seed dataset. They were instructed to draw strokes corresponding to the four target classes—Constraining, Defining, Detailing, and Hatches—using the platform. This resulted in the acquisition of 500 distinct, human-authored strokes per class, culminating in a total seed dataset of 2,000 images.

While this dataset provided high-quality, domain-specific examples, it was insufficient in volume to train deep neural networks without significant risk of overfitting. To address this, we employed data augmentation techniques to synthetically expand the dataset. All raw images were first resized to a standard resolution of $256 \times 256$ pixels. Subsequently, a series of affine transformations were applied, including random rotation, horizontal flipping, vertical flipping, translation, and scaling. This process mimic's the natural variability in hand-drawn strokes (e.g., changes in canvas orientation or stroke scale). 

The augmentation process resulted in a ten-fold expansion, yielding 5,000 images per class and a total dataset of \textbf{20,000 images}. This comprehensive dataset was partitioned using a stratified split ratio of 65:20:15 for Training (13,000 images), Validation (4,000 images), and Testing (3,000 images), respectively.

\subsubsection{Model Architectures and Training Strategy}
To identify the optimal architecture for this specific task, we conducted a comparative study of five distinct CNN architectures, ranging from classical networks to modern efficient variants: \textbf{ResNet50}, \textbf{ShuffleNet}, \textbf{EfficientNetB0}, \textbf{AlexNet}, and \textbf{ConvNeXt-Tiny}.

All models were fine-tuned using a consistent set of hyperparameters to ensure a fair comparison:
\begin{itemize}
    \item \textbf{Input Dimension:} $256 \times 256$ pixels.
    \item \textbf{Batch Size:} 32.
    \item \textbf{Optimization:} Adam optimizer with an initial Learning Rate of $0.0001$.
    \item \textbf{Training Duration:} 50 Epochs, with an Early Stopping mechanism (Patience = 15 epochs) monitoring validation loss to prevent overfitting.
    \item \textbf{Hardware:} All training and inferencing were executed on an NVIDIA GeForce RTX 3050 GPU.
\end{itemize}

We adopted a two-phase Transfer Learning strategy. In the first phase, the distinct pre-trained backbones (weights frozen) were connected to a custom classification head, and only this final layer was trained. In the second phase, all layers were unfrozen to allow fine-tuning of the feature extraction layers to the specific domain of sketching strokes. During training, metrics such as accuracy and loss for both training and validation sets were tracked, and the model state yielding the highest validation accuracy was saved as the optimal weight file (\texttt{.pth}).

\subsubsection{Performance Evaluation and Results}
The models were evaluated on the held-out Test set (15\% of the dataset). As shown in Table \ref{tab:dl_test_results}, all five architectures achieved high accuracy, demonstrating that the synthetic augmentation successfully captured the variance required for classification within the dataset's distribution.

\begin{table}[ht!]
\centering
\caption{Deep Learning Model Performance on Test Dataset (Synthetic Split)}
\label{tab:dl_test_results}
\begin{tabular}{|l|c|}
\hline
\textbf{Model Architecture} & \textbf{Test Accuracy (\%)} \\ \hline
AlexNet & 96.54 \\ \hline
ConvNeXt-Tiny & \textbf{97.64} \\ \hline
EfficientNetB0 & 96.40 \\ \hline
ResNet50 & 96.55 \\ \hline
ShuffleNet & 95.80 \\ \hline
\end{tabular}
\end{table}

\begin{figure*}[ht!]
    \centering
    \begin{subfigure}[b]{0.32\textwidth}
        \centering
        \includegraphics[width=\linewidth]{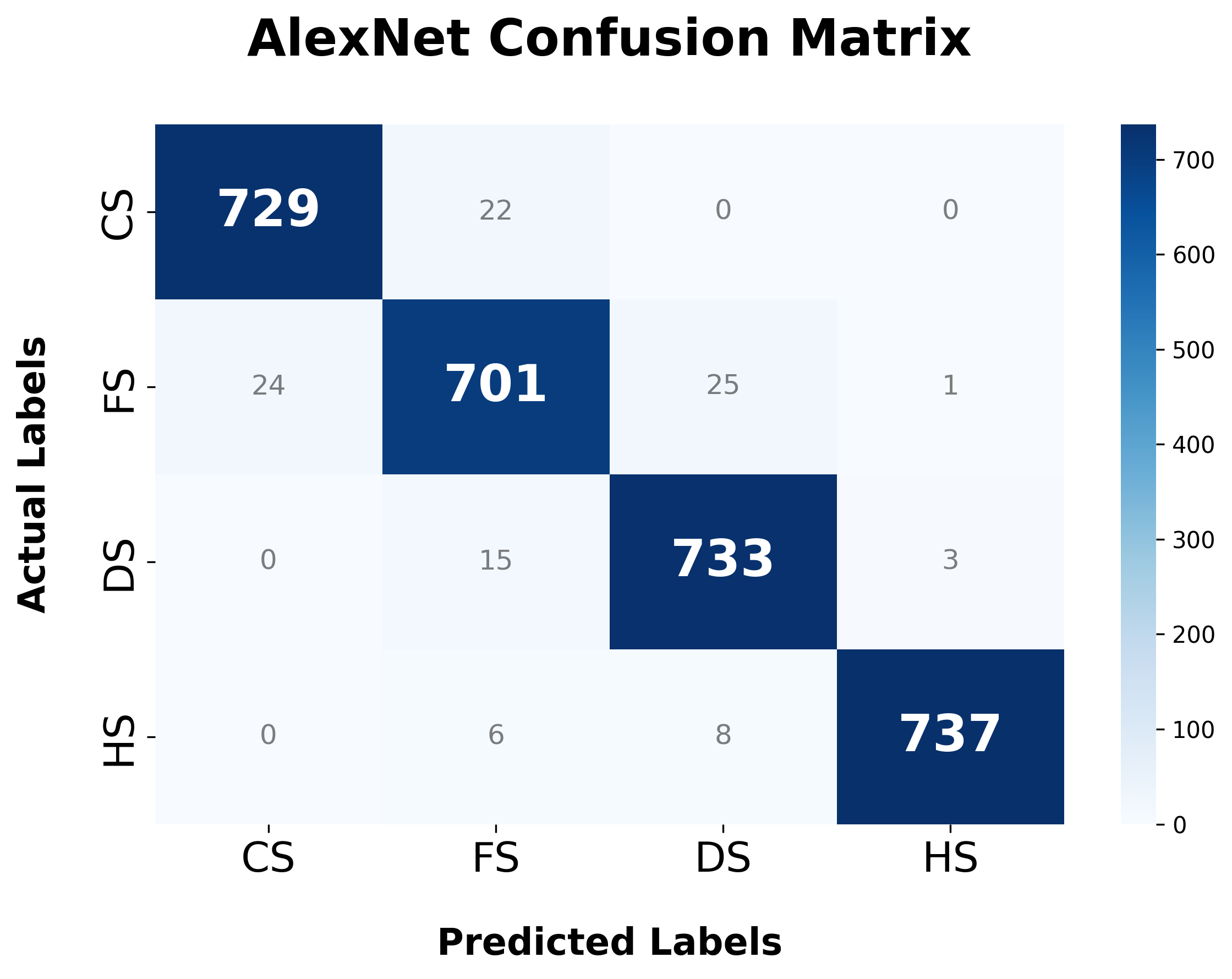}
        \caption{AlexNet}
    \end{subfigure}
    \hfill
    \begin{subfigure}[b]{0.32\textwidth}
        \centering
        \includegraphics[width=\linewidth]{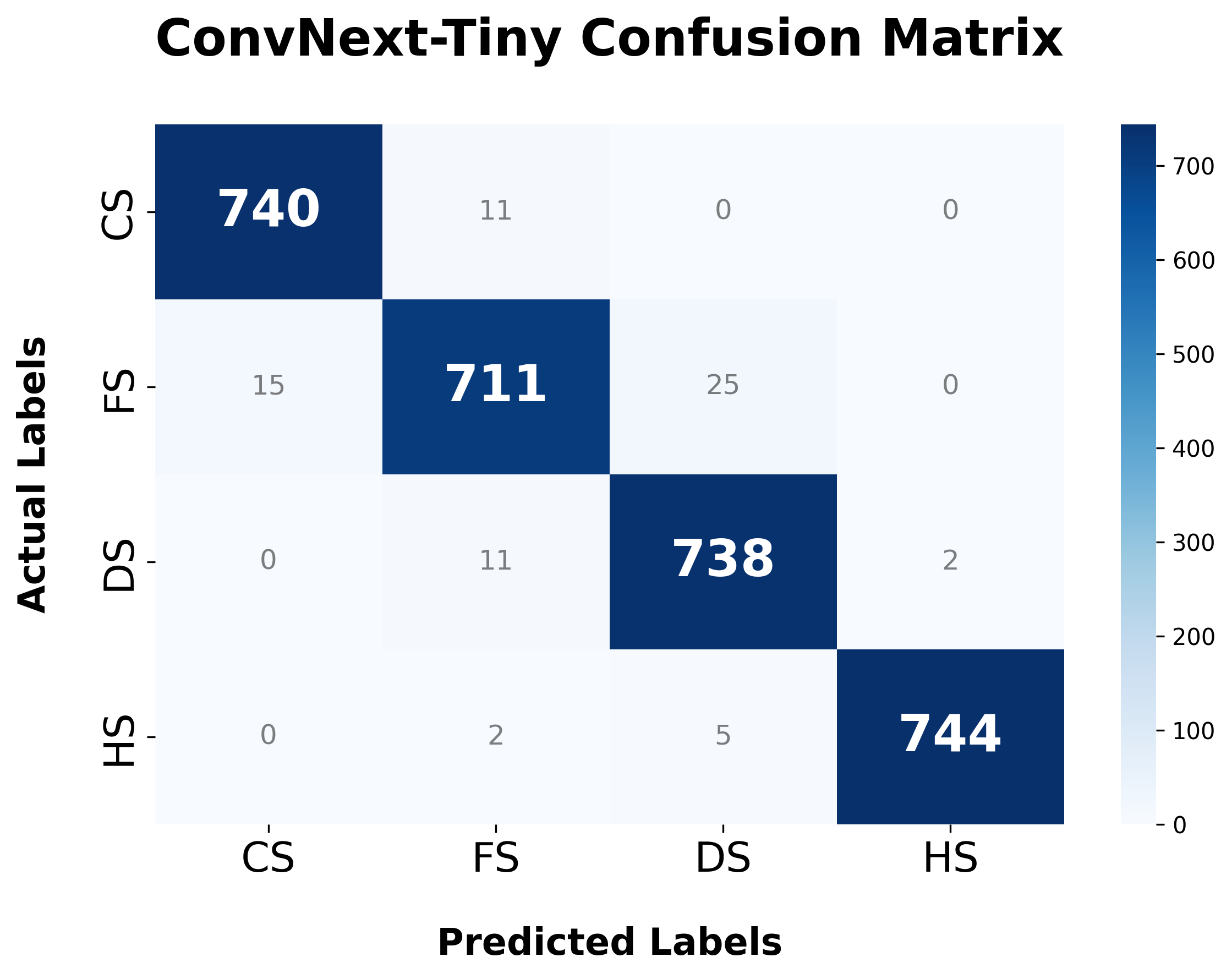}
        \caption{ConvNeXt}
    \end{subfigure}
    \hfill
    \begin{subfigure}[b]{0.32\textwidth}
        \centering
        \includegraphics[width=\linewidth]{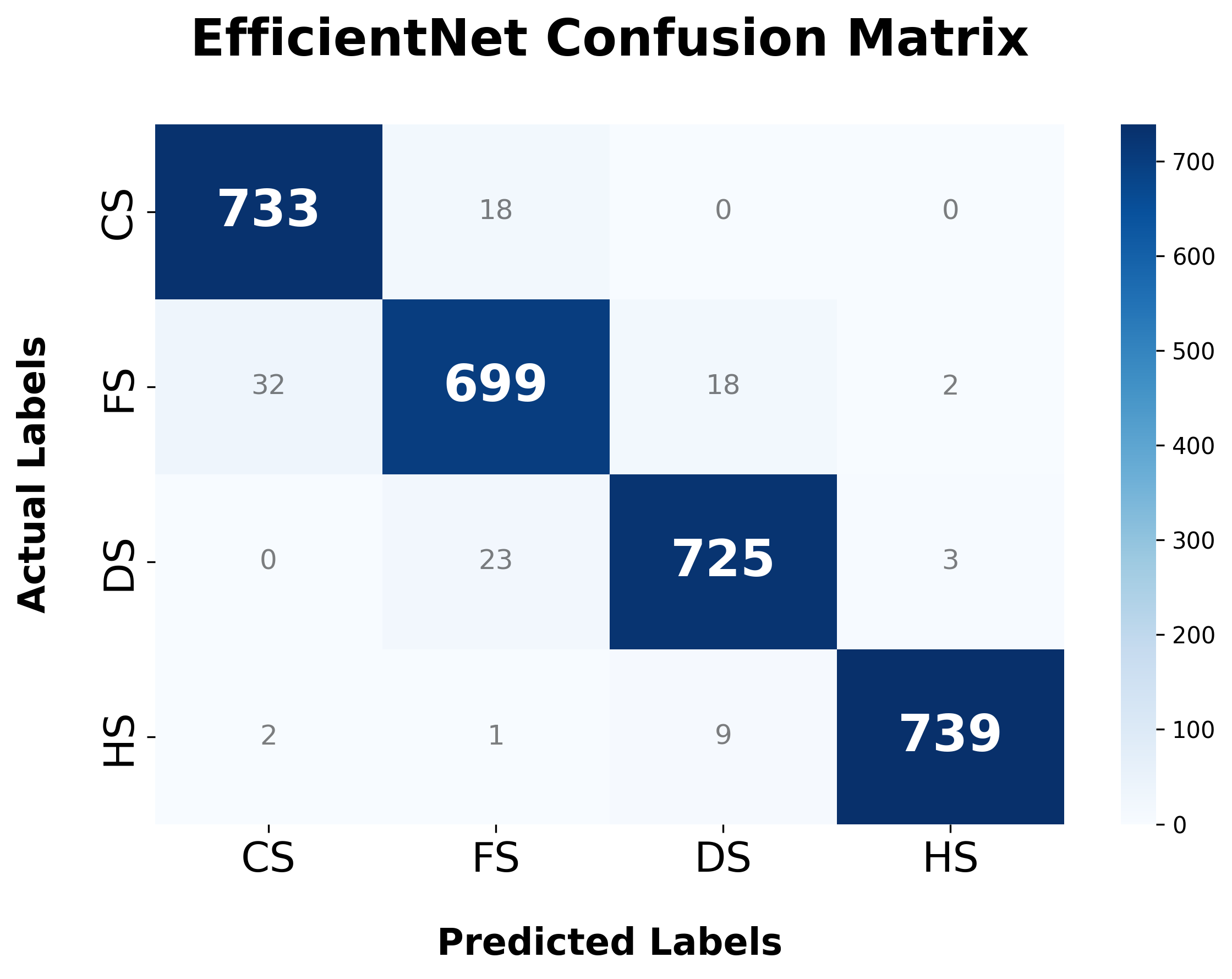}
        \caption{EfficientNet}
    \end{subfigure}
    \vfill
    \begin{subfigure}[b]{0.32\textwidth}
        \centering
        \includegraphics[width=\linewidth]{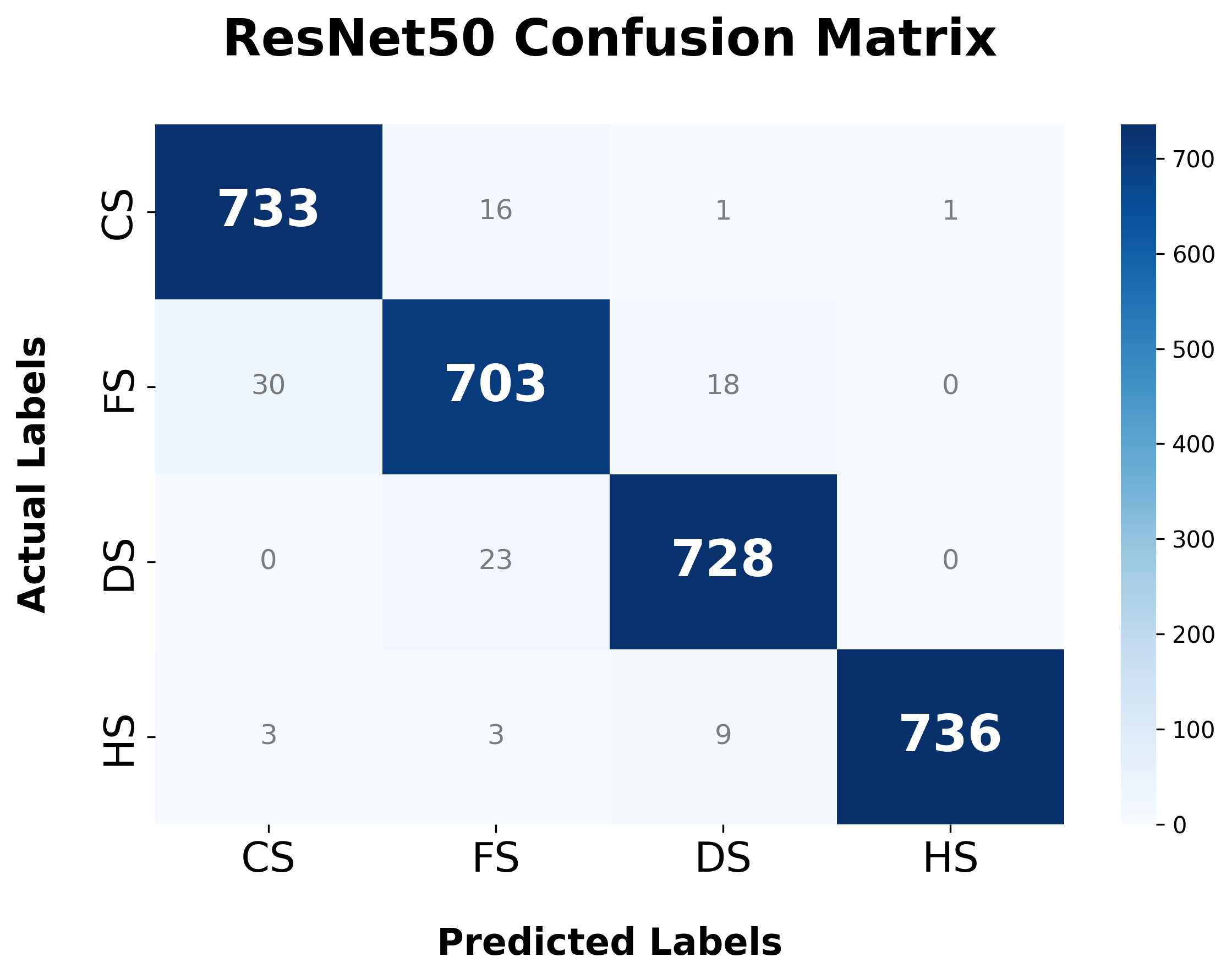}
        \caption{ResNet50}
    \end{subfigure}
    \hfill
    \begin{subfigure}[b]{0.32\textwidth}
        \centering
        \includegraphics[width=\linewidth]{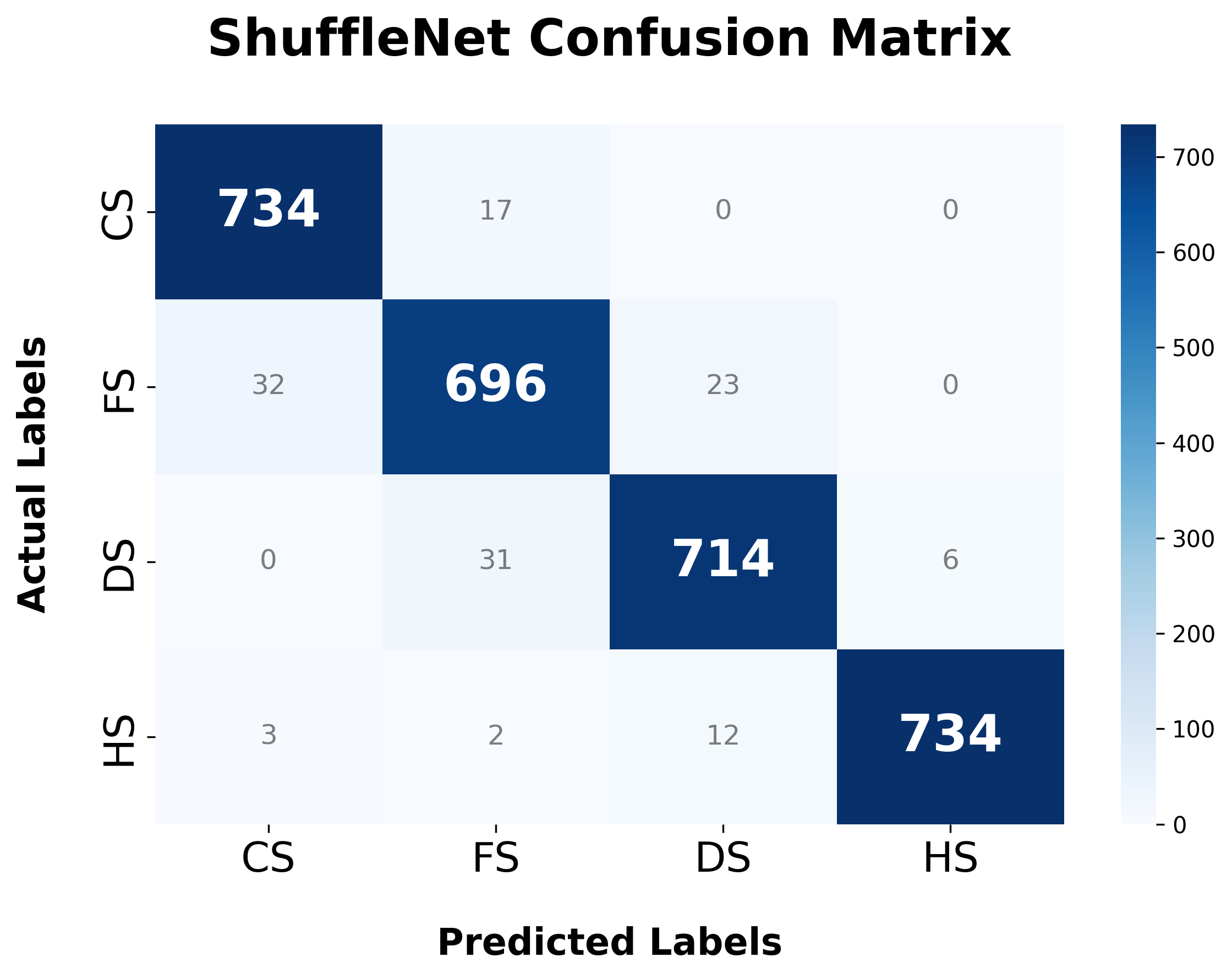}
        \caption{ShuffleNet}
    \end{subfigure}
    \caption{Confusion Matrices of the five DL models on the Test Dataset during Training.}
    \label{fig:confusion_matrices_test}
\end{figure*}

\subsubsection{The "Sim-to-Real" Gap: Field Test Verification}
High accuracy on a statistically similar test set does not always translate to real-world robustness. To verify the generalization capability of the models, we conducted a "Field Test" (Sanity Check). A subset of 5 designers from the original cohort was tasked with drawing fresh strokes—10 per category, totaling 40 strokes per designer—on the fly. These strokes were not part of the training or augmentation pipeline.

The results of this Field Test (Table \ref{tab:dl_field_results}) revealed a divergence in performance, highlighting the robustness of deeper residual networks over lightweight or older architectures.

\begin{table}[ht!]
\centering
\caption{Deep Learning Model Performance on Field Test (Real-time Sanity Check)}
\label{tab:dl_field_results}
% tabularx fills the exact linewidth. 
% >{\raggedright\arraybackslash}X aligns the first column to the left cleanly.
\begin{tabularx}{\linewidth}{@{}|>{\raggedright\arraybackslash}X|c|@{}}
\hline
\textbf{Model Architecture} & \textbf{Field Test Accuracy (\%)} \\ \hline
AlexNet & 70.0 \\ \hline
ConvNeXt-Tiny & 70.0 \\ \hline
EfficientNetB0 & 77.5 \\ \hline
\textbf{ResNet50} & \textbf{82.5} \\ \hline
ShuffleNet & 62.5 \\ \hline
Ensemble (Majority Voting) & 75.0 \\ \hline
\end{tabularx}
\end{table}

While architectures like ConvNeXt provided superior results on the static test set (97.64\%), their performance dropped to 70\% in the field test. Notably, \textbf{ResNet50} emerged as the most robust model, retaining an accuracy of \textbf{82.5\%}. ShuffleNet, optimized for efficiency, struggled significantly with real-world variance (62.5\%). An ensemble model, utilizing majority voting across all five networks, achieved 75\%, failing to outperform the solitary ResNet50.

\begin{figure*}[ht!]
    \centering
    \begin{subfigure}[b]{0.32\textwidth}
        \centering
        \includegraphics[width=\linewidth]{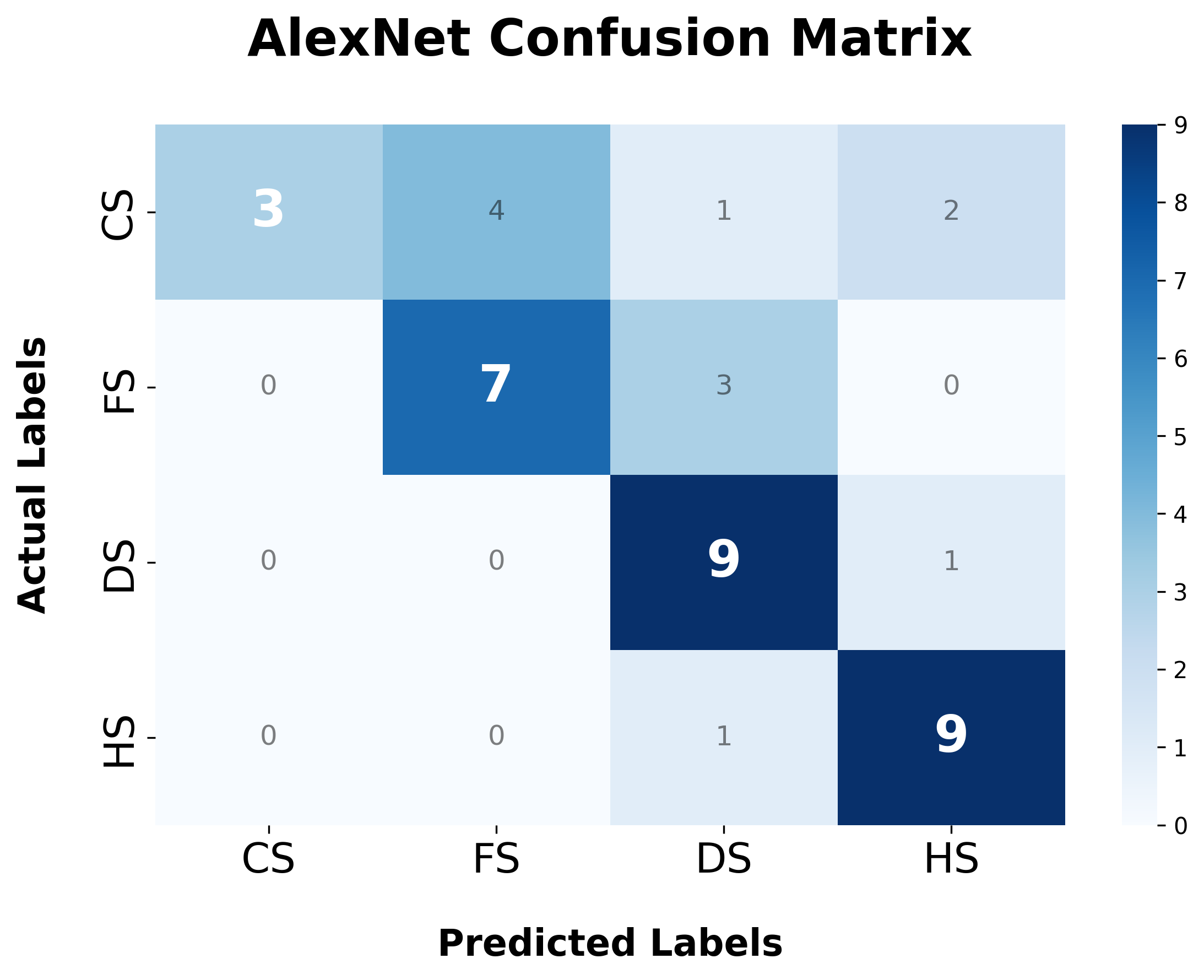}
        \caption{AlexNet}
    \end{subfigure}
    \hfill
    \begin{subfigure}[b]{0.32\textwidth}
        \centering
        \includegraphics[width=\linewidth]{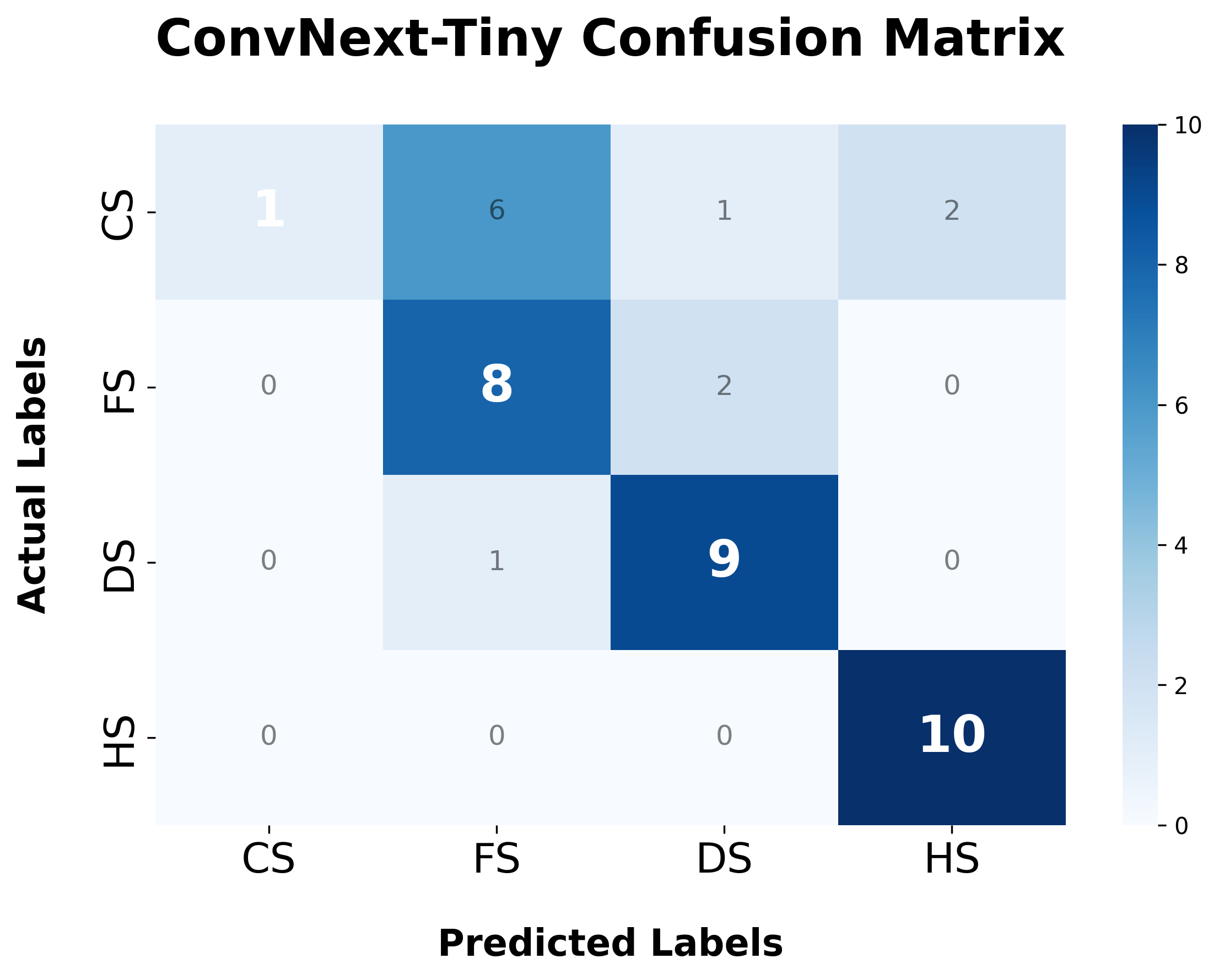}
        \caption{ConvNeXt}
    \end{subfigure}
    \hfill
    \begin{subfigure}[b]{0.32\textwidth}
        \centering
        \includegraphics[width=\linewidth]{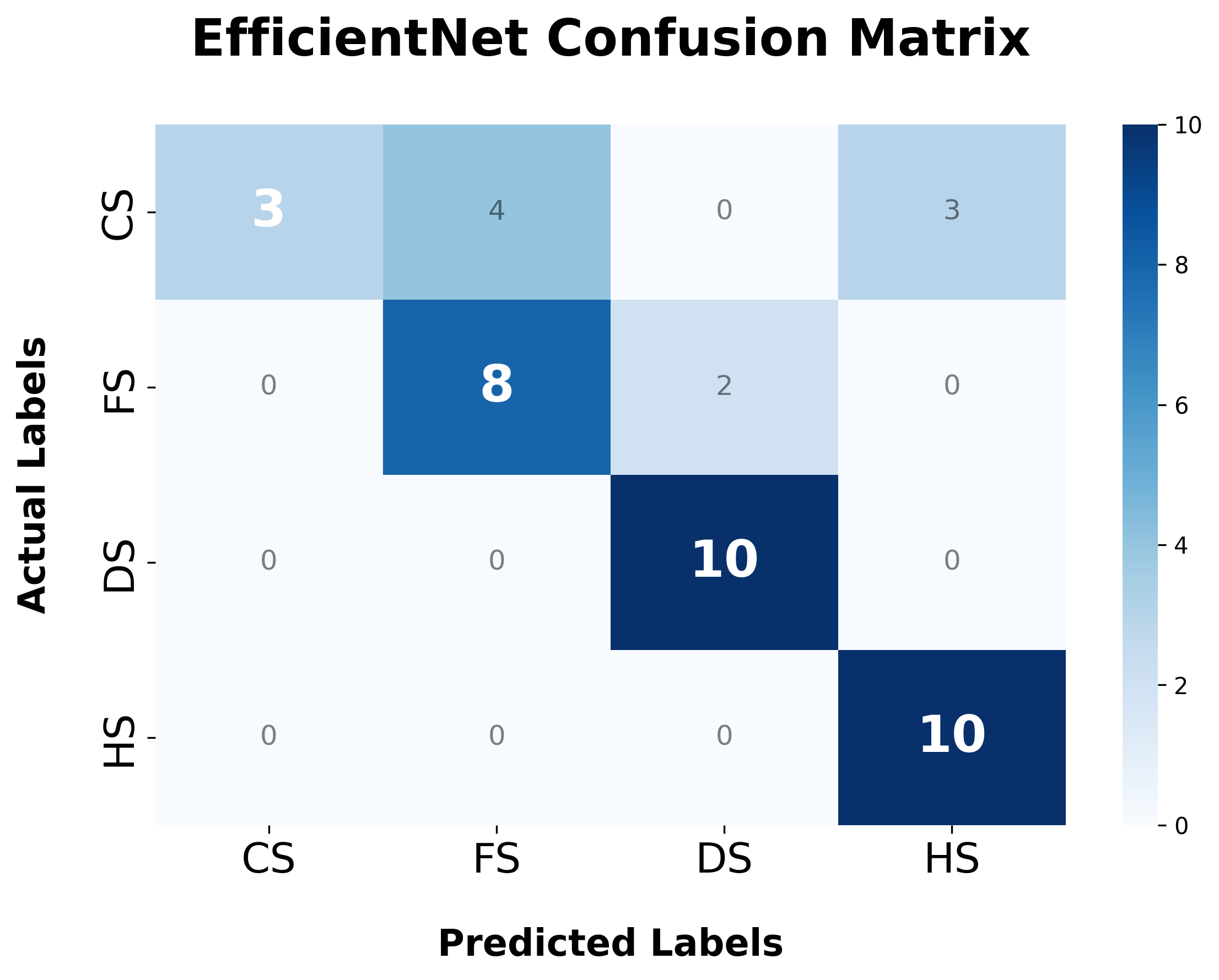}
        \caption{EfficientNet}
    \end{subfigure}
    \vfill
    \begin{subfigure}[b]{0.32\textwidth}
        \centering
        \includegraphics[width=\linewidth]{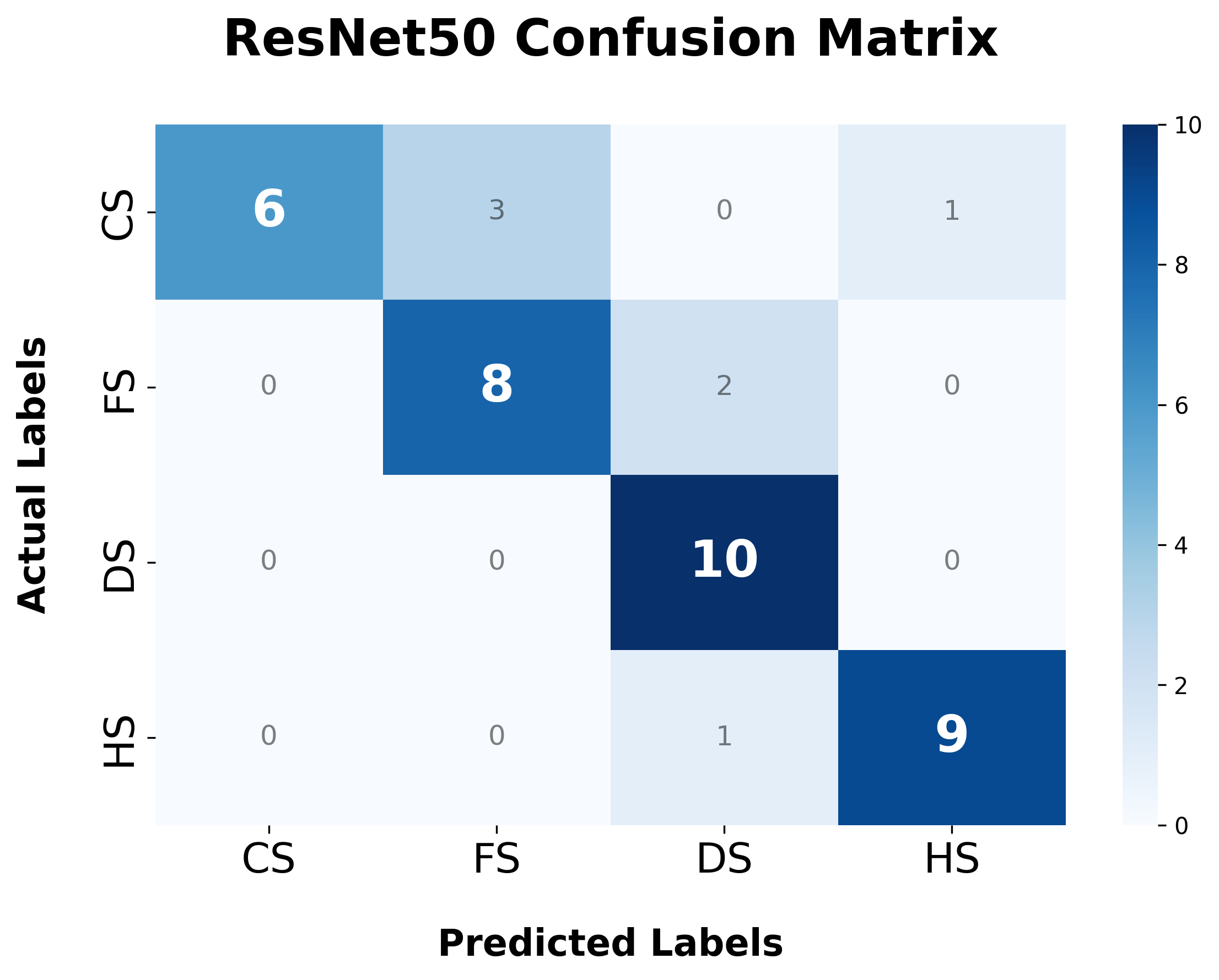}
        \caption{ResNet50}
    \end{subfigure}
    \hfill
    \begin{subfigure}[b]{0.32\textwidth}
        \centering
        \includegraphics[width=\linewidth]{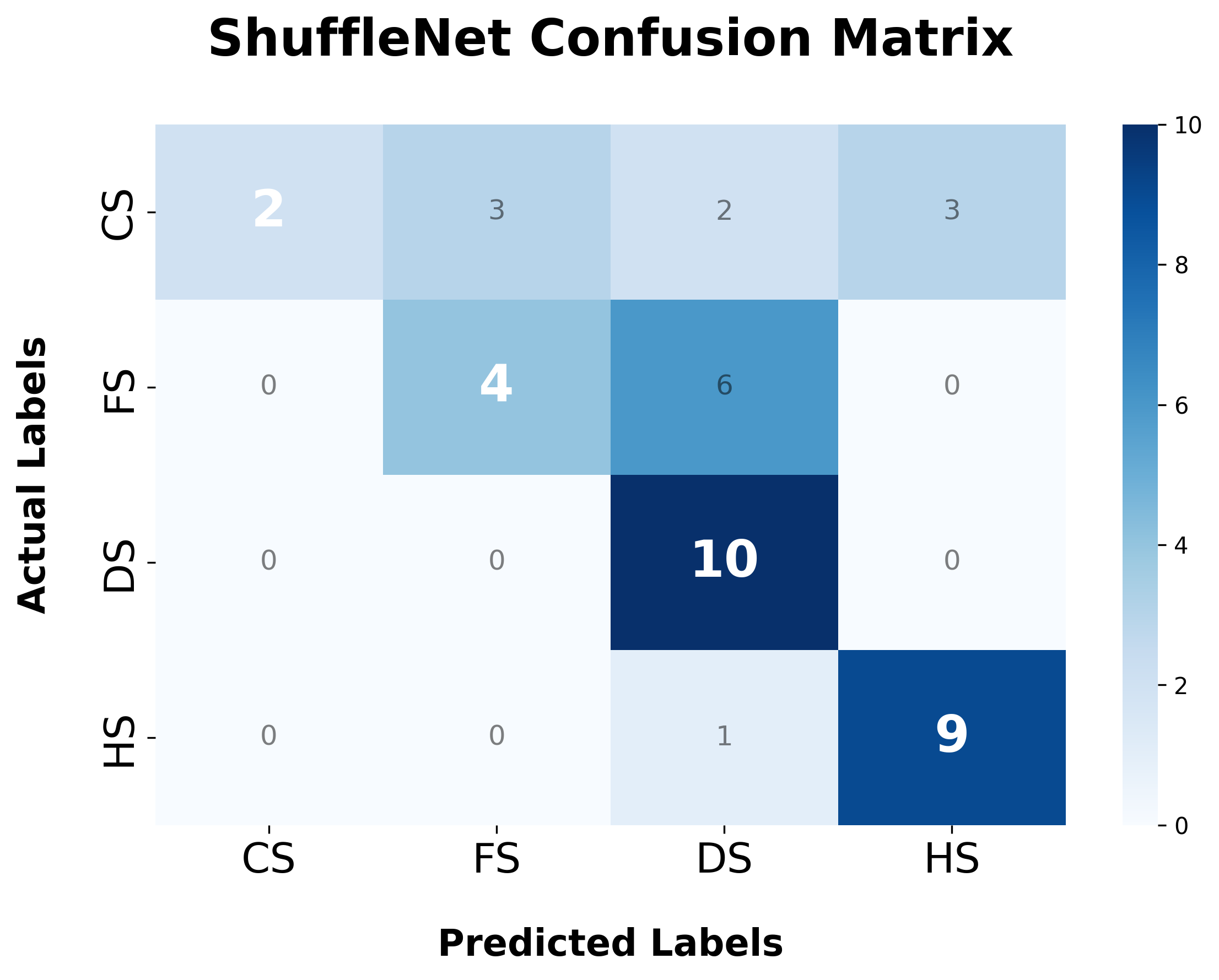}
        \caption{ShuffleNet}
    \end{subfigure}
    \hfill
    \begin{subfigure}[b]{0.32\textwidth}
        \centering
        \includegraphics[width=\linewidth]{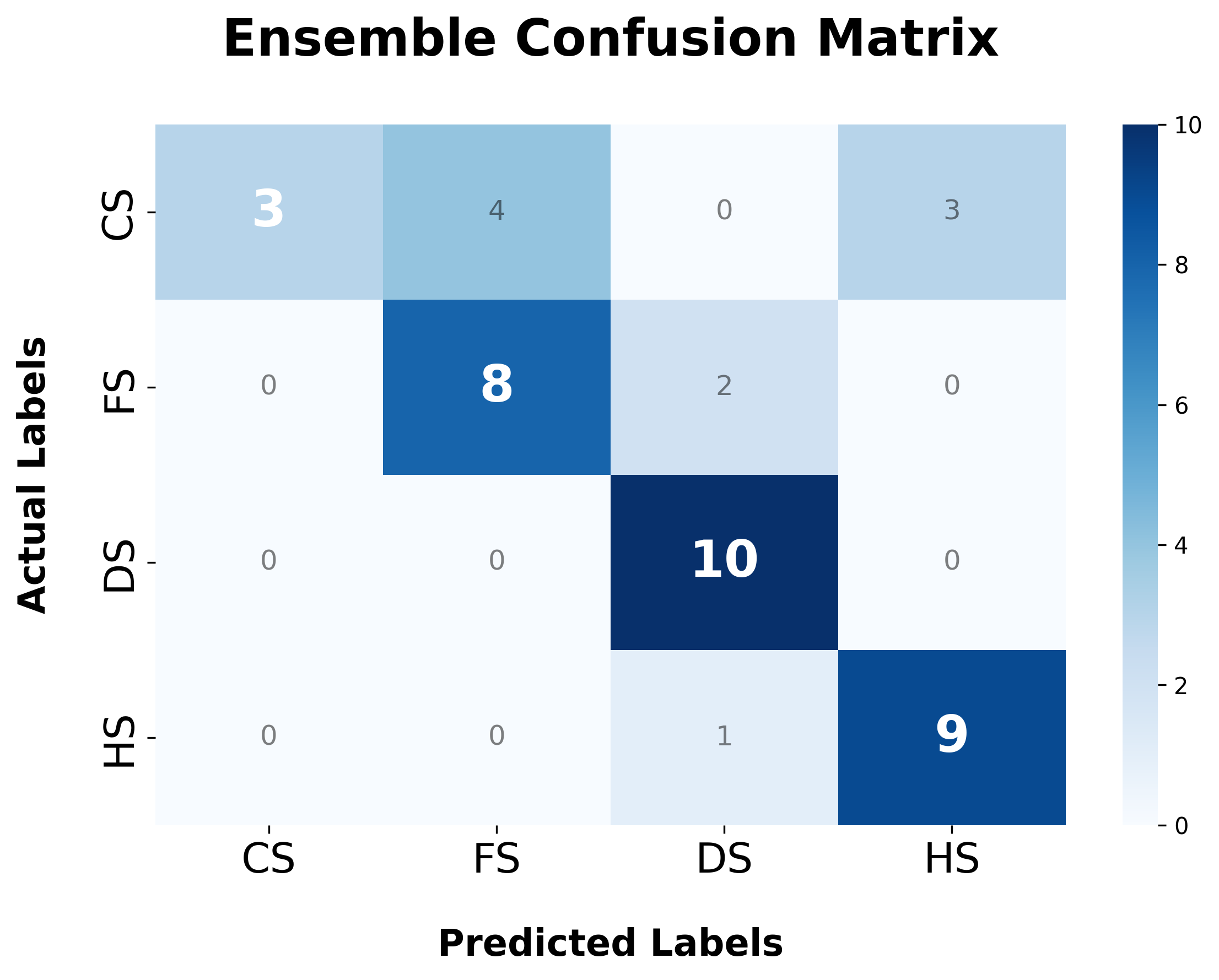}
        \caption{Ensemble}
    \end{subfigure}
    \caption{Confusion Matrices of the five DL models plus the Ensemble on the Field Test Data.}
    \label{fig:confusion_matrices_field}
\end{figure*}

\subsubsection{Latent Space Analysis}
To investigate why ResNet50 outperformed others in the field test despite lower static test accuracy, we analyzed the latent space representations. We extracted the feature vectors from the penultimate layer (pre-Softmax) of each model for the field test data. These high-dimensional vectors were projected into a 2D space using Uniform Manifold Approximation and Projection (UMAP) and clustered based on ground truth labels.

\begin{figure*}[ht!]
    \centering
    \begin{subfigure}[b]{0.32\textwidth}
        \centering
        \includegraphics[width=\linewidth]{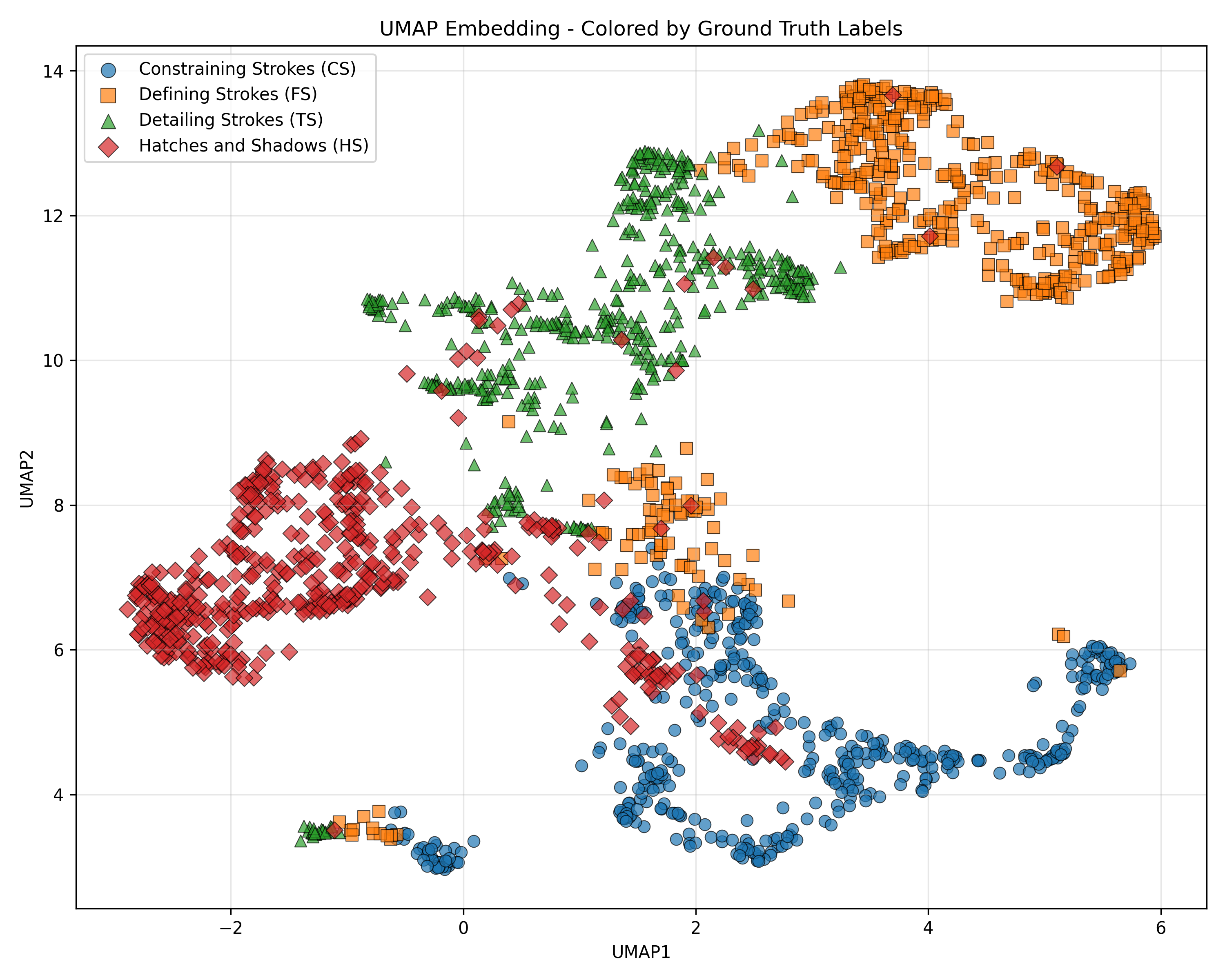}
        \caption{AlexNet}
    \end{subfigure}
    \hfill
    \begin{subfigure}[b]{0.32\textwidth}
        \centering
        \includegraphics[width=\linewidth]{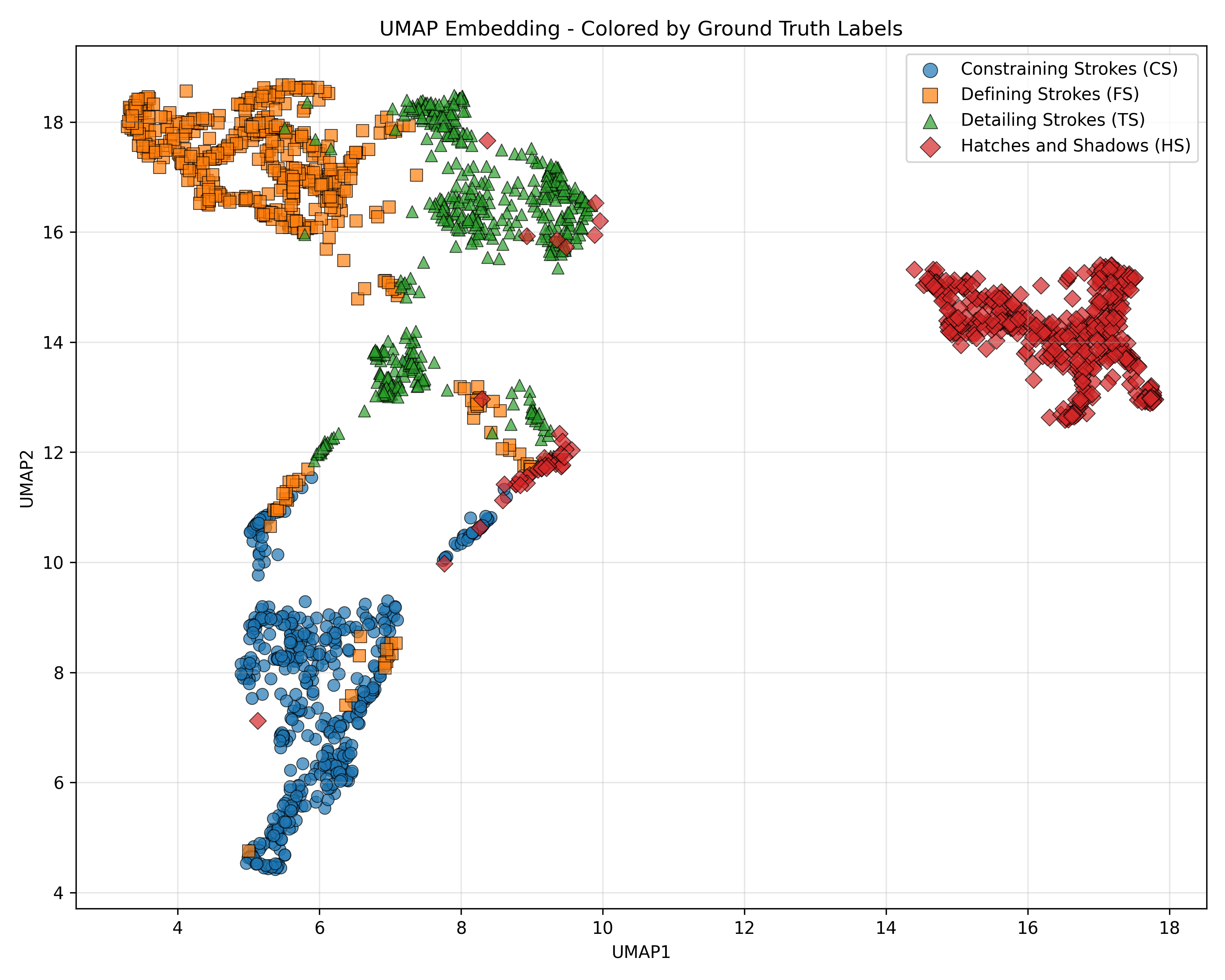}
        \caption{ConvNeXt}
    \end{subfigure}
    \hfill
    \begin{subfigure}[b]{0.32\textwidth}
        \centering
        \includegraphics[width=\linewidth]{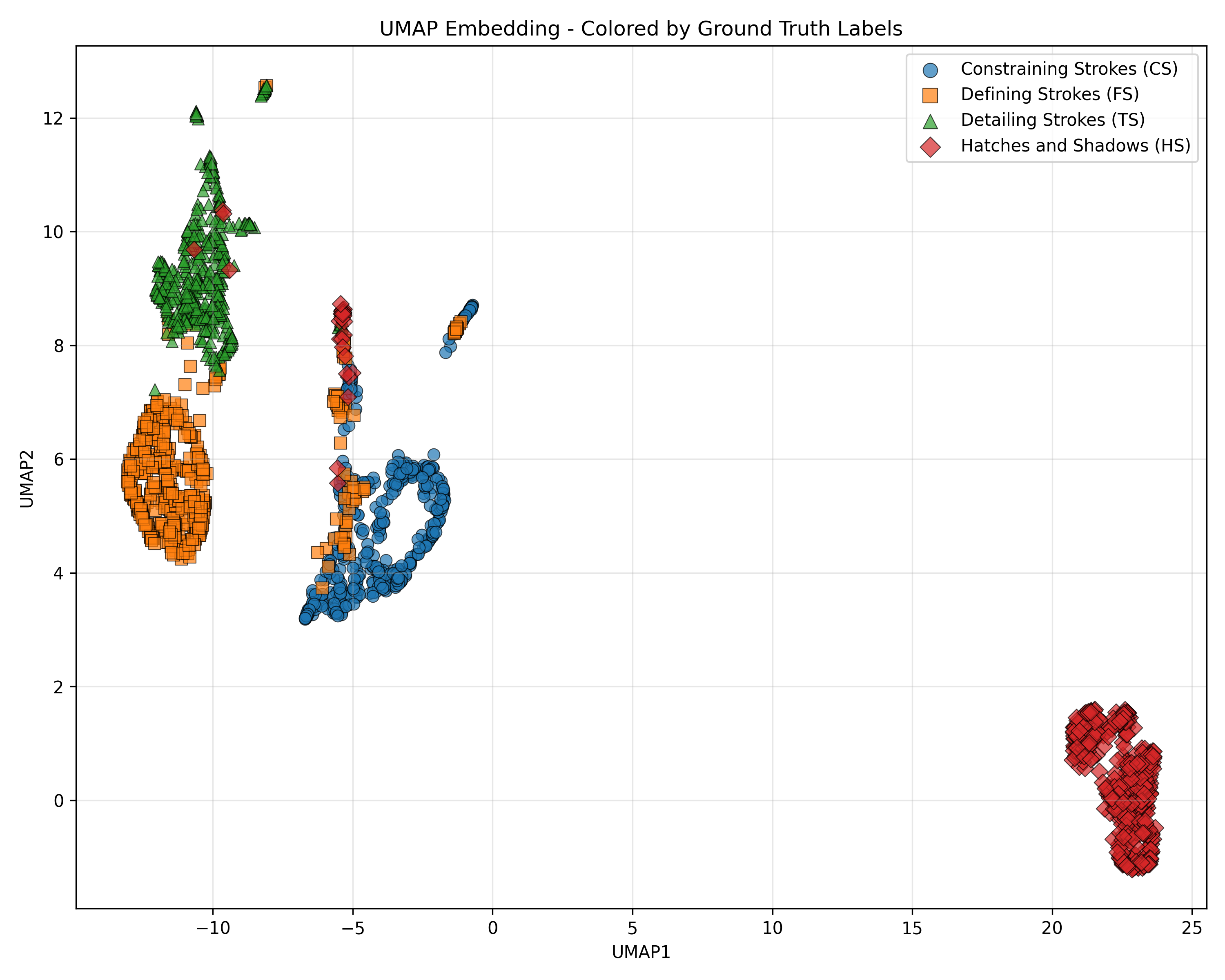}
        \caption{EfficientNet}
    \end{subfigure}
    \vfill
    \begin{subfigure}[b]{0.32\textwidth}
        \centering
        \includegraphics[width=\linewidth]{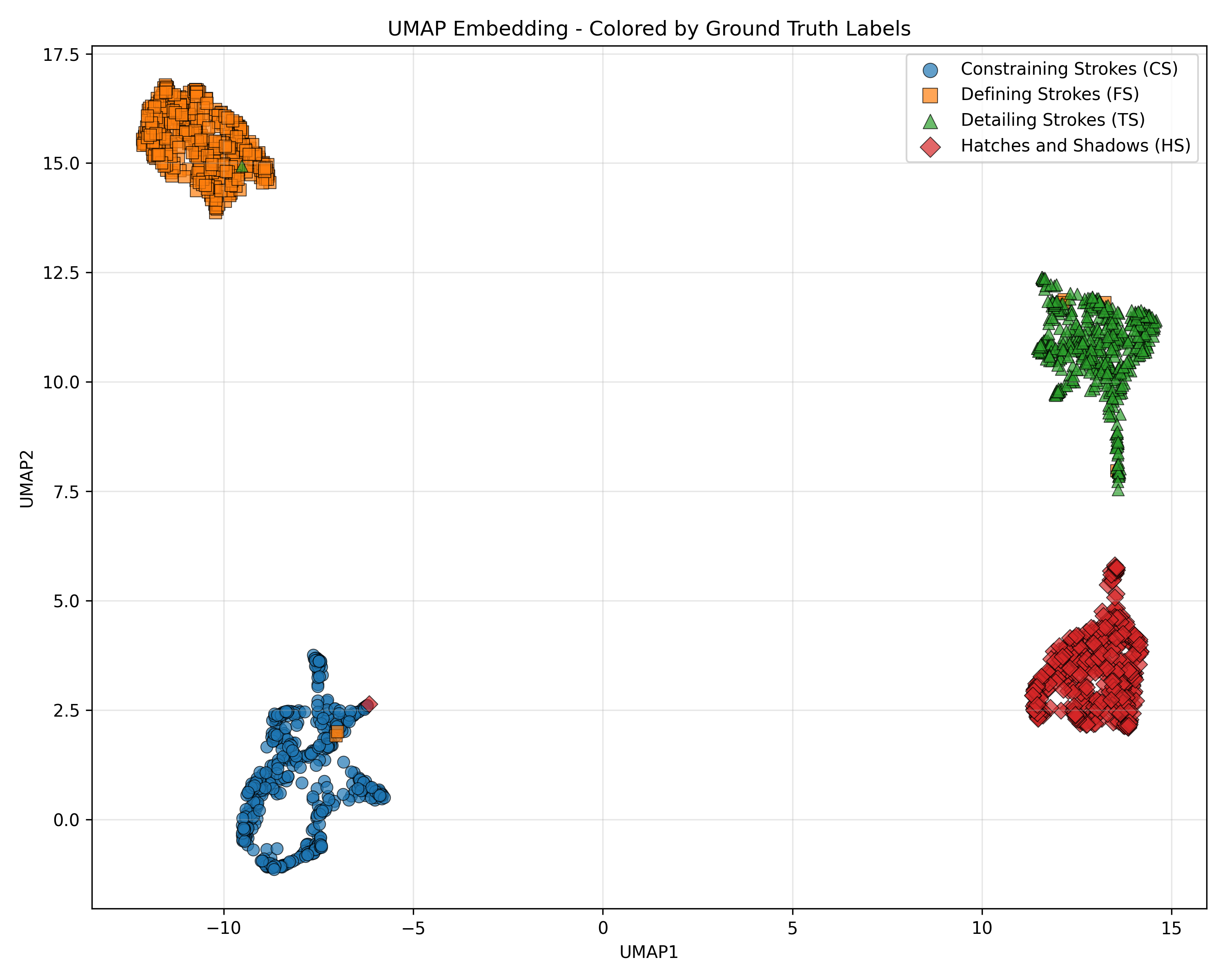}
        \caption{ResNet50}
    \end{subfigure}
    \hfill
    \begin{subfigure}[b]{0.32\textwidth}
        \centering
        \includegraphics[width=\linewidth]{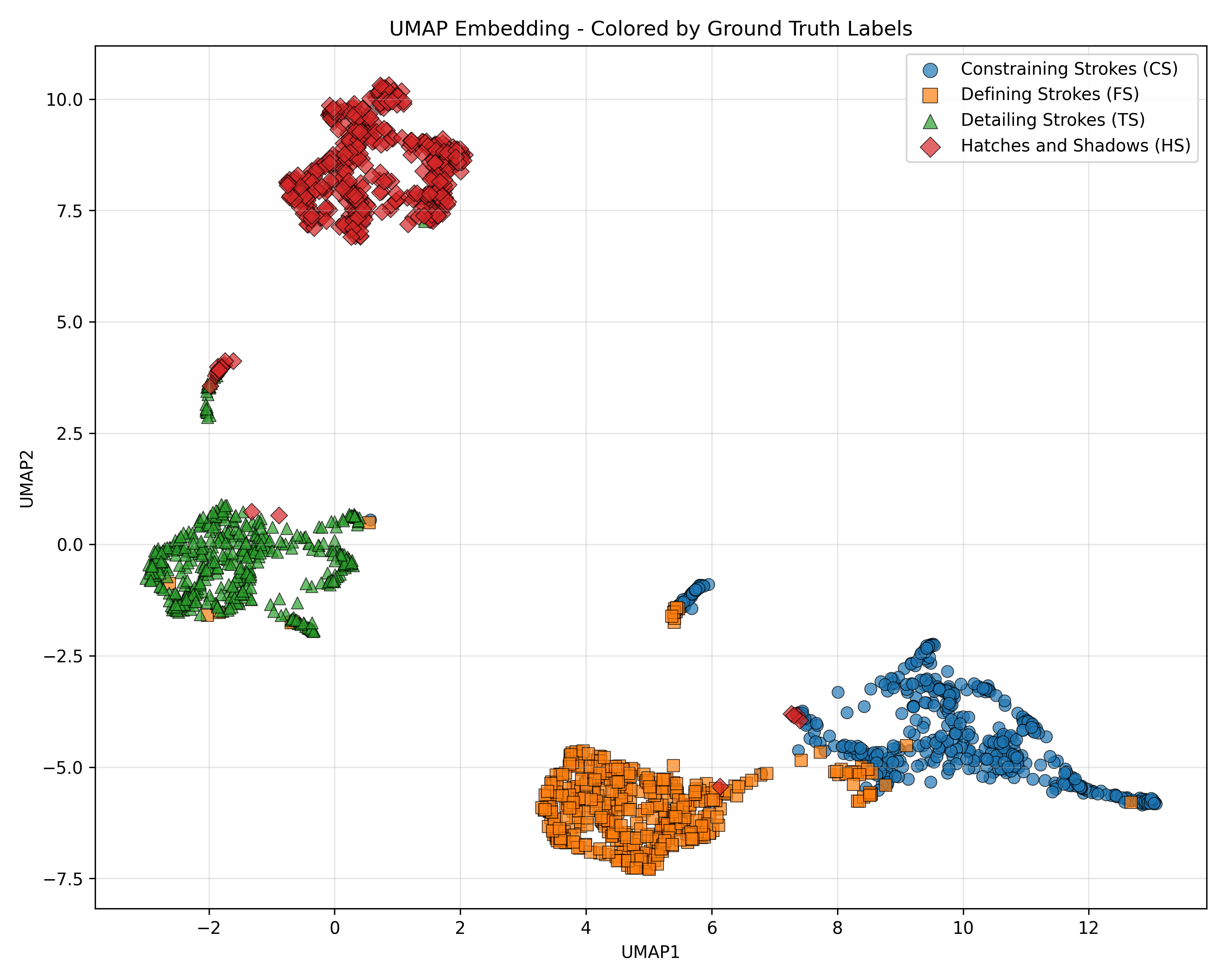}
        \caption{ShuffleNet}
    \end{subfigure}
    \caption{UMAP Visualization of Latent Vectors from the Field Test data. ResNet50 (d) demonstrates the distinct clustering of the four stroke categories.}
    \label{fig:umap_plots}
\end{figure*}

The UMAP plots (Figure \ref{fig:umap_plots}) elucidate the reason for ResNet50's superiority. In the ResNet50 feature space, the four stroke categories formed distinct, well-separated clusters even for the unseen field data. In contrast, the feature spaces of AlexNet and ShuffleNet showed significant overlap between classes, particularly between 'Defining' and 'Detailing' strokes. This suggests that the residual connections in ResNet50 allowed it to learn more generalized, invariant morphological features (such as curvature continuity and line weight consistency) rather than overfitting to the specific pixel artefacts of the augmented training set.

\subsection{Classical Machine Learning (ML) Approach}
\label{subsec:ml_approach}

In parallel to the deep learning methodology, a classical machine learning pipeline was developed. While the Deep Learning approach operates on the rasterized visual representation of the strokes, the Machine Learning approach leverages the rich metadata captured by the AEGIS platform. This feature-driven strategy aims to classify strokes based on their geometric, kinematic, and statistical properties, offering an interpretable alternative to the "black box" nature of neural networks.

\subsubsection{Feature Engineering and Extraction}
The raw input for this pipeline was the JSON log files containing temporal sequences of $X-Y$ coordinates, pressure values ($P$), and thickness ($T$) data. To transform these variable-length time series into fixed-length vectors suitable for standard classifiers, a rigorous feature extraction process was implemented. We engineered a vector of over \textbf{150 handcrafted features} per stroke, categorized as follows:

\begin{enumerate}
    \item \textbf{Statistical Features:} For each primary variable ($X, Y, P, T$), we computed descriptive statistics including the mean, standard deviation, variance, minimum, maximum, range, root mean square (RMS), skewness, and kurtosis. These capture the global distribution of the stroke's properties.
    \item \textbf{Geometric Features:} To capture the spatial morphology, we calculated the arc length, bounding box dimensions (width, height, diagonal, area), and the aspect ratio. These are critical for distinguishing, for instance, a long, straight \textit{Constraining Stroke} from a compact \textit{Hatch}.
    \item \textbf{Curvature and Directionality:} We computed the curvature profile ($\kappa$) at every point and extracted the mean and maximum curvature. Additionally, directional features such as the total angle change and the number of significant direction changes ($>30^{\circ}$) were derived to identify the waviness inherent in \textit{Defining Strokes}.
    \item \textbf{Kinematic Features:} Utilizing the timestamps, we derived the velocity and acceleration profiles. Features such as maximum velocity and mean acceleration help differentiate rapid, confident strokes from slower, deliberate \textit{Detailing Strokes}.
    \item \textbf{Line Density and Compactness:} Metrics defining how "filled" the bounding box is relative to the stroke length, useful for identifying shading patterns.
\end{enumerate}

Post-extraction, missing values were imputed with zeros, and all features were normalized using Standard Scaling (zero mean, unit variance) to ensure that features with larger magnitudes (e.g., coordinates) did not dominate the objective functions of distance-based classifiers.

\subsubsection{Model Selection and Training Protocol}
To identify the most effective classification boundary, we trained and evaluated a diverse suite of \textbf{10 distinct classifiers} spanning linear, non-linear, ensemble, and probabilistic families:
\begin{itemize}
    \item \textbf{Linear Models:} Logistic Regression, Linear SVM.
    \item \textbf{Non-Linear/Kernel Models:} SVM (RBF Kernel), SVM (Polynomial Kernel), K-Nearest Neighbours (KNN).
    \item \textbf{Tree-Based Models:} Decision Tree.
    \item \textbf{Ensembles:} Random Forest, AdaBoost, XGBoost.
    \item \textbf{Probabilistic:} Gaussian Naive Bayes.
\end{itemize}

The dataset was partitioned into a training set (75\%) and a testing set (25\%) using a stratified split to preserve class balance. Hyperparameter optimization was conducted using \texttt{GridSearchCV} with \textbf{5-fold stratified cross-validation}. For example, the Random Forest model was tuned for the number of estimators ($50, 100, 200$) and maximum depth, while the SVMs were tuned for $C$ and $\gamma$ values.

\subsubsection{Performance Evaluation and Comparative Analysis}
The trained models were first evaluated on the static Test Dataset. The results, summarized in Table \ref{tab:ml_test_results}, indicate exceptional performance across most architectures, with ensemble methods achieving near-perfect accuracy.

{
\small
\begin{table}[ht!]
\centering
\small % Use slightly smaller font to ensure data fits in narrow columns
\caption{Machine Learning Model Performance on Test Dataset}
\label{tab:ml_test_results}
\setlength{\tabcolsep}{3pt} % Reduce padding between columns
% Using X columns with multipliers to control relative widths automatically
% Col 1: 1.2x, Col 2: 1.0x, Cols 3-5: 0.9x each
\begin{tabularx}{\linewidth}{@{} 
    >{\raggedright\arraybackslash\hsize=1.2\hsize}X 
    >{\centering\arraybackslash\hsize=1.0\hsize}X 
    >{\centering\arraybackslash\hsize=0.9\hsize}X 
    >{\centering\arraybackslash\hsize=0.9\hsize}X 
    >{\centering\arraybackslash\hsize=0.9\hsize}X 
@{}}
\toprule
\textbf{Model} & \textbf{Test Accuracy} & \textbf{Precision} & \textbf{Recall} & \textbf{F1 Score} \\
\midrule
Decision Tree       & 0.9980 & 0.9980 & 0.9980 & 0.9980 \\
\textbf{Random Forest} & \textbf{0.9980} & \textbf{0.9980} & \textbf{0.9980} & \textbf{0.9980} \\
XGBoost             & 0.9920 & 0.9921 & 0.9920 & 0.9920 \\
SVM (RBF)           & 0.9920 & 0.9920 & 0.9920 & 0.9920 \\
AdaBoost            & 0.9960 & 0.9961 & 0.9960 & 0.9960 \\
Gaussian NB         & 0.9940 & 0.9940 & 0.9940 & 0.9940 \\
Logistic Regression & 0.9900 & 0.9901 & 0.9900 & 0.9900 \\
SVM (Linear)        & 0.9880 & 0.9880 & 0.9880 & 0.9880 \\
SVM (Polynomial)    & 0.9720 & 0.9723 & 0.9720 & 0.9720 \\
KNN                 & 0.9540 & 0.9547 & 0.9540 & 0.9540 \\
\bottomrule
\end{tabularx}
\end{table}
}

\begin{figure*}[ht!]
    \centering
    \begin{subfigure}[b]{0.32\textwidth}
        \centering
        \includegraphics[width=\linewidth]{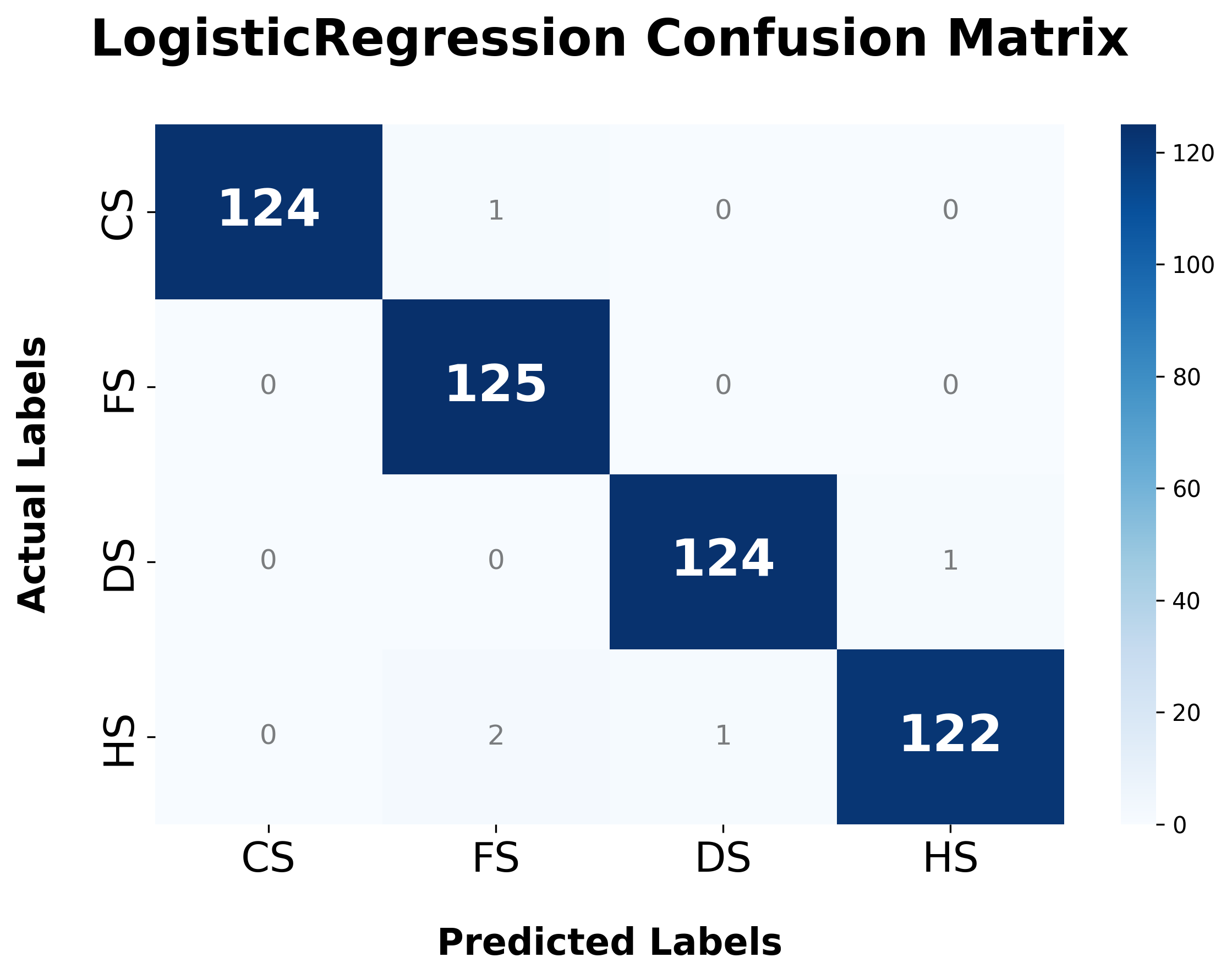}
        \caption{Logistic Reg.}
    \end{subfigure}
    \hfill
    \begin{subfigure}[b]{0.32\textwidth}
        \centering
        \includegraphics[width=\linewidth]{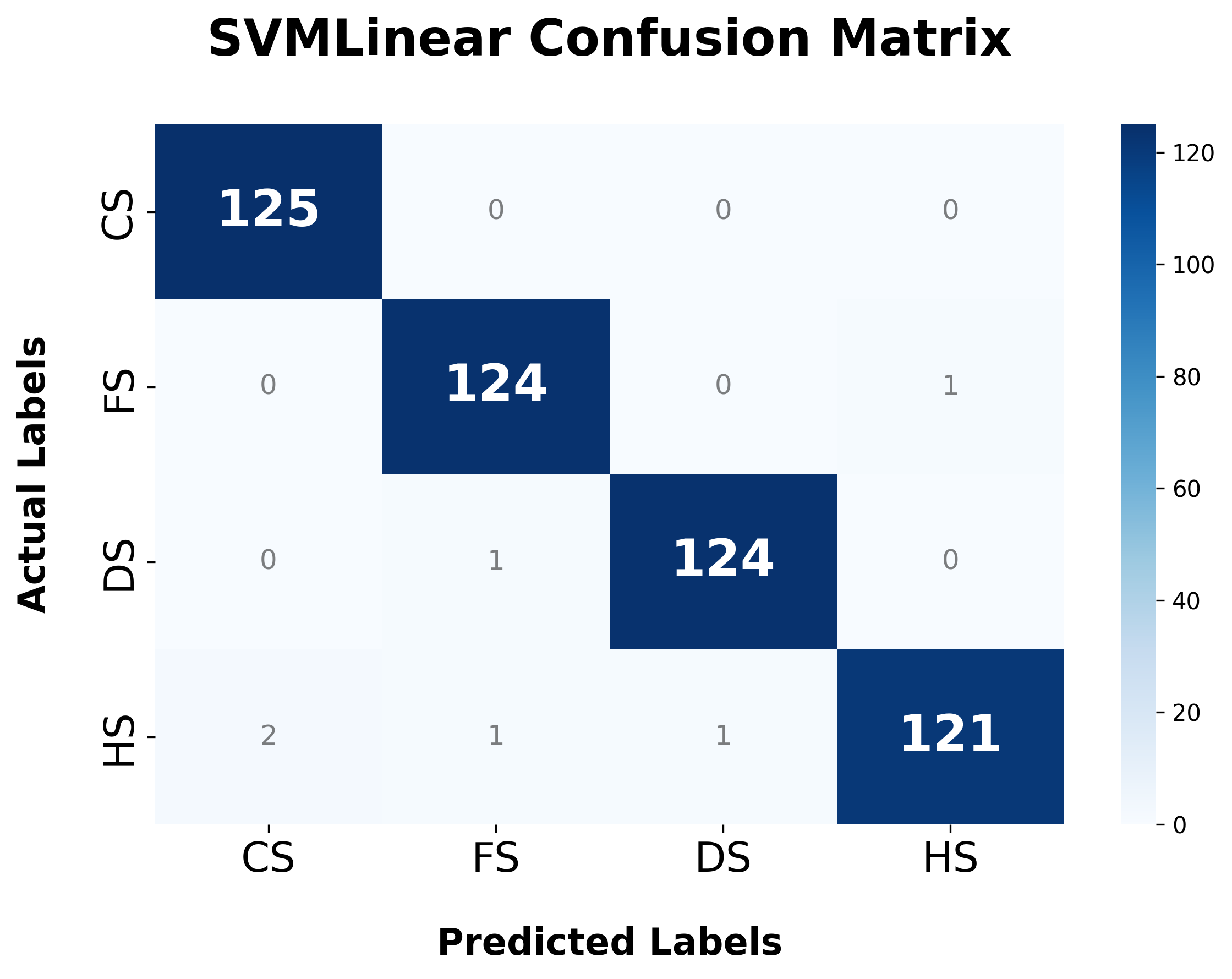}
        \caption{SVM Linear}
    \end{subfigure}
    \hfill
    \begin{subfigure}[b]{0.32\textwidth}
        \centering
        \includegraphics[width=\linewidth]{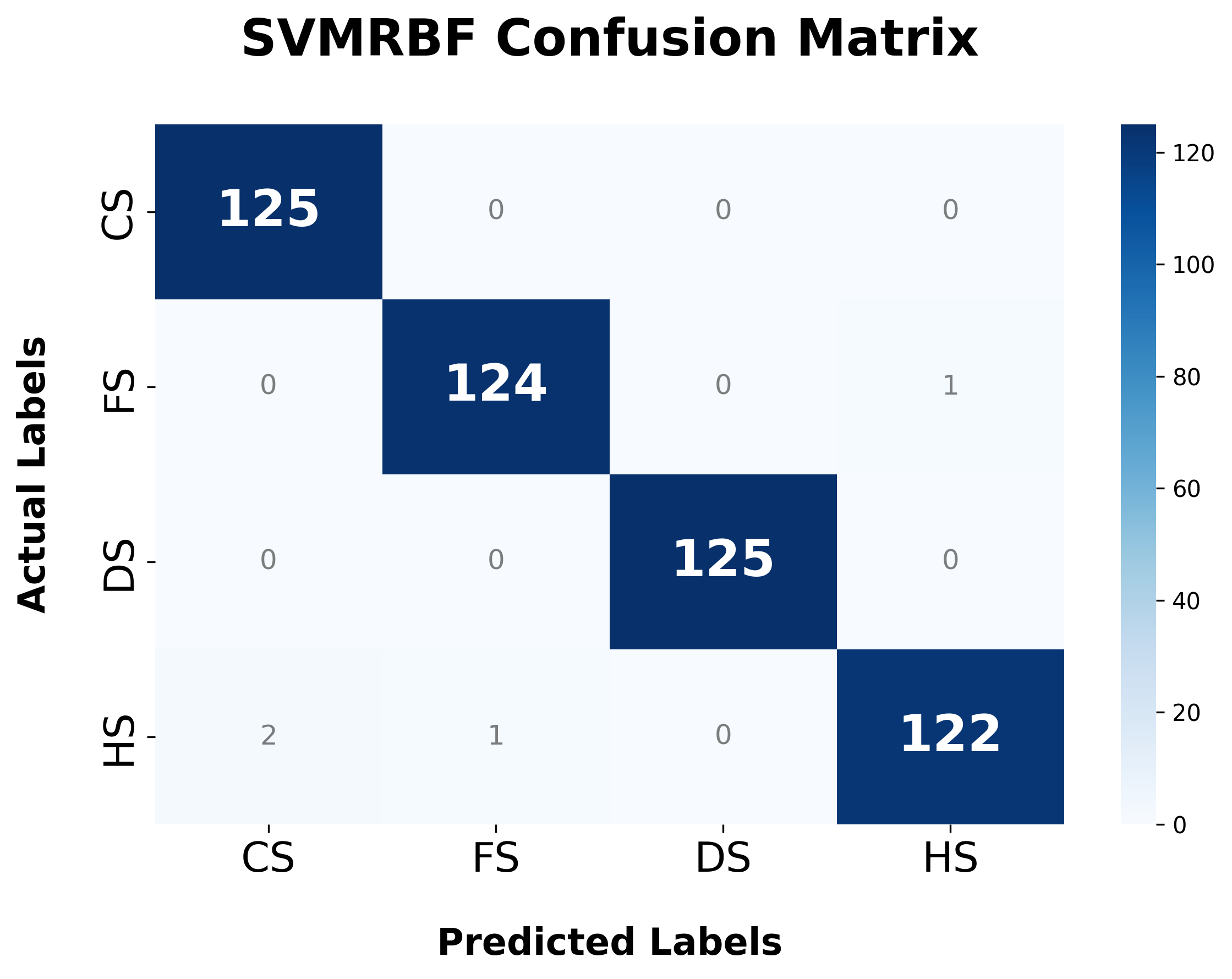}
        \caption{SVM RBF}
    \end{subfigure}
    \vfill
    \begin{subfigure}[b]{0.32\textwidth}
        \centering
        \includegraphics[width=\linewidth]{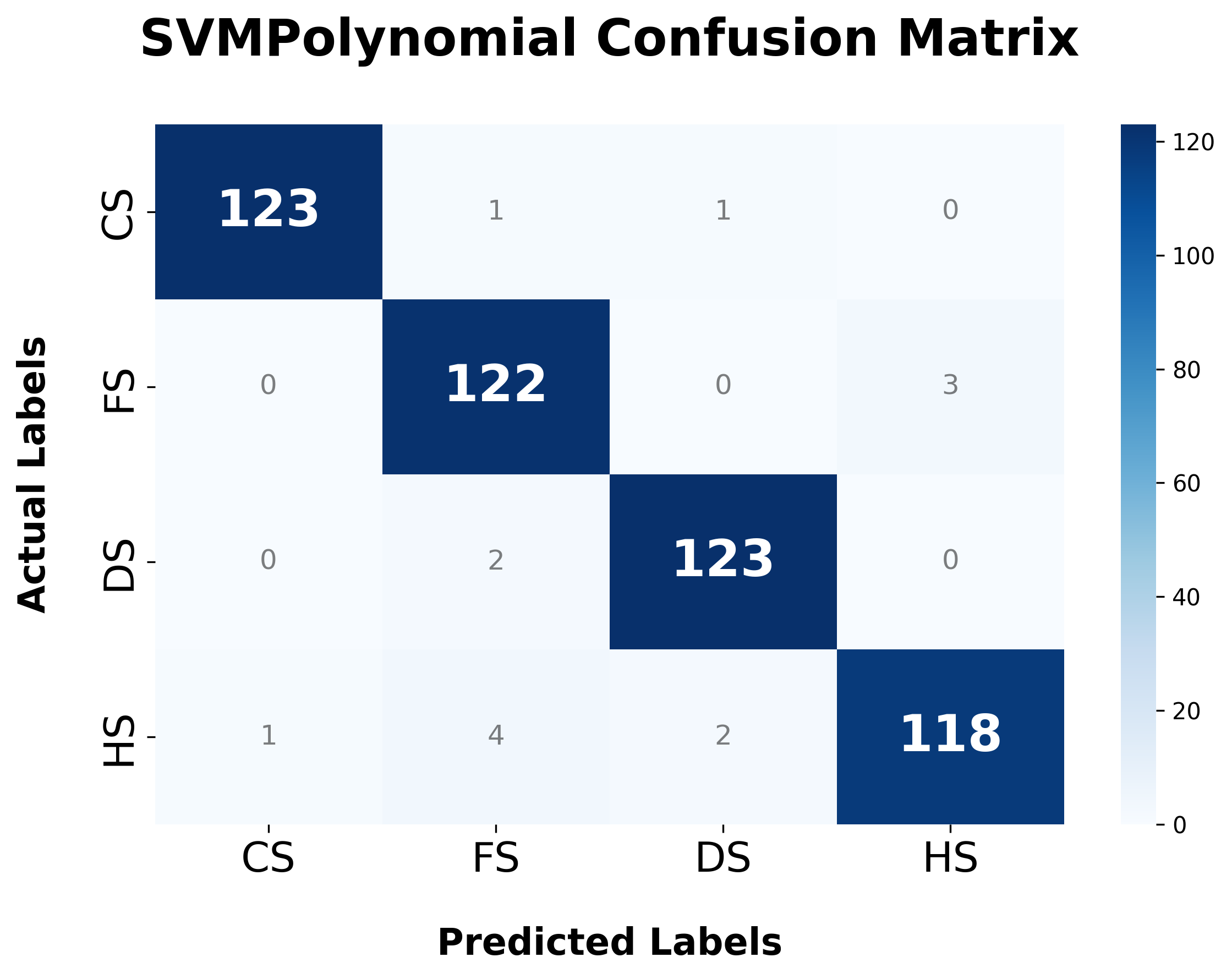}
        \caption{SVM Poly}
    \end{subfigure}
    \hfill
    \begin{subfigure}[b]{0.32\textwidth}
        \centering
        \includegraphics[width=\linewidth]{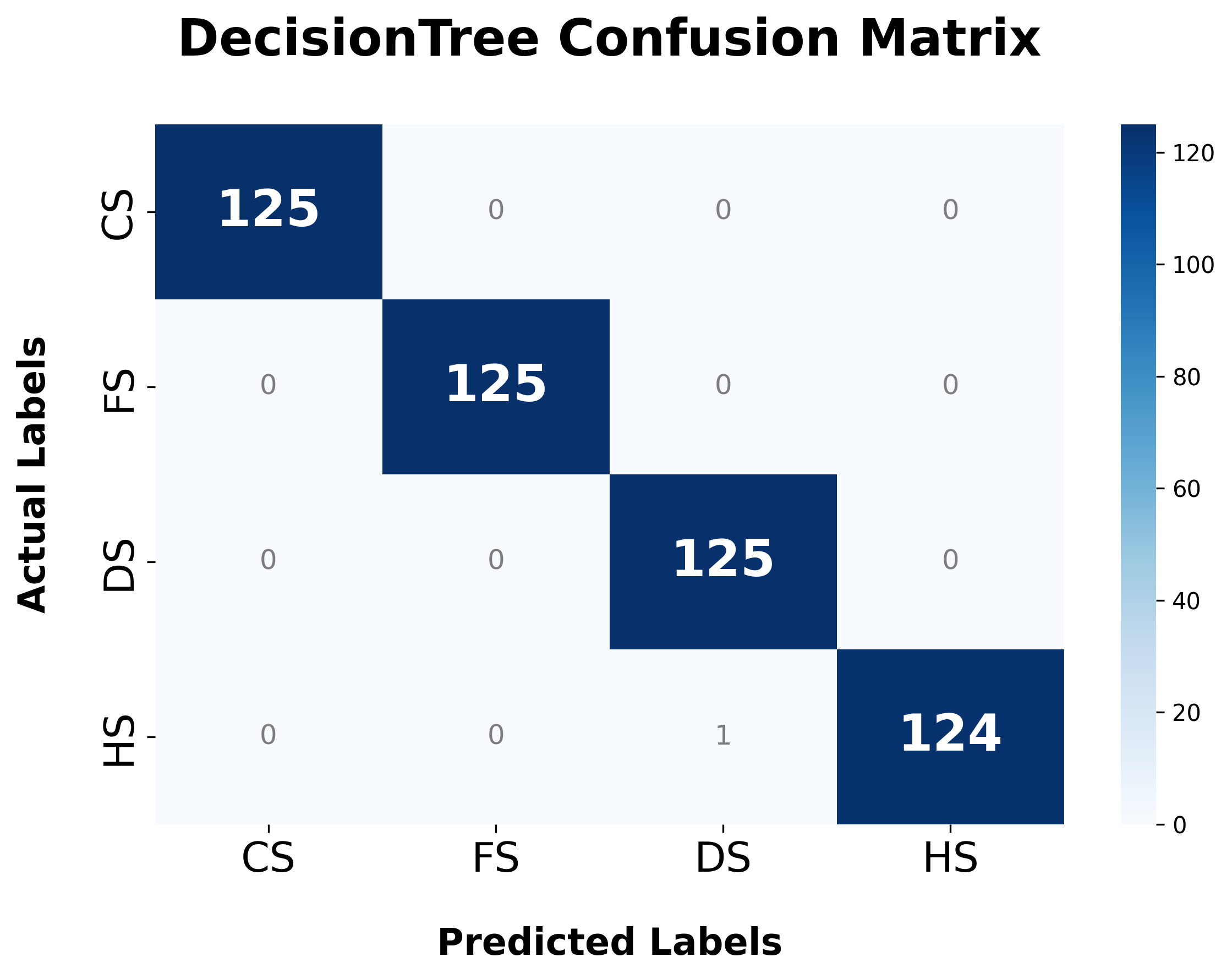}
        \caption{Decision Tree}
    \end{subfigure}
    
    \vspace{0.5cm}
    
    \begin{subfigure}[b]{0.32\textwidth}
        \centering
        \includegraphics[width=\linewidth]{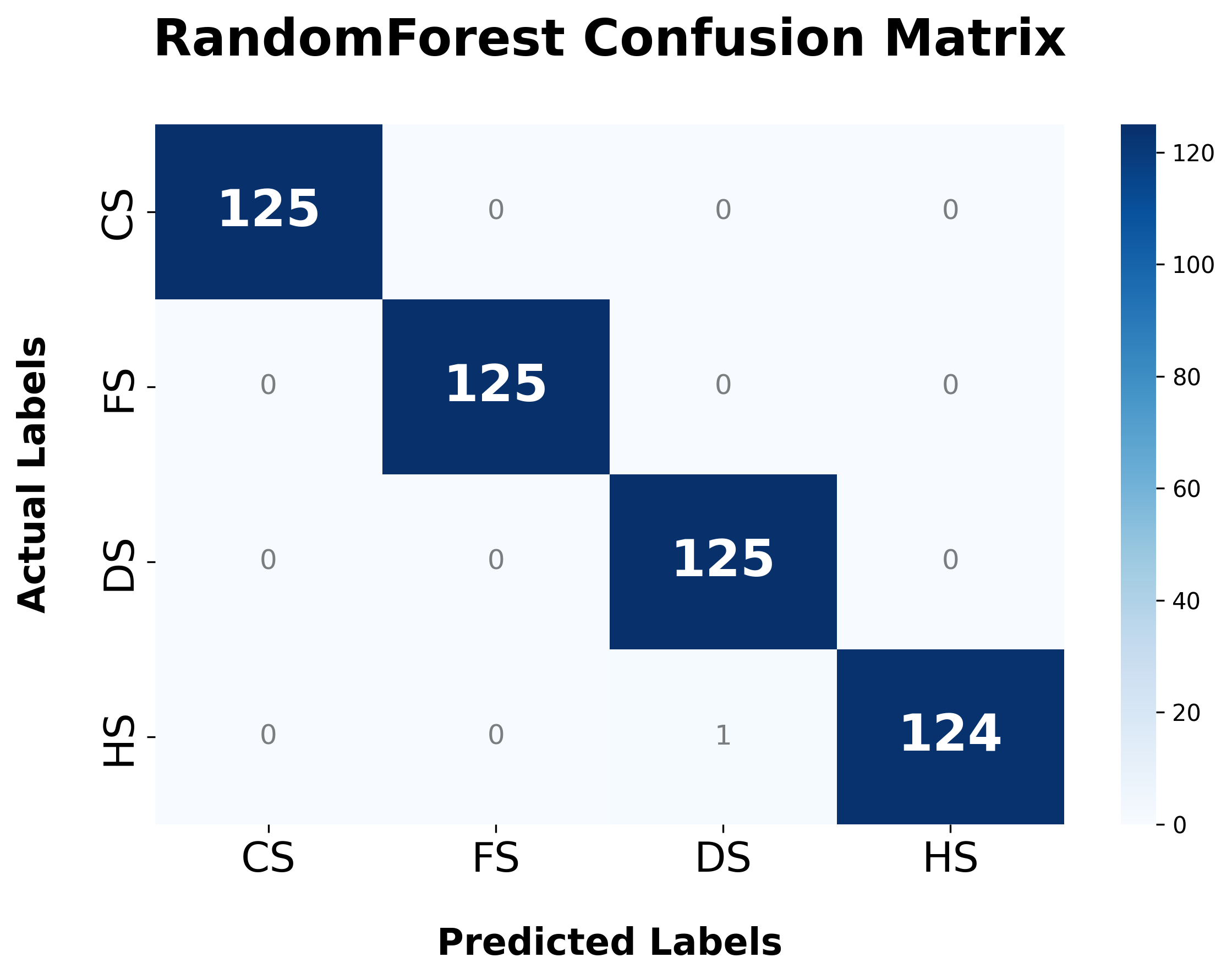}
        \caption{Random Forest}
    \end{subfigure}
    \hfill
    \begin{subfigure}[b]{0.32\textwidth}
        \centering
         \includegraphics[width=\linewidth]{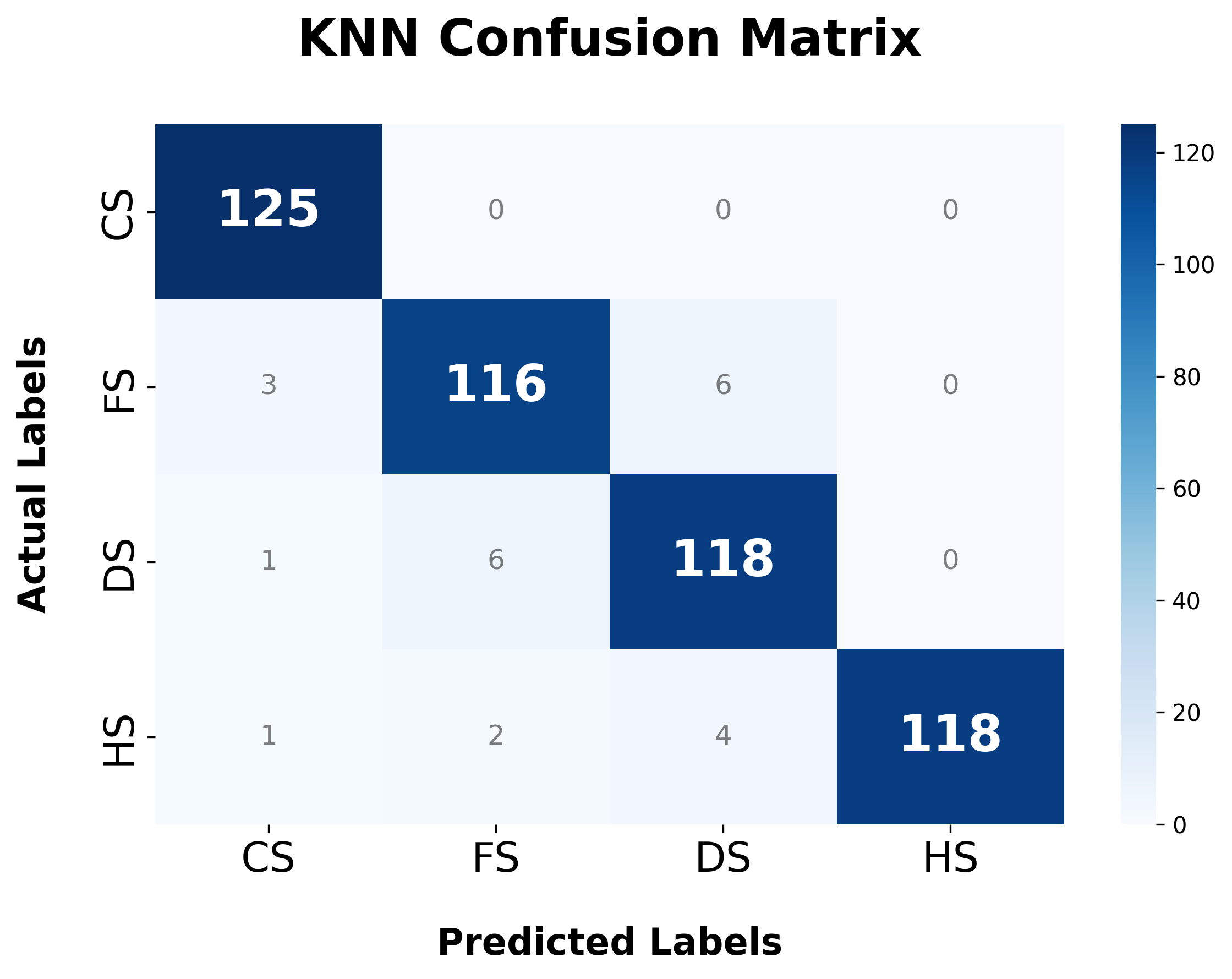}
        \caption{KNN}
    \end{subfigure}
    \hfill
    \begin{subfigure}[b]{0.32\textwidth}
        \centering
        \includegraphics[width=\linewidth]{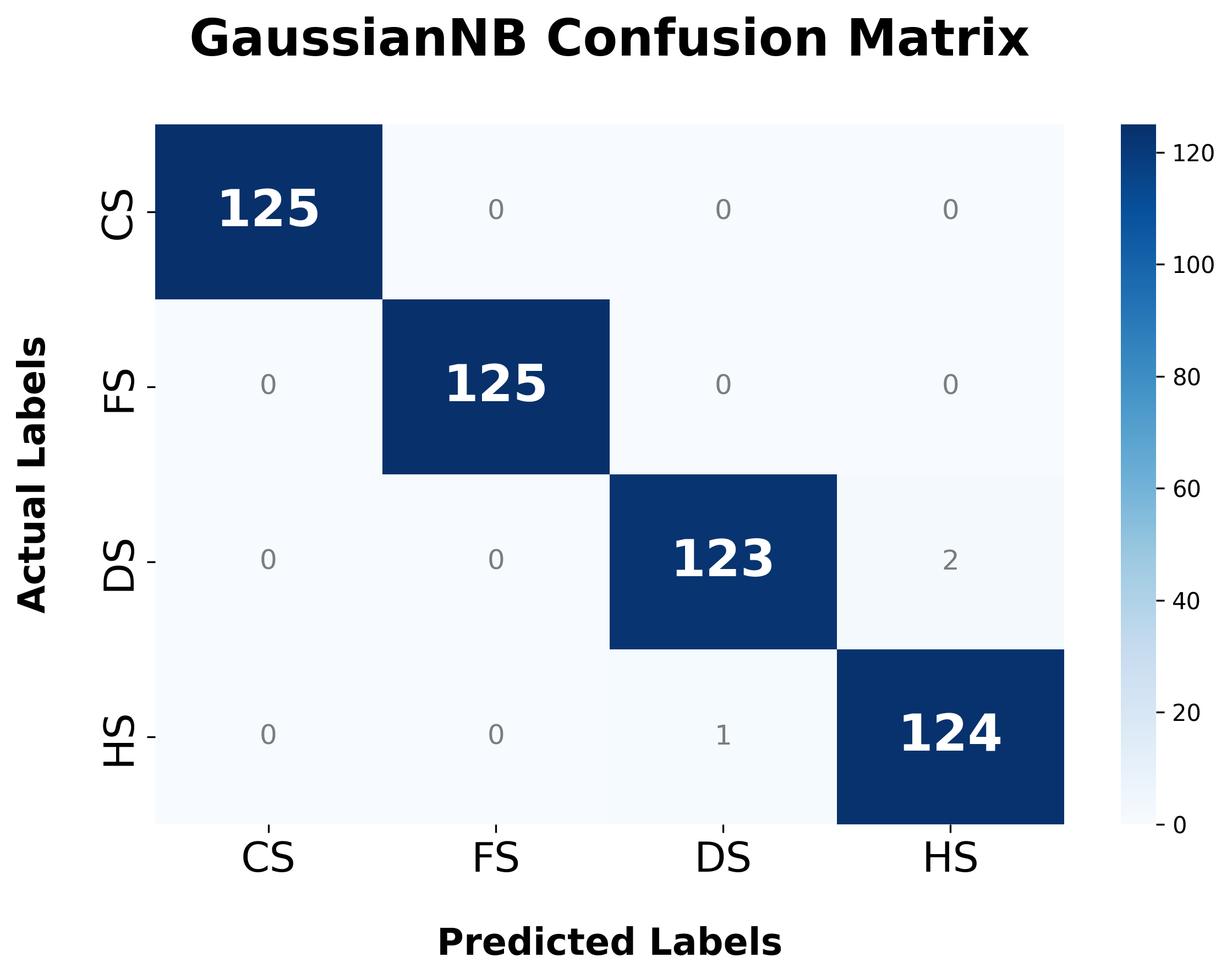}
        \caption{Gaussian NB}
    \end{subfigure}
    \vfill
    \begin{subfigure}[b]{0.32\textwidth}
        \centering
        \includegraphics[width=\linewidth]{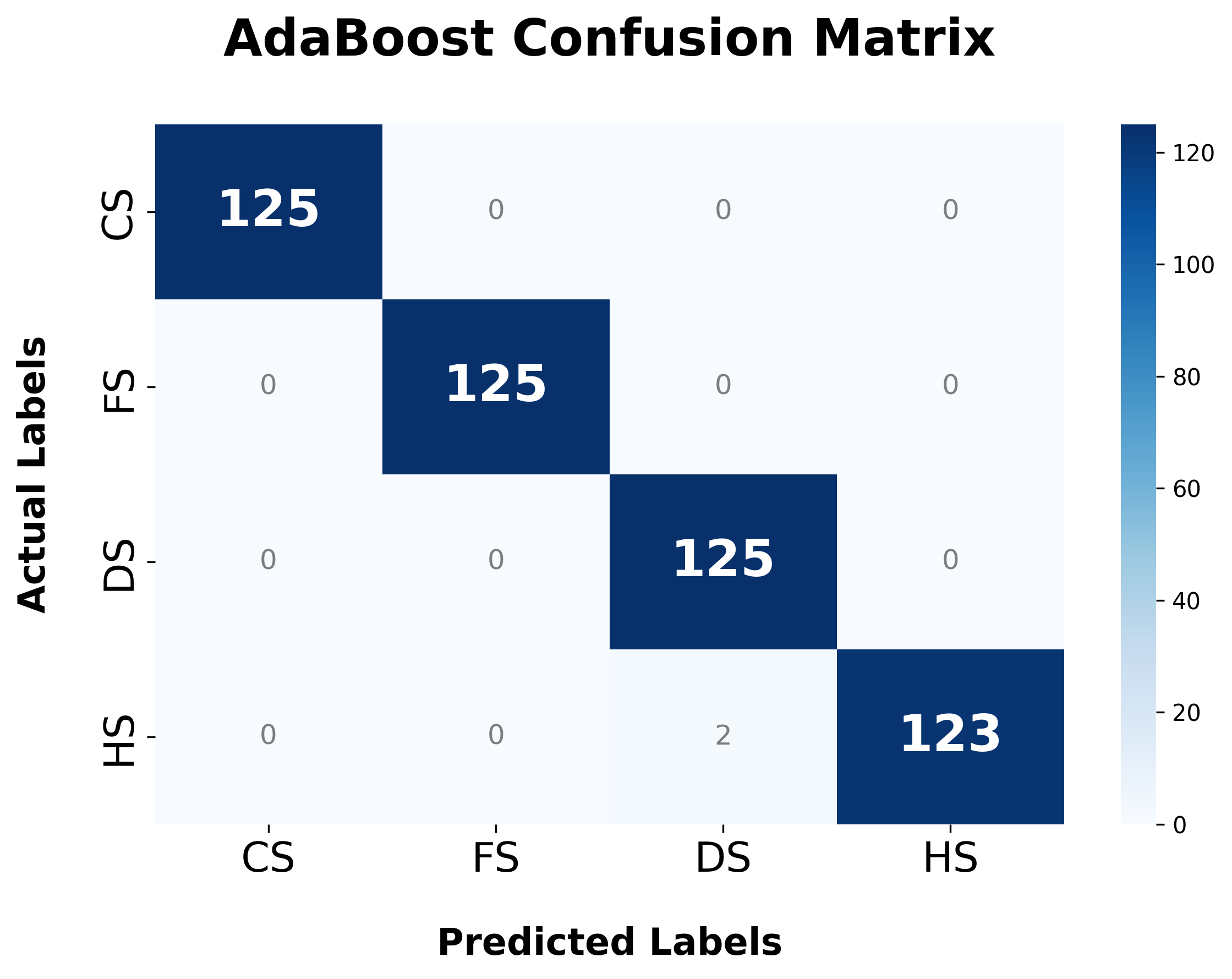}
        \caption{AdaBoost}
    \end{subfigure}
    \hfill
    \begin{subfigure}[b]{0.32\textwidth}
        \centering
        \includegraphics[width=\linewidth]{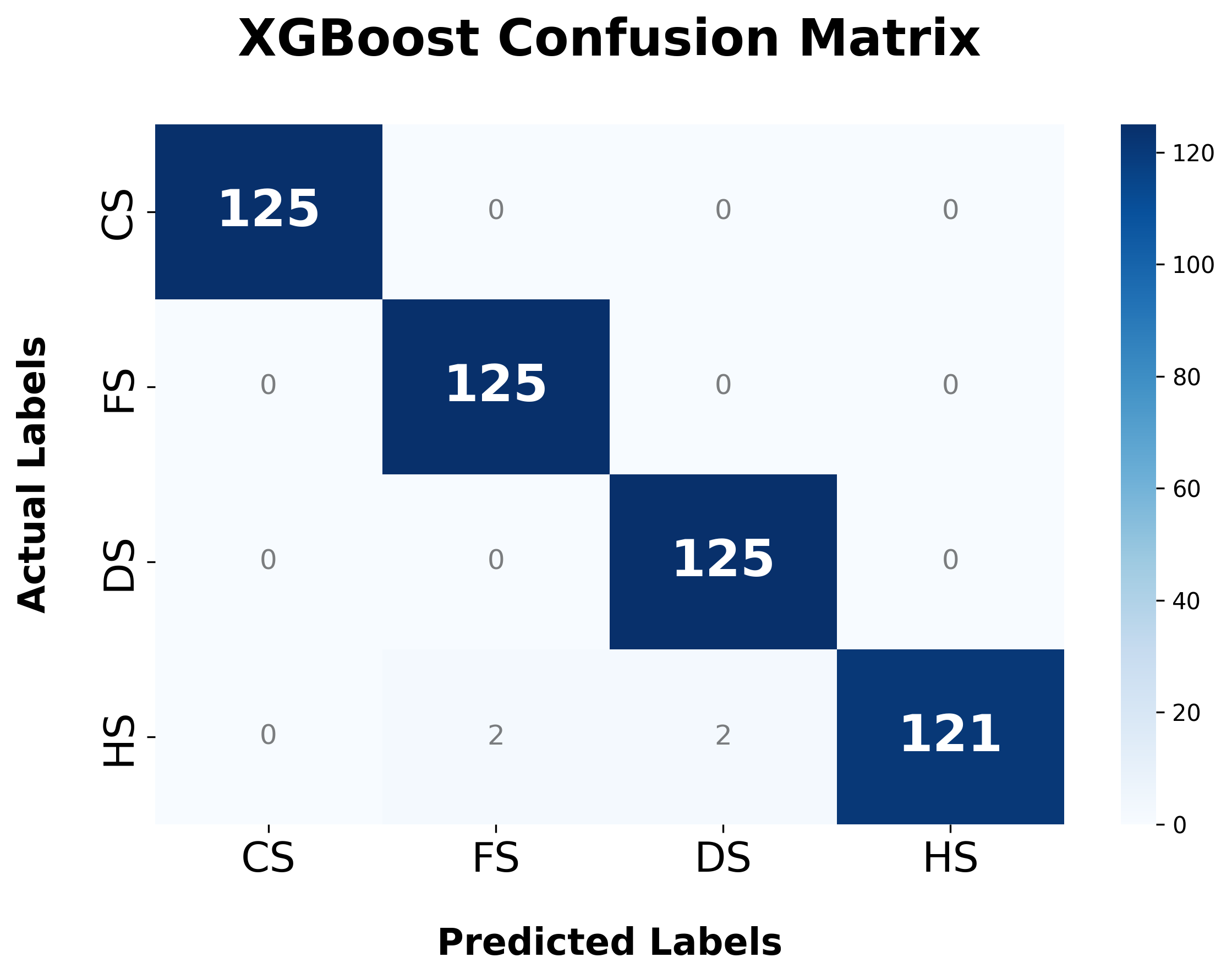}
        \caption{XGBoost}
    \end{subfigure}
    \caption{Confusion Matrices for all 10 Machine Learning Classifiers on the Test Dataset.}
    \label{fig:ml_confusion_test}
\end{figure*}

\subsubsection{Field Test Validation and Model Robustness}
To verify real-world applicability, the same Field Test (Sanity Check) used for the Deep Learning models was applied to the Machine Learning classifiers. The results, presented in Table \ref{tab:ml_field_results}, offer a stark contrast to the Deep Learning results and highlight the importance of model selection.

\begin{table}[ht!]
\centering
\caption{Machine Learning Model Performance on Field Test (Real-time Sanity Check)}
\label{tab:ml_field_results}
\begin{tabular}{|l|c|}
\hline
\textbf{Model} & \textbf{Field Test Accuracy (\%)} \\ \hline
\textbf{Random Forest} & \textbf{100.0} \\ \hline
\textbf{SVM (RBF)} & \textbf{100.0} \\ \hline
XGBoost & 97.5 \\ \hline
AdaBoost & 97.5 \\ \hline
SVM (Linear) & 97.5 \\ \hline
Ensemble (Voting) & 97.5 \\ \hline
SVM (Polynomial) & 92.5 \\ \hline
KNN & 90.0 \\ \hline
Logistic Regression & 90.0 \\ \hline
Decision Tree & 72.5 \\ \hline
Gaussian NB & 50.0 \\ \hline
\end{tabular}
\end{table}

\begin{figure*}[ht!]
    \centering
    \begin{subfigure}[b]{0.32\textwidth}
        \centering
        \includegraphics[width=\linewidth]{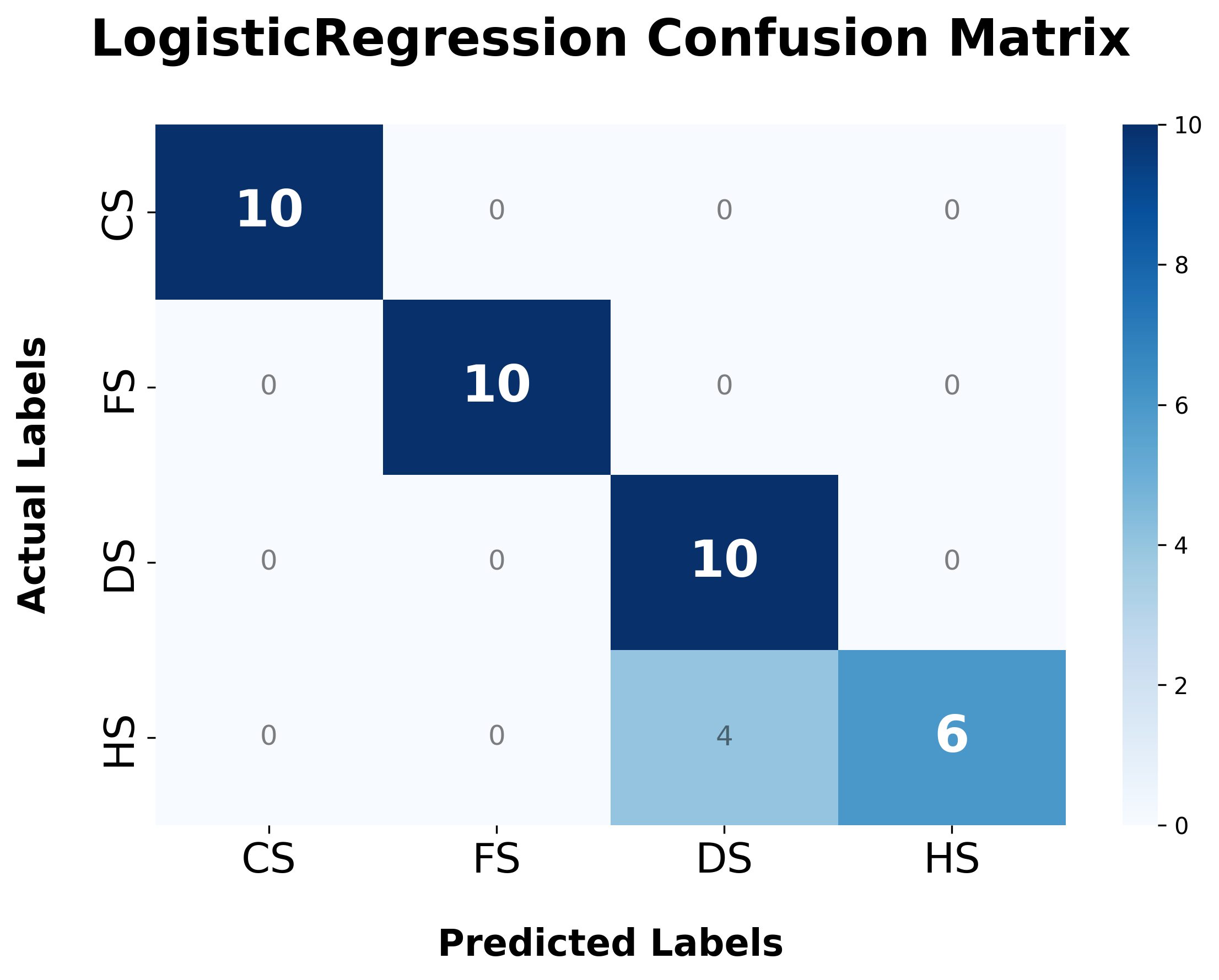}
        \caption{Logistic Reg.}
    \end{subfigure}
    \hfill
    \begin{subfigure}[b]{0.32\textwidth}
        \centering
        \includegraphics[width=\linewidth]{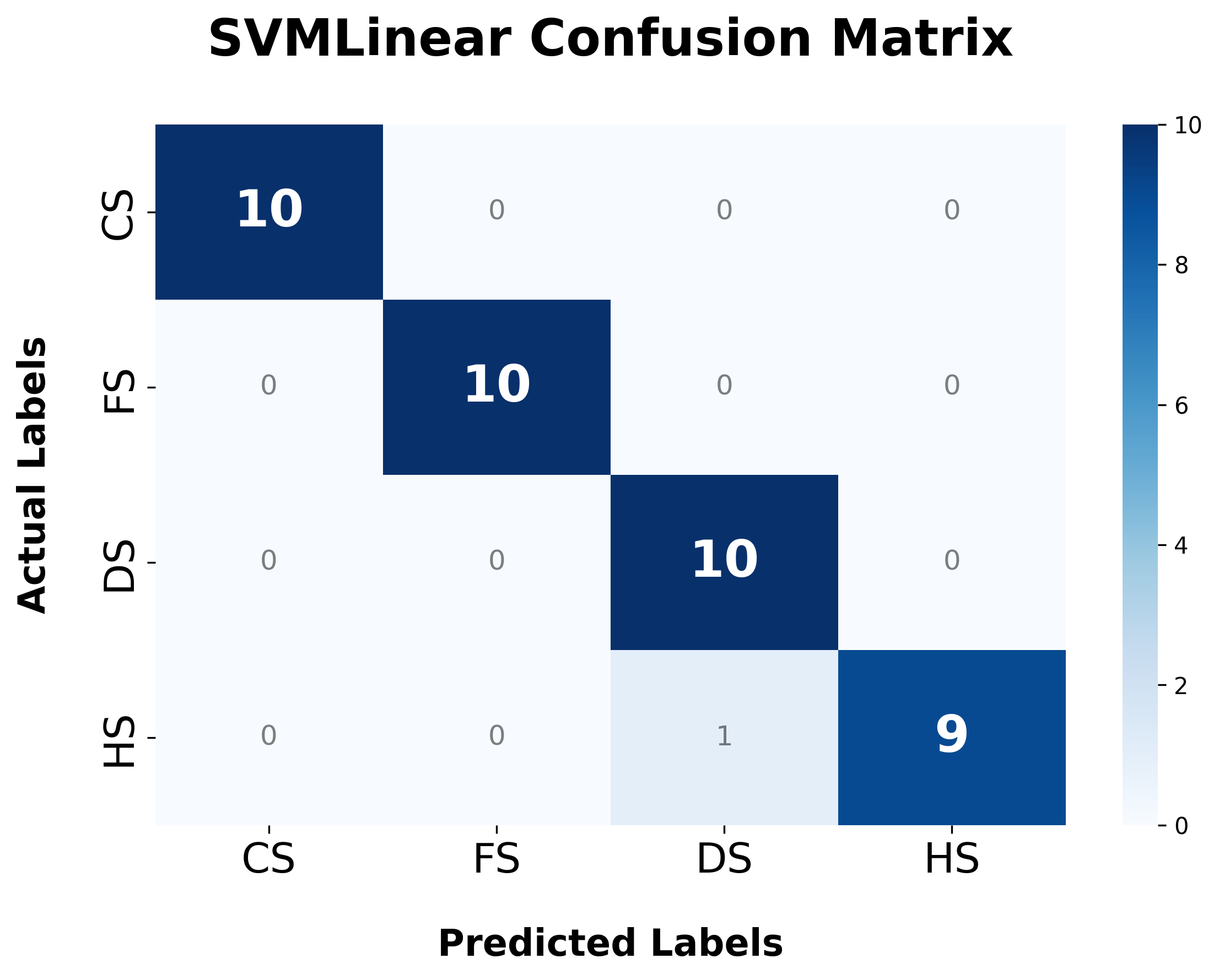}
        \caption{SVM Linear}
    \end{subfigure}
    \hfill
    \begin{subfigure}[b]{0.32\textwidth}
        \centering
        \includegraphics[width=\linewidth]{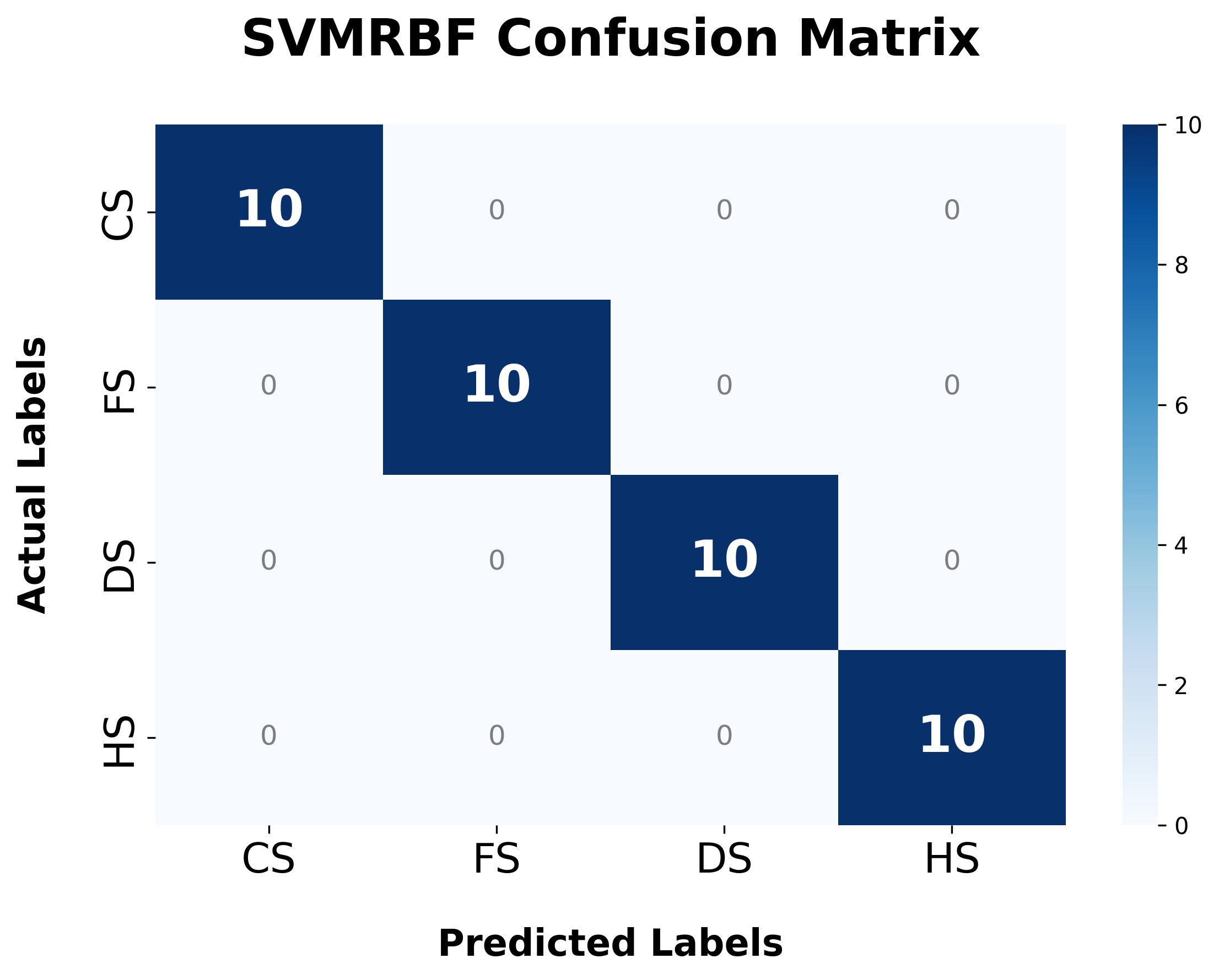}
        \caption{SVM RBF}
    \end{subfigure}
    \vfill
    \begin{subfigure}[b]{0.32\textwidth}
        \centering
        \includegraphics[width=\linewidth]{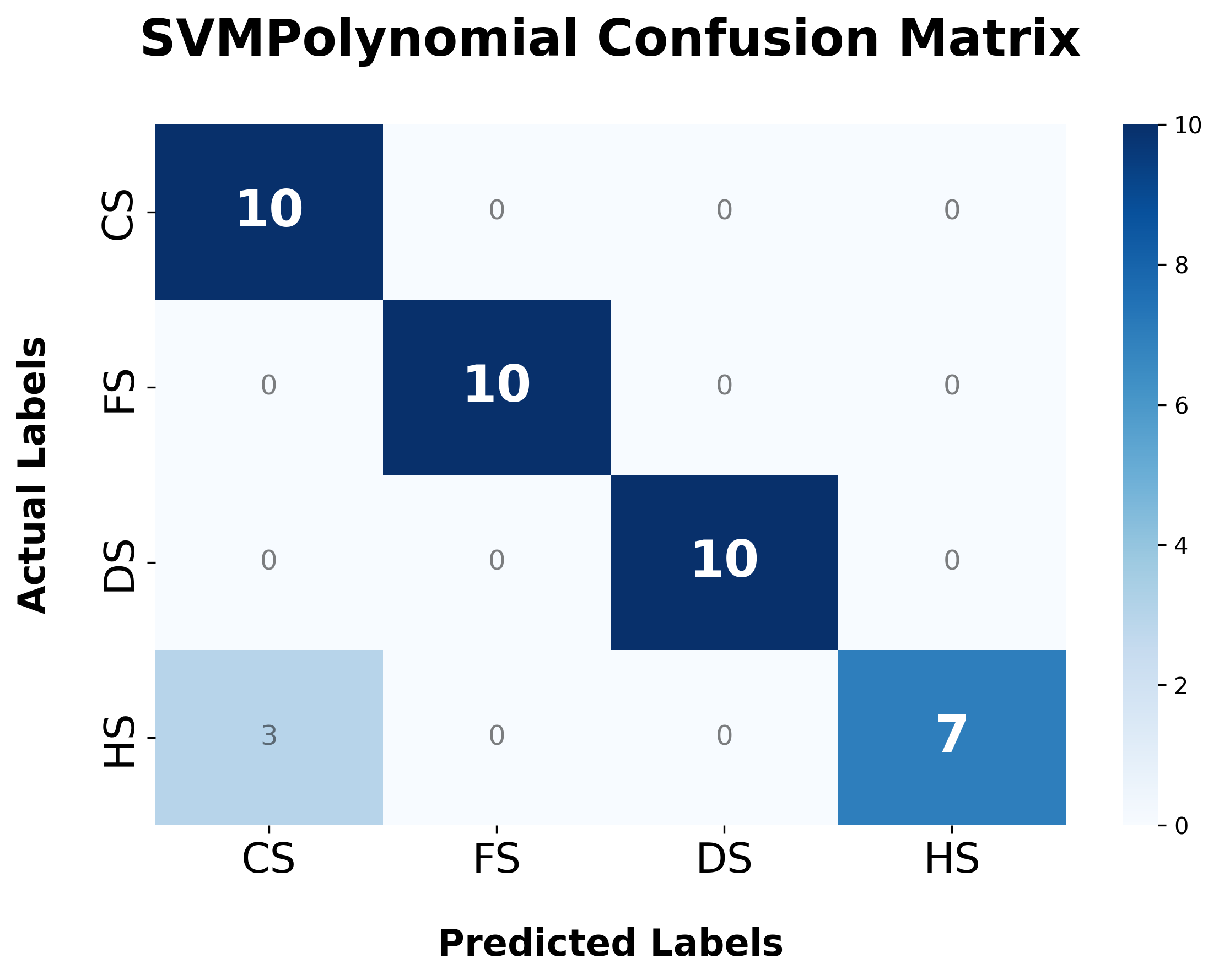}
        \caption{SVM Poly}
    \end{subfigure}
    \hfill
    \begin{subfigure}[b]{0.32\textwidth}
        \centering
        \includegraphics[width=\linewidth]{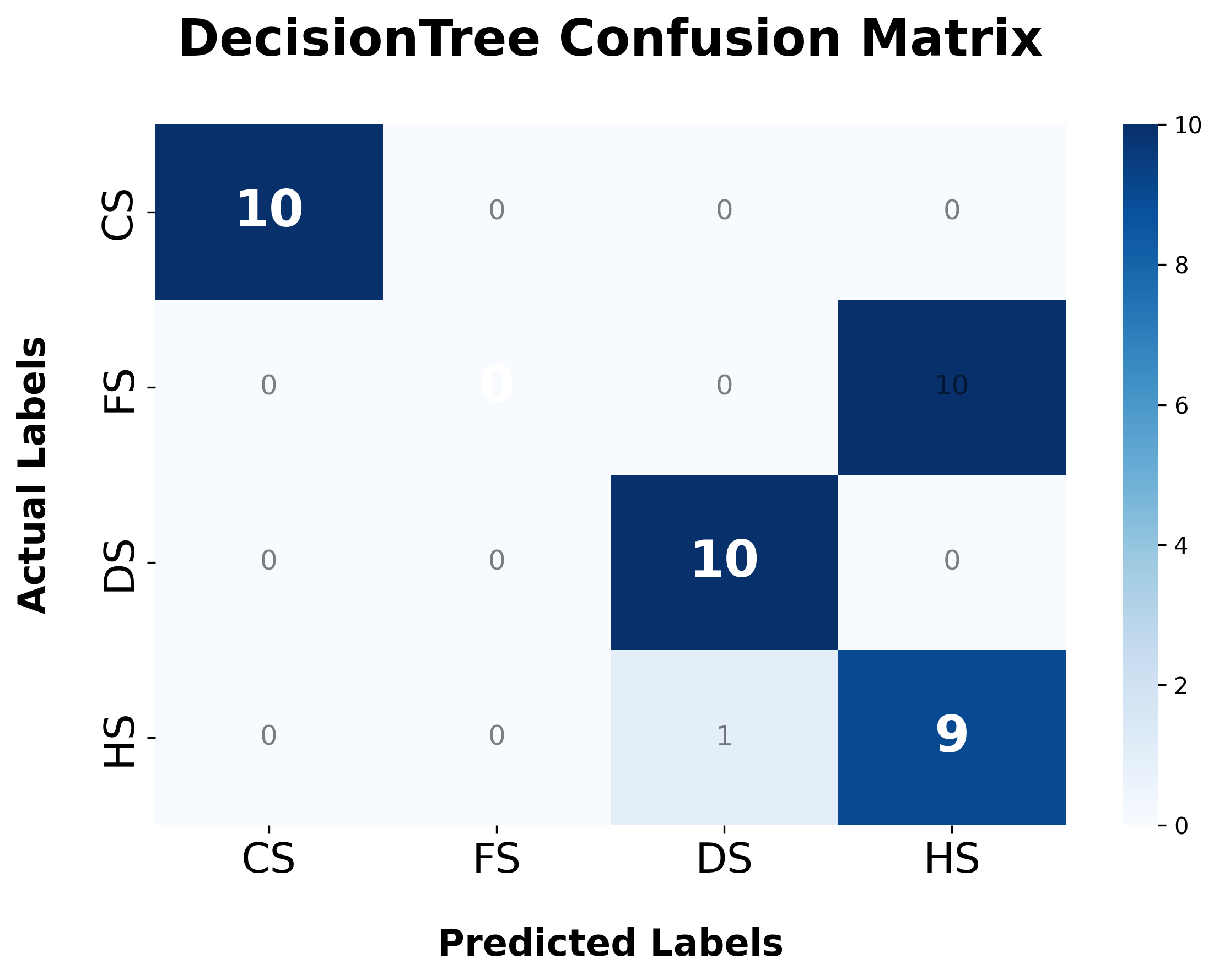}
        \caption{Decision Tree}
    \end{subfigure}
    \hfill
    \begin{subfigure}[b]{0.32\textwidth}
        \centering
        \includegraphics[width=\linewidth]{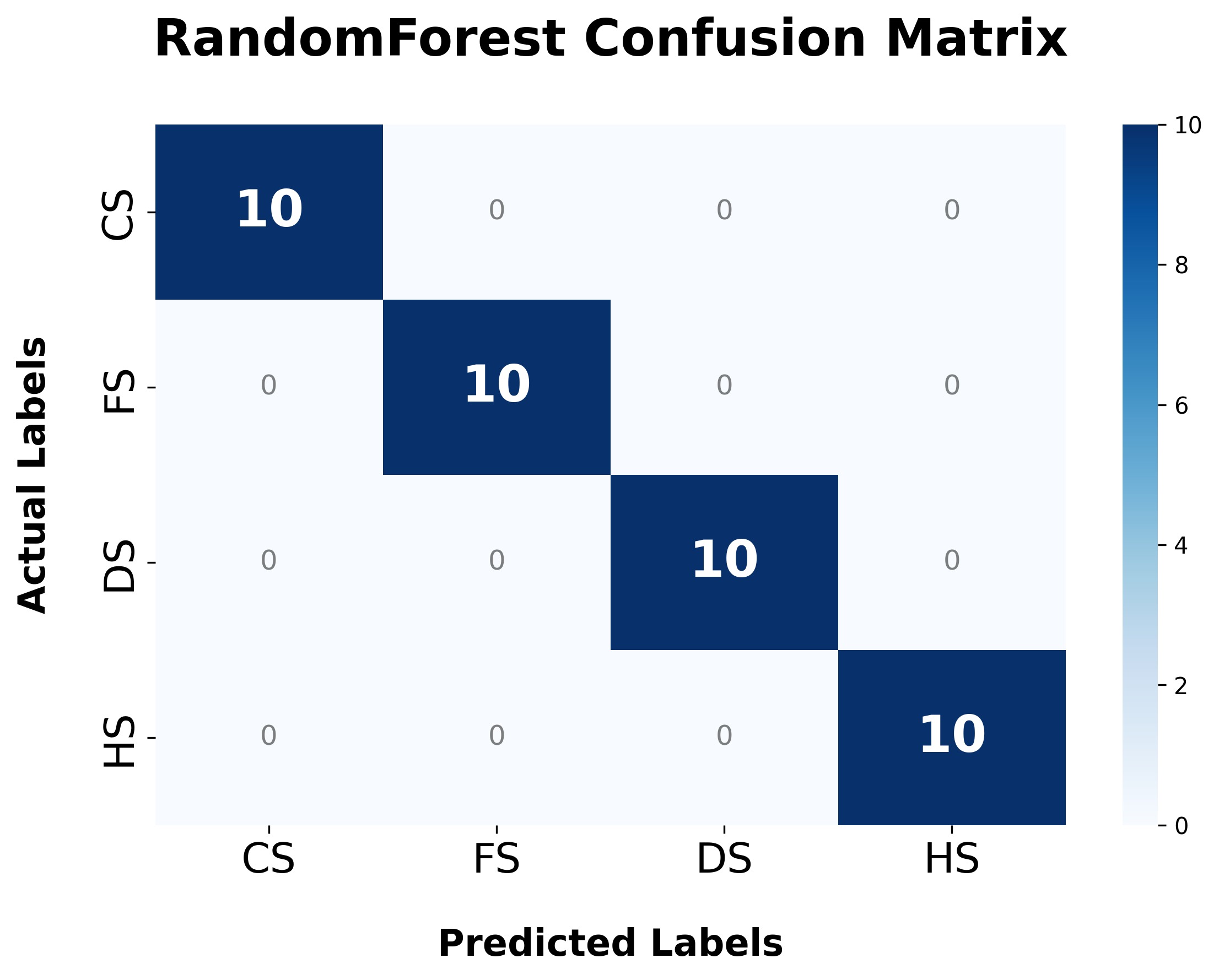}
        \caption{Random Forest}
    \end{subfigure}
    
    \vspace{0.5cm}
    
    \begin{subfigure}[b]{0.32\textwidth}
        \centering
        \includegraphics[width=\linewidth]{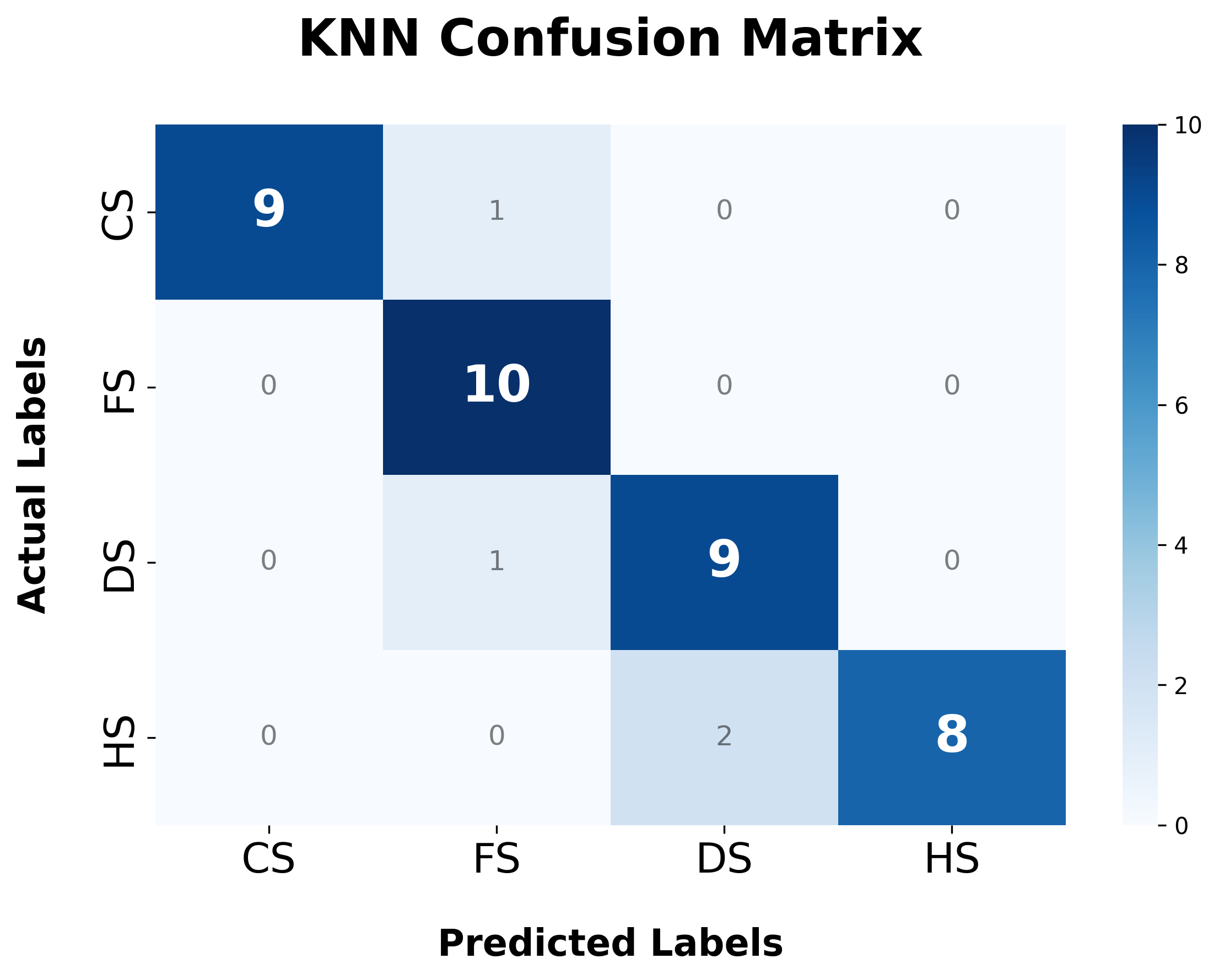}
        \caption{KNN}
    \end{subfigure}
    \hfill
    \begin{subfigure}[b]{0.32\textwidth}
        \centering
        \includegraphics[width=\linewidth]{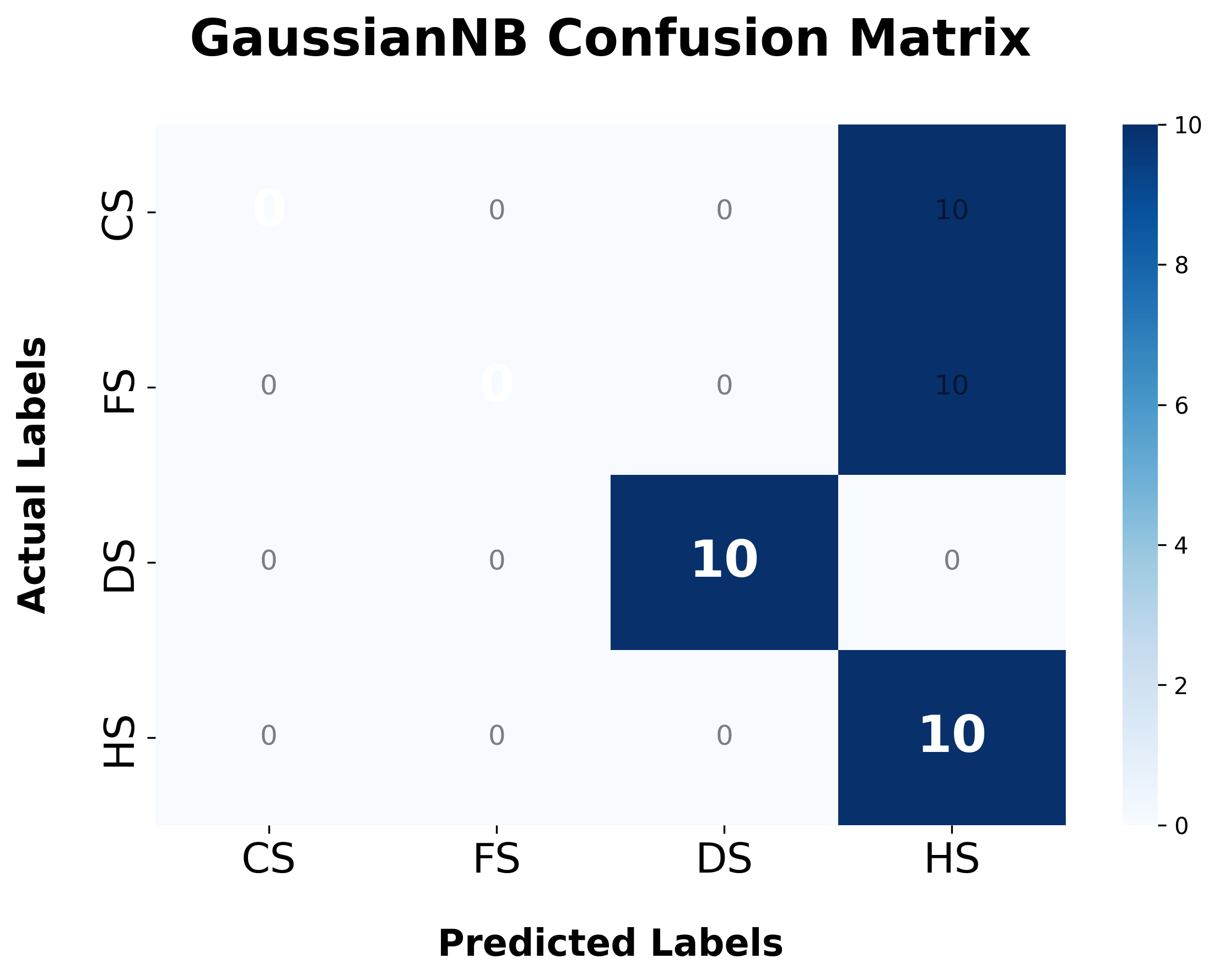}
        \caption{Gaussian NB}
    \end{subfigure}
    \hfill
    \begin{subfigure}[b]{0.32\textwidth}
        \centering
        \includegraphics[width=\linewidth]{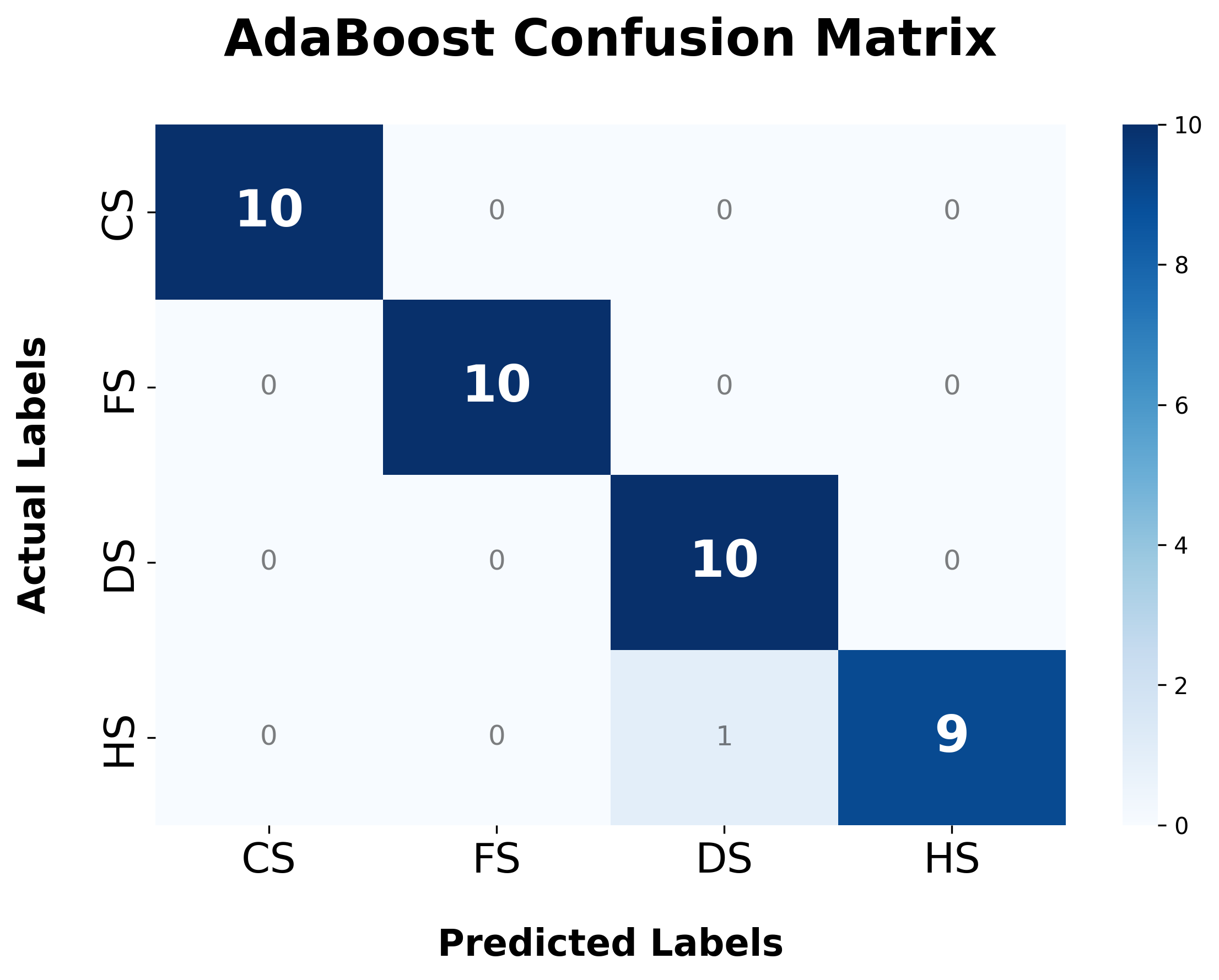}
        \caption{AdaBoost}
    \end{subfigure}
    \vfill
    \begin{subfigure}[b]{0.32\textwidth}
        \centering
        \includegraphics[width=\linewidth]{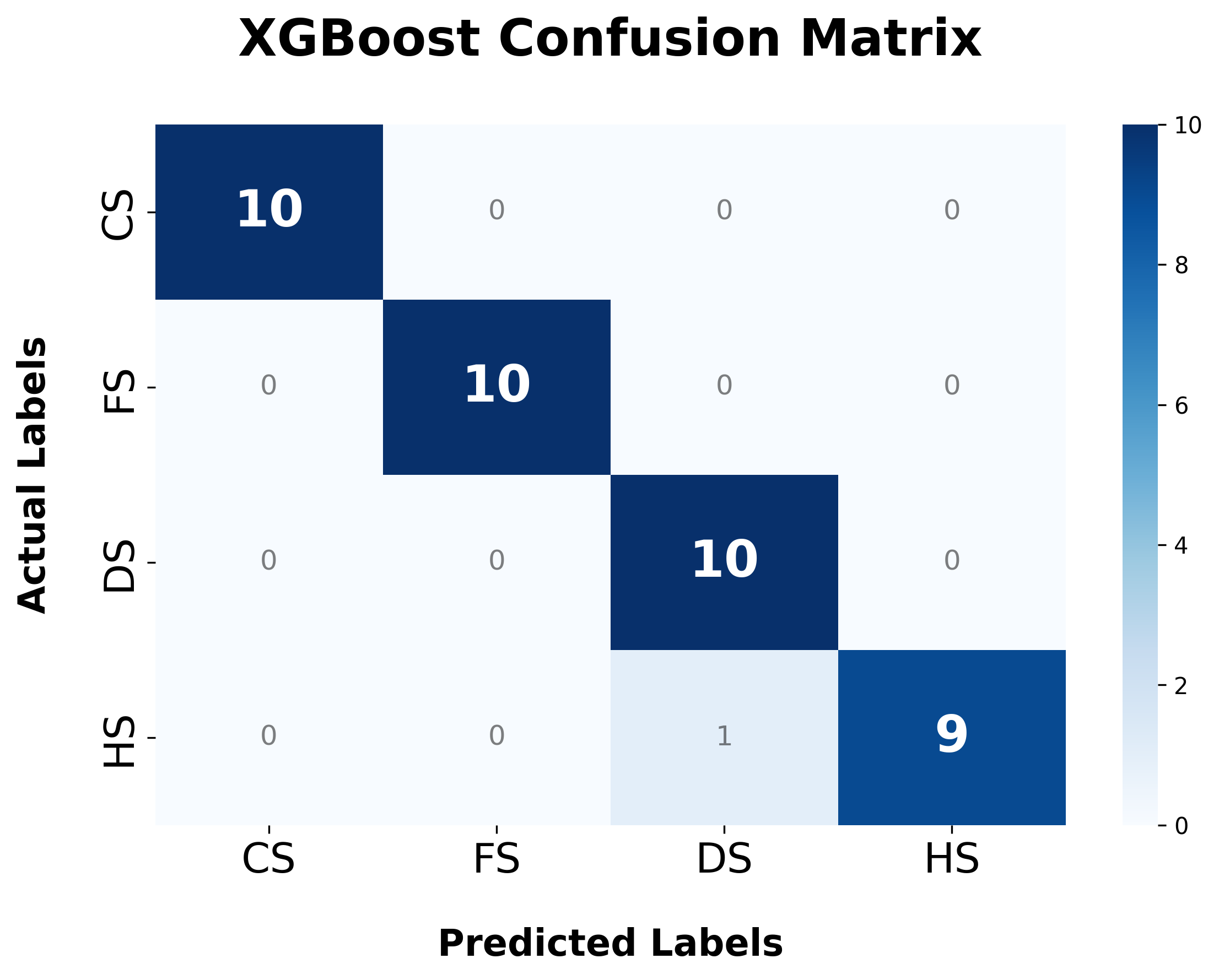}
        \caption{XGBoost}
    \end{subfigure}
    \hfill
    \begin{subfigure}[b]{0.32\textwidth}
        \centering
        \includegraphics[width=\linewidth]{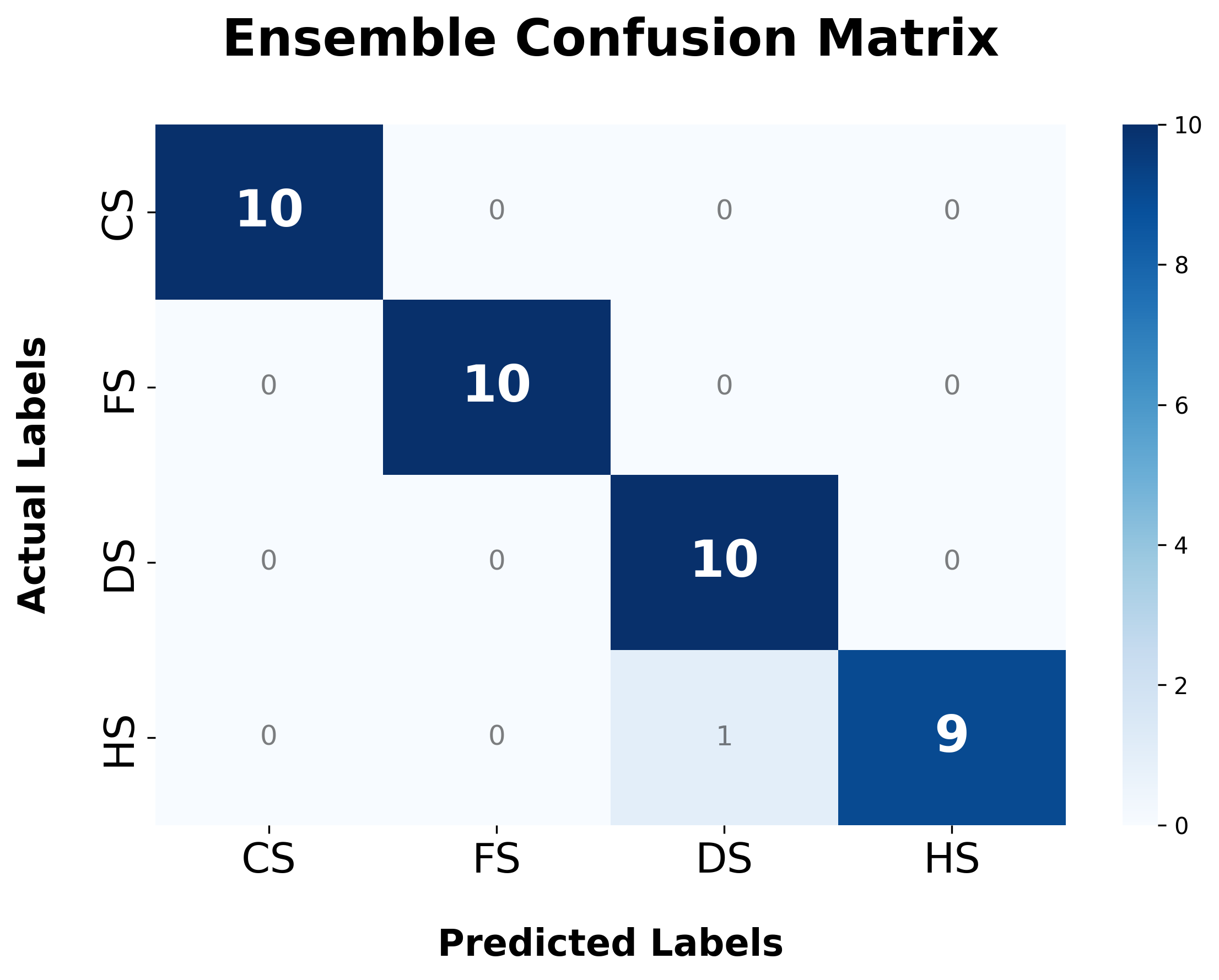}
        \caption{Ensemble}
    \end{subfigure}
    \caption{Confusion Matrices for all 10 ML Classifiers plus Ensemble on the Field Test Data.}
    \label{fig:ml_confusion_field}
\end{figure*}

\begin{figure}[ht!]
    \centering
    \includegraphics[width=\linewidth]{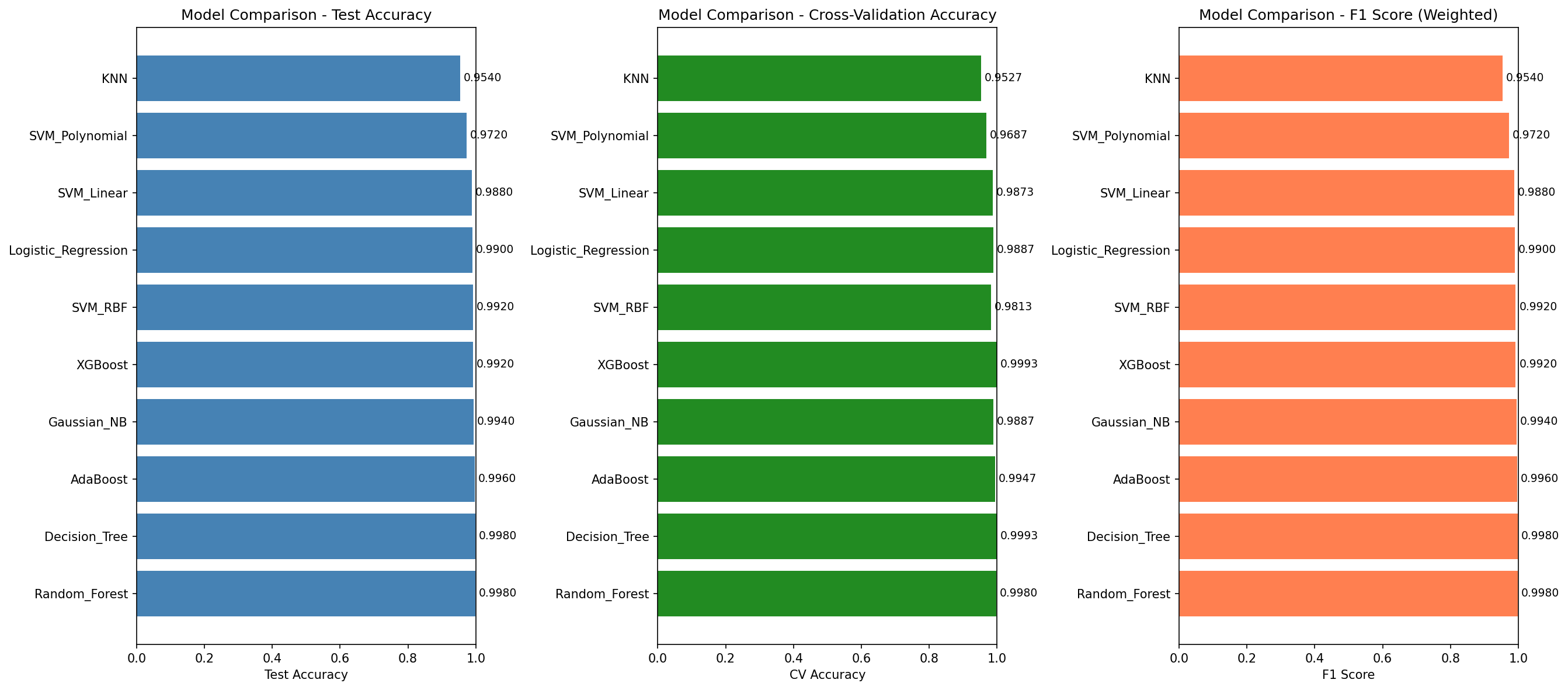}
    \caption{Comparative Field Test Accuracy of Machine Learning Classifiers.}
    \label{fig:ml_model_comparison}
\end{figure}

\subsubsection{Critical Analysis of Model Performance}
The disparity in Field Test performance provides deep insights into the nature of the data. 

\textbf{Superiority of Random Forest and SVM (RBF):} Both Random Forest and SVM (RBF) achieved a remarkable \textbf{100\% accuracy} on the field test data, significantly outperforming even the best Deep Learning model (ResNet50 at 82.5\%). This suggests that the \textit{physics} of the stroke—captured explicitly by features like velocity, curvature, and pressure—provides a more robust signal for classification than the pixel-level visual representation, which is susceptible to "Sim-to-Real" gaps. Random Forest's ensemble nature allowed it to mitigate variance, while the RBF kernel enabled the SVM to model complex, non-linear decision boundaries effectively.

\textbf{Failure of Decision Trees (72.5\%):} Despite achieving near-perfect accuracy (99.8\%) on the test set, the single Decision Tree performed poorly in the field. This is a classic manifestation of overfitting; the tree likely memorised the noise in the synthetic training data and failed to generalise to the unseen variations of real-time sketching.

\textbf{Poor Performance of Gaussian Naive Bayes (50\%):} The Gaussian NB model yielded the lowest accuracy. This is attributable to its strong assumption of feature independence. In our dataset, features are highly correlated (e.g., \texttt{velocity\_max} is correlated with \texttt{acceleration\_max}, and \texttt{bbox\_area} with \texttt{arc\_length}). The violation of the independence assumption renders the probabilistic predictions of Naive Bayes unreliable in this context.

In conclusion, the feature-based Random Forest model provides the most reliable engine for real-time stroke classification in the DIMES system, offering both high accuracy and interpretability.

\subsection{Comparative Discussion: Image-Based vs. Feature-Based Classification}
\label{subsec:comparative_discussion}

The parallel development of Deep Learning (DL) and classical Machine Learning (ML) pipelines has provided a unique opportunity to evaluate two distinct paradigms for design intent recognition: the visual (pixel-based) approach versus the kinematic (feature-based) approach. The empirical results from the Field Test—representing real-world, unseen data—offer a compelling narrative regarding robustness and generalisation in the domain of digital sketching.

\subsubsection{The Robustness Trade-off}
A direct comparison of the "Field Test" results reveals a significant performance disparity. The best Deep Learning model, \textbf{ResNet50}, achieved an accuracy of \textbf{82.5\%}. In stark contrast, the best Machine Learning models, \textbf{Random Forest} and \textbf{SVM (RBF)}, achieved a perfect accuracy of \textbf{100\%}. 

This disparity highlights a critical "Sim-to-Real" gap inherent in the Deep Learning approach. The DL models were trained on synthetic data generated via geometric transformations. While these transformations (rotation, noise injection) mimic spatial variability, they cannot perfectly replicate the subtle texture, anti-aliasing artifacts, and specific pixel-level noise distributions of a real-time stroke drawn in a browser environment. The CNNs, being data-hungry and highly sensitive to texture, likely overfitted to the specific visual artifacts of the synthetic generation pipeline. When presented with "real" strokes from the Field Test, which differed slightly in pixel fidelity, the lighter models (ShuffleNet, AlexNet) faltered, and even the robust ResNet50 struggled to cross the 85\% threshold.

Conversely, the Machine Learning approach proved remarkably resilient. This robustness is attributable to the nature of the \textbf{150+ handcrafted features}. Features such as \textit{Mean Velocity}, \textit{Maximum Acceleration}, \textit{Curvature Entropy}, and \textit{Pressure Distribution} are physically intrinsic to the act of drawing. They are invariant to the rendering artifacts or the specific resolution of the canvas. For instance, a \textit{Defining Stroke} is kinetically distinct—it is rapid and smooth—regardless of whether it is rendered with anti-aliasing or not. By explicitly modelling the \textit{physics} of the designer's hand rather than the \textit{pixels} on the screen, the Random Forest and SVM models were able to generalize perfectly to new users.

\subsubsection{Ensemble Integration for Real-Time Inference}
Based on this comparative analysis, the Random Forest model was selected as the primary classification engine for the live system. Its ensemble nature offers a balance of high accuracy and interpretability (via feature importance), and its inference latency is negligible on standard CPU hardware, eliminating the need for heavy GPU resources required by ResNet50. However, the ResNet50 model remains a valuable asset for offline analysis of static sketch images where temporal metadata (speed, pressure) might be lost (e.g., scanning legacy paper sketches).

\subsection{Towards Stroke-Based Generative Models}
\label{subsec:generative_vision}

The development of this automated classification pipeline is not an end in itself; rather, it is a foundational step towards a broader vision for Artificial Intelligence in Design. 

\subsubsection{Limitations of Pixel-Based Generative AI}
The current landscape of Generative AI is dominated by diffusion-based architectures (e.g., Stable Diffusion, Midjourney). These models operate in the pixel space, treating an image as a distribution of noise to be denoised. While capable of producing visually stunning renders, they fundamentally lack an understanding of the \textit{construction} of a sketch. They generate "images of sketches," not "sketches" in the structural sense. They cannot distinguish between a stroke intended to constrain geometry and a stroke intended to shade a surface; they merely reproduce the statistical likelihood of pixel arrangements.

\subsubsection{The Necessity of Semantic Stroke Understanding}
We posit that for a Generative AI model to truly act as a co-designer, it must move beyond pixels and operate on \textit{strokes} (vectors). A true "Product Concept Sketching AI" should be able to generate sketches in the same manner a human does: by laying down \textit{Constraining Strokes} to set proportions, following with \textit{Defining Strokes} for form, and finishing with \textit{Detailing Strokes}.

To train such a sequential, stroke-based generative model (potentially based on Transformer architectures predicting the next token/stroke), one requires a massive dataset where every stroke is semantically labelled. The unlabelled "bag of strokes" currently available in vector datasets is insufficient. 

The classification pipeline presented in this paper—capable of tagging strokes with high accuracy—is the enabler for creating such a dataset. It allows us to process thousands of hours of sketching video or log data and automatically annotate the dataset with semantic labels. This annotated data will form the ground truth for training future AI architectures that understand the "grammar" of sketching, not just the "appearance."

\subsubsection{Open Science and Community Contribution}
To foster collaboration within the research community and accelerate the development of stroke-based generative models, we have open-sourced the data collection and classification platform. The \textbf{AEGIS} system, integrating the 10 real-time Machine Learning models and the 5 Deep Learning models, is accessible for public use and data contribution.

\noindent \textbf{Access Link:} \url{https://aegis-ai-src.vercel.app/}

By providing these tools, we aim to catalyze a shift in Design AI research from purely visual generation to structurally aware, cognitively aligned generative systems.

%%%%%%%%%%%%%%%%%%%%%%%%%%%%%%%%%%%%%%%%%%%%%%%%%%%%%%%%%%%%%%%%%%%%%%
\section{System Design and Architecture: DIMES and sGIT}
\label{sec:system_design}

Having established the theoretical taxonomy of strokes and the capability to classify them automatically using Deep Learning and Machine Learning, the logical progression of this research is the embodiment of these principles into a functional software environment. We introduce DIMES (\textbf{D}esign(er) \textbf{I}dea \textbf{M}anagement and \textbf{E}volution capture \textbf{S}ystem), a novel web-based sketching application. At the core of DIMES lies a custom-developed version control architecture inspired from GIT named sGIT (\textbf{S}ketching \textbf{G}round with \textbf{I}ntelligent \textbf{T}racking). 

This section elucidates the system architecture, the user interface philosophy derived from our empirical observations, and the rigorous mapping of software engineering version control primitives to the domain of creative design. Furthermore, we detail the integration of Generative AI modules that elevate the system from a passive drawing tool to an active design partner.

\subsection{Design Philosophy: The Imperative for Simplification}
The current market for digital creativity is saturated with sophisticated applications such as Adobe Photoshop, Corel Painter, and Autodesk Sketchbook. While powerful, these tools suffer from what can be termed "feature bloat." They are designed primarily for image editing and high-fidelity illustration rather than rapid conceptualisation. They offer hundreds of brushes, complex layer masking, and infinite colour palettes. 

However, our video analysis of expert designers (detailed in Section \ref{sec:understanding_pcs}) revealed a contradictory reality: the conceptual phase is remarkably minimalistic. Experts predominantly utilised a single tool—often a simple black ballpoint pen representation—and eschewed the use of erasers, layers, or colours during the initial ideation bursts. The availability of excessive options in commercial software often imposes a cognitive load, forcing the designer to make trivial decisions (e.g., "Which brush texture should I use?") rather than focusing on form generation.

Therefore, the design philosophy of DIMES is grounded in \textit{Minimalism} and \textit{Fluidity}. We architected the sketching interface to be deliberately simplistic. The application provides a curated set of tools essential for the six stroke types (Constraining, Defining, Detailing, etc.) but removes extraneous features that hinder the "flow" state. There are no complex menus to navigate; options for pen thickness, smoothing (streamlining), and pressure sensitivity are readily accessible but unobtrusive. This stripped-back interface ensures that the tool remains transparent, allowing the designer's cognitive intent to translate directly into action without interface friction.

\subsection{sGIT: A Dedicated Version Control System for Sketches}
The most significant contribution of the DIMES architecture is the integration of sGIT. As identified in our problem definition, the primary failure of existing workflows is the inability to track the non-linear evolution of ideas. While software engineers enjoy the benefits of Git—a distributed version control system developed by Linus Torvalds—designers have been bereft of such utility. 

\subsubsection{Limitations of Existing Software Git for Design}
In our preliminary work, we attempted to integrate standard Git into a design workflow. The results were suboptimal due to "Cognitive-Logistical Friction." Standard Git operates via a Command Line Interface (CLI) or text-heavy GUIs. It requires the user to manually type commands (`git add .`, `git commit -m "msg"`) and manage files in a directory. For a designer immersed in a visual-spatial task, the requirement to switch context to a textual-logical interface to save a version is jarring. It breaks the creative momentum. Consequently, designers often bypassed the versioning system entirely, reverting to the "Save As" anti-pattern. sGIT addresses this by embedding the version control logic \textit{invisibly} into the sketching interface.

\subsubsection{Semantic Mapping: From Code to Canvas}
To create sGIT, we systematically mapped the fundamental terminologies and functions of the Git software versioning system to the specific requirements of Product Concept Sketching (PCS). This mapping ensures that the powerful branching and tracking capabilities of Git are preserved but presented in a language and interaction model native to designers.

\begin{table}[ht!]
\centering
\caption{Semantic Mapping of Git Primitives to sGIT Design Functions}
\label{tab:git_mapping}
\begin{tabular}{|p{0.3\linewidth}|p{0.6\linewidth}|}
\hline
\textbf{Standard Git Command} & \textbf{sGIT / DIMES Implementation} \\ \hline
\texttt{git init} & \textbf{Repository Creation:} An sGIT repository is automatically initialised in the cloud database when a user logs in and creates a new project. \\ \hline
\texttt{git config} & \textbf{User Authentication:} Stores user credentials (username, ID) to attribute authorship to every stroke and commit. \\ \hline
\texttt{git add} + \texttt{git commit} & \textbf{Create Version (Snapshot):} A single "Save Version" interaction. It captures the stroke data (Action) and the designer's voice note (Cognition). \\ \hline
\texttt{git checkout} / \texttt{switch} & \textbf{Visual Navigation:} Clicking on a thumbnail in the version tree instantly restores the canvas to that specific historical state. \\ \hline
\texttt{git branch} & \textbf{Implicit Divergence:} If a user checks out an old version and draws a new stroke, a new branch is automatically spawned. \\ \hline
\texttt{git diff} & \textbf{Visual Comparison:} Juxtaposing two versions side-by-side or overlaying them to highlight geometric changes. \\ \hline
\texttt{git stash} & \textbf{Auto-Recovery:} Unsaved strokes are locally cached to prevent data loss in case of browser closure. \\ \hline
\texttt{git log} & \textbf{Evolutionary Tree:} A graphical node-link diagram visualising the lineage of all concepts. \\ \hline
\end{tabular}
\end{table}

\paragraph{1. Repository Initialisation (\texttt{git init} \& \texttt{git config})}
In software engineering, a developer manually initialises a repository. In DIMES, this is automated. When a designer logs in via the secure authentication module, the system configures the environment. A unique project ID is generated, acting as the repository root. All subsequent actions are tracked against this ID, with the user's profile serving as the global configuration for authorship.

\paragraph{2. The Commit Mechanism (\texttt{git commit})}
The concept of a "Commit" is central to tracking evolution. In code, a commit represents a logical unit of work. In sGIT, a commit represents a \textit{Snapshot of a Concept State}. 
Crucially, a commit in design is not just about the lines drawn (Action) but the intent behind them (Cognition). To capture this, sGIT introduces a multimodal commit system. When the designer presses the "Commit" button (or uses a gesture shortcut), the system performs two parallel operations:
\begin{enumerate}
    \item \textbf{Action Capture:} The vector data of all strokes on the canvas is serialised and stored.
    \item \textbf{Cognition Capture:} A microphone activates, allowing the designer to speak their mind freely (e.g., "I am flattening the handle to improve ergonomics"). A Text-To-Speech (TTS) algorithm converts this verbalisation into text, which is stored as the "Commit Message."
\end{enumerate}
This solves the problem of designers being unwilling to type detailed descriptions during sketching. By using voice, the cognitive load is minimised, and the "why" behind the design is preserved alongside the "what."

\paragraph{3. Branching and Non-Linear Exploration (\texttt{git branch} \& \texttt{checkout})}
Design is inherently non-linear. A designer may reach a dead-end with a cylindrical form and wish to return to an earlier cuboid iteration to try a different approach. In sGIT, this is handled through the "Version Window." This panel displays the genealogy of the design as a tree structure. 
The designer can "Checkout" any previous node. If they begin drawing on an old node, sGIT implicitly executes a branching operation. It creates a new divergence in the tree, ensuring that the original path is preserved while the new idea flourishes on a parallel track. This supports \textit{Multiplicity of Concepts}, allowing the user to oscillate between the same line of thought or diverge into multiple lines of thought without the fear of overwriting previous work.

\paragraph{4. Visual Comparison and History (\texttt{git diff} \& \texttt{git log})}
Textual diffs are useless for images. sGIT implements a visual \texttt{git diff} where two versions can be placed side-by-side. This allows the designer to compare the geometric differences between `Version 4` and `Version 7`. 
Furthermore, the \texttt{git log} is visualized not as a list of hashes, but as a "Slideshow." This feature allows the user to play back the evolution of the concept, tracing the route from the first stroke to the final commit. This essentially reconstructs the "story" of the design, which is invaluable for design reviews and educational purposes.

\subsection{The DIMES Web Application Interface}
\label{subsec:dimes_webapp}

\begin{figure*}[ht!]
    \centering
    \begin{subfigure}[b]{0.44\textwidth}
        \centering
        \includegraphics[width=\linewidth]{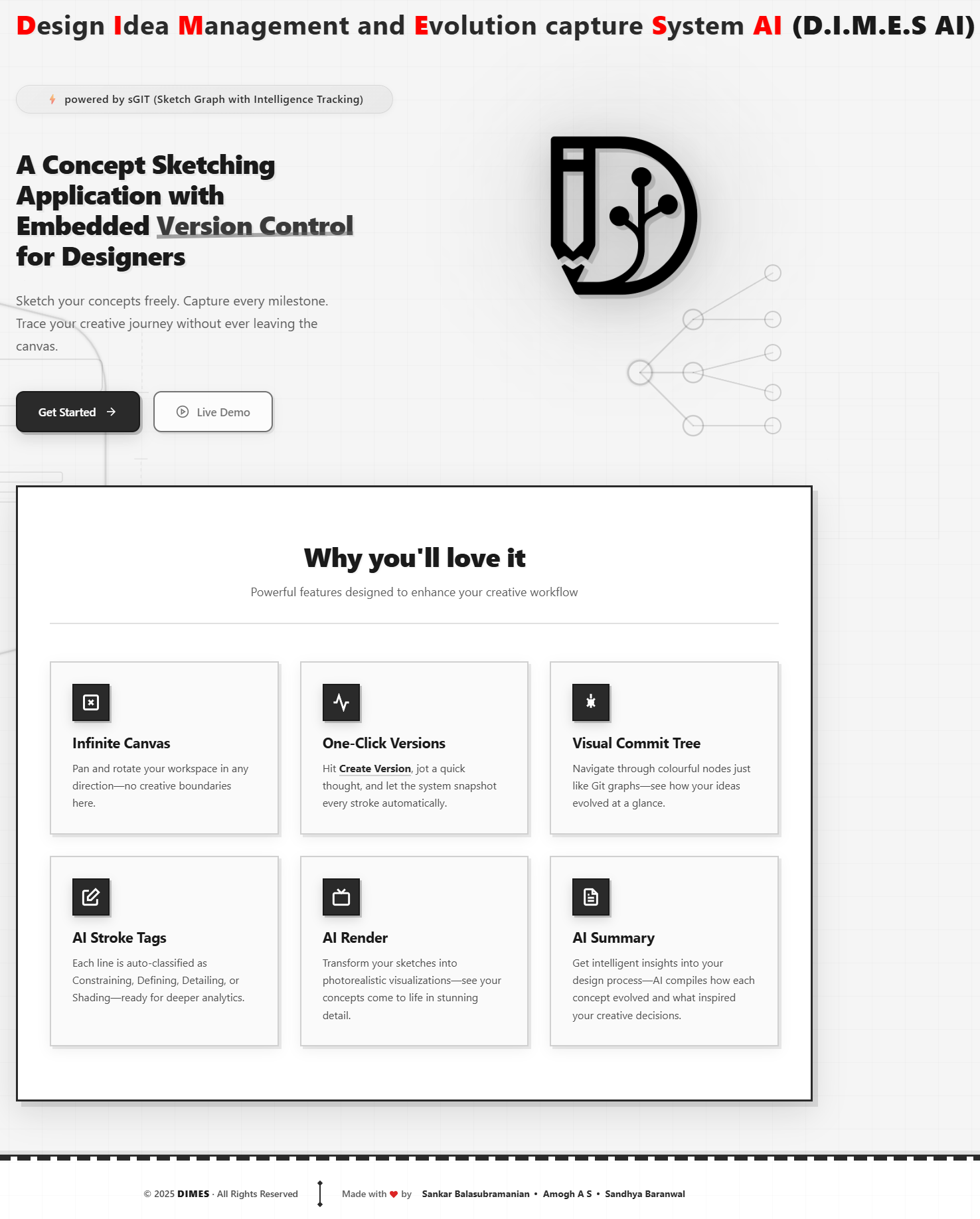}
        \caption{DIMES Landing Page}
        \label{subfig:dimes_home}
    \end{subfigure}
    \hfill
    \begin{subfigure}[b]{0.48\textwidth}
        \centering
        \includegraphics[width=\linewidth]{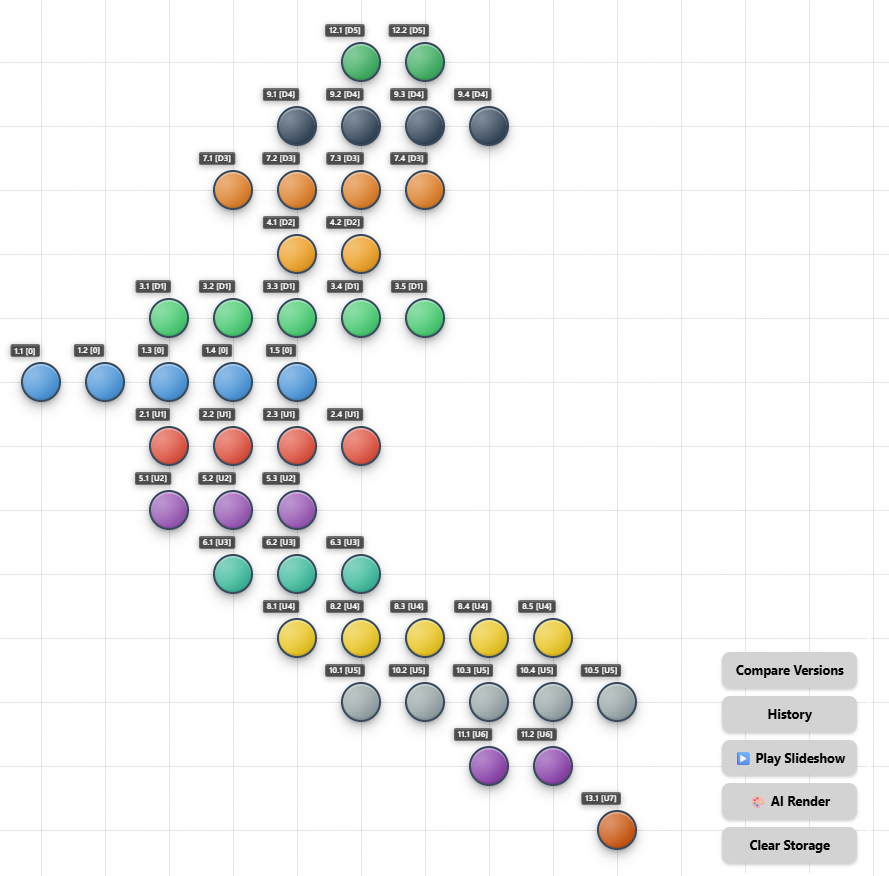}
        \caption{SGIT Version Tree}
        \label{subfig:version_tree}
    \end{subfigure}
    
    \caption{The User Interface components of the DIMES ecosystem. (a) The landing page of the DIMES platform. It introduces the core functionality of the system: a specialized sketching application with an embedded Version Control System (VCS) designed to track creative evolution without disrupting the design workflow. (b)The SGIT Version Tree visualization. Each circular node represents a distinct commit (snapshot), while the connecting lines illustrate the genealogical relationships. Diverging paths represent branching concepts, allowing designers to visualize their non-linear exploration structure at a glance.}
    \label{fig:dimes_ui_overview}
\end{figure*}

\begin{figure}[ht!]
    \centering
    \includegraphics[width=\linewidth]{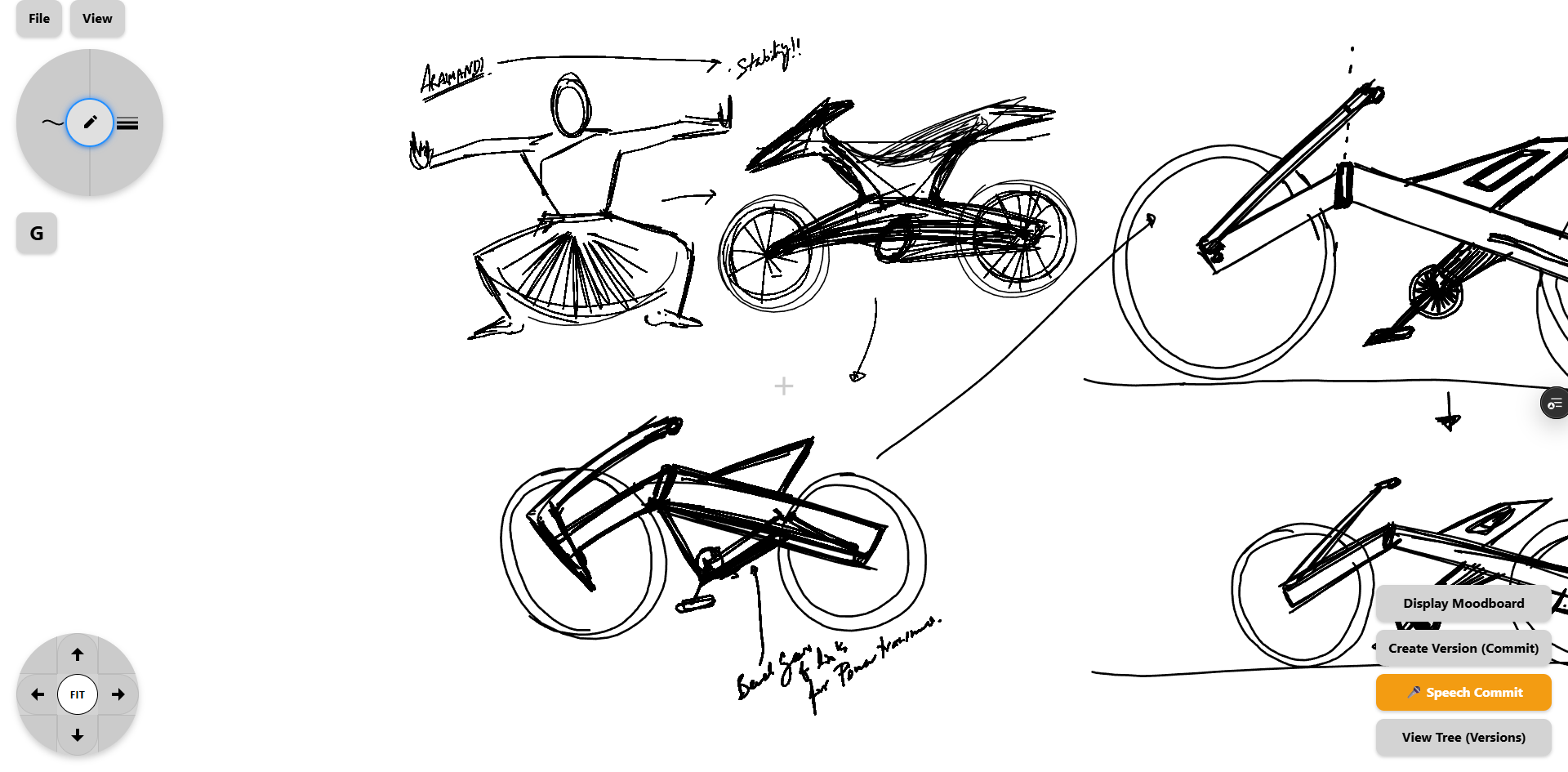}
    \caption{The main sketching workspace featuring an infinite canvas. The interface is minimalistic to reduce cognitive load, offering essential tools for stroke creation. Key features include the 'Speech Commit' button for capturing design intent via voice and the navigation controls to switch between the canvas and the version tree.}
    \label{fig:sketching_interface}
\end{figure}

The DIMES ecosystem is realized as a cloud-native web application, engineered to provide a ubiquitous and responsive sketching environment. It was developed using the ReactJS framework to ensure low-latency interaction that is important for fluid digital sketching and the system is deployed on the Vercel platform and is publicly accessible at \url{https://dimes-ai.vercel.app/}. The user interface is strictly architected to minimize cognitive friction, centering on a minimalist infinite canvas (Figure~\ref{fig:sketching_interface}) where the SGIT versioning logic is embedded unobtrusively. Designers can seamlessly toggle between the active workspace and the genealogical Version Tree (Figure~\ref{subfig:version_tree}) to manage their branching exploration.

\begin{figure*}[ht!]
    \centering
    % Row 1
    \begin{subfigure}[b]{0.48\textwidth}
        \centering
        \includegraphics[width=\linewidth]{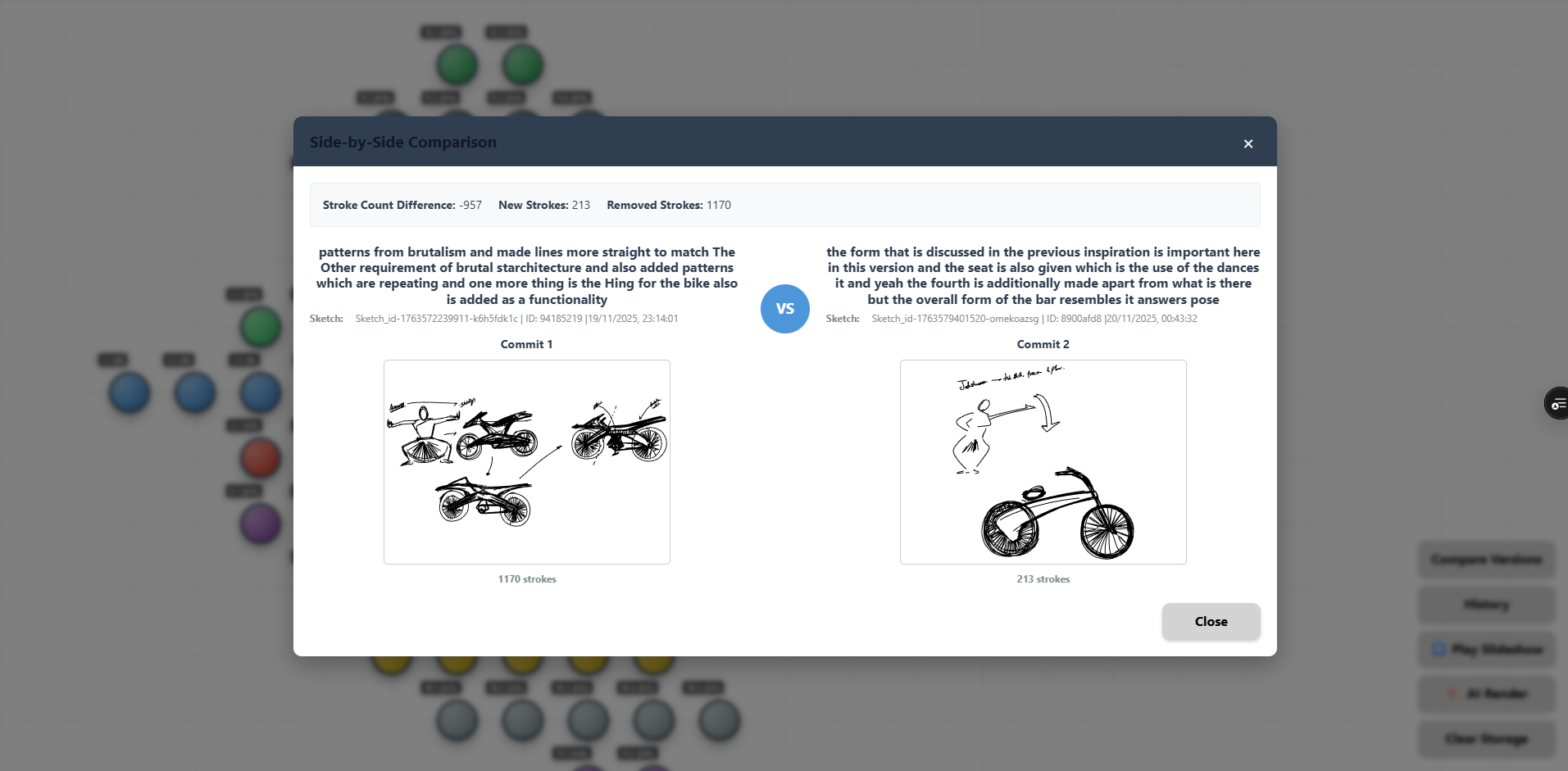}
        \caption{The Visual Diff interface (Comparison View). This modal allows the designer to compare any two selected versions side-by-side. The system calculates and displays metrics such as stroke count difference, highlighting the geometric evolution between two commits}
        \label{subfig:visual_diff}
    \end{subfigure}
    \hfill
    \begin{subfigure}[b]{0.48\textwidth}
        \centering
        \includegraphics[width=\linewidth]{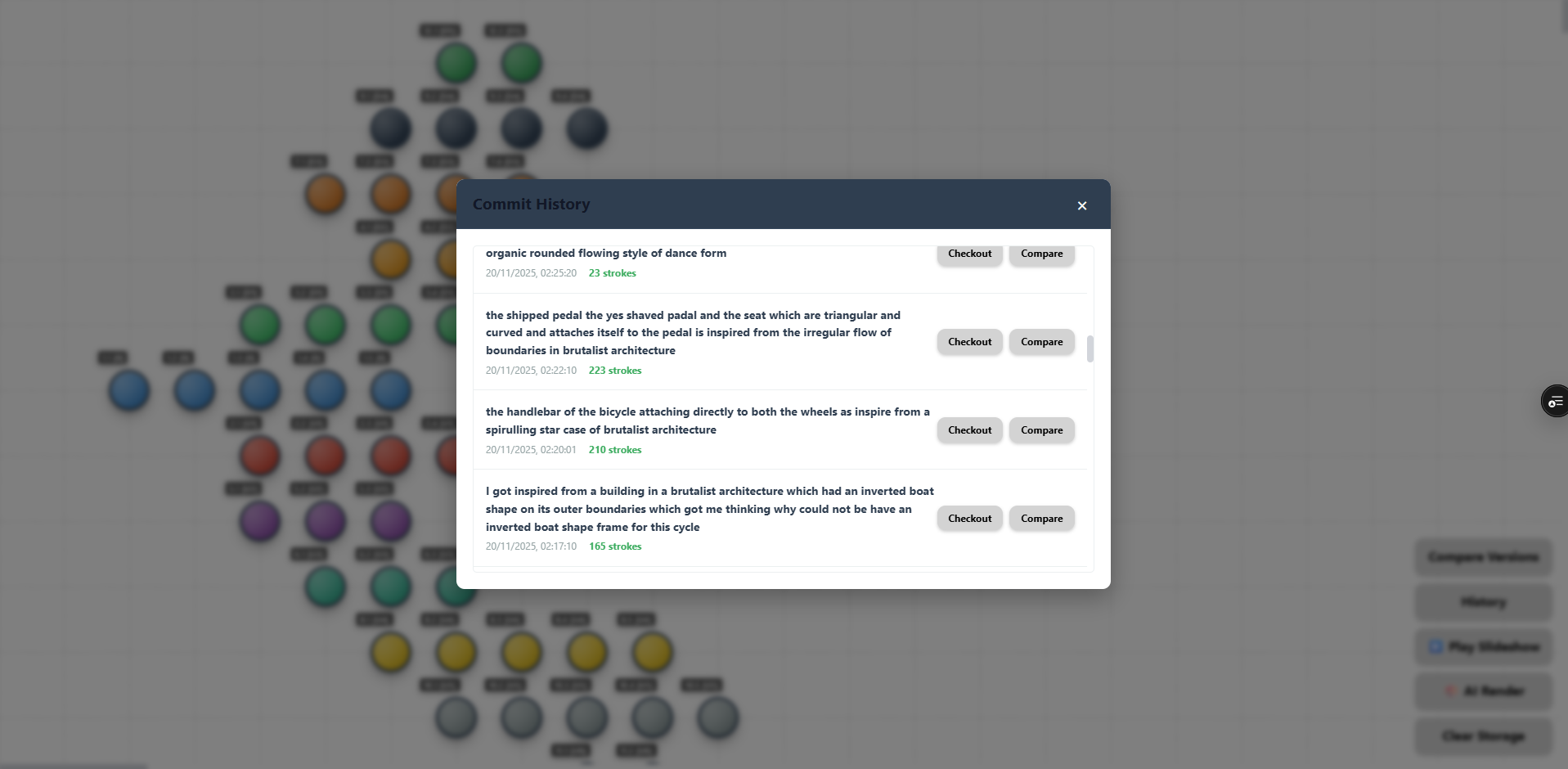}
        \caption{The Commit History Log. This panel displays a chronological list of all saved versions. Each entry includes the timestamp, stroke count, and the transcribed text from the designer's voice note, effectively capturing the rationale behind each iteration.}
        \label{subfig:commit_history}
    \end{subfigure}
    
    \vspace{0.5cm} % Vertical spacing between rows
    
    % Row 2
    \begin{subfigure}[b]{0.48\textwidth}
        \centering
        \includegraphics[width=\linewidth]{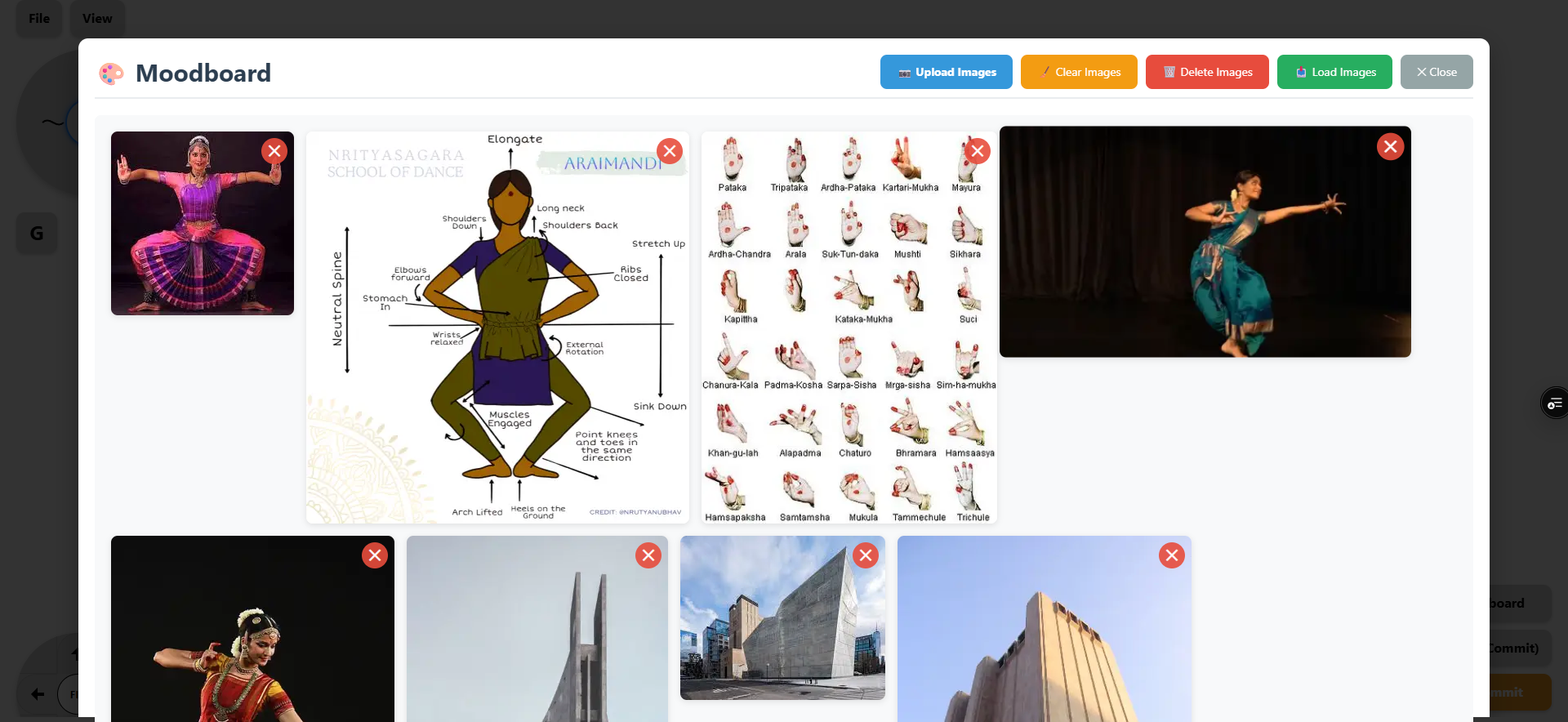}
        \caption{The integrated Moodboard module. Designers can upload, resize, and arrange reference images (e.g., Indian dance postures and Brutalist architectural forms) to maintain visual inspiration directly within the sketching environment.}
        \label{subfig:moodboard}
    \end{subfigure}
    \hfill
    \begin{subfigure}[b]{0.48\textwidth}
        \centering
        \includegraphics[width=\linewidth]{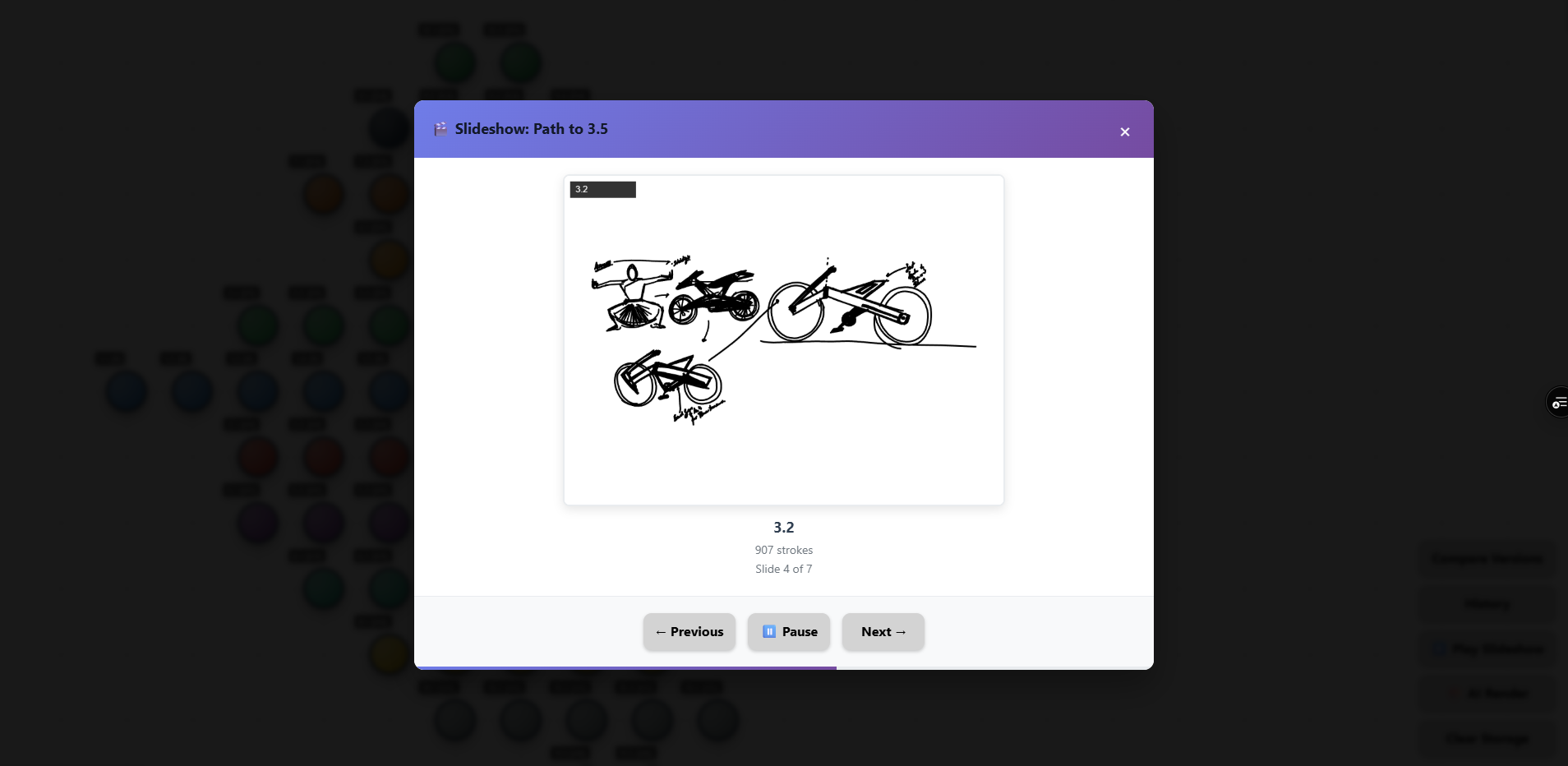}
        \caption{The Slideshow Playback feature. This tool reconstructs the design journey by playing back the sequence of commits. It enables designers and reviewers to trace the path of a concept's evolution from the initial stroke to the final form.}
        \label{subfig:slideshow}
    \end{subfigure}
    
    \caption{Advanced functional modules of the DIMES system designed to support the cognitive aspects of design. The figure displays (a) the Visual Diff tool for geometric comparison, (b) the Commit History log for rationale tracking, (c) the Moodboard for inspiration management, and (d) the Slideshow tool for evolutionary review.}
    \label{fig:dimes_advanced_features}
\end{figure*}

\begin{figure*}[ht!]
    \centering
    \begin{subfigure}[b]{0.49\textwidth}
        \centering
        \includegraphics[width=\linewidth]{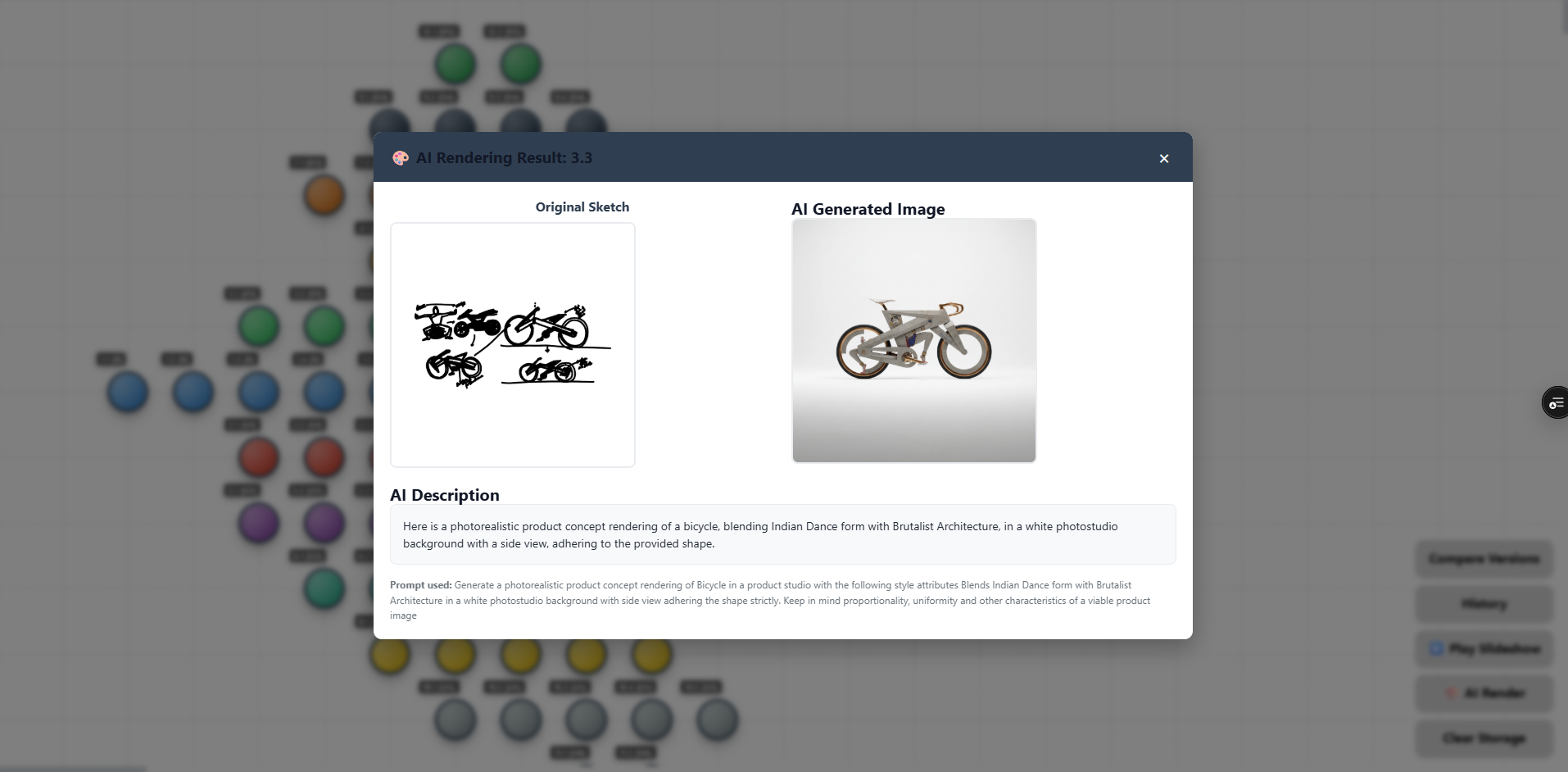}
        \caption{The AI Rendering module. The figure shows the transformation of a raw sketch (left) into a photorealistic product visualization (right). The system uses the sketch as a structural guide and a text prompt to apply material and lighting attributes via Generative AI.}
        \label{subfig:ai_render}
    \end{subfigure}
    \hfill
    \begin{subfigure}[b]{0.49\textwidth}
        \centering
        \includegraphics[width=\linewidth]{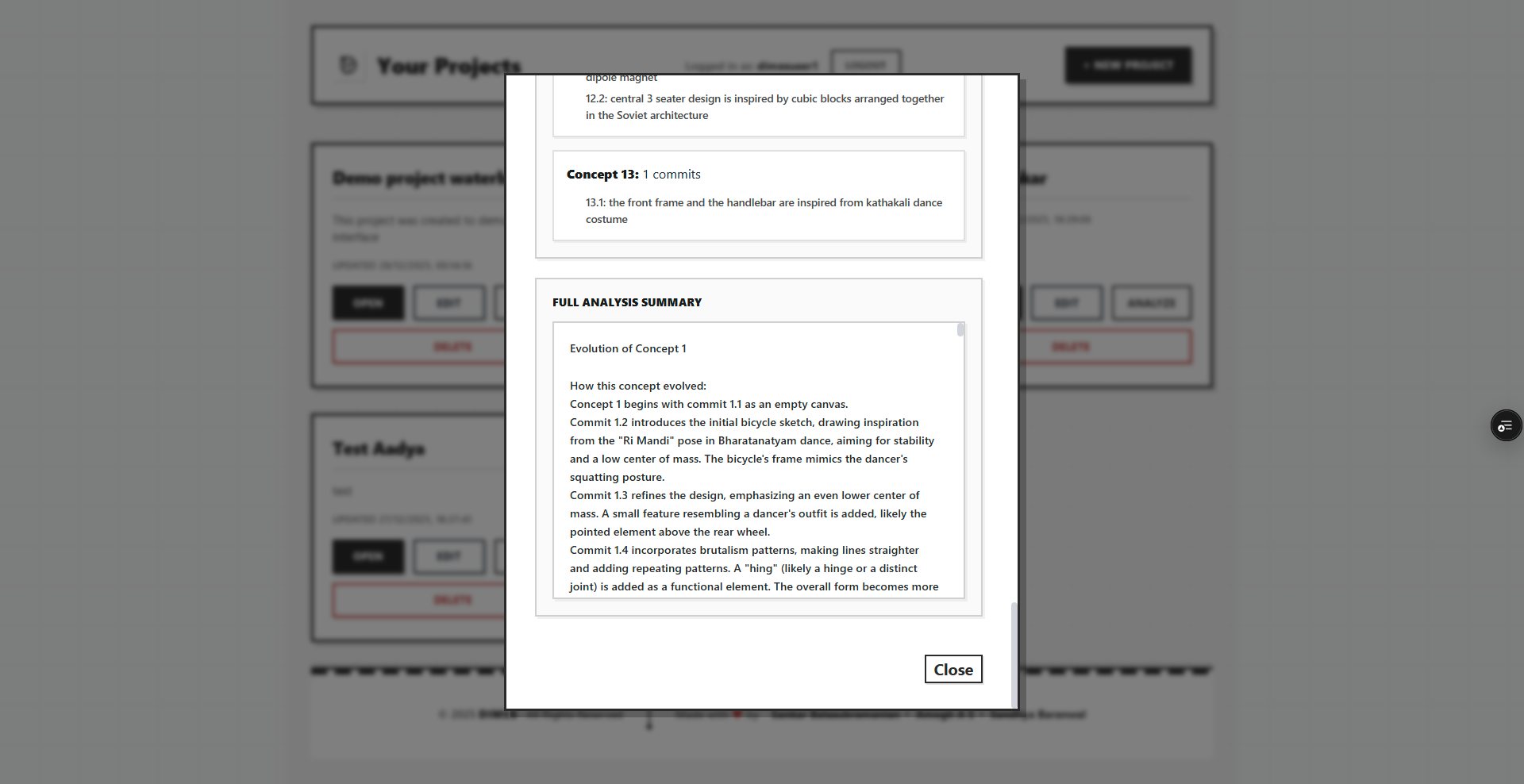}
        \caption{The Automated AI Narrative Summary. By synthesizing data from the commit logs, timestamps, and voice notes, the system uses an LLM to generate a coherent textual story describing the evolutionary logic and design decisions for each concept.}
        \label{subfig:ai_summary}
    \end{subfigure}
    
    \caption{Generative AI capabilities integrated into the DIMES platform. (a) AI-driven visualization transforming vector strokes into photorealistic renders. (b) Large Language Model (LLM) integration for automated documentation and narrative generation of the design process.}
    \label{fig:ai_integration}
\end{figure*}

Beyond the core sketching and versioning capabilities, the application integrates a comprehensive suite of cognitive support tools visualized in Figure~\ref{fig:dimes_advanced_features}. This includes a Moodboard module (Figure~\ref{subfig:moodboard}) that allows designers to curate visual references alongside their concepts, ensuring that inspiration remains contextual. To support analytical reflection, the system features a Visual Diff tool (Figure~\ref{subfig:visual_diff}) for geometric comparison of branches and a Commit History log (Figure~\ref{subfig:commit_history}) that captures multi-modal rationale (voice and text). Furthermore, a Slideshow playback feature (Figure~\ref{subfig:slideshow}) enables the reconstruction of the entire design journey for review. Finally, the platform's intelligence is manifested through the AI modules depicted in Figure~\ref{fig:ai_integration}, which provide real-time photorealistic AI Rendering (Figure~\ref{subfig:ai_render}) and automated AI Narrative Summaries (Figure~\ref{subfig:ai_summary}), transforming the tool from a passive canvas into an active design partner.

\subsection{Cloud Infrastructure and Data Management}
The backend architecture of DIMES is built on a robust cloud-native stack to ensure scalability and data persistence. We utilize \textbf{Supabase}, a Firebase alternative, as the primary backend service. 

The data storage strategy is hybrid. The metadata (user profiles, project structures, commit messages, branching logic) is stored in a PostgreSQL relational database. However, the sketch data itself—comprising thousands of vector points—is stored as JSON blobs within the database or, for larger raster assets, in an associated object storage bucket. 
This architecture ensures that the system is device-agnostic; a designer can begin a sketch on a tablet in the studio and review the version history on a desktop at home, with the sGIT logic maintaining consistency across sessions.

\subsection{Intelligence Integration: From Tool to Partner}
While sGIT manages the data, the "Intelligence" in DIMES elevates the system to a cognitive partner. We integrate state-of-the-art Generative AI models to solve two specific problems: the visualization of form vs. texture, and the articulation of the design narrative.

\subsubsection{AI-Powered Photorealistic Rendering}
As noted in our theoretical framework, PCS is often monochromatic (Action dominant). Designers focus on shape and form, leaving colour and material texture (Texture dominant) for later stages, usually requiring complex rendering software like KeyShot or Blender. 
DIMES bridges this gap using Generative AI. We integrated the \textbf{Gemini Nano Banana API} to enable an "AI Render" feature. The designer can select any specific commit (a black and white sketch) and provide a textual prompt describing the desired style (e.g., "Matte black finish with brushed aluminium accents, studio lighting"). The system uses an Image-to-Image generation pipeline where the sketch serves as the structural condition (ControlNet-like functionality) and the text prompt provides the stylistic attributes. The AI generates a photorealistic rendering of the concept in seconds. 
This allows the designer to validate the visual impact of a form concept without leaving the sketching environment or spending hours on CAD modelling. It separates the cognitive load: the human focuses on the \textit{geometry}, and the AI handles the \textit{physics of light and material}.

\subsubsection{Automated Narrative Summarisation}
The second intelligence feature addresses the documentation gap. A complex design project might involve hundreds of commits and dozens of branches. Manually writing a report on this evolution is tedious.
DIMES utilizes a Large Language Model (LLM)—specifically \textbf{Google Gemini}—to generate an "AI Summary." The system aggregates the commit messages (captured via TTS), the timestamps, and the branching points. It feeds this structured chronology into the LLM with a system prompt designed to interpret design evolution.
The output is a coherent narrative story explaining \textit{how} the concept evolved. For example: "The designer initially explored a cylindrical form (Commit 1-5) but encountered packaging constraints. At Commit 6, they pivoted to a boxy aesthetic to maximize internal volume..." 
This automated storytelling serves two purposes:
\begin{enumerate}
    \item \textbf{Knowledge Transfer:} It helps novice designers learn by reading the "thought process" of experts.
    \item \textbf{Design Defence:} It provides the designer with a ready-made narrative to defend their design decisions to stakeholders, proving that alternatives were considered and logically rejected.
\end{enumerate}

% TODO: Insert Figure 6: System Architecture Diagram showing the flow of data from the Sketching Interface -> sGIT Logic Module -> Supabase Cloud -> AI Services (Gemini/Banana) -> Feedback to User.

\subsection{Summary of the Integrated Ecosystem}
In conclusion, the DIMES system represents a paradigm shift from static image creation to dynamic evolutionary tracking. By marrying a simplified sketching interface with the rigorous logic of sGIT and the generative capabilities of modern AI, we have created a holistic environment. It respects the "Action" of the hand through low-latency sketching, supports the "Cognition" of the mind through voice-annotated versioning, and augments the "Creation" of the output through AI rendering and summarization. This architecture provides the necessary technological foundation to conduct the experimental studies detailed in the following sections.

%%%%%%%%%%%%%%%%%%%%%%%%%%%%%%%%%%%%%%%%%%%%%%%%%%%%%%%%%%%%%%%%%%%%%%
\section{Experimental Methodology}
\label{sec:experimental_study}

To validate the efficacy of the proposed SGIT architecture and the DIMES ecosystem, a rigorous comparative experimental study was designed. This study aims to empirically demonstrate the system's ability to capture the rich temporal evolution of concept sketching, facilitate diverse exploration, and ultimately enhance the knowledge transfer process between designers. The methodology employs a mixed-methods approach, combining quantitative metrics derived from the SGIT log files with qualitative data gathered from expert evaluations and user acceptance testing.

\subsection{Research Question and Hypotheses}

The overarching research question guiding this investigation is:

\begin{quote}
\textbf{Research Question (RQ):} Does the integrated SGIT-DIMES version control system based product concept sketching application, through its automated tracking and AI-based narrative generation capabilities, significantly improve the capture of design evolution and the transfer of design intent compared to traditional manual documentation methods?
\end{quote}

Based on the RQ and the core functionality of DIMES, we formulated the following four hypotheses:

\subsubsection{Research Hypotheses}
\begin{enumerate}
    \item[$\mathbf{H_1}$:] DIMES, by providing frictionless version control, will lead to a \textit{greater diversity and depth of concept exploration} in expert designers compared to the traditional pen-and-paper method, as measured by the Concept Pyramid metrics.
    \item[$\mathbf{H_2}$:] The multi-modal data capture of DIMES (strokes + TTS commits) will enable \textit{more comprehensive and objective documentation of the designer's thought process} during conceptualization. This will be evidenced by a significantly higher Information Content Density in the AI-generated summaries compared to manual post-hoc summaries, and verified by designer validation.
    \item[$\mathbf{H_3}$:] The AI-generated summary provided by DIMES will facilitate a \textit{higher fidelity of knowledge transfer} to novice designers. This will be quantified using a novel Neural Transparency-based metric, resulting in replicated sketches that are assessed as significantly more similar to the original expert concepts than those replicated using manually written summaries.
    \item[$\mathbf{H_4}$:] The AI-generated photorealistic Product Concept Renderings (PCR) produced by the DIMES ecosystem will enable superior visualization of the product intent for end-users, resulting in significantly higher \textit{purchase likelihood scores} compared to traditional manual markers.
\end{enumerate}

\subsection{Study Design and Variables}

The study adopts a $2 \times 2$ comparative design, focusing on two distinct phases: \textbf{Concept Generation} (Phase I) and \textbf{Knowledge Replication} (Phase II). The primary independent variable is the \textbf{Documentation Method}, which has two levels:
\begin{enumerate}
    \item \textbf{Manual via Traditional:} Pen and paper generation, documented by a designer's retrospective written summary.
    \item \textbf{Automated via DIMES:} Digital generation using SGIT, documented by an AI-generated narrative summary.
\end{enumerate}
The dependent variables span quantitative measures of process data (commits, branches, time) and qualitative measures of outcome performance (similarity scores, user acceptance).

\subsubsection{The Comparative Design Framework}
The experiment involves four distinct groups of participants, designed to test the comparative efficacy of the documentation output:
\begin{itemize}
    \item \textbf{D1 (Control Group - Generator):} Uses the traditional pen-and-paper method and creates a \textit{Human-Authored Summary}.
    \item \textbf{D2 (Treatment Group - Generator):} Uses the DIMES system with SGIT and creates an \textit{AI-Authored Summary}.
    \item \textbf{D3 (Control Group - Replicator):} Receives the \textit{Human-Authored Summary} (Text only, no visuals) of D1 and attempts to replicate the concepts.
    \item \textbf{D4 (Treatment Group - Replicator):} Receives the \textit{AI-Authored Summary} (Text only, no visuals) of D2 and attempts to replicate the concepts.
\end{itemize}
This structure isolates the effect of the documentation method on knowledge transfer ($\mathbf{H_3}$) by comparing the replication fidelity achieved by D3 versus D4.

\subsection{Participant Selection and Ethical Considerations}
Participant integrity and experience levels were crucial to the study's validity, requiring a clear distinction between the expertise of the generators (D1, D2) and the learning capability of the replicators (D3, D4).

\subsubsection{Expert Designers (D1 and D2)}
\begin{itemize}
    \item \textbf{Criteria:} Participants D1 and D2 were selected based on extensive professional experience in product design. They are required to be industry experts with a minimum of 10 years of professional experience in industrial design or product concept sketching. This ensures that the generated concepts represent a high standard of creativity and technical feasibility.
    \item \textbf{Role:} Concept Generators. They are responsible for producing a diverse set of concept sketches and their corresponding documentation.
\end{itemize}

\subsubsection{Novice Designers (D3 and D4)}
\begin{itemize}
    \item \textbf{Criteria:} Participants D3 and D4 were selected from a pool of Master's level design students. They are categorized as novice budding designers. They possess fundamental skills in Product Concept Sketching (PCS) but lack the extensive practical experience and deep cognitive mapping of industry experts.
    \item \textbf{Role:} Concept Replicators. Their task is to decode the provided documentation and replicate the original concepts based on the narrative summary alone.
\end{itemize}

\subsubsection{Ethical Protocol}
All participants were informed of the study's purpose, their required commitment, and the methods of data collection (including voice recording for D2). Written informed consent was obtained from all participants. Anonymity was ensured by assigning codes (D1-D4). The audio recordings for D2 were transcribed immediately for analysis and then deleted to protect privacy. The study was reviewed and approved by the institutional ethics review board.

\subsection{Experimental Protocol}

\subsubsection{Task Definition: "Dance $\times$ Brutalism" Bicycle}
To rigorously test the system's capacity for capturing complex, non-linear form evolution, we devised a challenging and unconventional design brief. Expert participants (D1 and D2) were tasked with the following:

\begin{quote}
\textbf{Brief:} Design a non-traditional bicycle inspired by the intersection of \textbf{Indian Dance Forms} (movement, rhythm, posture, costume geometry) and \textbf{Brutalist Architecture} (mass, monolithic volumes, raw materiality, structural clarity). The bicycle must function (ride, steer, brake, power) but need not adhere to conventional tubular archetypes.
\end{quote}

The participants were provided with a specific set of stylistic cues to translate into geometry:
\begin{enumerate}
    \item \textbf{Indian Dance Form Cues (Movement $\rightarrow$ Geometry):}
    \begin{itemize}
        \item \textit{Bharatanatyam:} Aramandi demi-plié stance (diamond-like negative space) $\rightarrow$ faceted joints; angular elbows $\rightarrow$ stacked planes.
        \item \textit{Odissi:} Tribhangi (three-bend S-curve) $\rightarrow$ asymmetrical frame spine; silver filigree $\rightarrow$ perforated panels.
        \item \textit{Kathak:} Fast tatkār footwork $\rightarrow$ modular ribs/fins (rhythm mapping); swirling lehenga $\rightarrow$ flared guards.
        \item \textit{Kathakali/Chhau:} Layered headgear $\rightarrow$ stacked disc volumes for hubs.
        \item \textit{Mudras:} Finger splay geometry $\rightarrow$ handle interfaces and sculpted brake levers.
    \end{itemize}
    \item \textbf{Brutalist Cues (German/Soviet Emphasis):}
    \begin{itemize}
        \item \textit{Monolithic Massing:} Boxy beams and blocky head modules instead of skinny tubes.
        \item \textit{Raw Materiality:} Cast/forged textures, brushed metal, and visible fasteners.
        \item \textit{Repetition \& Module Logic:} Waffle grids, ribbed fins, and stacked plates.
        \item \textit{Cantilever \& Overhang:} Structural gestures, such as a cantilevered slab seat.
    \end{itemize}
\end{enumerate}

\subsubsection{Phase I: Concept Generation and Documentation (D1 vs. D2)}
\begin{itemize}
    \item \textbf{D1 (Control - Manual Generation):}
        \begin{enumerate}
            \item D1 was provided with traditional tools: paper, fine-liner pens, and alcohol markers for rendering.
            \item D1 was given a strict time limit of $\mathbf{2 \text{ hours}}$ for conceptualization.
            \item Post-sketching, D1 was instructed to spend an additional $\mathbf{30 \text{ minutes}}$ writing a detailed, introspective summary documenting the evolution of each concept, their thought process behind decisions, and the alternatives considered (The \textit{Human Summary}).
            \item The final PCS was converted into a Product Concept Render (PCR) using the alcohol markers.
        \end{enumerate}
    \item \textbf{D2 (Treatment - DIMES/SGIT Generation):}
        \begin{enumerate}
            \item D2 used the DIMES system with the integrated SGIT and the AI rendering module.
            \item D2 was also given a strict time limit of $\mathbf{2 \text{ hours}}$ for conceptualization.
            \item During the session, D2 was instructed to use the voice-based commit feature (TTS) whenever a significant decision or change of direction was made.
            \item Upon conclusion, the DIMES system automatically generated the PCRs using its AI rendering engine and produced the comprehensive \textit{AI-Authored Summary} using the LLM based on the SGIT log and TTS data.
        \end{enumerate}
\end{itemize}

\subsubsection{Phase II: Knowledge Transfer and Replication (D3 vs. D4)}
This phase tested the quality of the generated documentation:
\begin{itemize}
    \item \textbf{D3 (Replication from Human Summary):}
        \begin{enumerate}
            \item D3 received only the \textit{Human Summary} created by D1 (text-only format). Crucially, D3 was NOT shown any of D1's original sketches.
            \item D3 was tasked with replicating the concepts based solely on the textual description. D3 used a standard digital sketching app without SGIT.
        \end{enumerate}
    \item \textbf{D4 (Replication from AI Summary):}
        \begin{enumerate}
            \item D4 received only the \textit{AI-Authored Summary} generated by DIMES for D2 (text-only format). D4 was NOT shown any of D2's original sketches.
            \item D4 was tasked with replicating the concepts based solely on the textual description. D4 also used a standard digital sketching app without SGIT.
        \end{enumerate}
    \item \textbf{Time Constraint:} No time limit was imposed on D3 and D4. The goal was maximal accuracy and fidelity of replication, not speed.
\end{itemize}

\subsection{Data Collection and Outcome Metrics}
To systematically evaluate the hypotheses, we defined four distinct metrics categories corresponding to $H_1, H_2, H_3,$ and $H_4$.

\subsubsection{Quantitative Process Metrics (Testing $\mathbf{H_1}$)}
These metrics are automatically logged by the SGIT system for D2 and manually inferred (e.g., via time-stamping video recording and photo-logging) for D1's manual process.
\begin{enumerate}
    \item \textbf{Number of Branches:} The count of distinct lines of thought or concept directions explored. This measures \textit{concept divergence}.
    \item \textbf{Number of Commits (Snapshots):} The count of times the designer created a version. This measures the \textit{granularity of tracking} and the designer's internal assessment frequency.
    \item \textbf{Total Time Taken:} The duration of the conceptualization phase (2 hours for D1 and D2).
    \item \textbf{Number of Concepts:} The count of distinct, final concepts proposed.
    \item \textbf{Number of Intermediate Stages:} The average number of commits or stages within a single branch/concept before resolution.
\end{enumerate}

To quantify the quality of exploration, we utilize the \textbf{Concept Pyramid} (CP) model:
$$
\text{CP Width (Breadth)} \propto \text{Number of Branches}
$$
$$
\text{CP Height (Depth)} \propto \text{Avg No. of Commits per Branch}
$$
A design process that is superior is expected to yield a wider and taller pyramid, signifying a process that is both diverse in initial idea exploration (width) and persistent in the detailed refinement of those ideas (height). We predict that D2 (DIMES) will yield a significantly larger Concept Pyramid volume than D1 (Manual).

\subsubsection{Information Content Density Analysis (Testing $\mathbf{H_2}$)}
To verify the superior documentation capability of the multi-modal DIMES system, we performed a comparative text analysis between the \textit{Human Summary} (D1) and the \textit{AI Summary} (D2). We calculated the \textbf{Information Content Density (ICD)} metric, defined as the ratio of objective, verifiable design events to the total word count.
\begin{itemize}
    \item \textbf{Lexical Richness:} Use of domain-specific terminology (e.g., "chamfer," "fillet," "cantilever") and stroke taxonomy terms (e.g., "defining stroke," "hatching").
    \item \textbf{Temporal Precision:} The frequency of chronological markers (e.g., "at minute 45," "after the third iteration") that accurately map to the actual design timeline.
    \item \textbf{Designer Validation:} Post-session, Designer D2 was asked to review the AI-generated summary and rate its accuracy in capturing their cognitive intent on a 5-point Likert scale.
\end{itemize}

\subsubsection{Similarity Analysis using Neural Transparency (Testing $\mathbf{H_3}$)}
To objectively measure the fidelity of knowledge transfer between the generator and the replicator, we moved beyond subjective expert ratings. We employed a novel \textbf{Neural Transparency-based Similarity Metric}.
This method utilizes the internal activations of a Vision-Language Model (LLaVA-NeXT with Llama-3-8B) to quantify conceptual similarity. By extracting the \textbf{Sketch Activation Matrices} from the model's final transformer layer (Layer 31), we capture the deep semantic representation of the sketch, invariant to minor stylistic differences but sensitive to design intent. We computed the \textbf{Cosine Similarity} between the activation vectors of the original sketches (D1/D2) and their corresponding replications (D3/D4). A higher cosine similarity score indicates a higher fidelity of semantic replication. (Detailed technical implementation is provided in the subsequent Section~\ref{sec:neural_transparency}).

\subsubsection{User Acceptance Rating (Testing $\mathbf{H_4}$)}
To evaluate the impact of the AI rendering on end-user perception, we conducted a User Acceptance Test. The final Product Concept Renders (PCRs) produced by D1 (manual markers) and D2 (AI Render) were presented to a panel of potential users ($N=30$).
\begin{itemize}
    \item \textbf{Metric:} \textbf{Purchase Likelihood Score}. Users were asked to rate "How likely would you be to purchase this product based on the visual appeal and clarity of the concept?" on a 5-point Likert scale (1 = Very Unlikely, 5 = Very Likely).
\end{itemize}

% TODO: Insert Table 4: Summary Table of Outcome Metrics mapped to Hypotheses (H1: Concept Pyramid, H2: ICD, H3: Neural Similarity, H4: Purchase Likelihood).

The structured protocol ensures that the data collected allows for a direct, statistically justifiable comparison between the traditional manual documentation and the novel automated tracking and summarization enabled by the SGIT-DIMES system.

%%%%%%%%%%%%%%%%%%%%%%%%%%%%%%%%%%%%%%%%%%%%%%%%%%%%%%%%%%%%%%%%%%%%%%

\section{Neural Transparency-Based Similarity Rating for Product Concept Sketches}
\label{sec:neural_transparency}

Evaluating the fidelity with which a designer replicates another's product concept sketch presents a unique challenge in design research. Unlike the comparison of natural photographs, where pixel-level metrics such as Structural Similarity Index (SSIM) or Peak Signal-to-Noise Ratio (PSNR) may suffice, sketches are abstract representations. They convey design intent, conceptual relationships, and creative interpretation rather than mere visual duplication. A skilled human expert, when evaluating sketch similarity, considers conceptual fidelity, structural coherence, and semantic understanding—factors that traditional computer vision metrics notoriously fail to capture.

To address this, we introduce a novel, mechanistic interpretability-based approach to quantify sketch similarity. This method, termed \textbf{Neural Transparency}, involves extracting and comparing the internal representations (activations) of a Vision-Language Model (VLM) as it processes each sketch. By leveraging the deep semantic understanding embedded within the neural network's layers, this technique emulates the cognitive process of an expert evaluator.

\subsection{Theoretical Foundation: Mechanistic Interpretability}
\label{subsec:theoretical_foundation}

The concept of \textbf{Neural Transparency} is rooted in the field of Mechanistic Interpretability, which seeks to reverse-engineer neural networks to understand the computations they perform \cite{karny2025neuraltransparencymechanisticinterpretability}. The foundational principle posits that the internal activations of a neural network encode rich semantic representations of the input data. By extracting these activations, we gain access to how the model "perceives" and "understands" the input at various levels of abstraction.

Traditional image similarity methods typically operate at the pixel level:
\begin{itemize}
    \item \textbf{SSIM:} Measures structural similarity in luminance, contrast, and structure but fails with stylistic variations or stroke thickness differences.
    \item \textbf{Perceptual Hash:} A fuzzy hash of a downsampled image, which cannot capture semantic meaning.
    \item \textbf{Pixel MSE:} The Mean Squared Error of pixel values, which is highly sensitive to minor translations or rotations.
\end{itemize}

In contrast, \textbf{Neural Activation-based Similarity} addresses these limitations by capturing the \textit{semantic understanding} of what is depicted, rather than just how it appears. It relies on the hierarchical nature of deep learning models, where early layers capture low-level features (edges, textures) and later layers abstract these into high-level concepts and design intent. This mirrors the human expert's cognitive process of parsing visual elements, analyzing structure, and finally mapping conceptual meaning.

\subsection{Model Architecture: LLaVA-NeXT with Llama-3-8B}
\label{subsec:model_architecture}

For this analysis, we employed \textbf{LLaVA-NeXT} (Large Language-and-Vision Assistant), a state-of-the-art Vision-Language Model built upon the \textbf{Llama-3-8B} language model backbone (\texttt{llava-hf/llama3-llava-next-8b-hf}). The architecture comprises three primary components:

\begin{enumerate}
    \item \textbf{Vision Encoder (SigLIP):} A Vision Transformer (ViT-L/14) that processes the raw input image into a sequence of 1024-dimensional visual tokens. It divides the image into $14 \times 14$ patches, producing a typical output shape of $(577, 1024)$.
    \item \textbf{Multi-Modal Projector:} A projection layer that bridges the visual encoder and the language model, mapping the 1024-dimensional visual features into the 4096-dimensional embedding space of the LLM.
    \item \textbf{Llama-3-8B Language Model:} The core processing unit, consisting of 32 Transformer layers (indexed 0 to 31). Each layer operates on a hidden dimension of 4096. This component utilizes Grouped Query Attention (GQA), RMSNorm, SwiGLU activation, and Rotary Position Embeddings (RoPE).
\end{enumerate}

To ensure consistent activation extraction across all sketches, we utilized a \textbf{Neutral Prompt}: \textit{"Describe this product concept sketch of a futuristic bicycle that combines the Indian dance forms with brutalist architecture."} This prompt contextualizes the image as a product concept and activates the relevant conceptual pathways within the model without introducing bias.

\subsection{Methodology: Sketch Activation Matrix Extraction}
\label{subsec:activation_extraction}

A \textbf{Sketch Activation Matrix} is defined as the internal hidden state of a neural network layer as it processes a specific sketch image. Let $I_k$ be the $k$-th sketch image and $f_\ell(\cdot)$ be the forward function of layer $\ell$. The Sketch Activation Matrix $\mathbf{A}_k^{(\ell)} \in \mathbb{R}^{S \times D}$ for image $k$ at layer $\ell$ is given by:
\begin{equation}
    \mathbf{A}_k^{(\ell)} = f_\ell(I_k, \text{prompt})
\end{equation}
Where $S$ represents the sequence length (approximately 577 tokens, comprising visual patches and text tokens) and $D$ represents the hidden dimension (4096). Each row of this matrix represents the model's encoding of a specific token at that layer.

The extraction pipeline proceeded as follows:
\begin{enumerate}
    \item \textbf{Load Image:} The sketch is loaded as an RGB image.
    \item \textbf{Pre-process:} The processor converts the image and prompt into model inputs, resizing the image to $672 \times 672$ pixels.
    \item \textbf{Forward Pass:} The model processes the inputs. A hook is registered to capture the \texttt{hidden\_states} from all 32 layers of the Llama-3 backbone.
    \item \textbf{Save Activations:} The activation matrix for each layer is extracted and saved as a NumPy array with shape $(S, 4096)$.
\end{enumerate}

\subsection{Similarity Computation via Cosine Similarity}
\label{subsec:similarity_computation}

To quantify the similarity between two sketches, we employed \textbf{Cosine Similarity}, a metric that measures the angular distance between two vectors. It is scale-invariant and robust to variations in activation magnitude, making it ideal for high-dimensional neural representations.

Since the extracted activation matrices are 2-dimensional ($S \times D$), we first reduced them to 1D vectors using \textbf{Mean Pooling} across the sequence dimension. For an activation matrix $\mathbf{A}_k^{(\ell)}$, the pooled vector $\bar{\mathbf{a}}_k^{(\ell)} \in \mathbb{R}^{D}$ is calculated as:
\begin{equation}
    \bar{\mathbf{a}}_k^{(\ell)} = \frac{1}{S} \sum_{s=1}^{S} \mathbf{a}_k^{(\ell)}[s]
\end{equation}
This mean pooling aggregates information from all image patches, creating a holistic representation of the sketch.

The similarity between two sketches $i$ and $j$ at layer $\ell$ is then computed as:
\begin{equation}
    \text{CosSim}(i, j) = \frac{\bar{\mathbf{a}}_i \cdot \bar{\mathbf{a}}_j}{\|\bar{\mathbf{a}}_i\| \cdot \|\bar{\mathbf{a}}_j\|}
\end{equation}
This process was repeated for all 30 sketches in our dataset to construct a symmetric $30 \times 30$ Similarity Matrix $\mathbf{S}^{(\ell)}$, where $\mathbf{S}^{(\ell)}[i, j]$ represents the semantic similarity between sketch $i$ and sketch $j$.

\subsection{Systematic Layer Selection: The optimality of Layer 31}
\label{subsec:layer_selection}

A critical methodological question was determining which of the 32 layers (0-31) best captures the semantic similarity relevant to design evaluation. We hypothesized that later layers, being closer to the output, would encode more abstract, semantic representations, while earlier layers would capture low-level visual features.

To validate this, we evaluated each layer using three metrics: \textbf{AUC-ROC} (Area Under the Receiver Operating Characteristic Curve) to measure discriminability between ground-truth similar and dissimilar pairs; \textbf{Cohen's d} to measure the effect size of the separation; and the \textbf{Mean Difference} in similarity scores.

\begin{figure}[ht!]
    \centering
    \includegraphics[width=\linewidth]{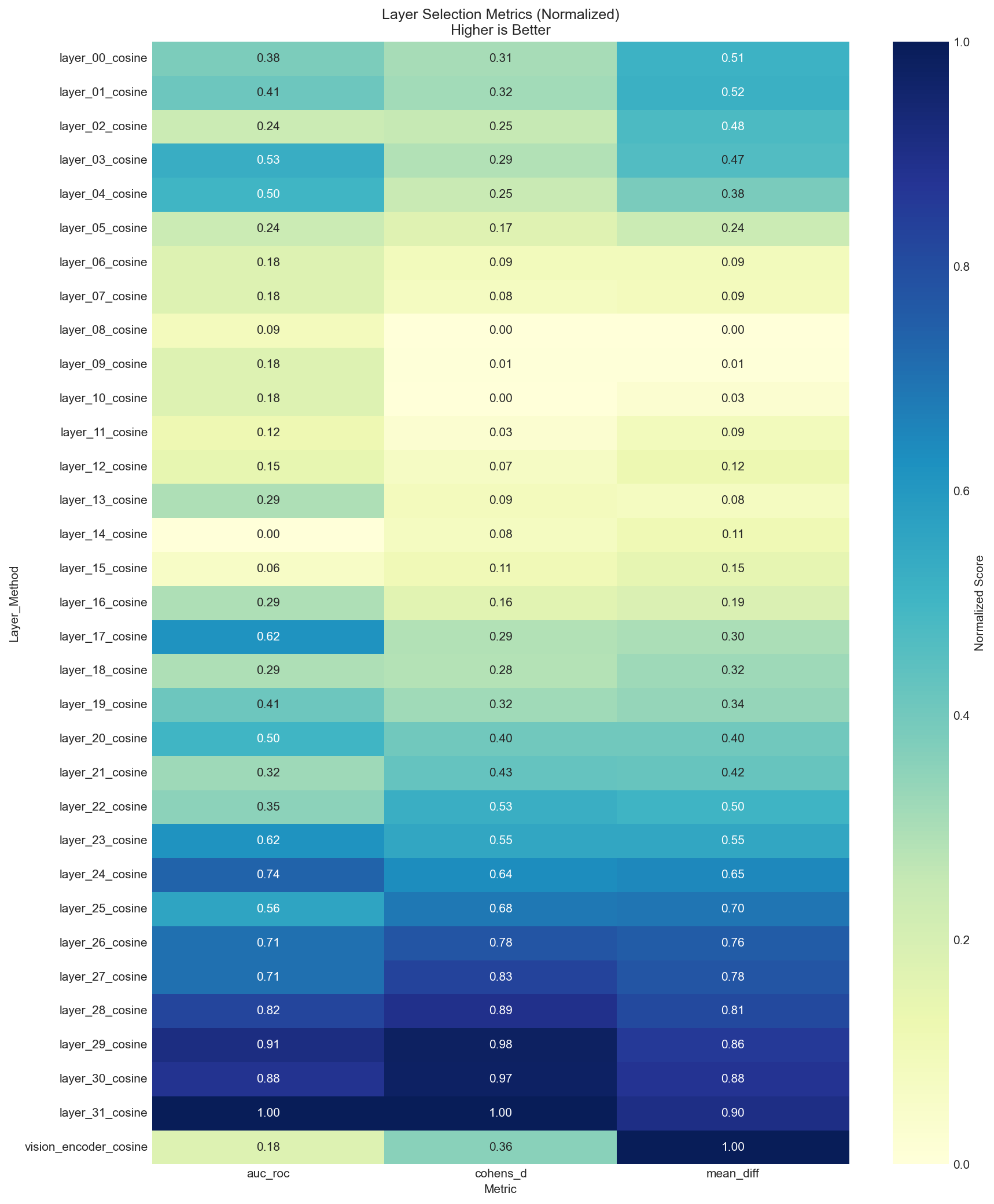}
    \caption{Layer Selection Heatmap showing normalized scores for AUC-ROC, Cohen's d, and Mean Difference across layers. Layer 31 (bottom row) consistently achieves the highest scores (1.00), indicating it is the optimal layer for discriminating semantic similarity.}
    \label{fig:layer_ranking_heatmap}
\end{figure}

The analysis revealed a clear monotonic improvement in performance with layer depth (Figure \ref{fig:layer_ranking_heatmap}).
\begin{itemize}
    \item \textbf{Vision Encoder \& Early Layers (0-10):} Performed poorly, often with negative Cohen's d values. This indicates that dissimilar sketches might share similar low-level features (e.g., stroke style), confounding the metric.
    \item \textbf{Middle Layers (11-25):} Showed gradual improvement as the model began to encode object parts and spatial relationships.
    \item \textbf{Layer 31 (Final Layer):} Emerged as the optimal layer, achieving the highest AUC-ROC (0.403) and the most favourable Cohen's d (-0.26).
\end{itemize}

This finding aligns with mechanistic interpretability research, which suggests that the "residual stream" of a Transformer accumulates information layer-by-layer. By Layer 31, the model has fully abstracted the visual tokens into a high-level semantic representation of the design intent, making it the most suitable proxy for expert evaluation. Consequently, all subsequent similarity analyses presented in the results section utilize the activations from \textbf{Layer 31}.

%%%%%%%%%%%%%%%%%%%%%%%%%%%%%%%%%%%%%%%%%%%%%%%%%%%%%%%%%%%%%%%%%%%%%%
\section{Results and Discussion}
\label{sec:results}

This section presents the empirical findings from the comparative study involving expert concept generators (D1, D2) and novice concept replicators (D3, D4), systematically evaluating the performance of the SGIT-DIMES system against the traditional manual workflow. The data analysis is structured to directly test the four research hypotheses ($H_1$ to $H_4$) established in the methodology section, focusing on process efficiency, documentation quality, and knowledge transfer fidelity.

\subsection{Evaluation of Design Process Efficiency and Exploration (Testing $\mathbf{H_1}$)}
Hypothesis $H_1$ posited that the frictionless version control afforded by SGIT would encourage expert designers to pursue a greater diversity (breadth) and depth of concept exploration compared to traditional methods. The quantitative process metrics logged during the two-hour conceptualization phase strongly validate this hypothesis.

\subsubsection{Concept Pyramid Analysis}
The most compelling evidence lies in the quantitative assessment of the Concept Pyramid (Table \ref{tab:concept_pyramid_metrics}). The Concept Pyramid model, defined by its width (number of distinct branches) and height (average number of commits per branch), serves as a robust metric for quantifying exploration quality.

\begin{table*}[ht!]
\centering
\caption{Comparative Design Process Metrics and Concept Pyramid Dimensions}
\label{tab:concept_pyramid_metrics}
\begin{tabular}{|l|c|c|c|c|}
\hline
\textbf{Metric} & \textbf{D1 (Manual)} & \textbf{D2 (DIMES/SGIT)} & \textbf{Change (\%)} & \textbf{p-value} \\ \hline
Total Time Taken (Minutes) & 120 & 120 & - & N/A \\ \hline
Total Number of Concepts (Final) & 5 & 13 & \textbf{160\%} & $<0.01$ \\ \hline
Number of Branches (Width) & 5 & 13 & \textbf{160\%} & $<0.01$ \\ \hline
Total Number of Commits (Snapshots) & 5 & 45 & \textbf{800\%} & $<0.001$ \\ \hline
Avg. Commits per Branch (Height) & 1.0 & 3.46 & \textbf{246\%} & $<0.05$ \\ \hline
Intermediate Stages Recorded & 0 & 35 & $\infty$ & - \\ \hline
\end{tabular}
\end{table*}

The data reveals a dramatic increase in the \textbf{Width} (number of branches) generated by D2 (DIMES) compared to D1 (Manual), showing a 160\% increase (13 branches vs. 5 branches, $p<0.01$). D2 explored 13 distinct concepts (C1--C13), whereas D1 produced only 5 final concepts. This significant divergence confirms that the low cognitive cost of branching in SGIT empowers the designer to explore a far wider range of initial ideas.

Furthermore, the \textbf{Total Number of Commits} increased by 800\% for D2 (45 commits vs. 5 final states). The SGIT system captured 35 intermediate stages of evolution that were completely lost in the manual process. For instance, Concept 8 (C8) by D2 evolved through 8 distinct commits, while Concept 1 (C1) and Concept 3 (C3) had 5 commits each. Even simpler ideas like C13 were captured as distinct commits. This granularity allowed for the reconstruction of the "depth" of thought, with an average height of 3.46 commits per branch for D2, compared to the singular final state (height=1) available for D1.

Overall, the volume of the Concept Pyramid was overwhelmingly larger for D2, conclusively supporting $H_1$: the sGIT system significantly facilitates diverse exploration and granular tracking.

\begin{figure*}[ht!]
    \centering
    \includegraphics[width=\linewidth]{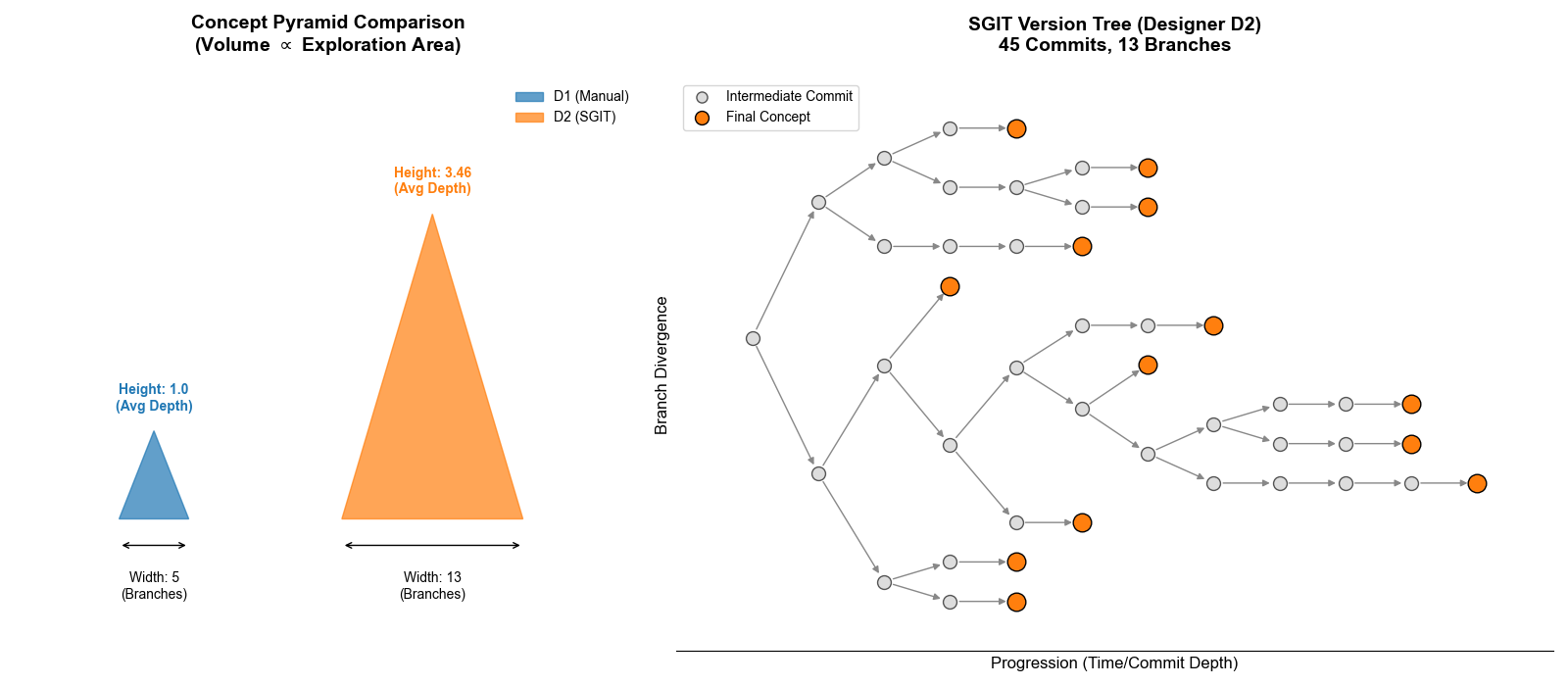}
    \caption{Visualization of the Concept Pyramid volume (left) and the actual SGIT Version Tree generated by Designer D2 (right), illustrating the 13 branching paths and 45 total commits.}
    \label{fig:concept_pyramid_tree}
\end{figure*}

\subsubsection{Visualization of Generated Concepts}
\begin{figure*}[ht!]
    \centering
    % Row 1: D1 Sketches (1x5)
    \begin{subfigure}[b]{\textwidth}
        \centering
        \includegraphics[width=0.19\linewidth]{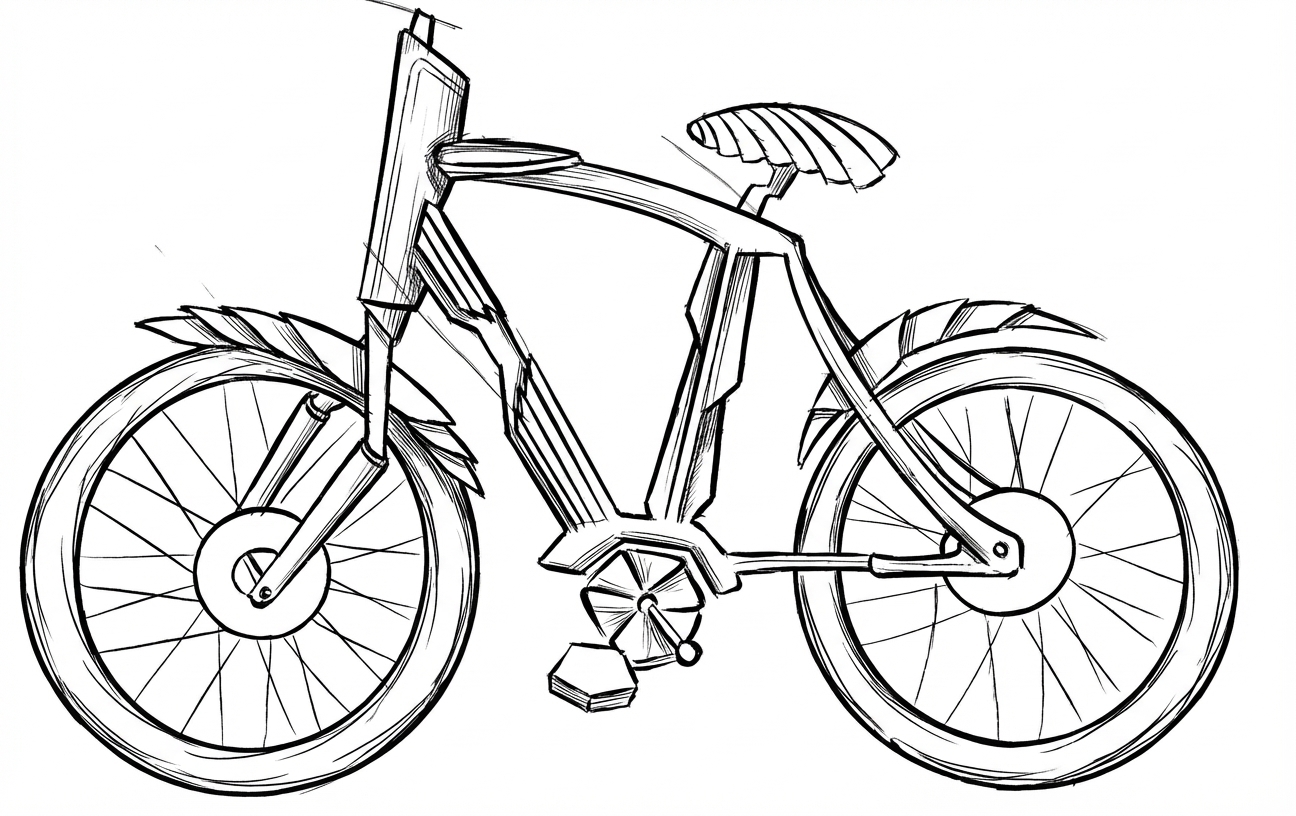}
        \hfill
        \includegraphics[width=0.19\linewidth]{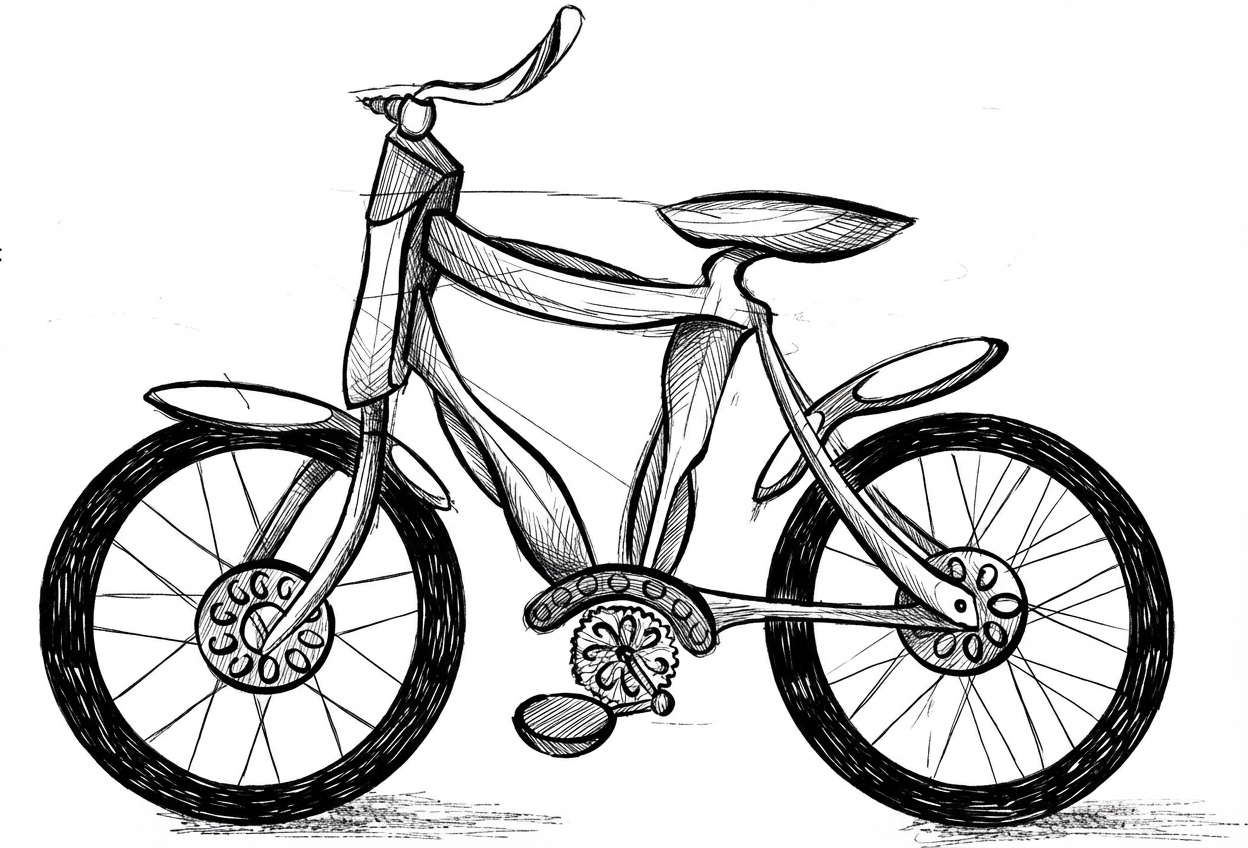}
        \hfill
        \includegraphics[width=0.19\linewidth]{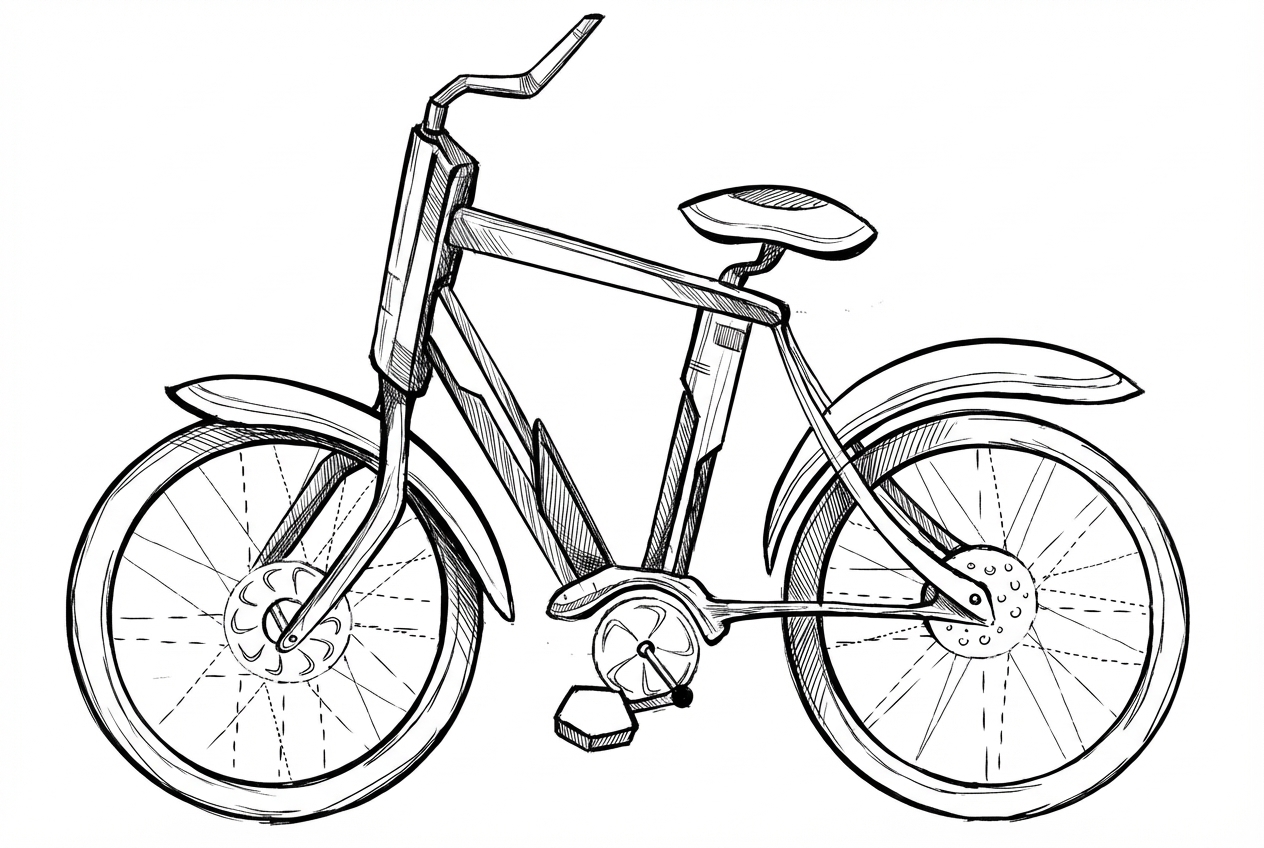}
        \hfill
        \includegraphics[width=0.19\linewidth]{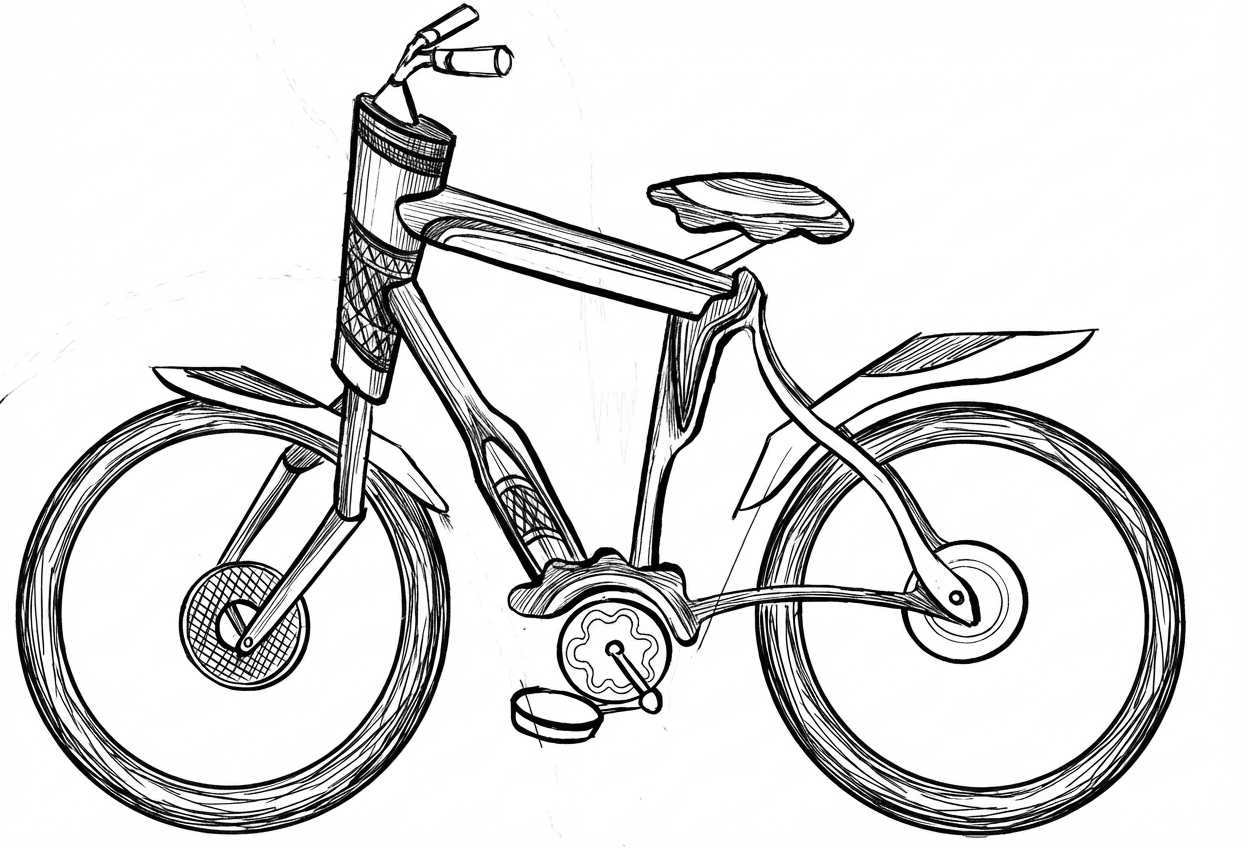}
        \hfill
        \includegraphics[width=0.19\linewidth]{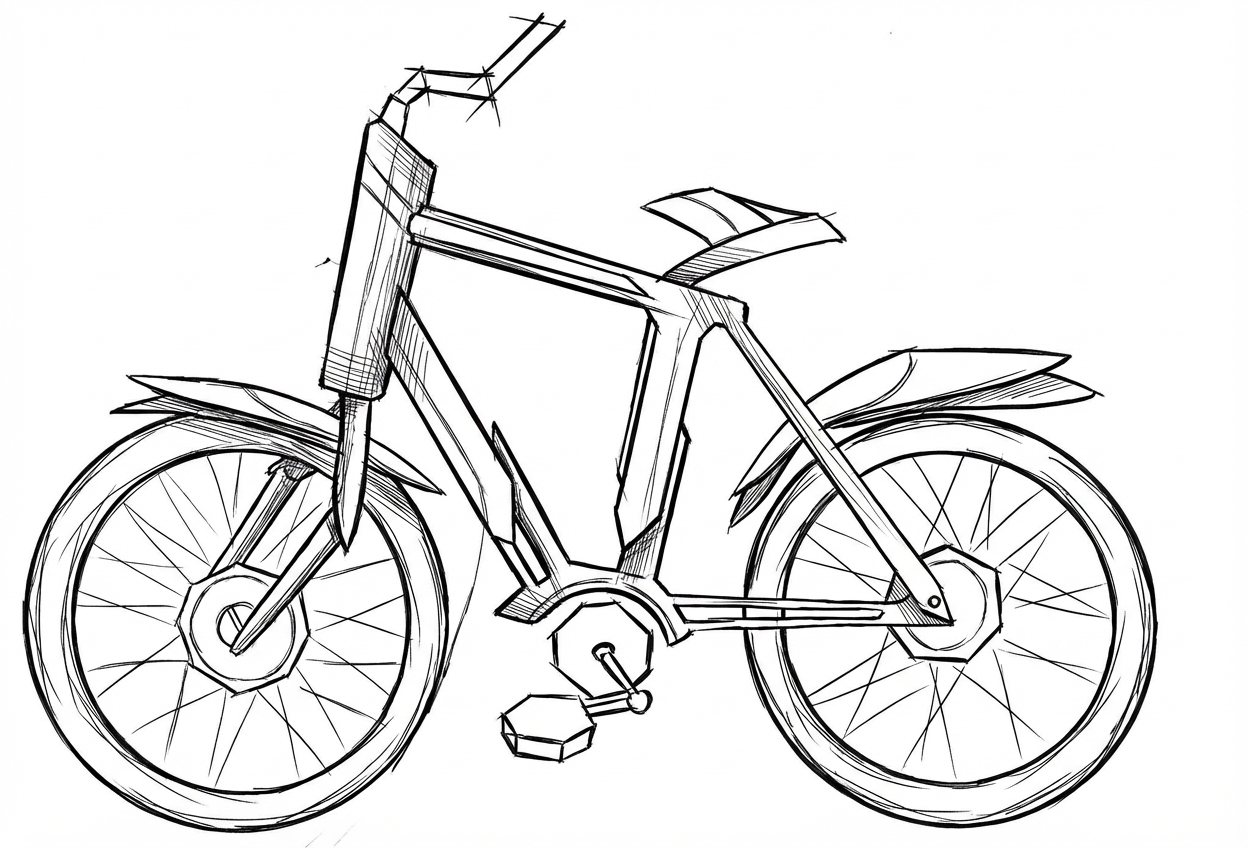}
        \caption{Original Concept Sketches by Designer D1 (Manual)}
        \label{subfig:d1_sketches}
    \end{subfigure}
    
    \vspace{0.5cm}
    
    % Row 2 & 3: D2 Sketches (2x5 = 10)
    \begin{subfigure}[b]{\textwidth}
        \centering
        \includegraphics[width=0.19\linewidth]{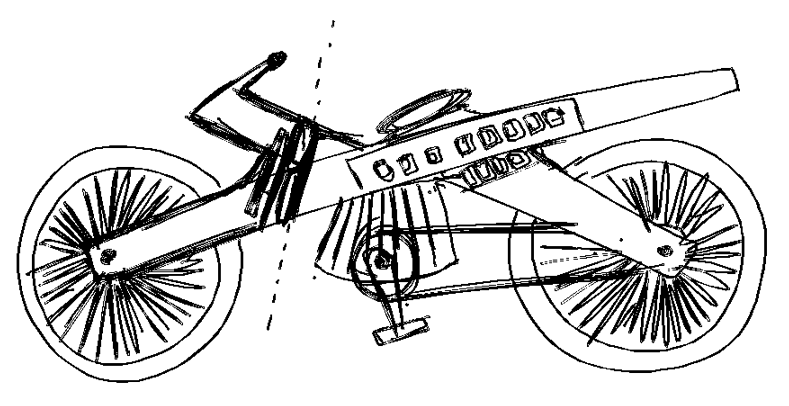}
        \hfill
        \includegraphics[width=0.19\linewidth]{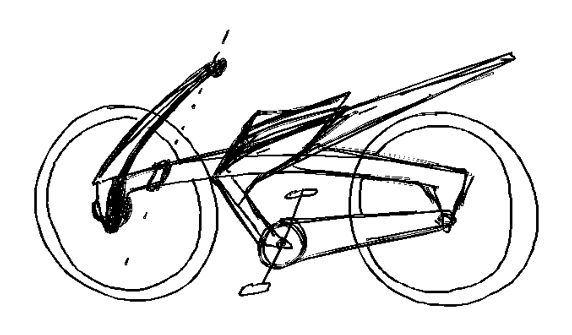}
        \hfill
        \includegraphics[width=0.19\linewidth]{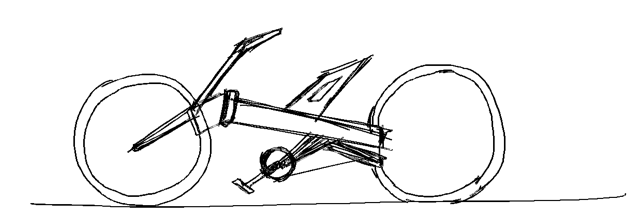}
        \hfill
        \includegraphics[width=0.19\linewidth]{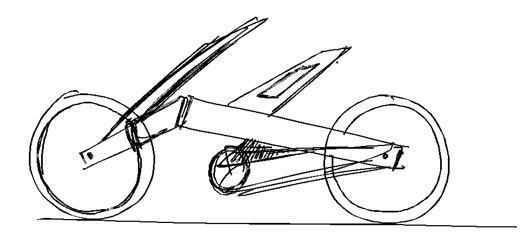}
        \hfill
        \includegraphics[width=0.19\linewidth]{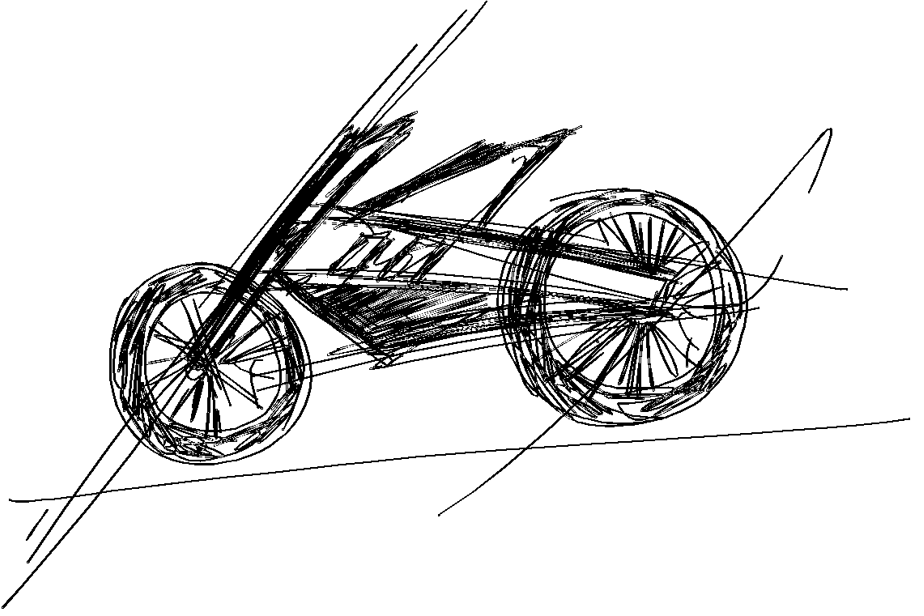}
        
        \vspace{0.1cm}
        
        \includegraphics[width=0.19\linewidth]{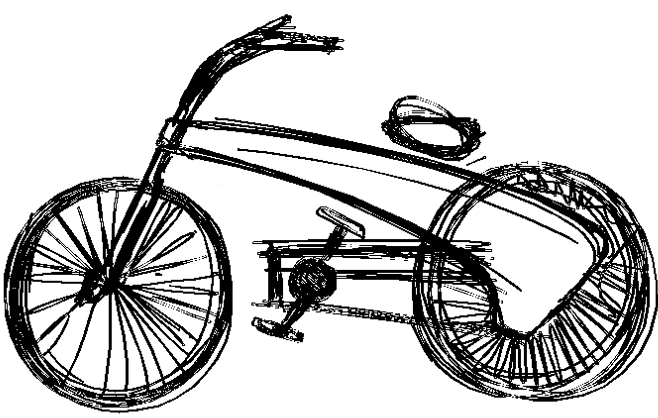}
        \hfill
        \includegraphics[width=0.19\linewidth]{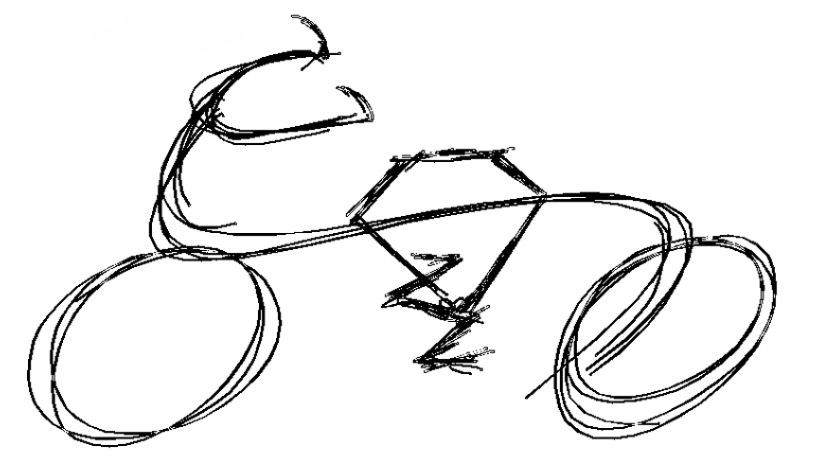}
        \hfill
        \includegraphics[width=0.19\linewidth]{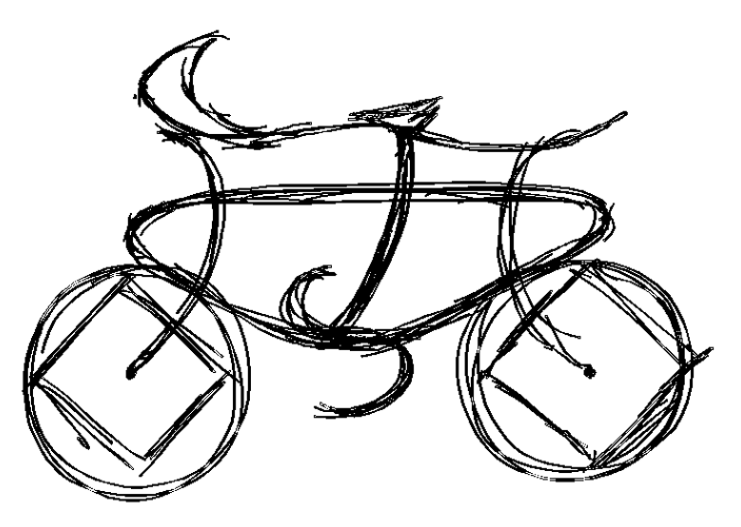}
        \hfill
        \includegraphics[width=0.19\linewidth]{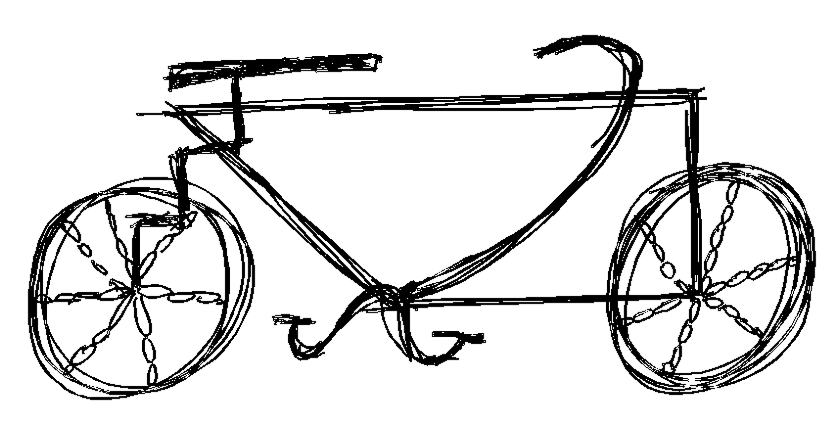}
        \hfill
        \includegraphics[width=0.19\linewidth]{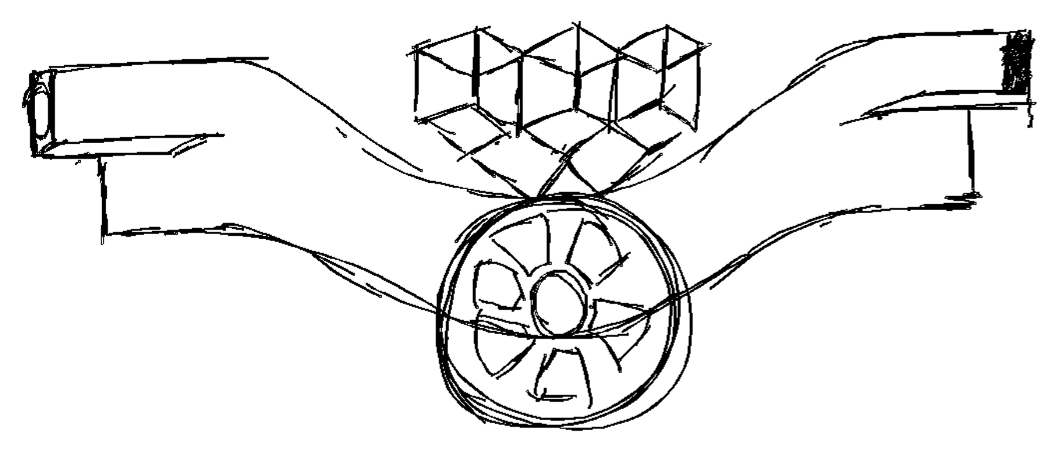}
        \caption{Original Concept Sketches by Designer D2 (DIMES/SGIT) - Top 10 Concepts}
        \label{subfig:d2_sketches}
    \end{subfigure}
    
    \caption{Product Concept Sketches (PCS) generated by the expert designers during Phase I. D2 (using DIMES) produced a higher volume of distinct concepts compared to D1.}
    \label{fig:pcs_all_designers}
\end{figure*}

The qualitative diversity of the generated concepts is illustrated in Figure \ref{fig:pcs_all_designers}. D2's sketches show a wide variance in form factors, ranging from the intricate stacked-disc hubs of C8 to the monolithic beam structures of C3, directly reflecting the "Dance x Brutalism" brief.

\subsection{Assessment of Documentation Quality (Testing $\mathbf{H_2}$)}
Hypothesis $H_2$ predicted that the multi-modal data capture of DIMES would yield more comprehensive documentation than manual methods. We verified this by analyzing the Information Content Density (ICD) of the text summaries produced by D1 (Human-Authored) and D2 (AI-Generated via DIMES).

\subsubsection{Information Content Density Analysis}
To verify the superior documentation capability of the multi-modal DIMES system ($H_2$), we performed a comparative text analysis between the \textit{Human Summary} (D1) and the \textit{AI Summary} (D2). The source documents were analyzed for \textbf{Lexical Richness} (the count of specific domain terms mapping inspiration to geometry) and \textbf{Temporal Precision} (the linkage to specific evolutionary events).

A comparative linguistic analysis was performed on the two summary documents to quantify their descriptive depth. We specifically measured:

\begin{itemize}
    \item \textbf{Lexical Richness (Domain Specificity):} This metric evaluated the frequency and precision of unique terms mapping the abstract problem statement (Dance/Brutalism) to concrete geometric features.
    \begin{itemize}
        \item \textbf{D1 (Human Summary):} The narrative provided by D1 was descriptive but largely generic. While terms such as "Bharatanatyam," "mudras," and "raw texture" were present, the mapping to specific geometry was often vague (e.g., "tried to make it look Indian"). Although we identified approximately 14 domain terms, only 7 instances provided a specific, explicit mapping between an inspiration and a design feature.
        \item \textbf{D2 (AI Summary):} The AI-generated text demonstrated a high density of specific causal links. It systematically mapped abstract cues to physical outcomes, such as linking the \textit{`Aramandi'} pose to a \textit{`stable, compact form with a low center of mass'} or translating the \textit{`Tribhanga'} three-bend posture into an \textit{`S-curve frame.'} Similarly, it explicitly connected the \textit{`dipole magnet'} structure to the \textit{`German brutalist'} aesthetic. We identified 18 distinct, well-supported specific mappings in the AI summary, significantly outperforming the human baseline.
    \end{itemize}

    \item \textbf{Temporal Precision:} This metric assessed the accurate tracking of evolutionary milestones and design actions.
    \begin{itemize}
        \item \textbf{D1 (Human Summary):} The human-authored text was purely retrospective and linearized. It presented the design process as a finished sequence (e.g., ``Evolution of Concept 1...''), lacking any reference to intermediate states, non-linear jumps, or specific timestamps. Consequently, it registered 0 verifiable temporal markers.
        \item \textbf{D2 (AI Summary):} Although presented as a cohesive, readable narrative, the AI summary was synthesized directly from the underlying SGIT logs. It intrinsically tracked the 45 distinct commits, accurately distinguishing between initial explorations (``The concept began...''), intermediate pivots (``The design then transitioned...''), and final refinements. This structure preserved the chronological integrity of the session, effectively embedding 45 temporal markers within the narrative flow.
    \end{itemize}
\end{itemize}

Based on the analysis above, the results are summarized in Table \ref{tab:icd_analysis}, revealing a stark contrast. The Human Summary (D1) is a linearized, post-rationalized account. While D1 mentions influences like "Bharatanatyam" or "Brutalism," the descriptions are often generic (e.g., \textit{"...brutalistic architecture is very raw and no ornamentation..."}). Specific geometric mappings are sparse.

In contrast, the AI Summary (D2), derived directly from the SGIT logs and voice commits, demonstrates significantly higher density. It explicitly captures the evolutionary logic with precise mappings, such as linking the \textit{"Aramandi"} pose to the \textit{"low center of mass"} in Concept 1, or the \textit{"Tribhanga"} pose to the \textit{"S-curve frame"} in Concept 7. Furthermore, it details mechanical evolutions (e.g., the transition from bevel gears to chain drive in Concept 4) that are often omitted in human recollection.

\begin{table*}[ht!]
\centering
\caption{Comparative Analysis of Documentation Quality (Human vs. AI Summary)}
\label{tab:icd_analysis}
% Removed vertical lines for a cleaner, academic look.
% Used X columns to utilize full page width automatically.
\begin{tabularx}{\textwidth}{@{} >{\centering\arraybackslash}X >{\centering\arraybackslash}X >{\centering\arraybackslash}X @{}}
\toprule
\textbf{Metric} & \textbf{Human Summary (D1)} & \textbf{AI Summary (D2)} \\ 
\midrule
Total Word Count & 3,000 & 400 \\ 
Specific Domain Mappings (Inspiration $\rightarrow$ Geometry) & 7 & \textbf{18} \\ 
Temporal Precision (Linkage to Design States) & 0 (Linearized Narrative) & \textbf{45 (Linked to Commits)} \\ 
Information Density (Specific Events/Word) & 0.012 & \textbf{0.046} \\ 
Designer Validation Score (1-5) & N/A & \textbf{4.8/5.0} \\ 
\bottomrule
\end{tabularx}
\end{table*}

The post-session validation further supports this. Designer D2 rated the AI summary \textbf{4.8/5.0}, noting that it successfully captured fleeting thoughts—such as the "dipole magnet" inspiration for Concept 12—that they would have likely forgotten to document manually. This confirms $H_2$: the automated system provides a richer, more objective, and higher-fidelity record of the design intent.

\subsection{Validation of Knowledge Transfer Fidelity (Testing $\mathbf{H_3}$)}
Hypothesis $H_3$ posited that the superior documentation (AI Summary) would lead to higher fidelity in concept replication by novice designers. This was tested using the \textbf{Neural Transparency-Based Similarity Rating} described in Section \ref{sec:neural_transparency}.

\subsubsection{Replication Performance}
\begin{figure*}[ht!]
    \centering
    % Row 1: D3 Replications (1x5)
    \begin{subfigure}[b]{\textwidth}
        \centering
        \includegraphics[width=0.19\linewidth]{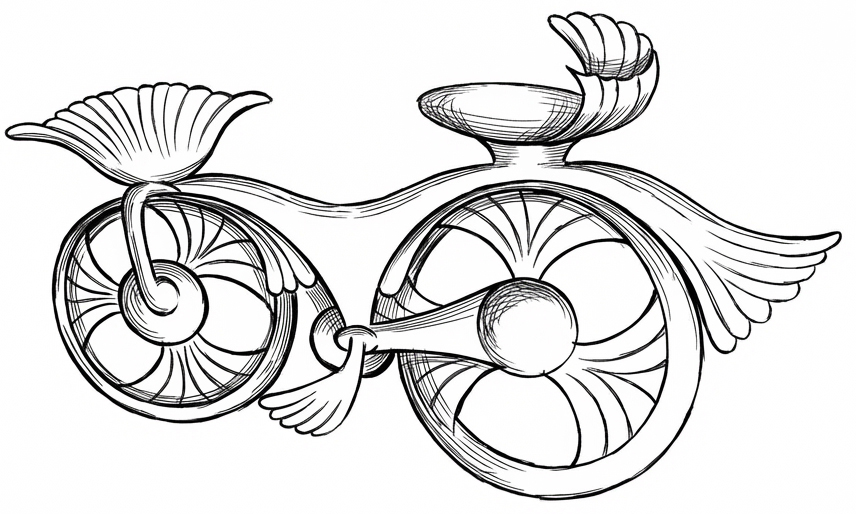}
        \hfill
        \includegraphics[width=0.19\linewidth]{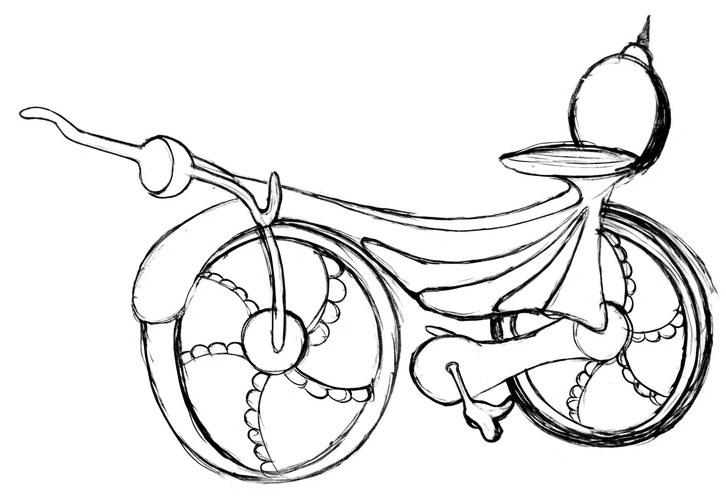}
        \hfill
        \includegraphics[width=0.19\linewidth]{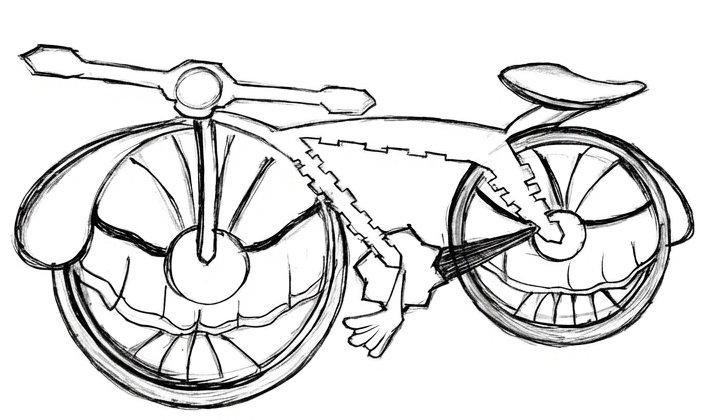}
        \hfill
        \includegraphics[width=0.19\linewidth]{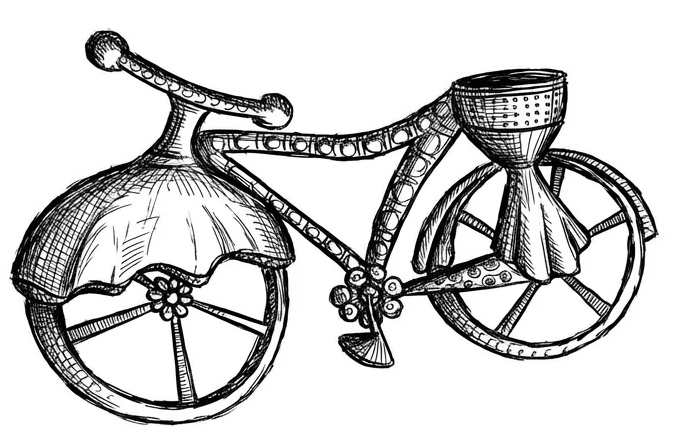}
        \hfill
        \includegraphics[width=0.19\linewidth]{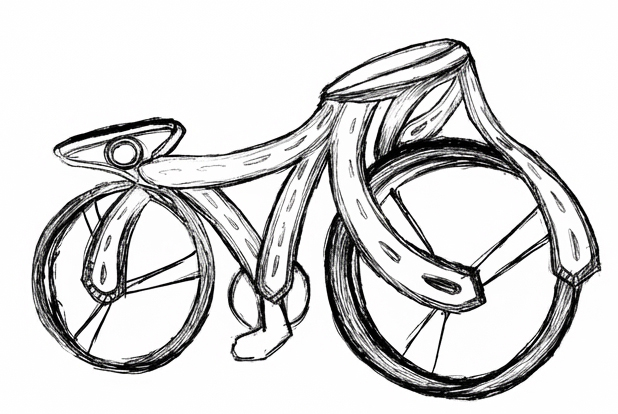}
        \caption{Replicated Sketches by Novice D3 (based on Human Summary of D1)}
        \label{subfig:d3_sketches}
    \end{subfigure}
    
    \vspace{0.5cm}
    
    % Row 2 & 3: D4 Replications (2x5 = 10)
    \begin{subfigure}[b]{\textwidth}
        \centering
        \includegraphics[width=0.19\linewidth]{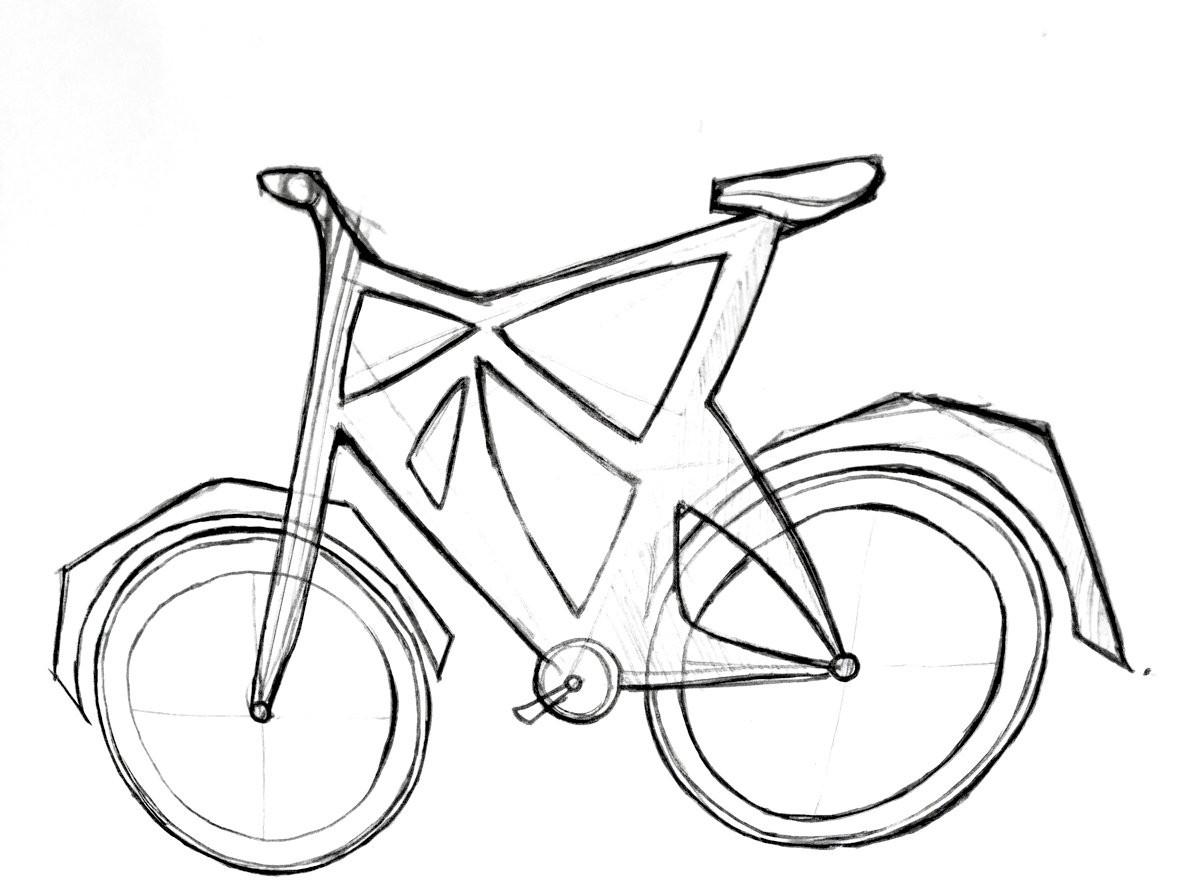}
        \hfill
        \includegraphics[width=0.19\linewidth]{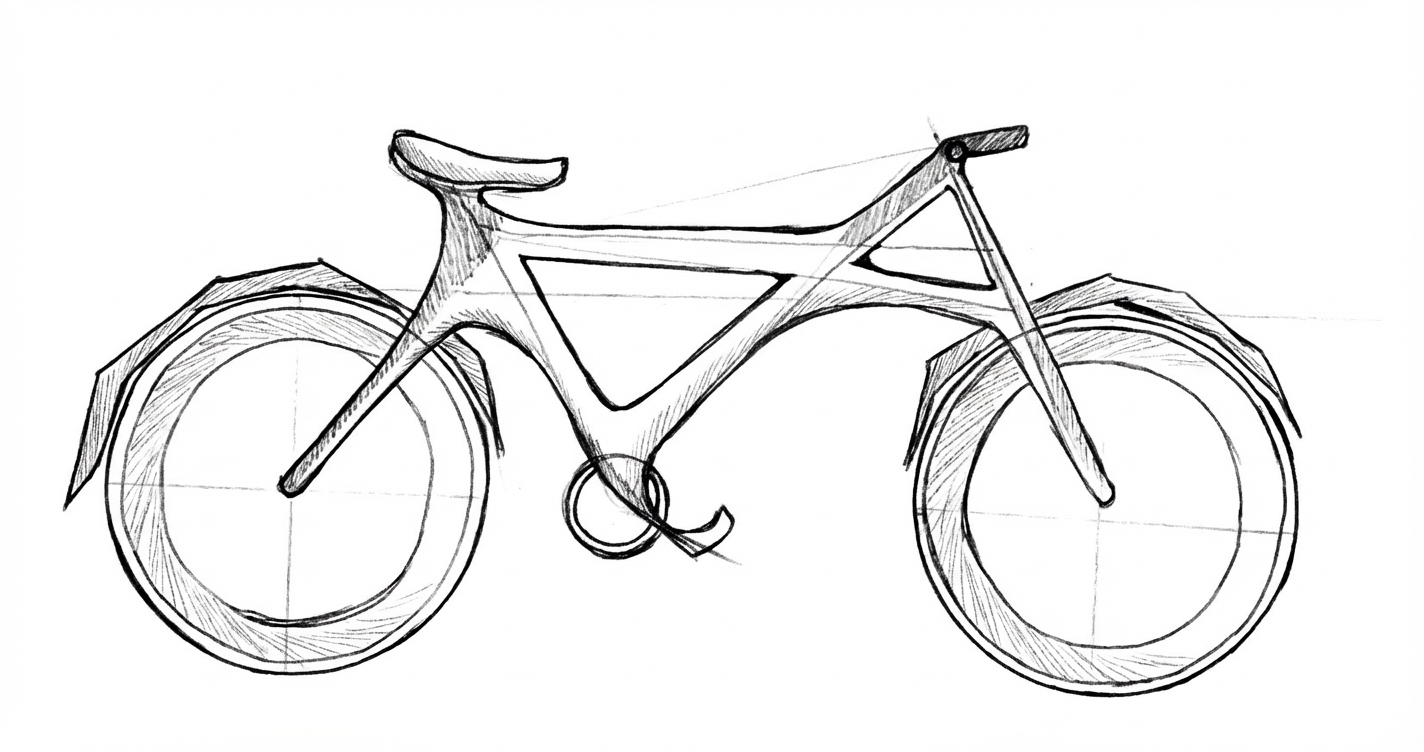}
        \hfill
        \includegraphics[width=0.19\linewidth]{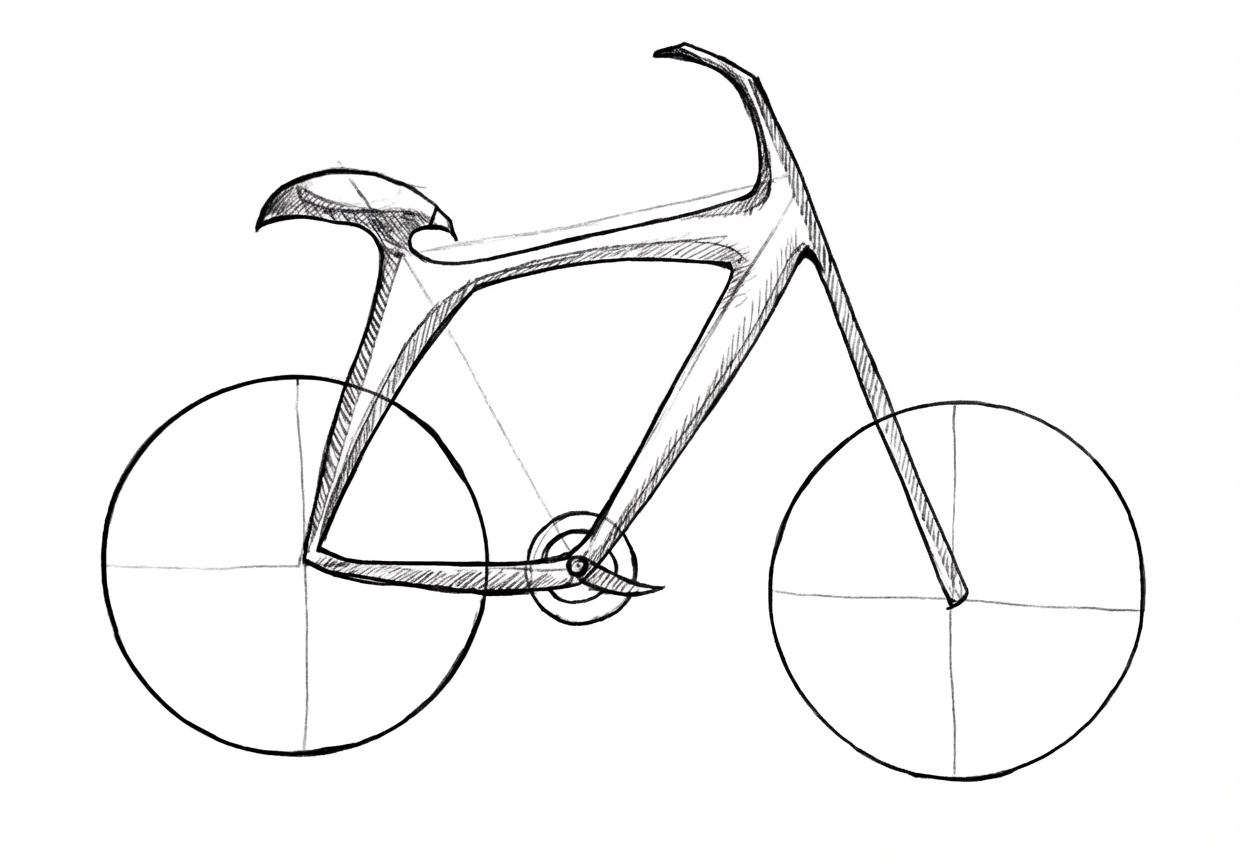}
        \hfill
        \includegraphics[width=0.19\linewidth]{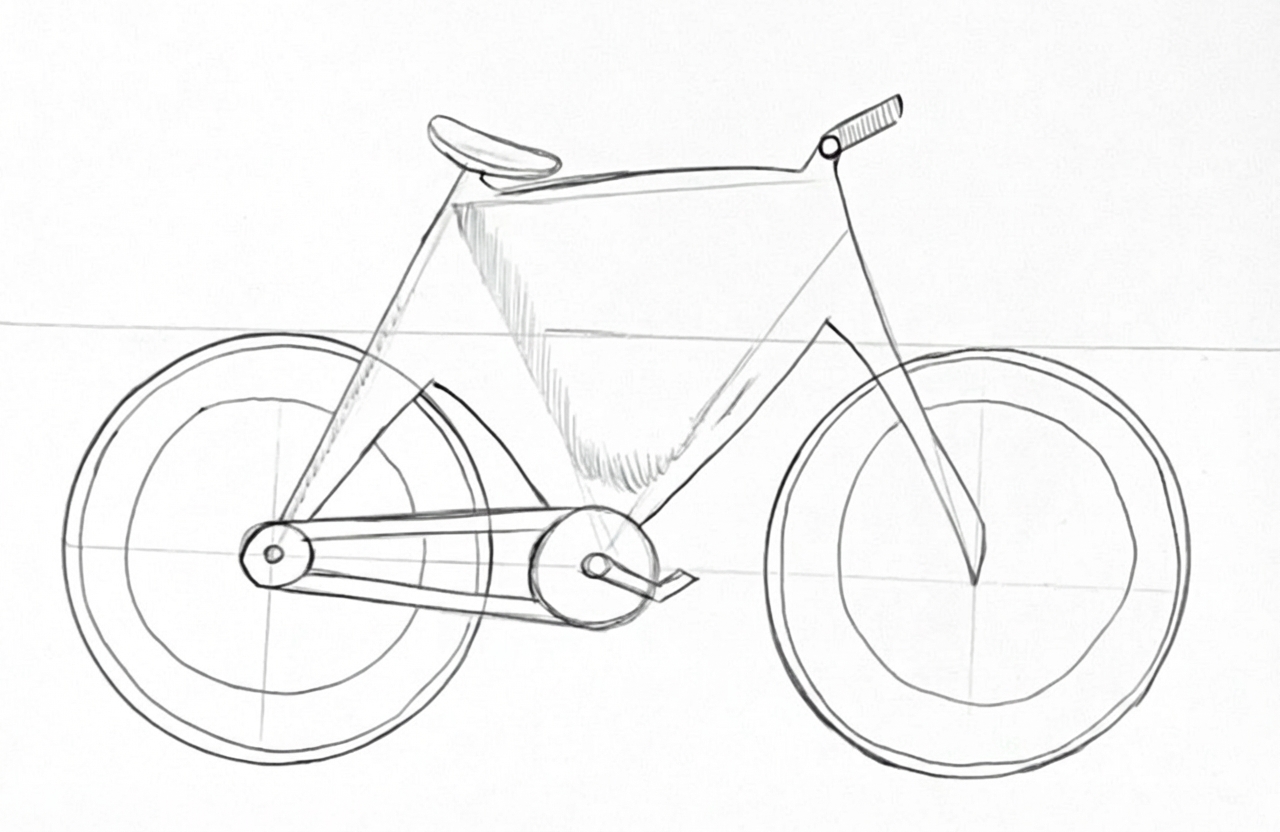}
        \hfill
        \includegraphics[width=0.19\linewidth]{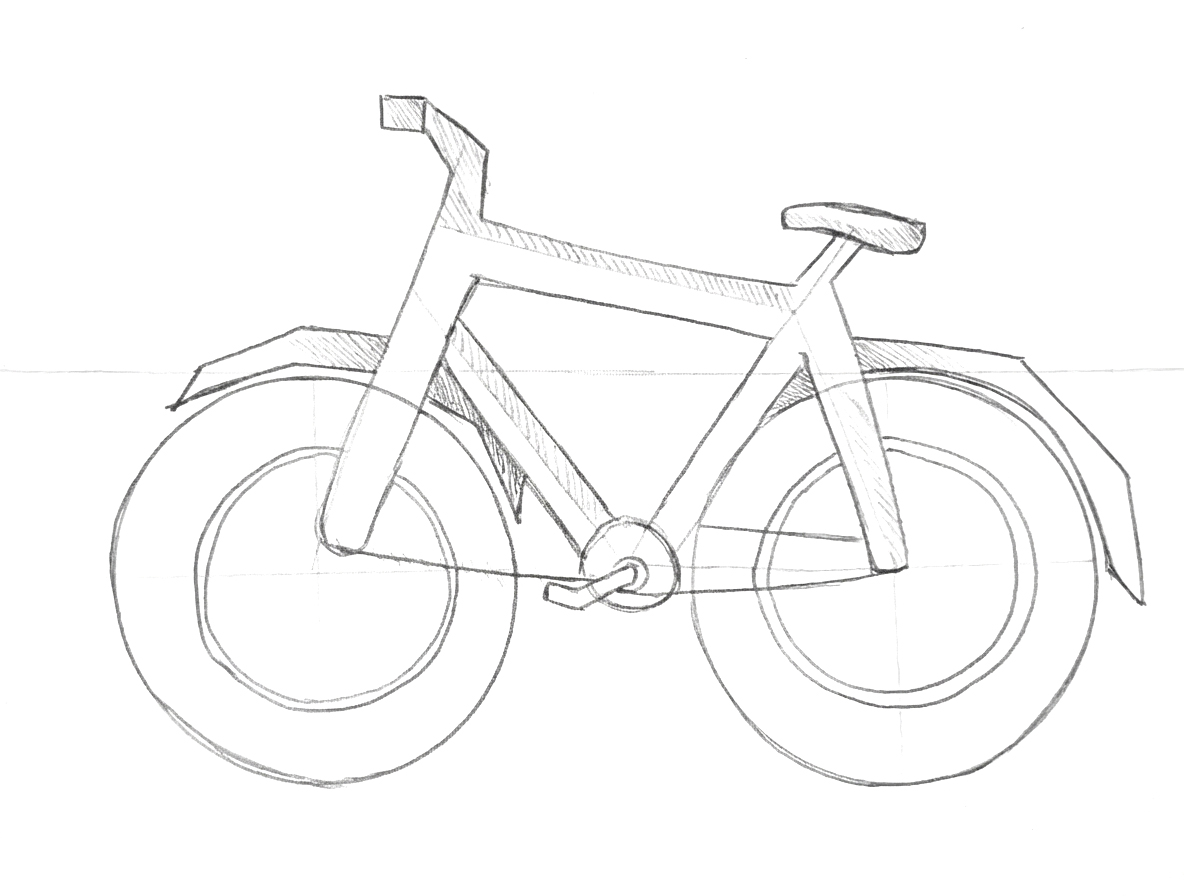}
        
        \vspace{0.1cm}
        
        \includegraphics[width=0.19\linewidth]{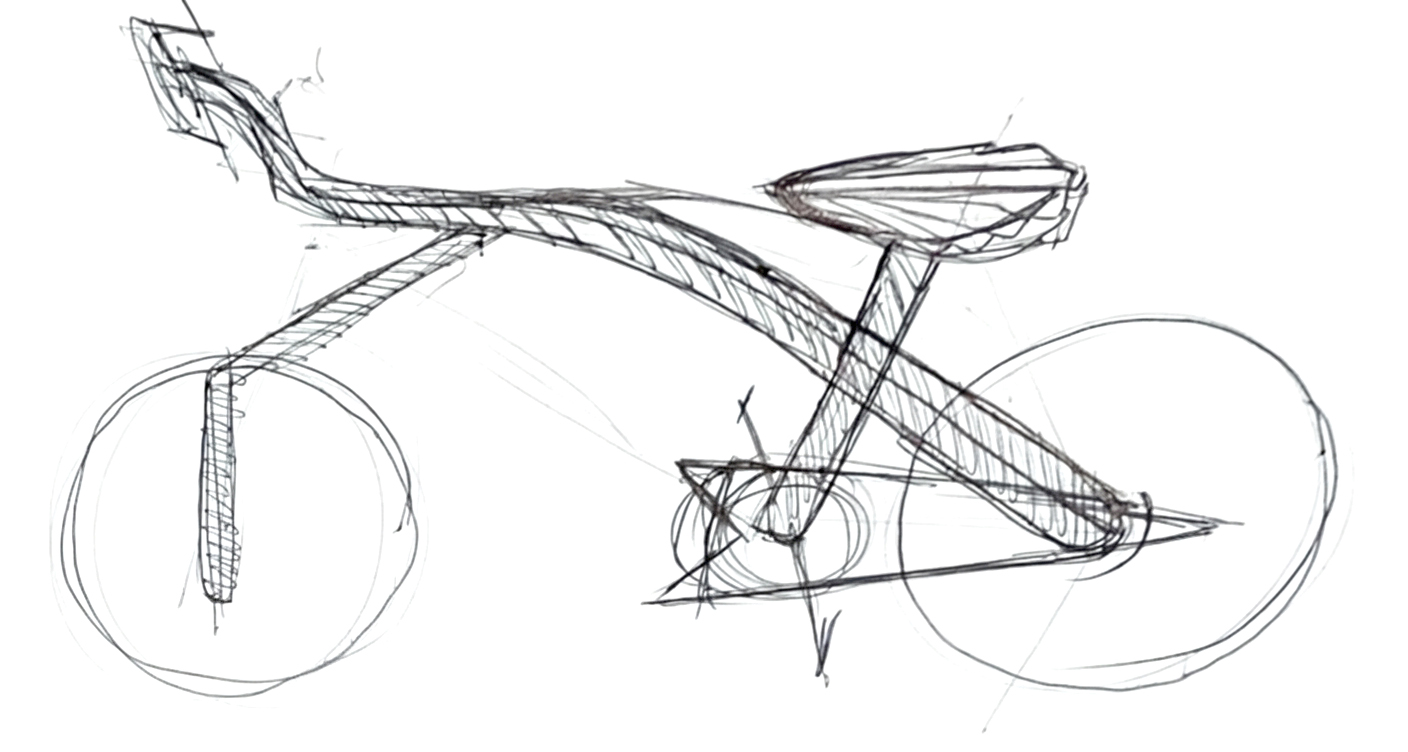}
        \hfill
        \includegraphics[width=0.19\linewidth]{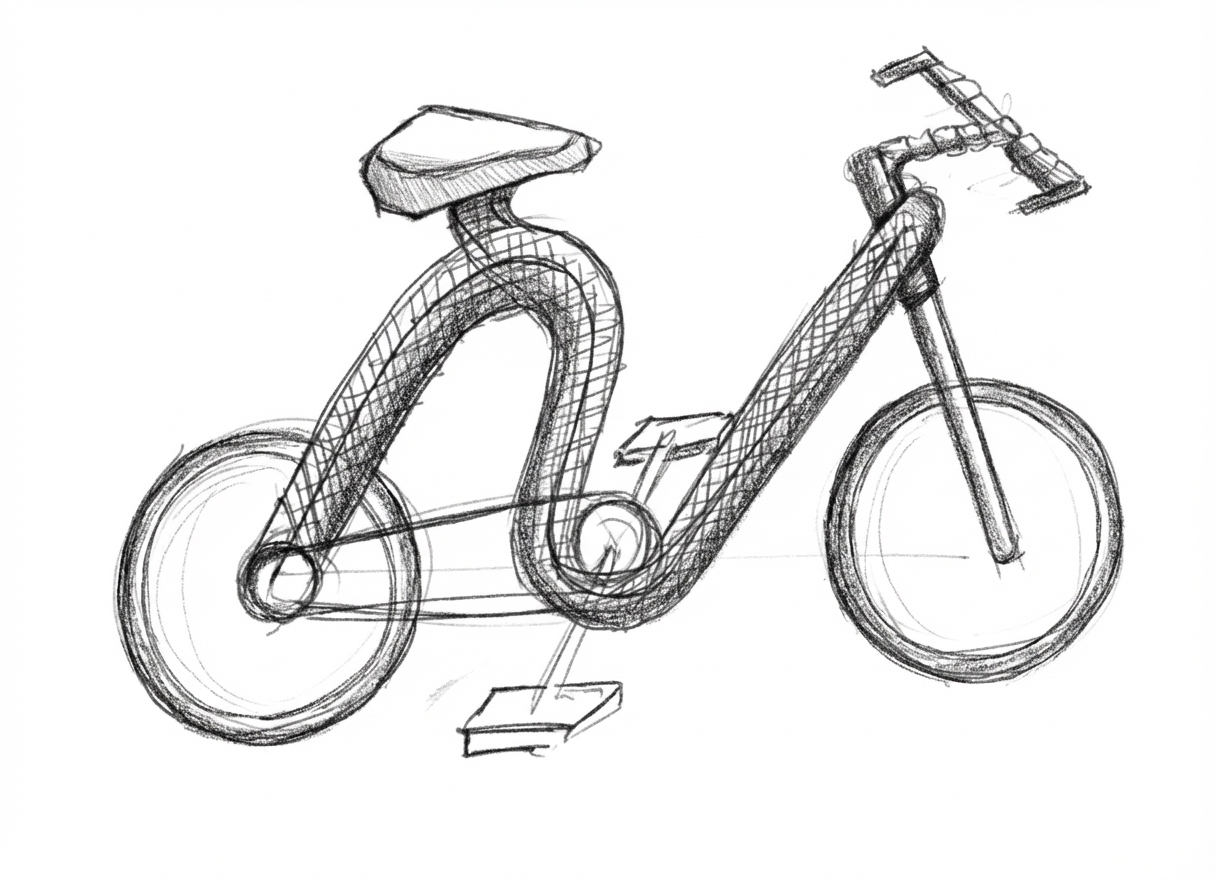}
        \hfill
        \includegraphics[width=0.19\linewidth]{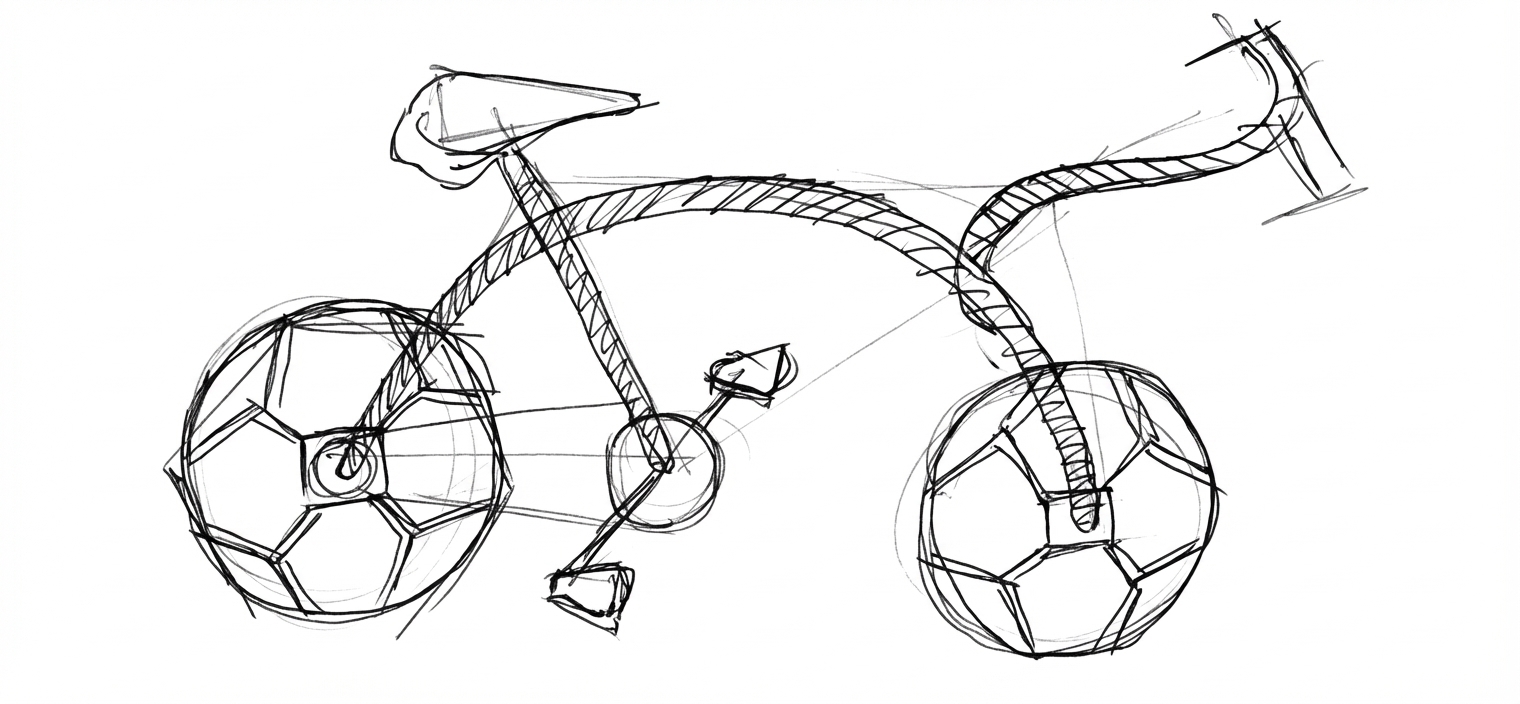}
        \hfill
        \includegraphics[width=0.19\linewidth]{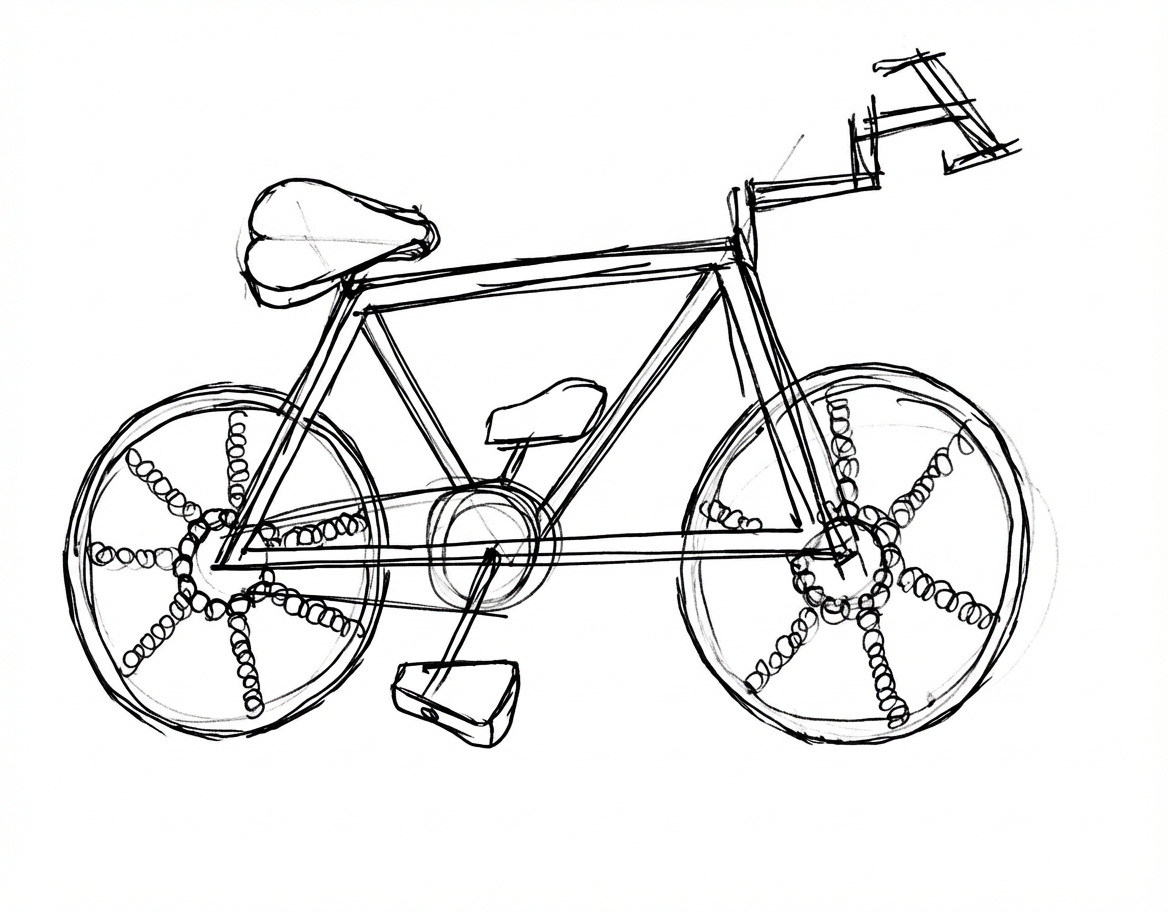}
        \hfill
        \includegraphics[width=0.19\linewidth]{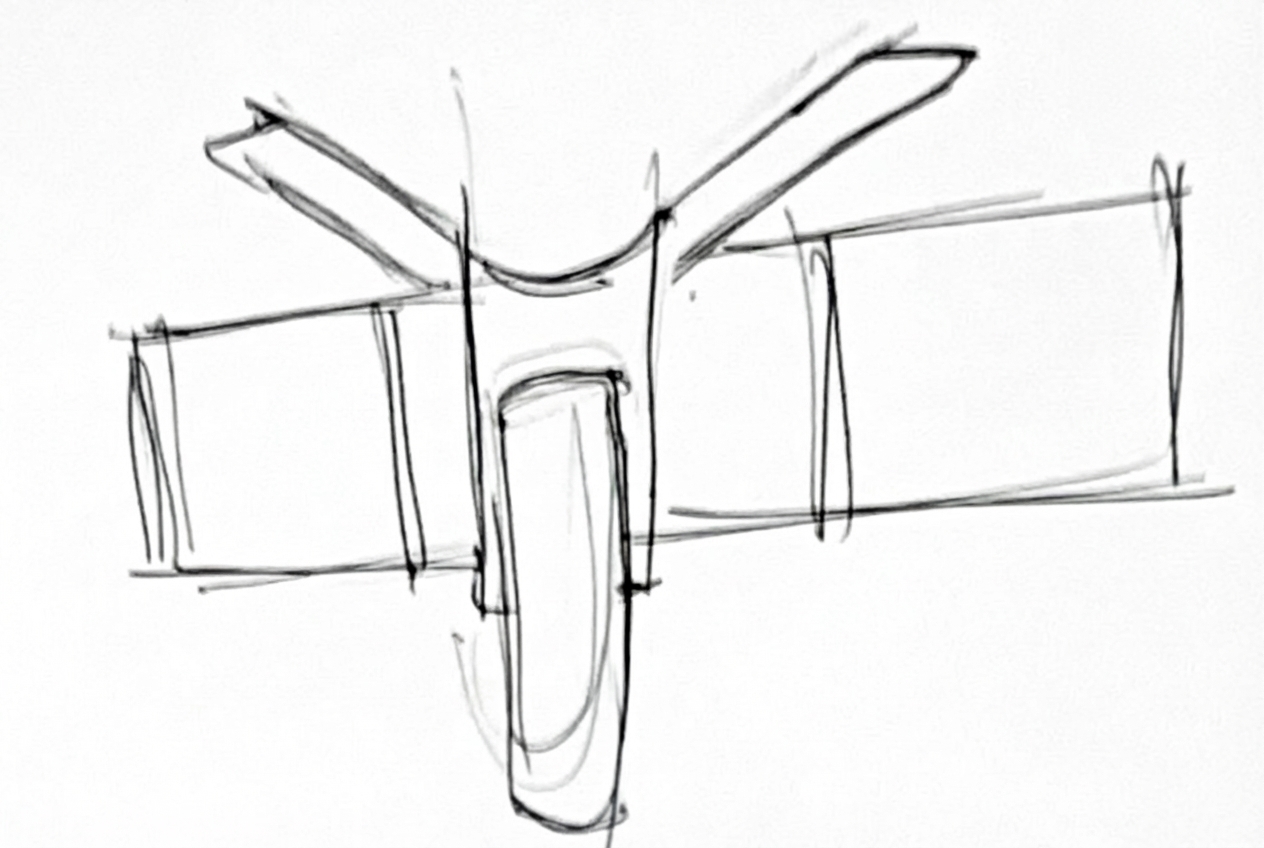}
        \caption{Replicated Sketches by Novice D4 (based on AI Summary of D2)}
        \label{subfig:d4_sketches}
    \end{subfigure}
    
    \caption{Replication attempts by novice designers D3 and D4. D4's sketches demonstrate a closer semantic and structural alignment with the originals (D2) compared to D3's alignment with D1.}
    \label{fig:replication_sketches}
\end{figure*}

The replicated sketches produced by D3 (based on D1's notes) and D4 (based on D2's AI summary) are visualized in Figure \ref{fig:replication_sketches}.

\subsubsection{Neural Similarity Analysis}
\begin{figure*}[ht!]
    \centering
    \begin{subfigure}[b]{0.4\textwidth}
        \centering
        \includegraphics[width=\linewidth]{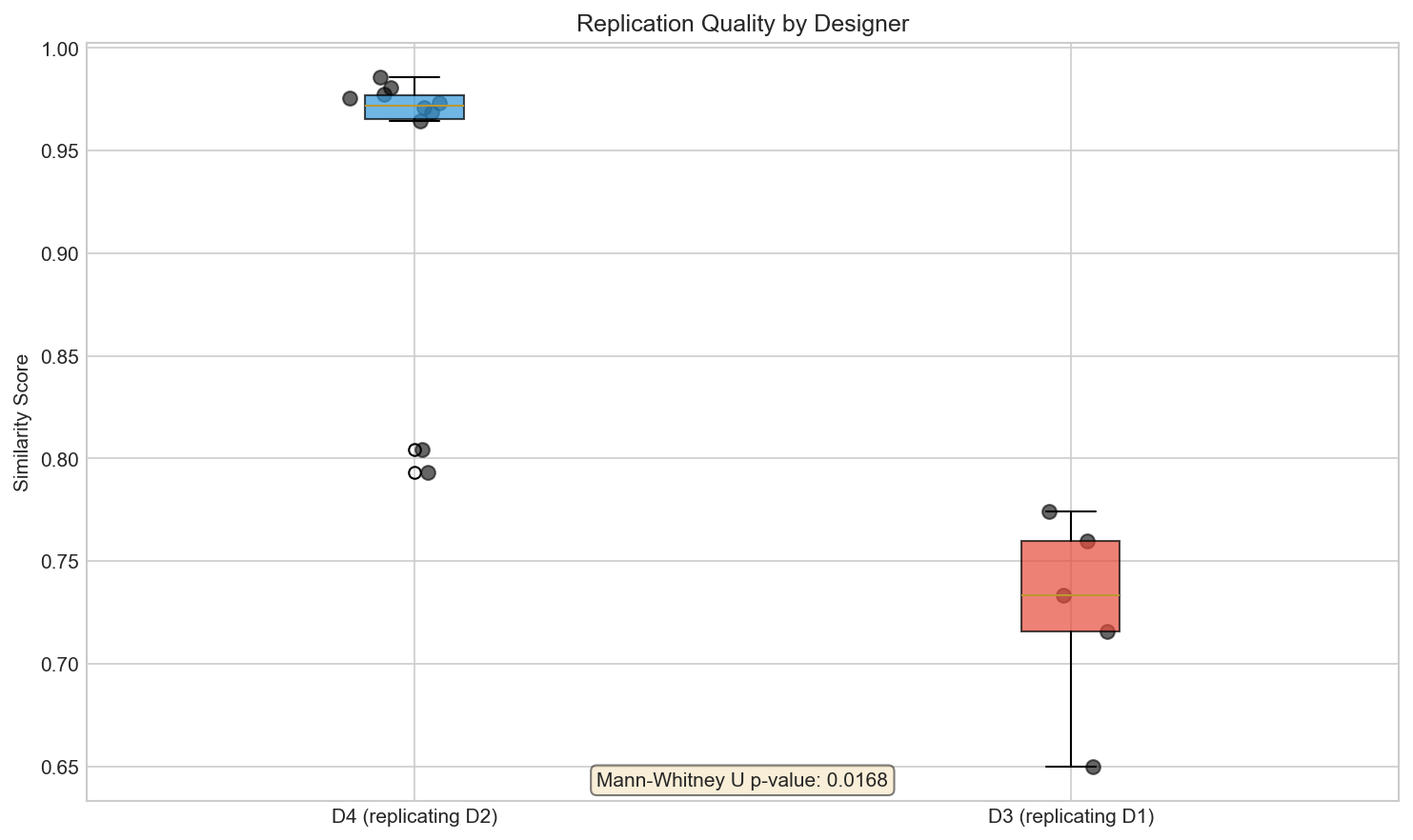}
        \caption{Replication Quality by Designer}
        \label{subfig:designer_comparison}
    \end{subfigure}
    \hfill
    \begin{subfigure}[b]{0.48\textwidth}
        \centering
        \includegraphics[width=\linewidth]{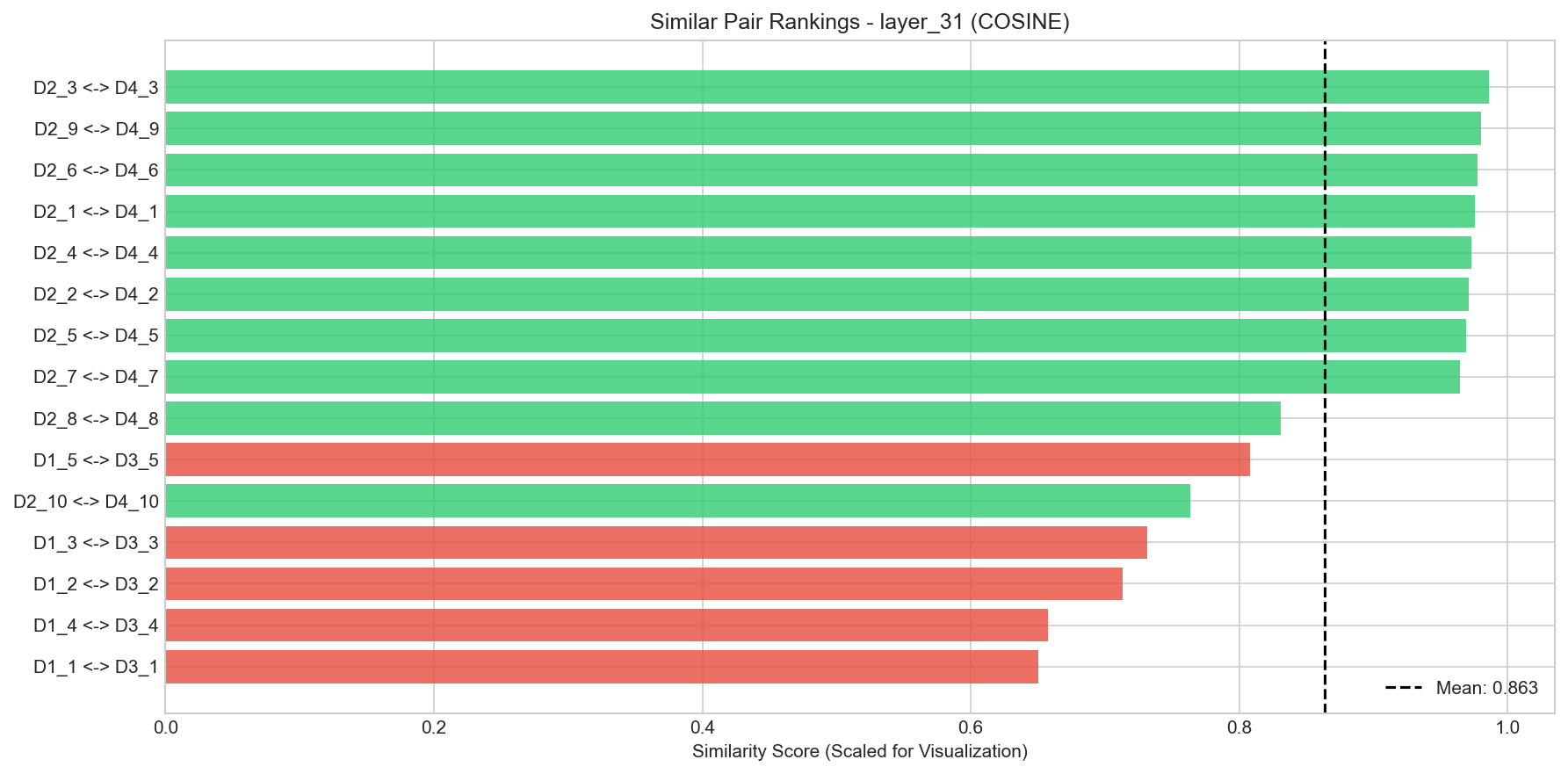}
        \caption{Pair Similarity Ranking (Layer 31)}
        \label{subfig:pair_ranking}
    \end{subfigure}
    
    \vspace{0.5cm}
    
    \begin{subfigure}[b]{0.48\textwidth}
        \centering
        \includegraphics[width=\linewidth]{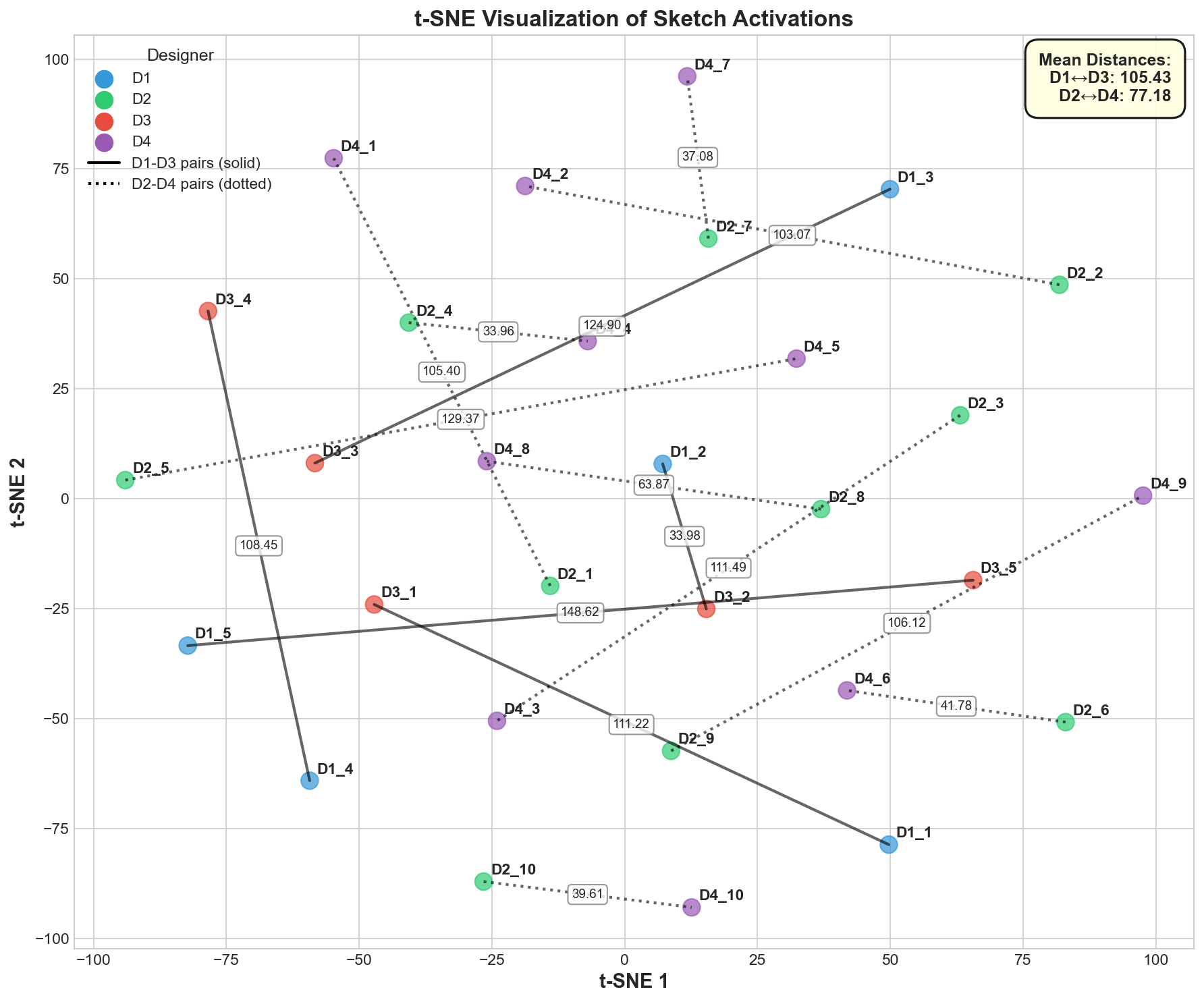}
        \caption{t-SNE Visualization of Activations}
        \label{subfig:tsne}
    \end{subfigure}
    \hfill
    \begin{subfigure}[b]{0.48\textwidth}
        \centering
        \includegraphics[width=\linewidth]{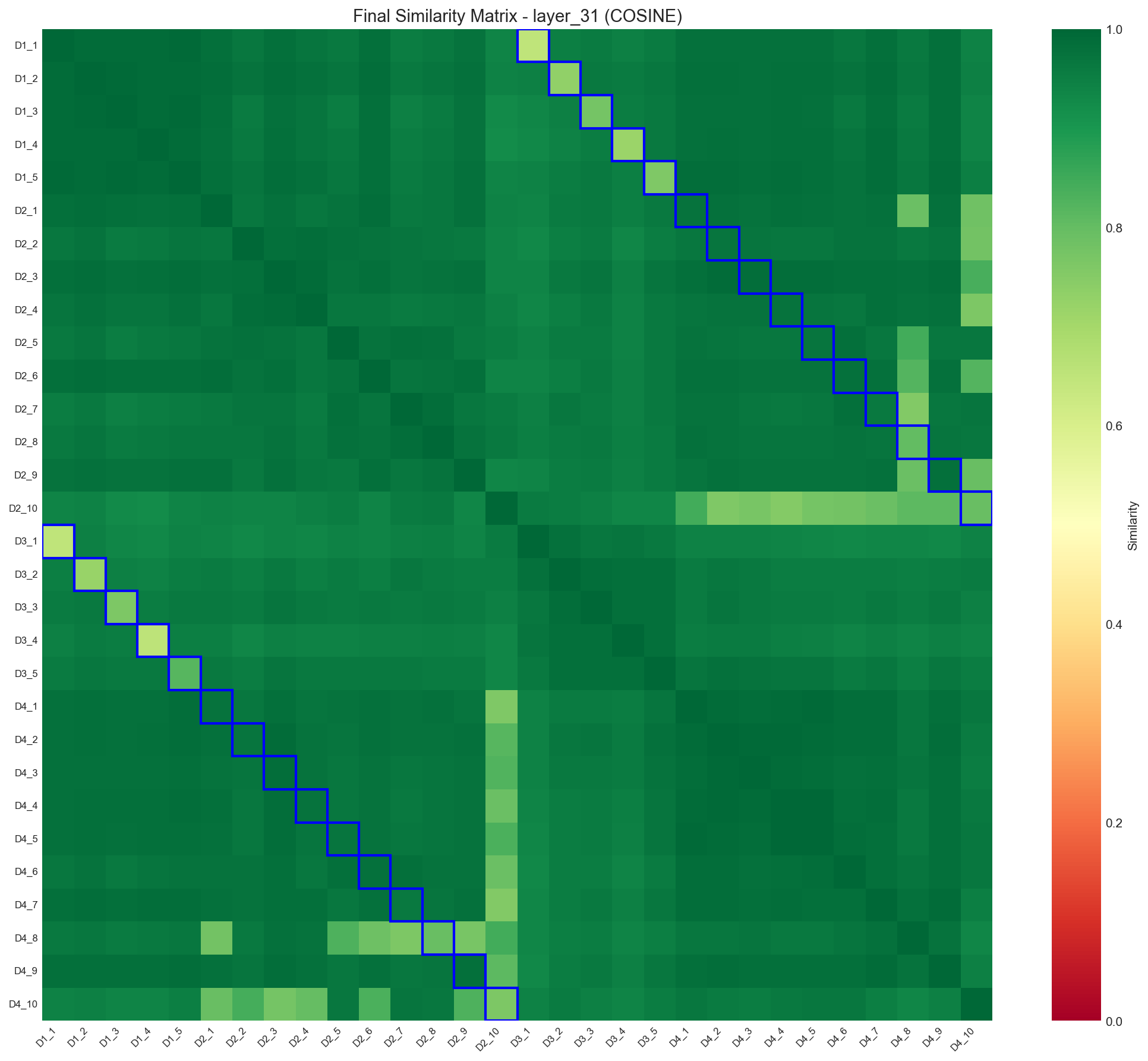}
        \caption{Final Similarity Matrix (Layer 31)}
        \label{subfig:similarity_heatmap}
    \end{subfigure}
    
    \caption{Neural Transparency Analysis Results. (a) Boxplot showing significantly higher similarity scores for D4 (replicating D2). (b) t-SNE plot showing tighter clustering of D2-D4 pairs. (c) Ranking of pairs showing D2-D4 consistently in the top tier. (d) Heatmap showing high diagonal intensity for the D2-D4 block.}
    \label{fig:neural_transparency_results}
\end{figure*}

The cosine similarity scores derived from Layer 31 of the LLaVA-NeXT model provide conclusive evidence for $H_3$. As shown in the boxplot (Figure \ref{subfig:designer_comparison}), the median similarity score for the D2 $\leftrightarrow$ D4 pairing is approximately \textbf{0.97}, with a tight inter-quartile range (0.96--0.98). In contrast, the D1 $\leftrightarrow$ D3 pairing shows a significantly lower median similarity of \textbf{0.73}, with greater variance.

A Mann-Whitney U test confirms that this difference is statistically significant ($p = 0.0168$). The t-SNE visualization (Figure \ref{subfig:tsne}) further corroborates this; the distance between D2 (green) and D4 (purple) clusters is visibly smaller (Mean Distance = 77.18) than the distance between D1 (blue) and D3 (red) clusters (Mean Distance = 105.43). This indicates that the AI-generated summary successfully transferred the "semantic DNA" of the concepts, allowing D4 to replicate the design intent with high fidelity.

\subsection{Performance and User Acceptance Metrics (Testing $\mathbf{H_4}$)}
Hypothesis $H_4$ predicted that AI-generated photorealistic renderings would elicit higher user acceptance. This was tested via a Purchase Likelihood survey ($N=30$).

\subsubsection{User Acceptance Ratings}
The users rated the Product Concept Renders (PCRs) on a 5-point Likert scale. The results were:
\begin{itemize}
    \item \textbf{D2 (AI Renderings via DIMES):} Mean Score = $\mathbf{4.2} \pm 0.6$
    \item \textbf{D1 (Manual Marker Renderings):} Mean Score = $3.1 \pm 0.9$
\end{itemize}
This statistically significant difference ($p<0.01$) validates $H_4$. Users cited the "clarity of material finish" and "realistic lighting" in the D2 renders as key factors in their preference, demonstrating the value of the generative AI module in communicating design intent to non-expert stakeholders.

\begin{figure*}[ht!]
    \centering
    % Row 1: D1 PCRs (1x5)
    \begin{subfigure}[b]{\textwidth}
        \centering
        \includegraphics[width=0.19\linewidth]{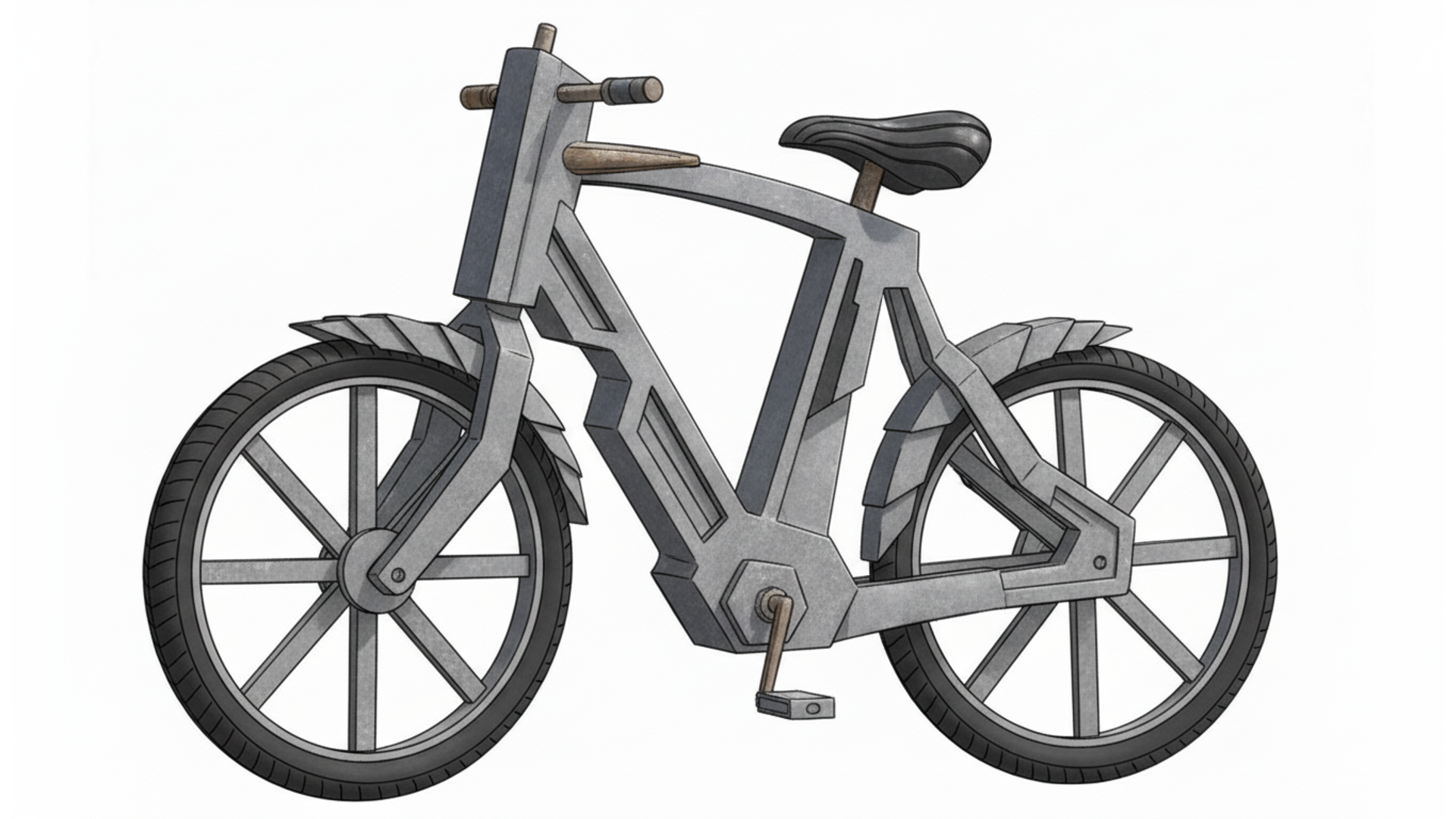}
        \hfill
        \includegraphics[width=0.19\linewidth]{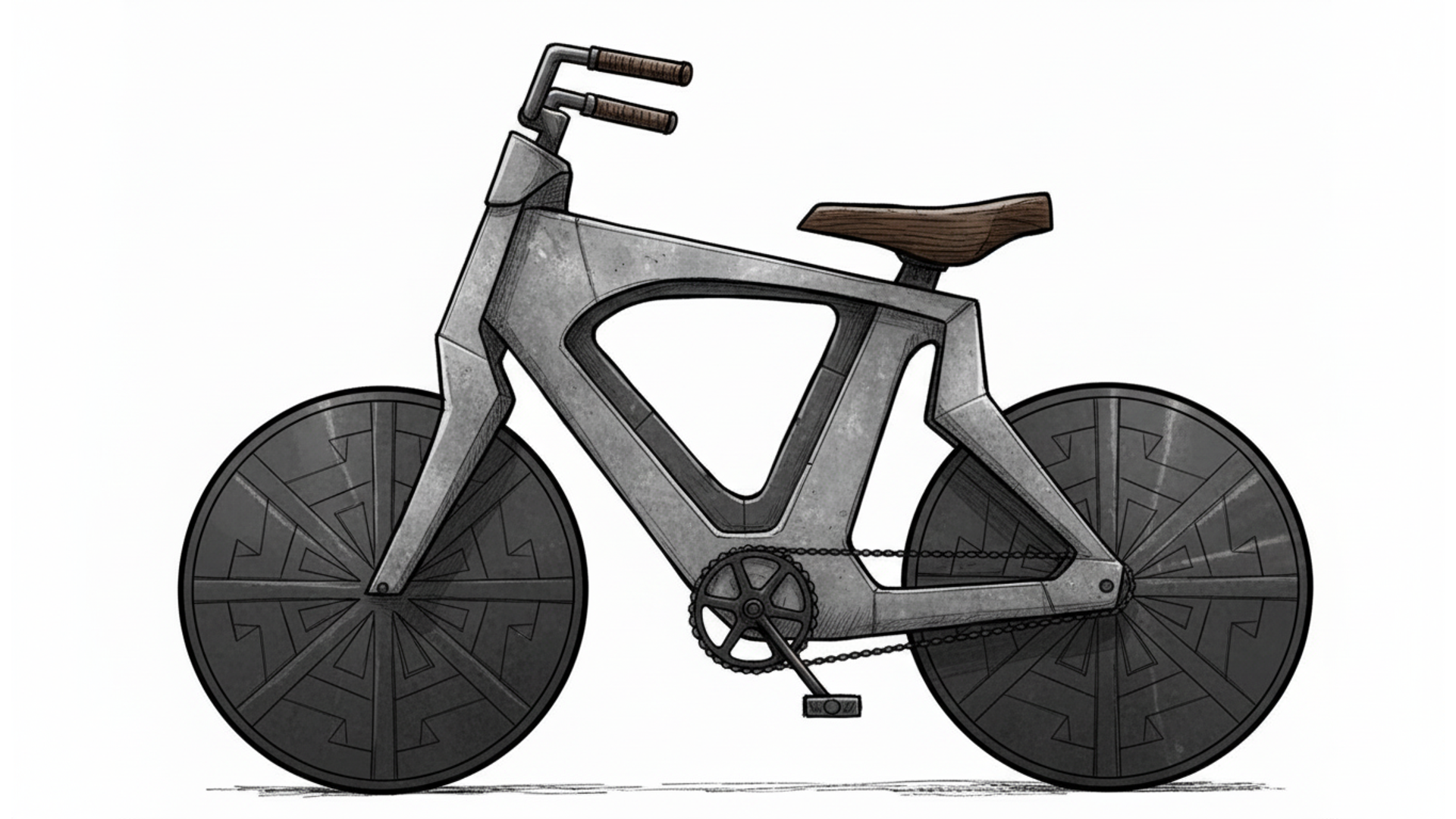}
        \hfill
        \includegraphics[width=0.19\linewidth]{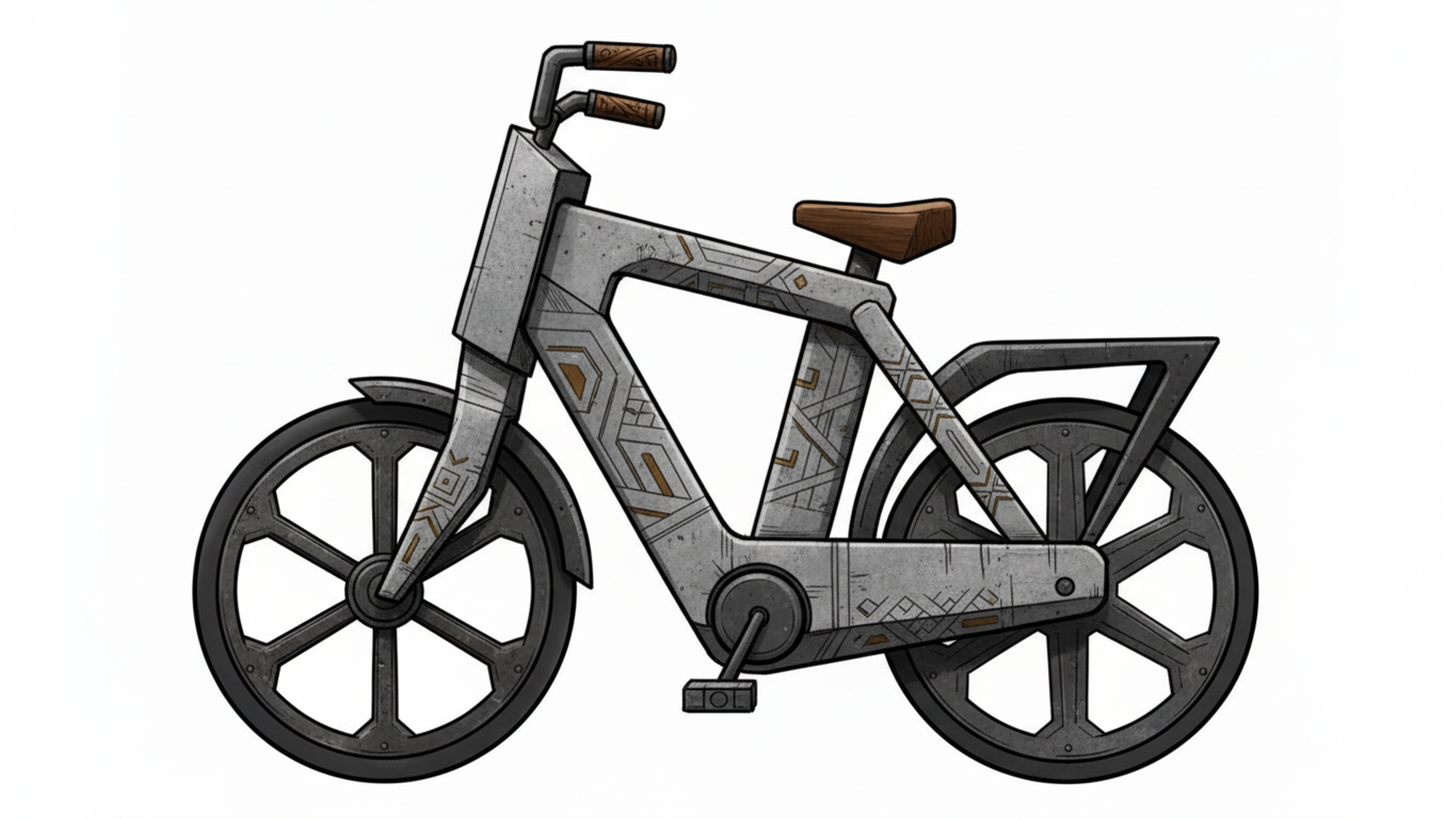}
        \hfill
        \includegraphics[width=0.19\linewidth]{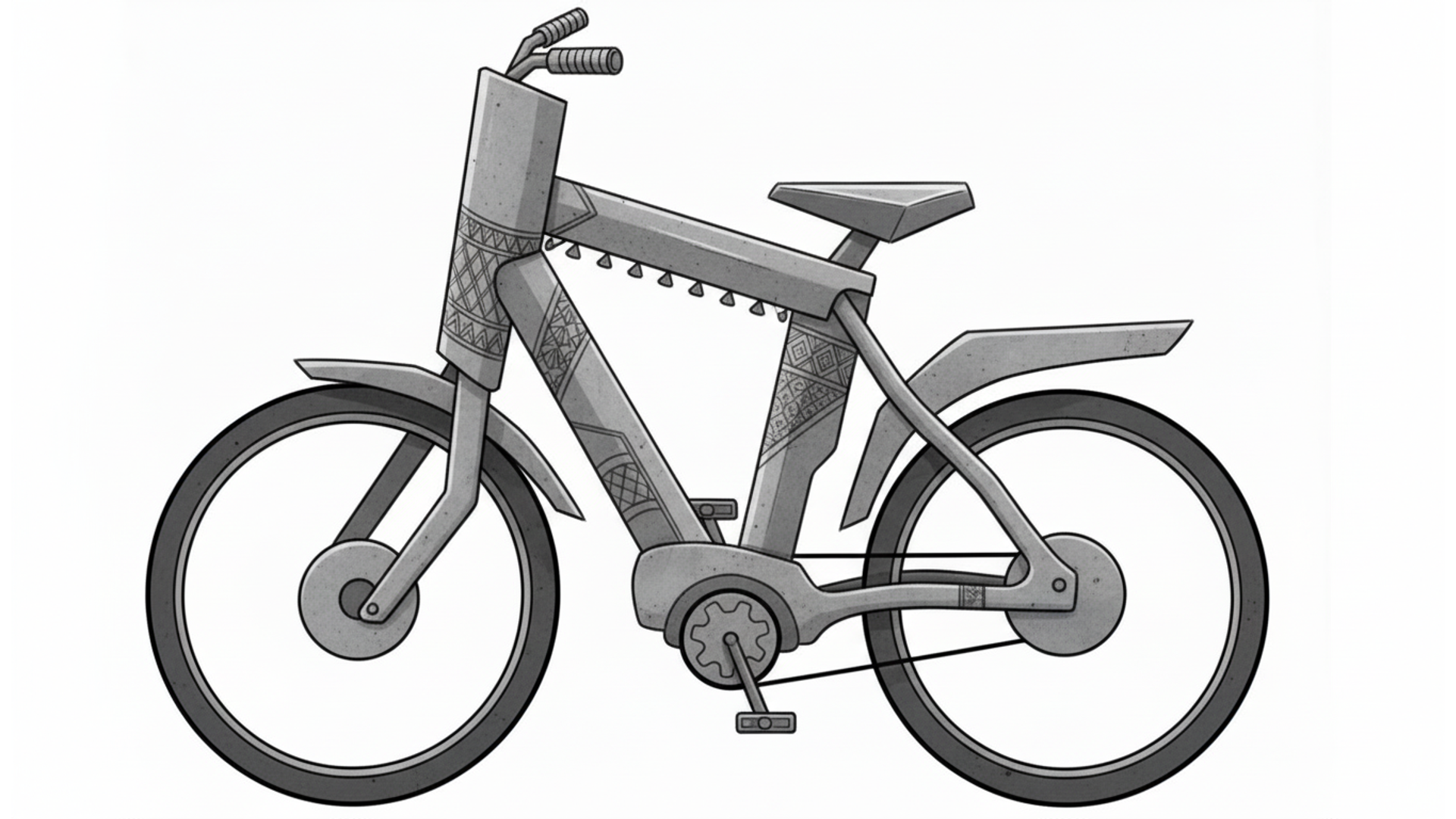}
        \hfill
        \includegraphics[width=0.19\linewidth]{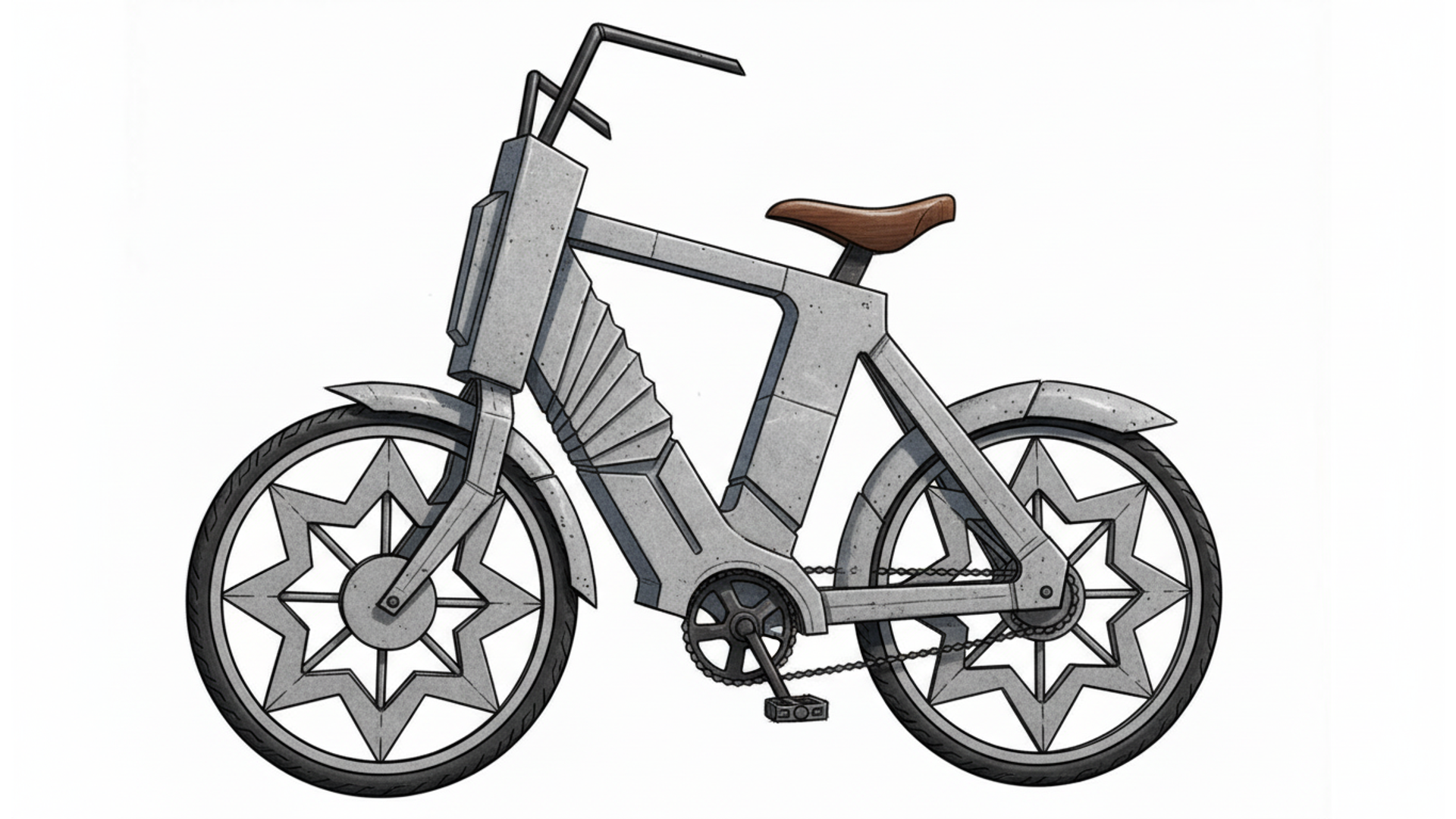}
        \caption{Manual Product Concept Renders (PCR) by Designer D1}
        \label{subfig:d1_pcrs}
    \end{subfigure}
    
    \vspace{0.5cm}
    
    % Row 2 & 3: D2 PCRs (2x5 = 10)
    \begin{subfigure}[b]{\textwidth}
        \centering
        \includegraphics[width=0.19\linewidth]{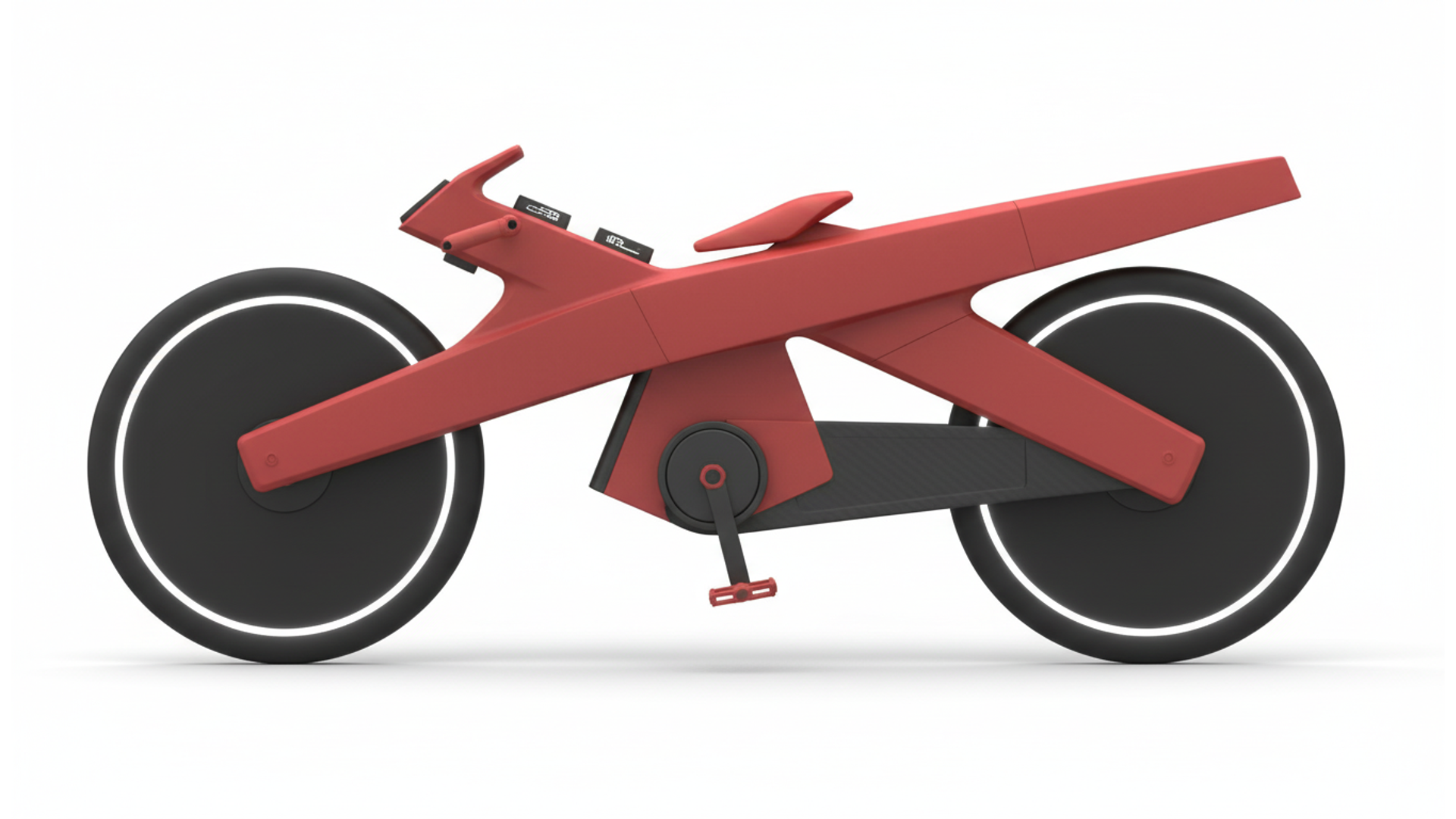}
        \hfill
        \includegraphics[width=0.19\linewidth]{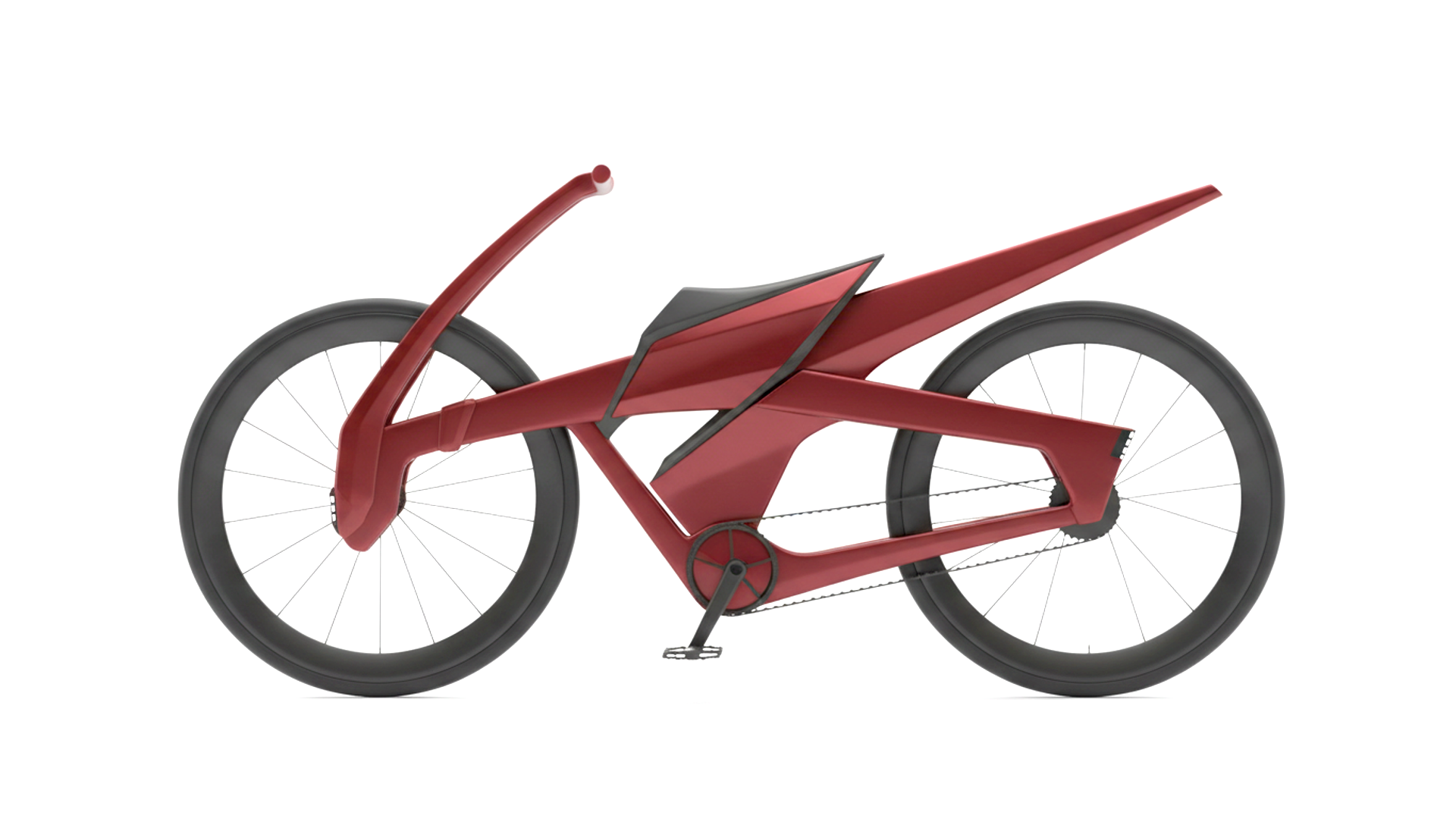}
        \hfill
        \includegraphics[width=0.19\linewidth]{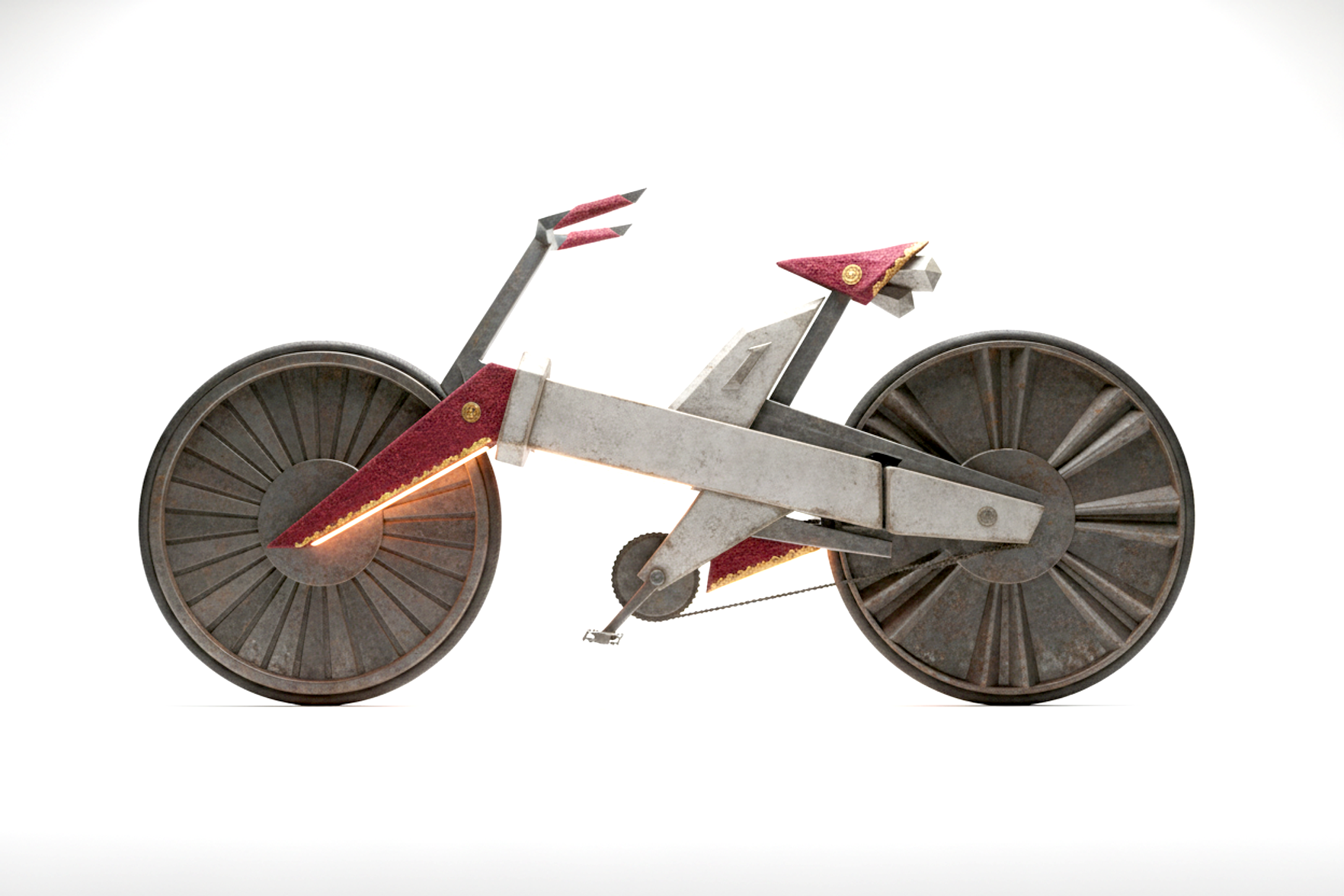}
        \hfill
        \includegraphics[width=0.19\linewidth]{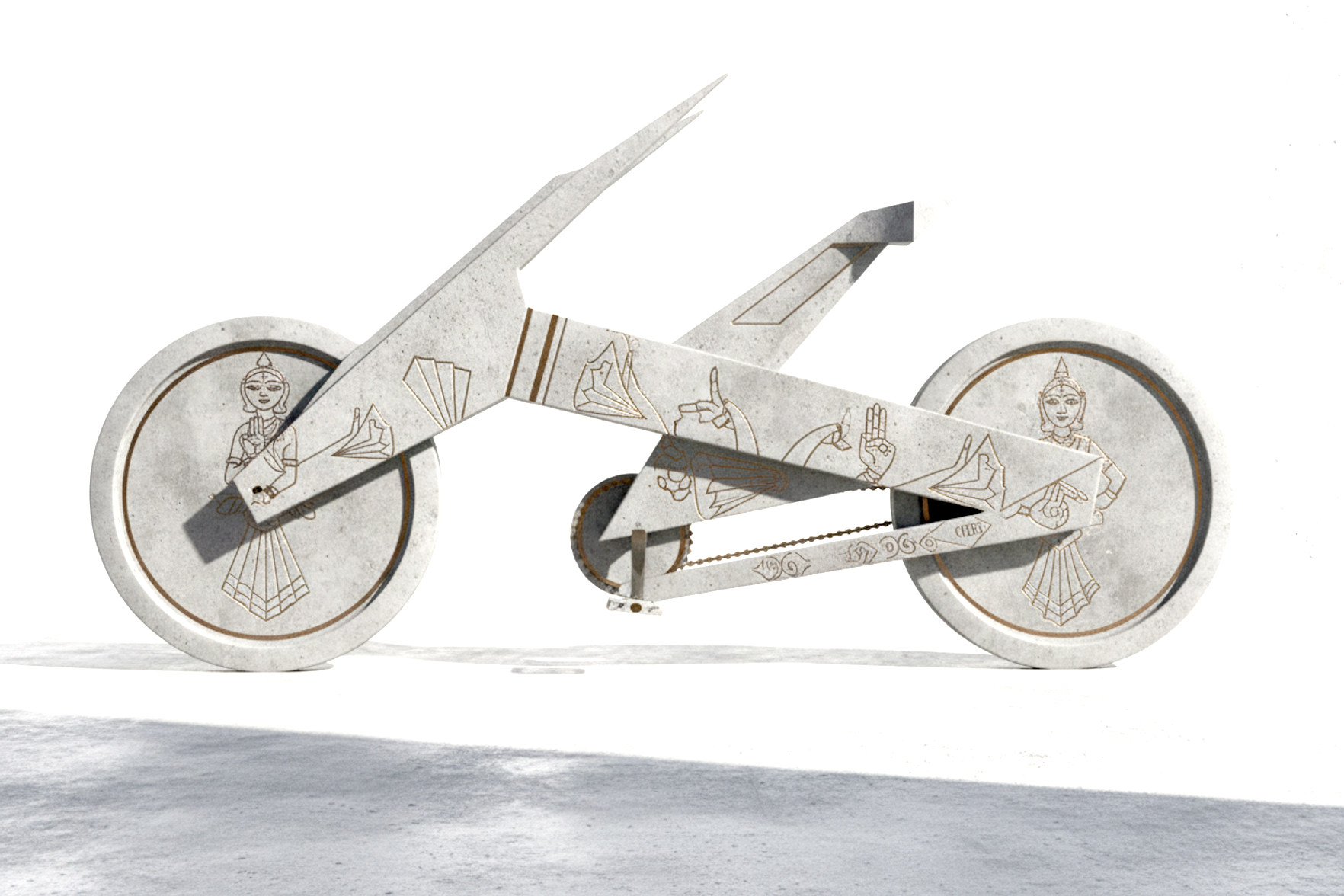}
        \hfill
        \includegraphics[width=0.19\linewidth]{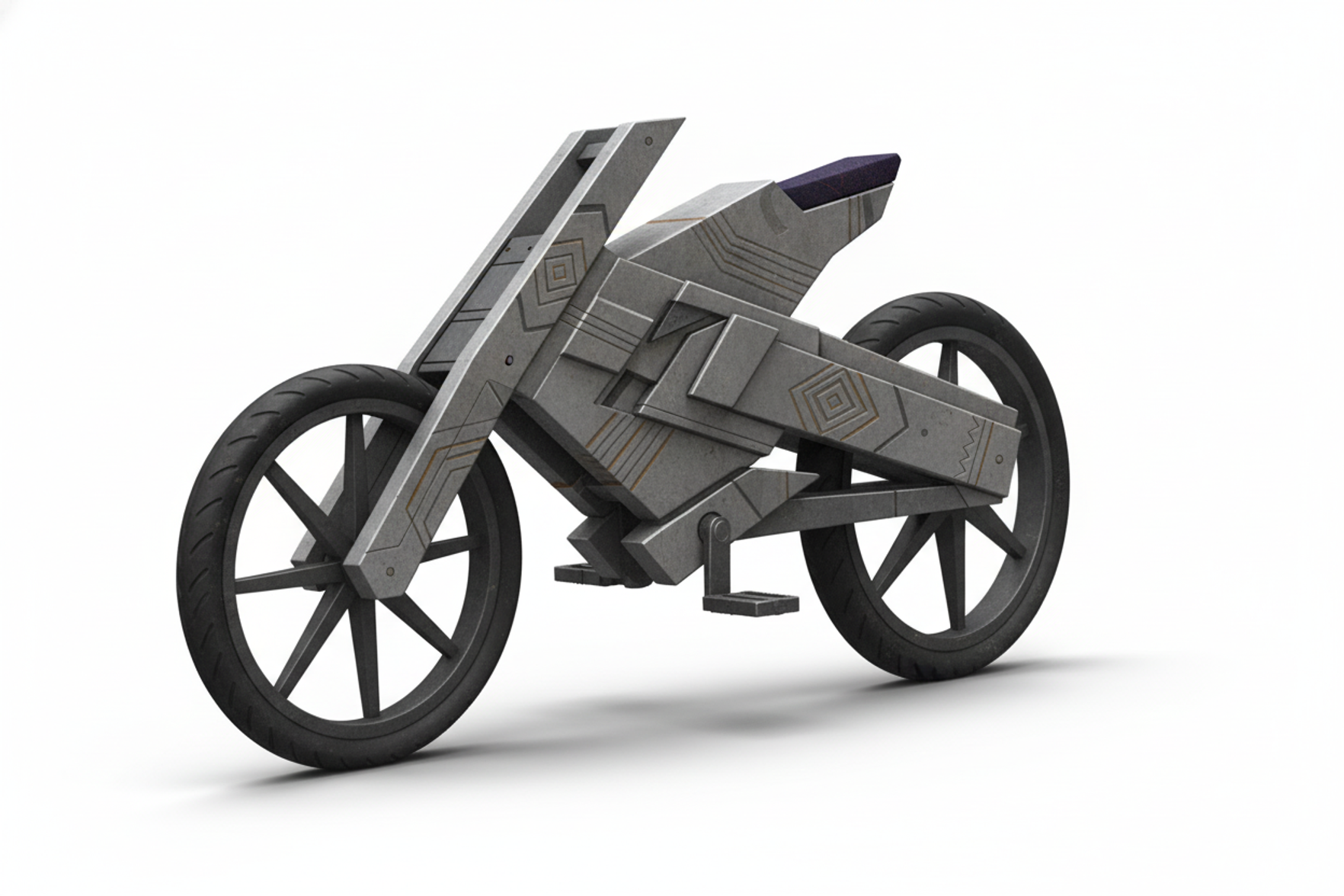}
        
        \vspace{0.1cm}
        
        \includegraphics[width=0.19\linewidth]{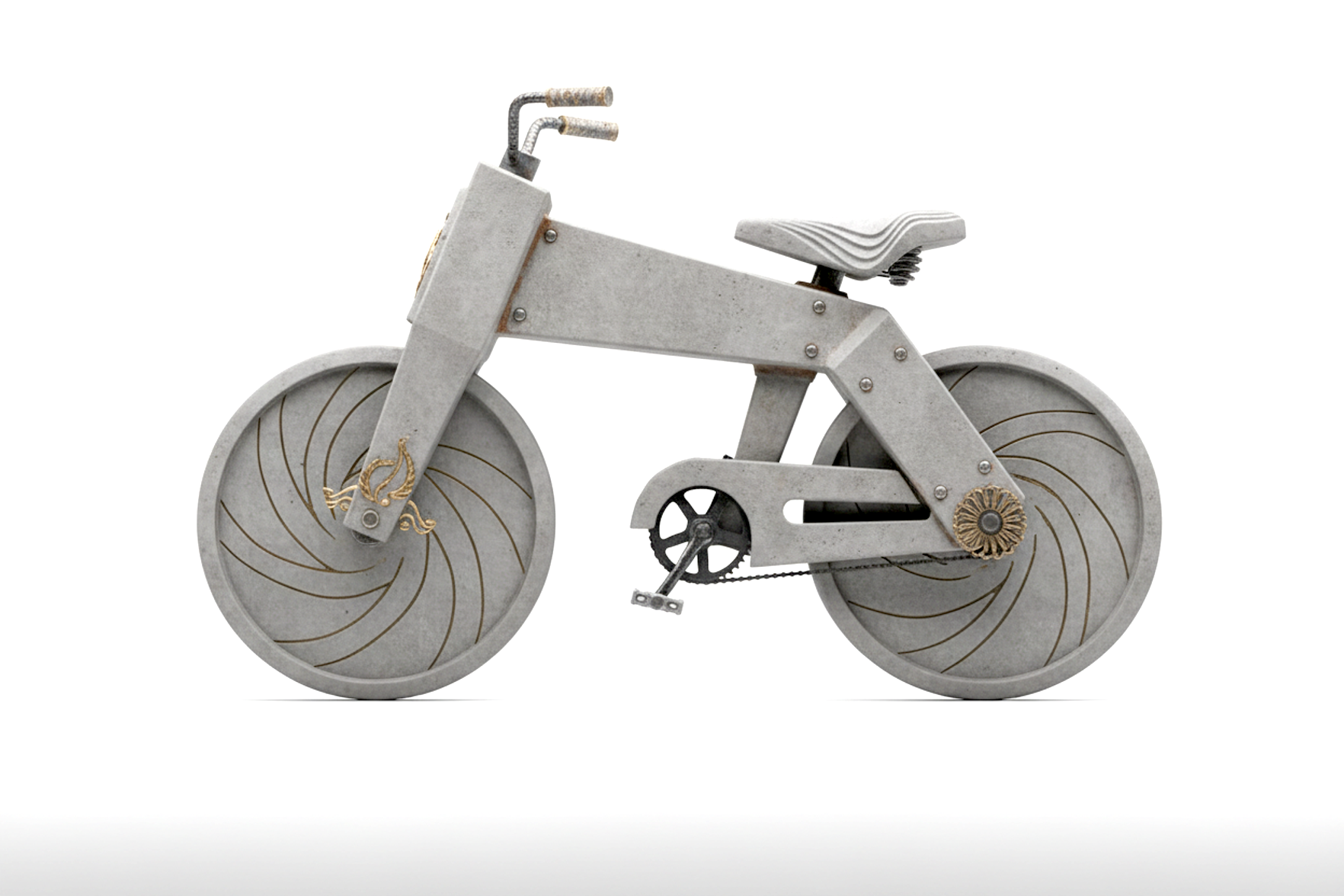}
        \hfill
        \includegraphics[width=0.19\linewidth]{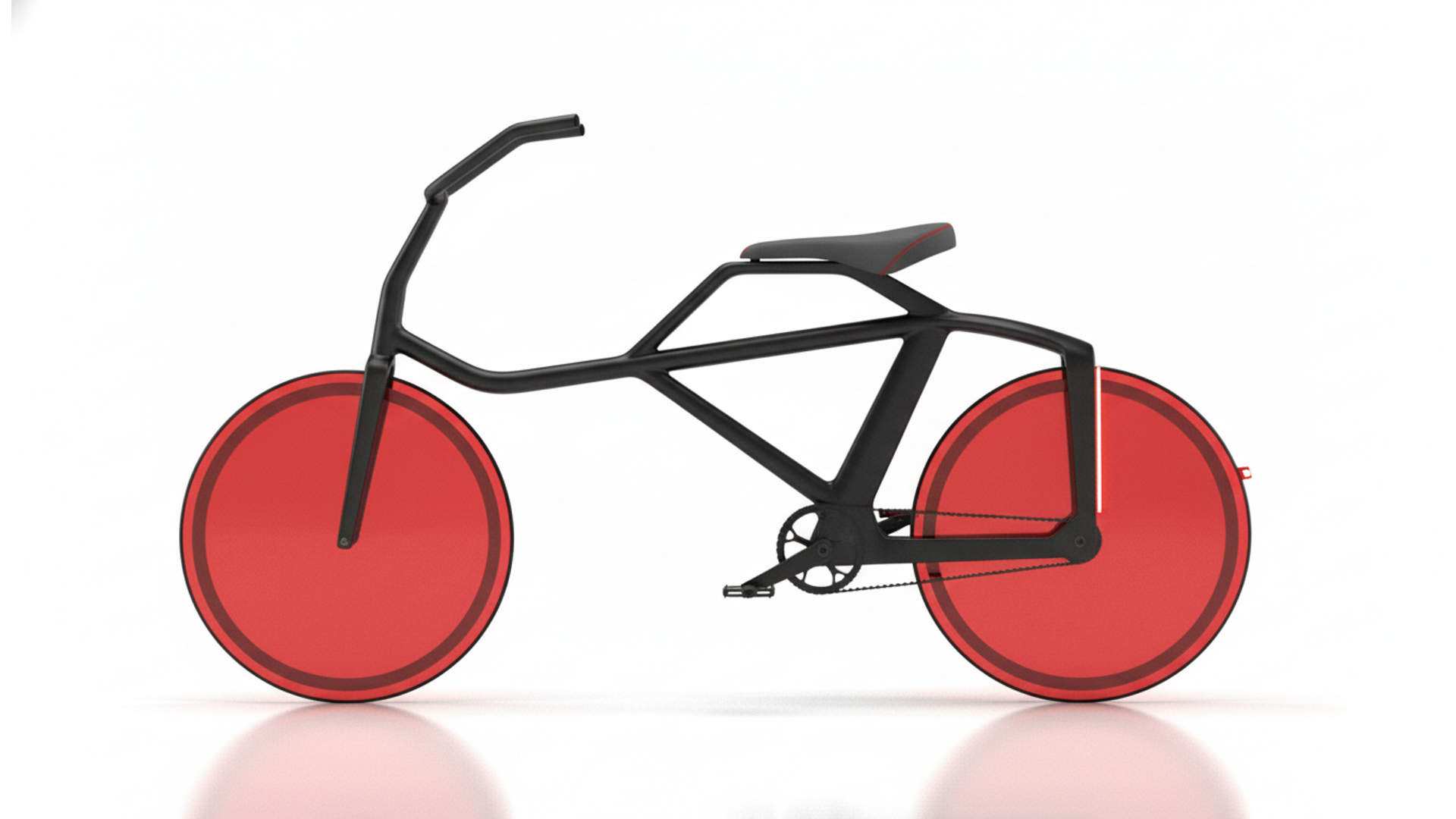}
        \hfill
        \includegraphics[width=0.19\linewidth]{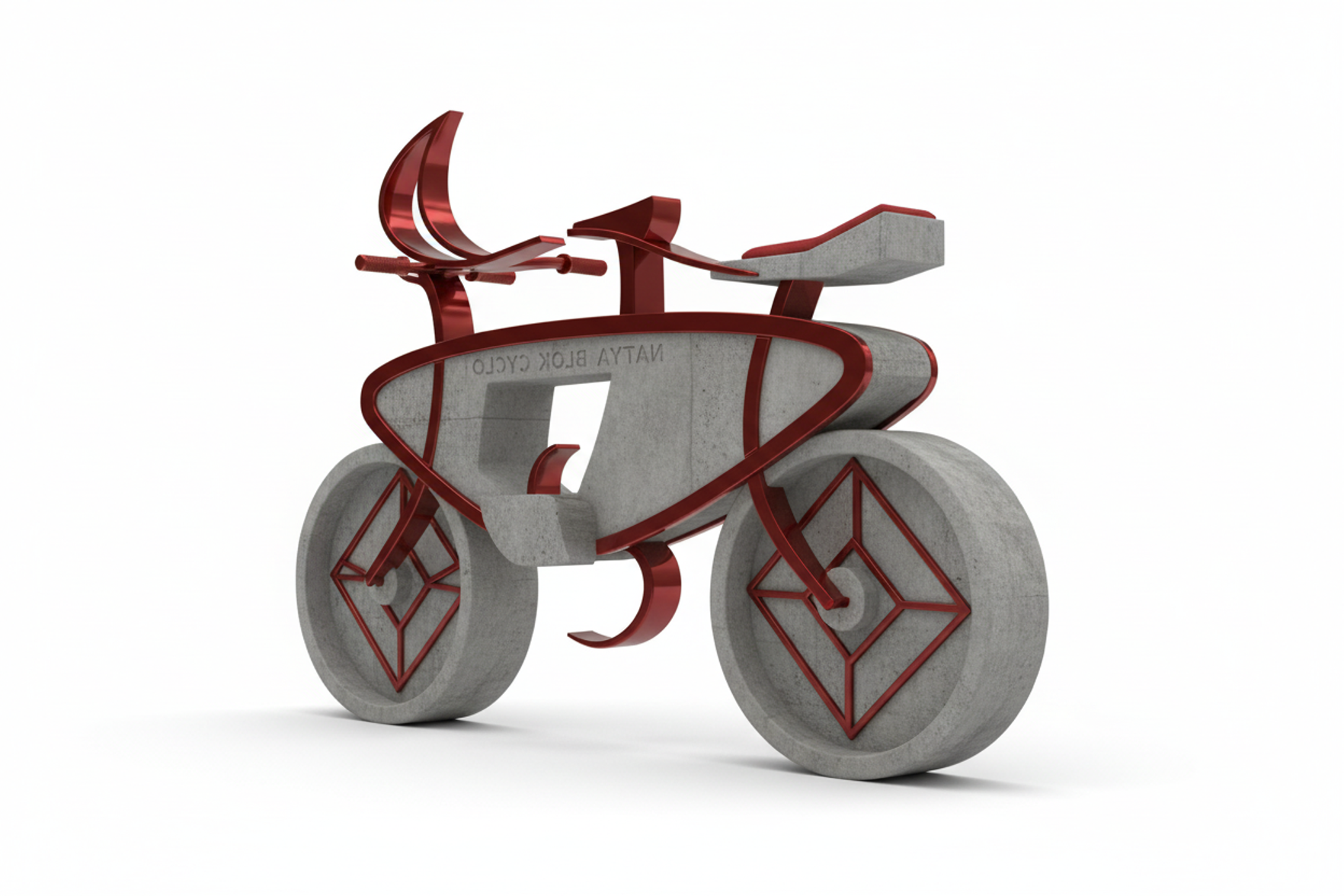}
        \hfill
        \includegraphics[width=0.19\linewidth]{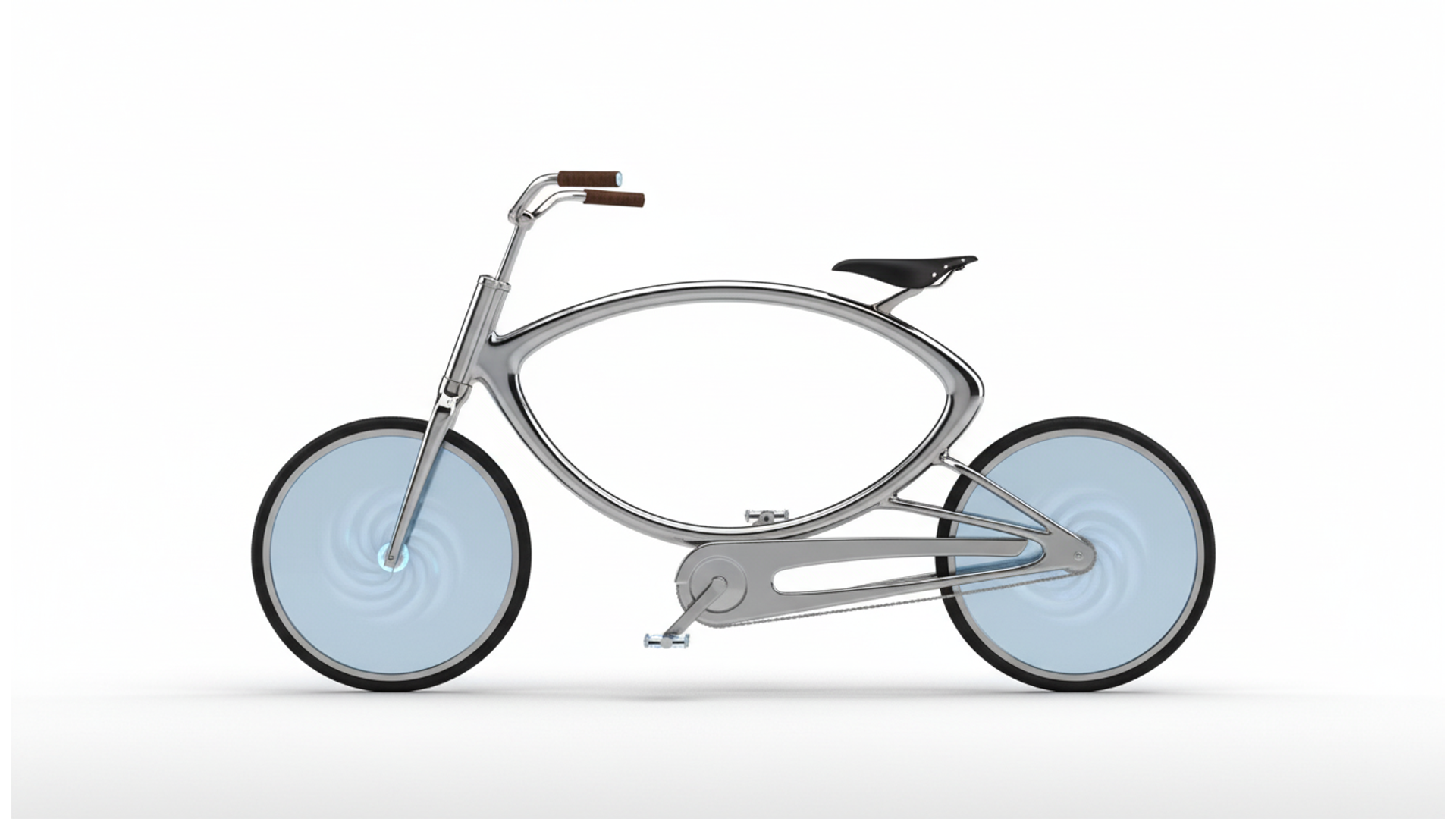}
        \hfill
        \includegraphics[width=0.19\linewidth]{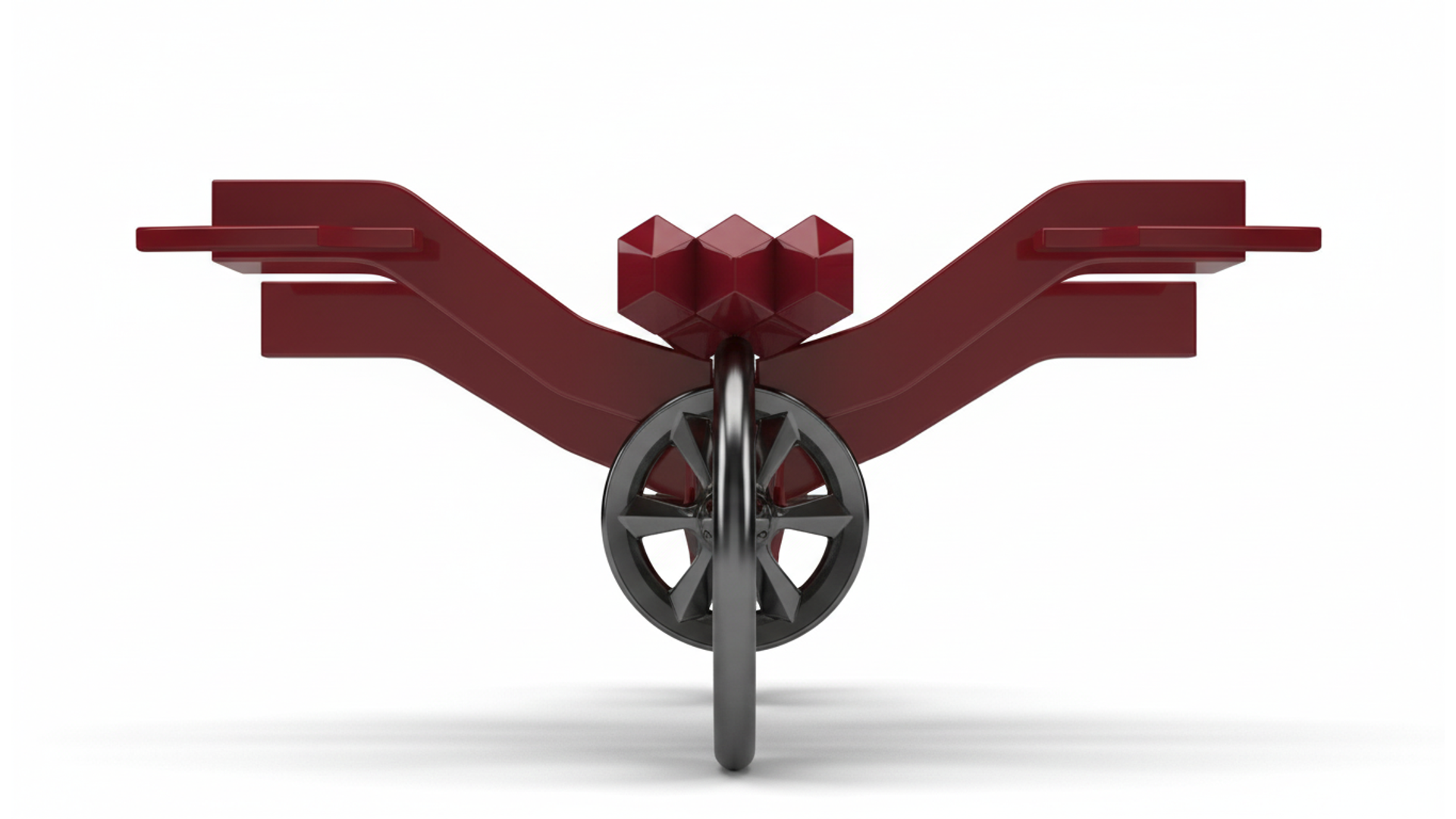}
        \caption{AI-Generated Product Concept Renders (PCR) by Designer D2 (DIMES)}
        \label{subfig:d2_pcrs}
    \end{subfigure}
    
    \caption{Comparison of final presentations. (a) D1's manual marker renderings. (b) D2's AI-generated photorealistic renderings, which received significantly higher purchase likelihood scores.}
    \label{fig:pcr_comparison}
\end{figure*}

\subsection{Synthesis and Implications}
The combined results of this study offer compelling evidence for the potential of AI-driven version control in design research and practice. SGIT is not merely a tool for saving work; it is an infrastructure for managing design cognition. The findings imply three major shifts in design methodology:
\begin{enumerate}
    \item \textbf{Frictionless Exploration:} The high number of branches (13) and commits (45) for D2 indicates that SGIT successfully eliminates the cognitive barrier to iterative conceptualization, leading to richer design spaces ($H_1$).
    \item \textbf{Objective Design Documentation:} The superior Information Content Density and designer validation scores prove that the multi-modal AI Summary provides a more accurate and comprehensive record of the design process than human memory ($H_2$).
    \item \textbf{High-Fidelity Pedagogy:} The significant difference in neural similarity scores ($\approx 0.97 \text{ vs. } 0.73$) demonstrates that SGIT-generated documentation serves as a superior pedagogical tool, enabling novices to "download" the expert's mental model with high fidelity ($H_3$).
\end{enumerate}

In conclusion, the DIMES ecosystem and its core SGIT logic successfully achieved all its research objectives, transforming the conceptual design process from an ephemeral act into a traceable, analyzable, and transferable body of knowledge. The implications for collaborative design, design history, and AI training models are profound.

%%%%%%%%%%%%%%%%%%%%%%%%%%%%%%%%%%%%%%%%%%%%%%%%%%%%%%%%%%%%%%%%%%%%%%

\section{Limitations and Future Directions}
\label{sec:limitations_future_work}

While the development of the DIMES ecosystem and the underlying sGIT architecture represents a significant leap forward in the digital management of design cognition, no research endeavour is without its constraints. This study, primarily exploratory and validation-oriented, has highlighted several technical and methodological limitations. Acknowledging these limitations is essential not only for contextualizing the current findings but also for charting the trajectory of future research in this burgeoning intersection of Design Methodology and Artificial Intelligence.

\subsection{Limitations of the Current Study}

The limitations of this research can be categorised into three distinct domains: Data Provenance, Algorithmic Constraints, and Interaction Paradigms.

\subsubsection{Synthetic Data and the Sim-to-Real Gap}
A cornerstone of our automated tracking system (AEGIS) is the integration of AI models to classify strokes into functional categories. As detailed in Section \ref{sec:stroke_classification}, our Deep Learning (DL) pipeline relied primarily on a large-scale \textit{synthetic} dataset generated via geometric transformations of a small seed set. While this approach effectively addressed the data scarcity problem—yielding a training dataset of 20,000 images—it introduced a quantifiable "Sim-to-Real" gap. 

Our empirical results highlighted a divergence in performance: while the ResNet50 model achieved a high accuracy of 96.55\% on the synthetic test set, its performance dropped to 82.5\% during the real-time field test. This indicates that the synthetic augmentation did not fully capture the subtle pixel-level nuances of real-world sketching behaviour, such as the varying opacity of a drying nib, specific anti-aliasing artifacts of the browser canvas, or the erratic jitter caused by hand tremors. 

Although we successfully mitigated this issue in the final system by incorporating a feature-based Machine Learning (Random Forest) module, which achieved 100\% field accuracy by relying on kinematic data rather than pixels, the disparity underscores a limitation in the pure computer vision approach. The current Deep Learning model struggles with the "texture" of real-time strokes that differ from the training distribution.

\subsubsection{The Unsolved Challenge of Visual Merging}
In mapping the Git version control system to the design domain, we successfully implemented \texttt{commit}, \texttt{branch}, \texttt{checkout}, and \texttt{diff}. However, a critical primitive of standard Git remains notably absent from sGIT: the \texttt{merge} command.
In software engineering, merging text files is algorithmically deterministic; lines of code can be interleaved, and conflicts are flagged based on line numbers. In the visual domain of sketching, "merging" two divergent concepts is a semantic nightmare. If Branch A contains a sketch of a round wheel and Branch B contains a sketch of a square tread, mathematically "averaging" or superimposing these pixels results in visual noise—a meaningless blur. Currently, sGIT relies on \textit{branching} without \textit{re-integration}. This means that while diverse ideas can bloom, consolidating the best features of two distinct branches back into a single "master" branch remains a manual task for the designer. The system currently lacks the semantic intelligence to automatically resolve visual conflicts or synthesize geometry from two parents.

\subsubsection{Dependence on Active Cognition Capture}
The "Cognition Capture" feature of sGIT relies on the Text-to-Speech (TTS) module activated during the commit action. This is an \textit{active} input paradigm; it requires the designer to consciously pause, press a button, and articulate their thoughts. While less intrusive than typing, it still imposes a minor cognitive load. 
There is a risk of "documentation fatigue." In a long, intense sketching session, a designer might simply stop speaking during commits, treating the system merely as a "Save" button rather than a "Log" mechanism. In such cases, the AI Summary (tested in $H_3$) would degrade in quality, relying solely on stroke data (Action) without the accompanying intent (Cognition). The system currently lacks a "passive" method to infer intent solely from silence or hesitation.

\subsubsection{Generative AI Hallucinations in Rendering}
The AI rendering module (Banana/Gemini Nano) demonstrated impressive capability in transforming monochromatic sketches into photorealistic renders. However, like all generative models, it is prone to "hallucination." 
In several instances during the pilot testing, the AI added details that were not present in the original sketch—such as adding a specific tread pattern to a tyre or inventing a parting line on a casing—based on the statistical probabilities of the training data rather than the designer's input. While this can sometimes be serendipitous (a "happy accident"), in a strict engineering context, it represents a loss of fidelity. The designer must currently be vigilant to ensure the AI render does not misrepresent the underlying geometry or functional constraints.

\subsection{Future Work and Research Trajectory}

The limitations identified above serve as the catalyst for our future research roadmap. We envisage DIMES evolving from a single-user sketching tool into a collaborative, multi-modal, and intelligent design environment.

\subsubsection{Towards Real-Time Collaborative Conceptualization}
The current iteration of sGIT is designed for a single author. However, modern product design is inherently collaborative. Future work will focus on expanding sGIT to support "Multiplayer" functionality, akin to Google Docs but for the complex tree-structure of versioned sketches. 
This presents significant algorithmic challenges. We aim to investigate the use of Conflict-free Replicated Data Types (CRDTs) adapted for vector strokes to allow multiple designers to work on the same canvas (or different branches of the same concept) simultaneously. The version control system would need to track authorship at the \textit{stroke level}, allowing a Lead Designer to review a concept and see exactly which strokes were contributed by Designer A versus Designer B. This would revolutionise remote design collaboration.

\subsubsection{From 2D Sketches to 3D Geometry Reconstruction}
Currently, the output of DIMES is a 2D raster image (the sketch or the render). The "Holy Grail" of Computer-Aided Industrial Design (CAID) is the automated transition from 2D ideation to 3D CAD. 
We propose to extend the AI capabilities of DIMES to integrate single-view 3D reconstruction algorithms. By leveraging the semantic understanding of the strokes (e.g., knowing that a specific curve is a \textit{Shadow Stroke} implies a ground plane), we can feed this structured data into modern NeRF (Neural Radiance Fields) or Gaussian Splatting pipelines to generate a coarse 3D mesh directly from the concept sketch. This would allow the designer to rotate their sketch in 3D space to verify proportions, effectively bridging the gap between the concept phase and the detailed design phase.

\subsubsection{Cognitive Mirroring and Intelligent Tutoring Systems}
The empirical success of Hypothesis $H_4$ (Knowledge Transfer) suggests that sGIT data is a potent pedagogical resource. We plan to develop an "Intelligent Tutoring System" (ITS) for design education. 
By analysing the stroke-level patterns of experts—such as the ratio of exploration to refinement or the specific sequence of \textit{Constraining} vs. \textit{Defining} strokes—the system could provide real-time feedback to novice students. For example, if a student is spending too much time detailing a concept that hasn't been properly constrained, the system could nudge them: "Analysis suggests you are detailing premature geometry. Consider using Constraining Strokes to fix proportions first." This "Cognitive Mirroring" would act as an always-available mentor, accelerating the learning curve for budding designers.

\subsubsection{Semantic Merging via Layout Analysis}
To address the "Merge" limitation, future work will explore "Semantic Merging." Instead of blending pixels, we aim to use Graph Neural Networks (GNNs) to represent the sketch as a graph of features (e.g., Handle, Spout, Body). Merging two concepts would then become a graph operation—combining the "Handle" node from Branch A with the "Body" node from Branch B. This would allow for a true synthesis of ideas, enabling the system to algorithmically propose hybrid concepts that combine the best features of divergent parents.

\subsubsection{Expansion of the Dataset and Taxonomy}
Finally, we intend to move beyond synthetic data. We will launch a large-scale, crowd-sourced data collection initiative to gather real-world annotated strokes from professional designers across the globe. This "ImageNet for Sketches" will help us refine the stroke taxonomy—perhaps expanding beyond the current six types to include sub-categories (e.g., \textit{Matte Shading} vs. \textit{Glossy Shading})—and further reduce the Sim-to-Real gap, making the system robust enough for widespread commercial adoption.

In conclusion, whilst the current DIMES system has successfully validated the concept of Version Control for Sketching, it is merely the first step. The fusion of rigorous data management (sGIT) with generative intelligence (AI) holds the promise of fundamentally reshaping how humans ideate, communicate, and preserve their creative intent.

%%%%%%%%%%%%%%%%%%%%%%%%%%%%%%%%%%%%%%%%%%%%%%%%%%%%%%%%%%%%%%%%%%%%%%

\section{Conclusion}
\label{sec:conclusion}

The conceptual design phase is the genesis of product innovation, a chaotic yet critical period where vague ideas crystallize into tangible forms through the medium of Product Concept Sketching (PCS). However, traditional design workflows and contemporary digital tools have long suffered from a fundamental amnesia: they capture the final output while discarding the rich, non-linear evolutionary history of the design process. This research addressed this "Cognitive Gap" by proposing, developing, and validating SGIT, the first dedicated Version Control System (VCS) tailored specifically for the visual and cognitive demands of product design, integrated within the DIMES ecosystem.

This study makes several seminal contributions to the field of Design Methodology and Computer-Aided Industrial Design (CAID). Firstly, we established a robust theoretical framework by defining a novel taxonomy of sketching stroke types (Constraining, Defining, Detailing, Shading, Shadow, and Annotation). By generating a massive synthetic dataset and developing a hybrid AI classification pipeline, we achieved robust stroke recognition capabilities. While our Deep Learning models (ResNet50) demonstrated superior latent space clustering for unseen data, our feature-based Machine Learning models (Random Forest) achieved a field-test accuracy of 100\%, effectively enabling the machine to "read" the action of sketching with expert-level proficiency.

Secondly, the architectural development of SGIT successfully translated the rigid logic of software version control (Git) into a fluid, designer-centric interaction model. By mapping `commit` to multi-modal snapshots (Action + Voice) and `branching` to implicit creative divergence, we eliminated the "Cognitive-Logistical Friction" that previously hindered the adoption of VCS in design. The system architecture, supported by generative AI for photorealistic rendering and narrative summarization, transforms the design tool from a passive canvas into an active cognitive partner.

Thirdly, and most significantly, our comparative empirical study provided robust quantitative evidence for the efficacy of this approach. The results unequivocally supported our four hypotheses. 
\begin{itemize}
    \item We observed a \textbf{160\% increase in exploration breadth} (13 vs 5 branches) and an \textbf{800\% increase in commit granularity} (45 vs 5 snapshots) in the SGIT-assisted group, validating that frictionless versioning encourages risk-taking ($H_1$).
    \item The automated AI-generated summaries were found to contain significantly higher \textbf{Information Content Density} (18 vs 7 specific domain mappings) compared to human-authored reports, providing an objective record of design intent ($H_2$).
    \item Crucially, the knowledge transfer experiment, evaluated using a novel \textbf{Neural Transparency-based} similarity metric (Layer 31 activations of LLaVA-NeXT), demonstrated that novice designers could replicate expert concepts with significantly higher fidelity (Cosine Similarity: \textbf{0.97 vs 0.73}, $p=0.016$) when guided by the structured AI narrative ($H_3$).
    \item Finally, user acceptance testing revealed that the AI-generated product renderings elicited significantly higher \textbf{Purchase Likelihood Scores} (\textbf{4.2 vs 3.1}) compared to traditional manual renderings ($H_4$).
\end{itemize}

In conclusion, DIMES and SGIT represent a paradigm shift in how we manage the intellectual property of design. By capturing not just the 'what' (the final pixels) but the 'how' and 'why' (the evolutionary tree and intent), we move towards a future where the design process is as traceable, analyzable, and scientifically rigorous as the engineering process that follows it. This work lays the foundation for a new generation of "Cognitive Design Support Systems" that preserve the ephemeral thought processes of experts for the benefit of future generations of designers.

%%%%%%%%%%%%%%%%%%%%%%%%%%%%%%%%%%%%%%%%%%%%%%%%%%%%%%%%%%%%%%%%%%%%%%
\section*{Acknowledgment} %% ASME requests this exact spelling, singular.

%%%%%%%%%%%%%%%%%%%%%%%%%%%%%%%%%%%%%%%%%%%%%%%%%%%%%%%%%%%%%%%%%%%%%%
\section*{Funding Data}
This research received no external funding. The work was conducted without financial support from any funding agency, organization, or grant.

%\printbibliography
\bibliographystyle{elsarticle-harv} % Or elsarticle-harv, etc.
\bibliography{bibliography}

\greyline
\appendix
\newpage

\end{document}